\documentclass[twocolumn]{aastex631}
\newcommand{\Mlow}{\log\,M_\star/M_\odot = 10-10.5}
\newcommand{\Mhigh}{\log\,M_\star/M_\odot = 10.5-11}
\newcommand{\delSSFR}{\Delta\,\mathrm{SSFR}}
\newcommand{\deltahalf}{\delta_\mathrm{0.5Mpc}}
\newcommand{\deltaone}{\delta_\mathrm{1Mpc}}
\newcommand{\deltatwo}{\delta_\mathrm{2Mpc}}
\newcommand{\deltafour}{\delta_\mathrm{4Mpc}}
\newcommand{\deltaeight}{\delta_\mathrm{8Mpc}}

\shorttitle{Environments of Starbursts and Post-starbursts}
\shortauthors{Yesuf, H.}

\begin{document}

\title{Quenching in the Right Place at the Right Time: Tracing the Shared History of Starbursts, AGNs, and Post-starburst Galaxies Using Their Structures and Multiscale Environments}
\author{Hassen M. Yesuf}
\affiliation{Kavli Institute for the Physics and Mathematics of the Universe (WPI), UTIAS, University of Tokyo, Kashiwa, Chiba 277-8583, Japan}
\affiliation{Kavli Institute for Astronomy and Astrophysics, Peking University, Beijing 100871, People's Republic of China}

\begin{abstract}
This work uses multiscale environments and structures of galaxies in the Sloan Digital Sky Survey as consistency checks of the evolution from starburst to quiescence at redshift $z < 0.2$. The environmental indicators include fixed aperture mass overdensities ($\delta_{x\mathrm{Mpc}}$, $x \in \{0.5, 1, 2, 4, 8\}\,h^{-1}$Mpc), $k$-nearest neighbor distances, the tidal parameter, halo mass ($M_h$), and satellite/central classification. The residuals of specific star formation rates ($\Delta\,\mathrm{SSFR}$) is used to select starbursts ($\delSSFR  > 0.6$\,dex, $N \approx 8,\,600$). Quenched post-starbursts (QPSBs) are selected using H$\alpha < 3$\,{\AA} emission and H$\delta_A > 4$\,{\AA} absorption ($N \approx 750$). The environments of starbursts and QPSBs are compared with those of active galactic nuclei (AGNs) and inactive galaxies of varying $\delSSFR$. The environments of starbursts, AGNs, and QPSBs are unlike the environments of most quiescent galaxies (QGs). About $70\%-90$\% of starbursts, AGNs with H$\delta_A > 4$\,{\AA}, and QPSBs are centrals, $\sim 80\%-90\%$ have $M_h < 10^{13}\,M_\odot$, and only $\sim 2\%-4\%$ have $M_h > 10^{14}\,M_\odot$ or live in clusters. Their $M_h$ and satellite fractions are also different from those of QGs. All QPSBs are matched to some SFGs, starbursts, AGNs, and QGs of similar $M_\star$, environments, concentration indices, and velocity dispersions. A significant fraction ($\sim 20\%-30\%$) of starbursts cannot be matched to QPSBs or QGs. The implications are: (1) some starbursts do not quench rapidly. (2) Satellite-quenching mechanisms operating in high density environments cannot account for most QPSBs. (3) The evolution from starbursts to QPSBs to QGs is not the dominant path at $z < 0.2$. (4) Starbursts are not mainly triggered by tidal interactions.
\end{abstract}


\section{Introduction} 

The details of how the star formation rate (SFR) is regulated in galaxies in not well understood. Evidently, star formation is fueled by cold gas in and around galaxies \citep{Kennicutt+12, Genzel+15, Saintonge+17,Tumlinson+17,Catinella+18}. External mechanisms that are governed by dark matter halos or the environment and/or internal processes such as blackhole activity may regulate the amount, the in and out flows, and/or the thermodynamic state of the cold gas in galaxies or their surrounding halos, thereby impacting SFR \citep{Schaye+15, Henriques+17,Pillepich+18, Matthee+19}. Consequently, some of the observable properties of galaxies correlate with the properties of halos and the environment while others reflect the baryonic physics within galaxies. Most observable galaxy properties depend strongly on stellar mass ($M_\star$), which is likely the primary parameter of galaxies characterizing the internal mechanisms. For example, star-forming galaxies (SFGs) follow a tight relationship (scatter of $\sim 0.3-0.4$\,dex) between $M_\star$ and SFR \citep[e.g.,][]{Brinchmann+04, Elbaz+07, Noeske+07,Speagle+14}. The residuals of this relation, $\delSSFR$, for SFGs depend weakly on the environment. In contrast, the fraction of quenched galaxies depends strongly on both $M_\star$ and environment \citep{Baldry+06, Bamford+09, Peng+10,Darvish+16, Kawinwanichakij+17,WangHuiyuan+18}. The environmental quenching mechanisms are effective in low-mass galaxies ($M_\star \lesssim10^{10}\,M_\odot$); high-mass galaxies quench by internal processes such as blackhole feedback \citep[e.g.,][]{Croton+06, Hopkins+08,Fabian+12, Kormendy+13}. The majority of low-mass galaxies are quenched as satellites of more massive galaxies by interacting with the hot intracluster media (ICM) of their host halos, being starved of cold gas, or merging/interacting with other satellites \citep[for reviews see][]{Boselli+06,Cortese+21}.  

Numerous studies have explored trends of galaxy properties with the environment to establish its importance in shaping galaxy mergers/interactions, morphology, star formation, black hole activity, and cold gas \citep[e.g.,][]{Dressler+80, Gomez+03, Balogh+04, Kauffmann+04, LinLihwai+10, Ellison+10, Wetzel+12, Kampczyk+13, vandeVoort+17, Donnari+21}. The following points list some noteworthy observational trends with the environment: (1) galaxy mergers/interactions happen in all environments. Those that induce SFR enhancements, however, are favored in galaxies in low-density environments, likely because of their high gas fractions \citep{Ellison+10}. (2) The morphology-environment relation is partly driven by $M_\star$, in the sense that higher mass galaxies live in denser environments \citep{Kauffmann+04}. (3) The impact of the environment on SFR is stronger than on morphology \citep{Kauffmann+04, Ball+08, Bamford+09,Skibba+09}; at fixed SFR, the dependence of morphology on the environment is very weak. (4) Both mass quenching and environmental quenching are associated with morphological transformation of galaxies \citep{Carollo+16}. (5) The effects of the environment on SFR and morphology are relatively local, within a single halo, $\sim 1-2$\,Mpc \citep{Kauffmann+04, Blanton+07, Park+07,Blanton+09}. The observed environmental dependencies can be understood in terms of the host halo mass, and the galaxy's position within that halo \citep{WooJoanna+13}. The large-scale environment probably has little influence on galaxy properties. (6) The properties of dark matter haloes such as their growth rates, concentrations, and interaction histories correlate with the environment in scale-dependent ways \citep[][]{Wechsler+18,Behroozi+22}. These correlations are imprinted in observables such as the two-point correlation functions and $k$-nearest neighbor distances of galaxies, more prominently at $\lesssim 2$\,Mpc scale \citep{Behroozi+22}.

It is not straightforward to attribute the quenching of typical quiescent galaxies (QGs) to their current dense environments, because they might have been quenched by internal processes or in a different environment at much earlier time. Either way, the correlation with their current environments may not be causal. Factors such as the star formation history, the assembly history of the dense environment/cluster, and the pre-processing that happens before a small group falls onto the cluster need to be accounted for \citep{Dressler+13, Mahajan13, Donnari+21}. Alternatively, studying environments and structures of galaxies that transition rapidly between SFGs and QGs may provide valuable insights on the causal relationship between environment and star formation quenching. These galaxies, namely, starbursts and their descendants, are the focus of this study. Because of their short evolution span, we expect these galaxies to have similar $M_\star$, structures, and environments. This work examines this expectation in detail.

Quenched post-starburst galaxies \citep[QPSBs, also known as k+a galaxies;][]{Dressler+83, Quintero+04, Goto05, French+21} are a rare class of galaxies in transition from the star-forming to quiescent phase. Their spectra reveal very little ongoing star formation (as evidenced by weak emission lines due to lack of ionizing O \& B stars) but a substantial recent star formation within the last $\lesssim $1\,Gyr (as indicated by strong Balmer absorption lines due to abundance of A stars). Because of this unusual combination of two spectral features, QPSBs are relatively easy to identify in large spectroscopic surveys of nearby galaxies. Starbursts are also relatively easy to identify because they have unusually high ongoing SFRs. Note that the progenitor-descendant relationship between these two classes is inferred from a stellar population analysis indirectly. Do all starbursts become QPSBs? Can some QPSBs form without prior bursts? Do they collectively pass through the AGN phase? These are the kinds of questions we want to answer by using additional information about their environments and structures. Unfortunately, identifying a complete and pure sample of transition PSBs with ongoing star formation or/and AGN activity is extremely challenging \citep{Wild+10,Yesuf+14, Alatalo+16,Baron+22}. Some uncertainty remains on how AGNs with strong Balmer absorption lines relate to the starburst and QPSB populations. This work provides some clarity on these AGNs. 

\subsection{The Scope and Contributions of This Paper}

The main aim of this work is to do thorough consistency checks of the evolution from SFGs $\rightarrow$ starbursts $\rightarrow$ AGNs $\rightarrow$ QPSBs $\rightarrow$ QGs at $z < 0.16$ using multiple environmental indicators and structural properties based on the Sloan Digital Sky Survey (SDSS). The new contributions are: (1) examining this evolution as a whole using the same data and measurements in four narrow $M_\star$ ranges. As later discussion will elucidate, previous studies rarely examined the environments and structures of starbursts and QPSBs together, let alone keeping their $M_\star$ fixed. (2) Unlike many previous studies of starbursts or QPSBs, this study uses several multiscale environmental indicators to study their evolution. This is crucial because various environmental indicators have different information content and meaning, and they often lead to different results \citep{Wilman+10, Haas+12, Muldrew+12,WooJoanna+13}. Therefore, this paper explores fixed aperture environmental density indicators ranging from scales of $0.5-8\,h^{-1}$Mpc, $k$-nearest neighbor distances/densities ($k=1, 3, 5$), the tidal parameter, two estimates of halo masses \citep{Lim+17,Tinker21}, satellite/central classification, and group/cluster membership \citep{Tempel+14}. Compared to related past studies, this work uses entirely new measurements or similar measurements with updated data. (3) To put this study in the broader context of galaxy evolution, the environments of starbursts and QPSBs are meticulously compared with those of galaxies in the upper SFMS, SFMS, lower SFMS, green valley, and those of QGs as well as AGNs. Due to the improvements or approaches described above, this study demonstrates that all QPSBs can be linked to some SFGs, starbursts, AGNs, and QGs that have similar $M_\star$, structures, and multiscale environments. However, not all starbursts can be linked to QPSBs of similar properties, primarily because these starbursts are disc-dominated. In addition, this paper also shows that some QGs plausibly originated from recent quenching of starbursts, but most QGs live in utterly different environments today and are unlikely to be descendants of recently quenched galaxies.

\subsection{Inconsistencies of Previous Studies of Environments of Starbursts and QPSBs}

As summarized in this section and also discussed in section~\ref{sec:disc}, the environments of starbursts and QSBs have been investigated by many previous studies. However, the results from these studies were not completely consistent because of differences in methodology (e.g., quantifying environment) and sample selection and characteristics (e.g., sample size, redshift, and $M_\star$). Nonetheless, the majority of the studies at $z \lesssim 0.3$ found that most QPSBs reside in the low-density/field environments similar to those of SFGs \citep{Zabludoff+96, Blake+04, Hogg+06, Goto05, Nolan+07, Yan+09,Wilkinson+17, Pawlik+18}. The existence of a large population of QPSBs in the low-density environments indicates that environmental mechanisms are not the dominant route by which QPSBs formed recently. On the other hand, there is evidence that at least some low-$z$ QPSBs have quenched due to environmental mechanisms \citep{Owers+19, Paccagnella+17, Paccagnella+19, Vulcani+20}. For example, \citet{Paccagnella+19} found that the QPSB fraction increases from the outskirts toward the cluster center and from the least to the most massive halos, suggesting that clusters are more efficient at producing QPSBs. As clusters are rare and most galaxies do not reside in clusters, it is still unclear how important the cluster environment is for QPSB formation.

Likewise, the environments of starbursts are not well established. The handful of previous studies on environments of starbursts at $z < 0.3$, with the exception of \citet{Owers+07}, are limited to a special class of galaxies known as ultraluminous infrared galaxies (ULIRGs) and LIRGs \citep[][]{Koulouridis+06, HwangHS+10,Tekola+12, Tekola+14}. \citet{Owers+07} found that starbursts are less clustered at $\sim 1-15$\,Mpc scales than the overall 2dFGRS\footnote{2dF Galaxy Redshift Survey} galaxy population, and they do not preferentially live in clusters. \citet{HwangHS+10} found that, at fixed $M_\star$, the fraction of ULIRGs and LIRGs and their infrared luminosities show weak or no dependance on the environmental density, which was estimated using the projected distance to the 5th nearest neighbor galaxy ($r_{p,5}$). However, they found that the probability of being (U)LIRG and its IR luminosity both increase with the decreasing distance to the nearest late-type galaxy. Other similar studies found LIRGs live preferentially in low-density ($\delta_{5}$) environments \citep[][]{Goto05b,Ellison+13, Burton+13}. In contrast, \citet{Tekola+12} found that LIRGs live in denser environments than those of non-LIRGs. These authors measured the overdensity around LIRGs by counting their numbers in a cylinder of $2\,h^{-1}$\,Mpc radius and $10\,h^{-1}$\,Mpc length ($|\Delta v| < 1000\,\mathrm{km\,s^{-1}}$) and comparing their distribution with a random catalog. They found that LIRGs show a strong correlation between this environmental density and the infrared luminosity, while non-LIRGs do not show such correlation. In short, galaxies with high infrared luminosities preferentially reside in low-mass ($\lesssim 10^{13}\,M_\odot$) haloes \citep{Tekola+14} or low-density environments on large scales. But the small-scale ($ \lesssim 2 \,h^{-1}$\,Mpc) environments of LIRGs may be relatively higher than those of normal SFGs \citep{Koulouridis+06, Tekola+12}. The environmental studies of local LIRGs and ULIRGs are limited by small sample sizes; therefore, their conclusions are tentative. This study complements them by selecting $\sim 8,600$ starbursts using SFR based on UV-optical-mid IR spectral energy density (SED) fitting \citep{Salim+16,Salim+18}. 

Similarly, the environments of starbursts and QPSBs at $z \gtrsim 0.3$ are also not well constrained. Some studies found that the fractions of QPSBs/k+a are higher in high-density environments than in low-density environments \citep[e.g.,][]{Tran+04, Poggianti+09, Yan+09, Vergani+10, Muzzin+12, Dressler+13, Socolovsky+18, McNab+21} and others did not \citep{Balogh+99, Yan+09, Lemaux+17}. \citet{Yan+09} found that the distribution of $k$-nearest neighbor densities of QPSBs at $z \sim 0.8$ is similar to that of QGs, whereas \citet{Lemaux+17} found that it is similar to that of SFGs. Discrepancies between these studies are caused primarily by differences in their sample selection and/or methodology \citep[for a detailed discussion, see][]{Yan+09, Lemaux+17}. The typical environment of distant starbursts is also unknown. There is a long-standing debate about the SFR-density relation (the sign of the correlation or lack thereof) of distant galaxies \citep{Elbaz+07,Cooper+08, Patel+11,Sobral+11, Ziparo+14}.

\subsection{The Merger Origin and Structures of Starbursts and PSBs}

 A merger/interaction-driven galaxy evolution is one of the hypotheses that have been proposed to explain the observed link between structure and SFR. Major galaxy mergers in simulations lead to a central mass concentration and black hole growth \citep{Sanders+88, Mihos+96,Hopkins+08}. During galaxy mergers, the gas is perturbed and funneled to the center to power a nuclear starburst and an AGN activity. The gas is then depleted quickly by the starburst or removed or heated by stellar and/or AGN feedback, eventually forming a quiescent, early-type galaxy. For example, QPSBs in Magneticum cosmological simulations were shut down rapidly by a merger-triggered AGN feedback, which redistributed and heated gas in the PSBs. About 89\% of the simulated QPSBs at $z \approx 0$ had at least one merger within the last 2.5\,Gyr and about 65\% had major ($M_\star$ ratio greater than 1:3) mergers \citep{Lotz+21}. In EAGLE simulations, around 14\% of the simulated QPSBs have experienced mergers with $M_\star$ ratios greater than 1:10 and $\sim 3\%$ with ratios greater than 1:3 within 0.5\,Gyr of the onset of the PSB episode \citep{Davis+19}. While AGNs are important in removing star-forming gas from some galaxies, they are not the dominant cause of gas removal in PSBs in EAGLE; only $\sim 10\%$ of the QPSBs had enhanced AGN accretion/outburst. 

Consistent with the prevalence of mergers in simulated PSBs, a significant fraction of observed QPSBs show prominent signs of recent mergers or interactions \citep{Yamauchi+08, YangYujin+08, Pracy+09, Pawlik+16, Sazonova+21}. Most QPSBs are also more centrally concentrated than SFGs \citep{Nolan+07,YangYujin+08,Pracy+09, Sazonova+21}. Likewise, starbursts (including LIRGs and ULIRGs) are highly disturbed merger remnants \citep{Sanders+96, Veilleux+02, Luo+14, Larson+16, Cibinel+19, Shangguan+19,Yesuf+21}. Alternatively, minor mergers may produce a significant fraction of rejuvenated PSBs by triggering new cycles of starbursts in passive, bulge-dominated galaxies \citep{Dressler+13, Rowlands+18, Davis+19, Pawlik+18}. Although rare at $z \approx 0$, the compaction process, triggered by an intense gas inflow episode, involving mergers, counter-rotating streams, or recycled gas may also produce PSBs \citep{Dekel+14,Zolotov+15, Tacchella+16}. Several works have shown that PSBs at $z \sim 1-2$ are compact \citep{Yano+16, Almaini+17, Maltby+18, WuPo-Feng+20, Suess+21, Setton+22}.

Merger signatures are hard to identify in observations; they fade with time \citep{Lotz+08}, are hard to accurately quantify, and require deep imaging. Thus, disturbance parameters such as asymmetry cannot be measured reliably for typical starbursts and QPSBs using shallow SDSS images \citep{Pawlik+16,Pawlik+18}. Alternatively, we will demonstrate that easily measurable structural parameters such as $C$ and $\sigma_\star$ and multiscale environmental parameters provide valuable information for large samples of starbursts and their descendants. In fact, complementary to the asymmetry parameter, $C$ and $\sigma_\star$ are good predictors of $\delSSFR$ for central galaxies \citep[][]{Yesuf+21}.

The rest of the paper is structured as follows: Section~\ref{sec:data} presents the stellar, structural, and environmental data used in this study. Section~\ref{sec:res} presents the main results. To help interpret these results, Section~\ref{sec:disc} presents in-depth discussion and comparisons with previous studies of starbursts and QPSBs. The summary and conclusions of this work are given in Section~\ref{sec:conc}. We adopt a cosmology with $\Omega_\Lambda = 0.7$, $\Omega_m =  0.3$ and $h = 0.7$.

\section{Data and Methodology}\label{sec:data}

\subsection{Measurements of Galaxy Properties}

Our sample is selected from the Sloan Digital Sky Survey \citep[SDSS;][]{Aihara+11}. We use the publicly available Catalog Archive Server to retrieve some of the measurements used in this work such as the stellar velocity dispersion $\sigma_\star$, Petrosian radii, the spectral index H$\delta_A$, and emission-line fluxes \citep{Brinchmann+04}. The $M_\star$ and SFR measurements are taken from version 2 of the GALEX-SDSS-WISE Legacy Catalog \citep[GSWLC-2][]{Salim+16, Salim+18}. They are derived by spectral energy distribution (SED) fitting of UV-optical photometry with additional IR luminosity constraints using the Code Investigating GALaxy Emission \citep[CIGALE]{Noll+09}. \citet{Salim+18} estimated the total infrared luminosities from WISE 22\,$\mu$m or 12\,$\mu$m photometry using luminosity-dependent infrared templates of \citet{Chary+01} and calibrations derived from a subset of galaxies that have Herschel far-infrared data. For narrow-line AGNs, their IR luminosities are corrected using \ion{O}{3} 5007\,{\AA} emission-line equivalent width before using them in the SED fitting. Broad-line AGNs are excluded from the current sample because their SFRs estimates are not reliable. The SED fitting used template superposition of two exponential star formation histories (SFHs) of a younger population (100\,Myr to 5\,Gyr) and an old stellar population (formed 10\,Gyr ago), with the young mass fraction varying between zero and 50\%. The stellar population models were calculated for four stellar metallicities ($0.2-2.5\,Z_\odot$) using \citet{Bruzual+03} models and assuming a \citet{Chabrier03} stellar initial mass function.

\subsection{Sample Definitions}

From the SDSS main galaxy sample, we select $387, 258$ galaxies at $z = 0.02-0.16$, with $M_\star = 3\times 10^9 - 3\times 10^{11}\,M_\odot$,  with good SFR measurements ($\mathrm{flag\_SED=0}$), and which are at least 4\,$h^{-1}$Mpc away from the survey edge. We define the difference from the ridge line of the SFMS, $\delSSFR$, by fitting a simple linear relation of the form $\log\,(\mathrm{SSFR/Gyr^{-1}}) = (a - 1)[\log(M_\star /M_\odot) - 10.5] + b$ galaxies to the galaxies with SSFR $> 0.01\,\mathrm{Gyr^{-1}}$. We fix $\alpha=0.48$ based on the estimate of \citet{Speagle+14} and derive $\beta = -1.24$ from the median of the residuals. Based on their $\delSSFR$ we group galaxies into starbursts ($\Delta\,\mathrm{SSFR}  > 0.6$\,dex), upper SFMS ($\Delta\,\mathrm{SSFR}  = 0.2-0.6$\,dex), SFMS ($\Delta\,\mathrm{SSFR} = -0.2-0.2$\,dex), lower SFMS ($\Delta \,\mathrm{SSFR}  = -(0.2-0.5)$\,dex) , and green valley ($\Delta\,\mathrm{SSFR} =  -(0.5-1)$\,dex), and quiescent ($\Delta\,\mathrm{SSFR}  < -1$\,dex) galaxies. The total number of starbursts in our sample is 8,\,614.

We define QPSBs as galaxies with the equivalent width (EW) of H$\alpha< 3$\,{\AA} in emission and H$\delta_A > 4${\AA} in absorption \citep{Yesuf+14}. We only select QPSBs with reliable spectral measurements. Namely, we exclude galaxies whose median signal-to-noise ratio per pixel (S/N)  of the entire spectra are $\mathrm{S/N} < 10$ or have gaps around their H$\alpha$ continua. To automatically identify objects with bad H$\alpha$ measurements, we require the H$\alpha$ continuum fluxes to be non-zero and the EWs of H$\beta$ to be $< 3\,${\AA}. The total number of selected QPSBs is 752, which is about 10 times smaller than that of starbursts.

As mentioned before, selecting a complete and unbiased sample of the PSBs is not easy. Some attempts have been made to improve the definition of PSBs to include galaxies with ongoing star formation and/or AGN activity \citep{Wild+10,Yesuf+14, Alatalo+16}. We simply select a subsample of AGNs with H$\delta_A > 4$\,{\AA} and show that these galaxies have similar structural and environmental properties as QPSBs, whereas AGNs with H$\delta_A < 3$\,{\AA} are different from QPSBs. We identify ``pure" AGNs using the emission-line ratios \ion{O}{3}/H$\beta$ and \ion{N}{2}/H$\alpha$ \citep{Kewley+06} and H$\alpha > 3$\,{\AA}. We require $\mathrm{S/N} > 3$ for the emission line fluxes and $\mathrm{S/N} > 10$ for the entire continuum spectra. The total number of AGNs with H$\delta_A > 4$\,{\AA} (H$\delta_A < 3$\,{\AA}) is 1,\,006 (9,\,544).

\subsection{Measurements of Environmental Properties}

This paper uses several environmental indicators measured by its author or others \citep{Tempel+14, Lim+17, Tinker21}. The author calculates the mass density of galaxies within radii of 0.5, 1, 2, 4, and 8$\,h^{-1}\,\mathrm{Mpc}$ centered around each galaxy and the projected distances to its $k = {1, 3, 5}$ nearest neighbors ($r_{p,k}$). In these calculations only galaxies that have $|\Delta v| < 1000$\,km\,s$^{-1}$ relative to the primary galaxies are considered, thereby excluding unrelated foreground and background galaxies along the line of sight. The main results do not change if $|\Delta v| < 500$\,km\,s$^{-1}$ or $|\Delta v| < 1500$\,km\,s$^{-1}$ is used instead. The projected $k$-nearest neighbor surface density is defined as $\Sigma_{k} = k/(\pi r_{p,k}^2)$. All densities are normalized by the median densities of all galaxies in a given mass range and in a redshift bin of $\Delta z = 0.02$. This converts the densities into overdensities relative to the median density (e.g., $\delta_{k} \equiv \Sigma_{k}/\tilde{{\Sigma_{k}}}-1$) and effectively accounts for redshift variations in the selection rate of SDSS spectroscopy \citep{Cooper+09}. Subscripts of $\delta$ with or without units differentiate fixed radius densities from nearest neighbor densities. For example, $1+\delta_5$ denotes overdensity within the 5th nearest neighbor, and $1+\delta_\mathrm{8Mpc}$ denotes normalized mass density within 8$\,h^{-1}\,\mathrm{Mpc}$.

The environmental density indicators are significantly correlated with each other. For example, $\deltahalf$ is strongly correlated with $\deltaone$ ($\rho = 0.8$) and $\deltatwo$ ($\rho=0.6$) and it is moderately correlated with $\deltafour$ ($\rho =0.5$) and $\deltaeight$ ($\rho \approx 0.4$). The correlation between $\deltaeight$ and $\deltafour$ is also strong ($\rho \approx 0.8$). Likewise, $\delta_5$ is strongly correlated ($\rho \approx 0.8$) with $\deltaone$, $\deltatwo$, and $\deltafour$.

\begin{figure*}[hbt!]
\gridline{\fig{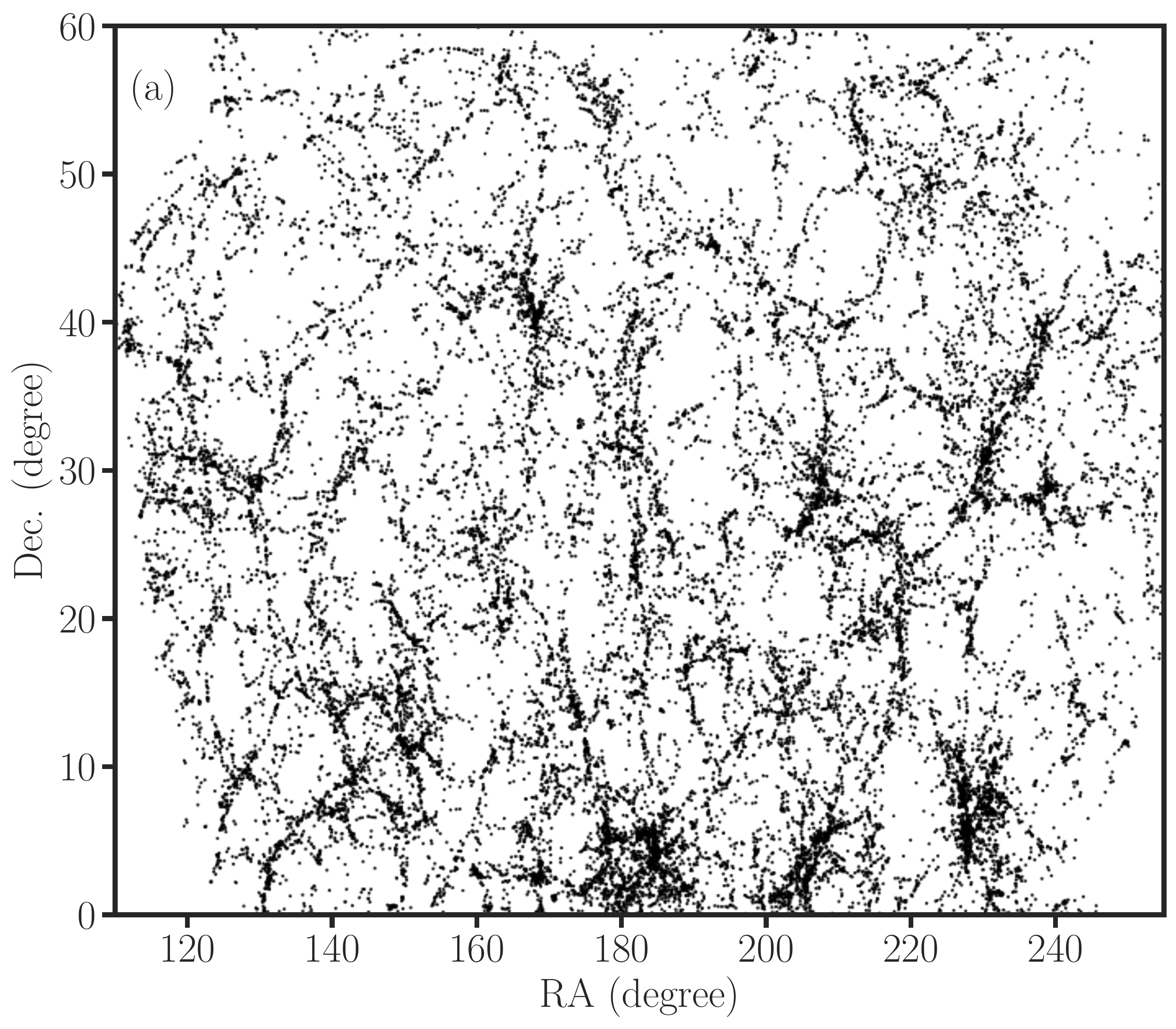}{0.40\textwidth}{}
          \fig{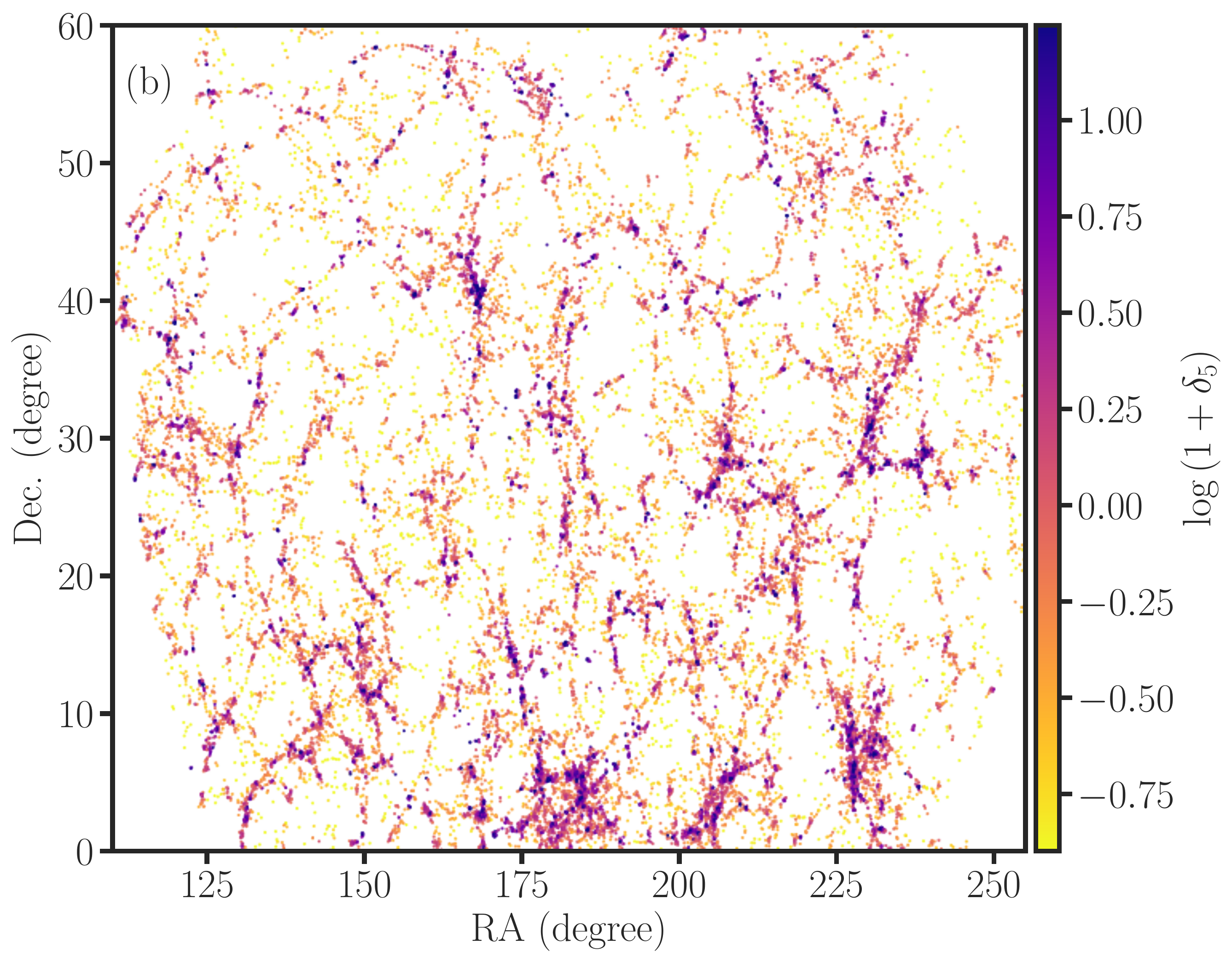}{0.45\textwidth}{}}
\gridline{\fig{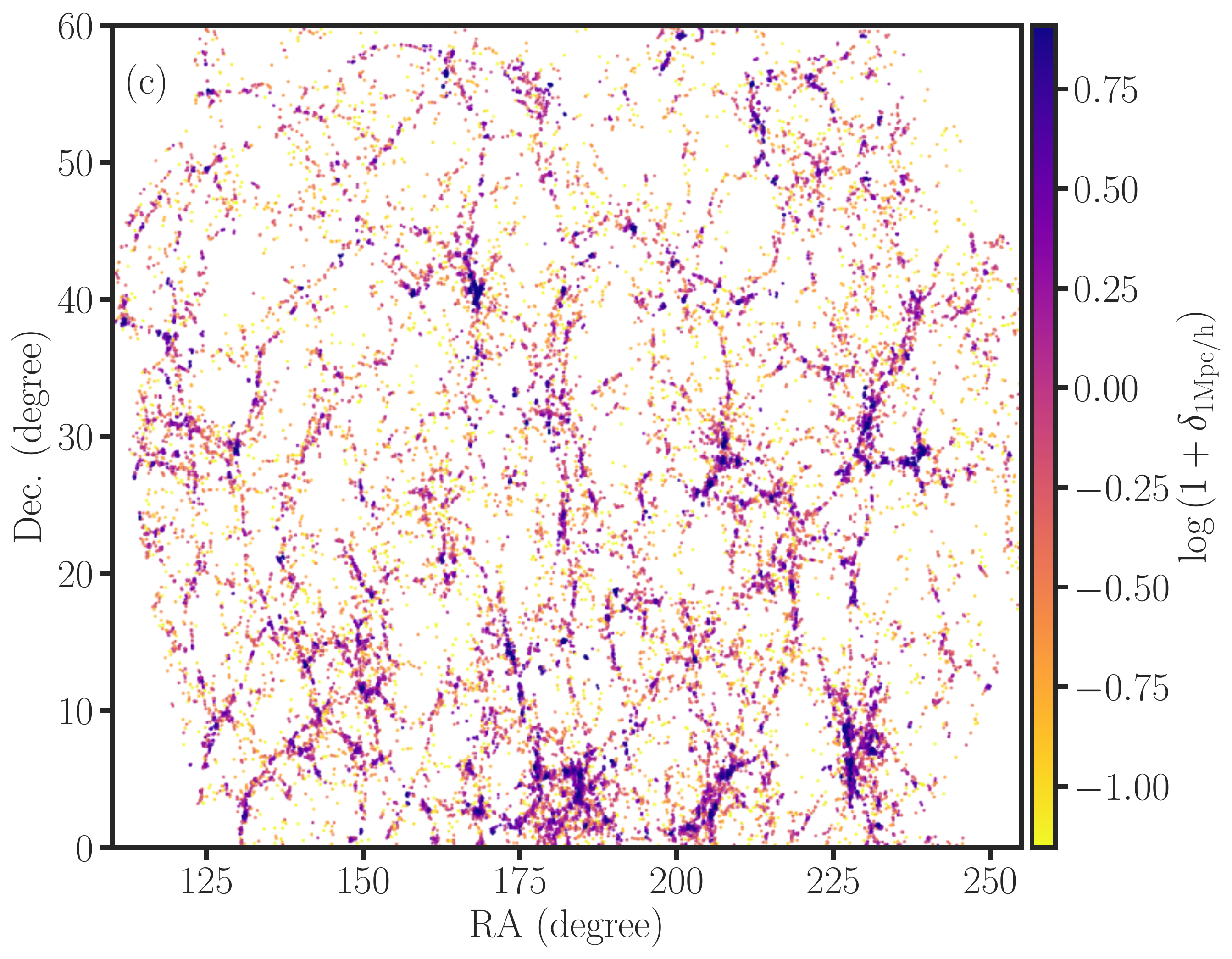}{0.45\textwidth}{}
         \fig{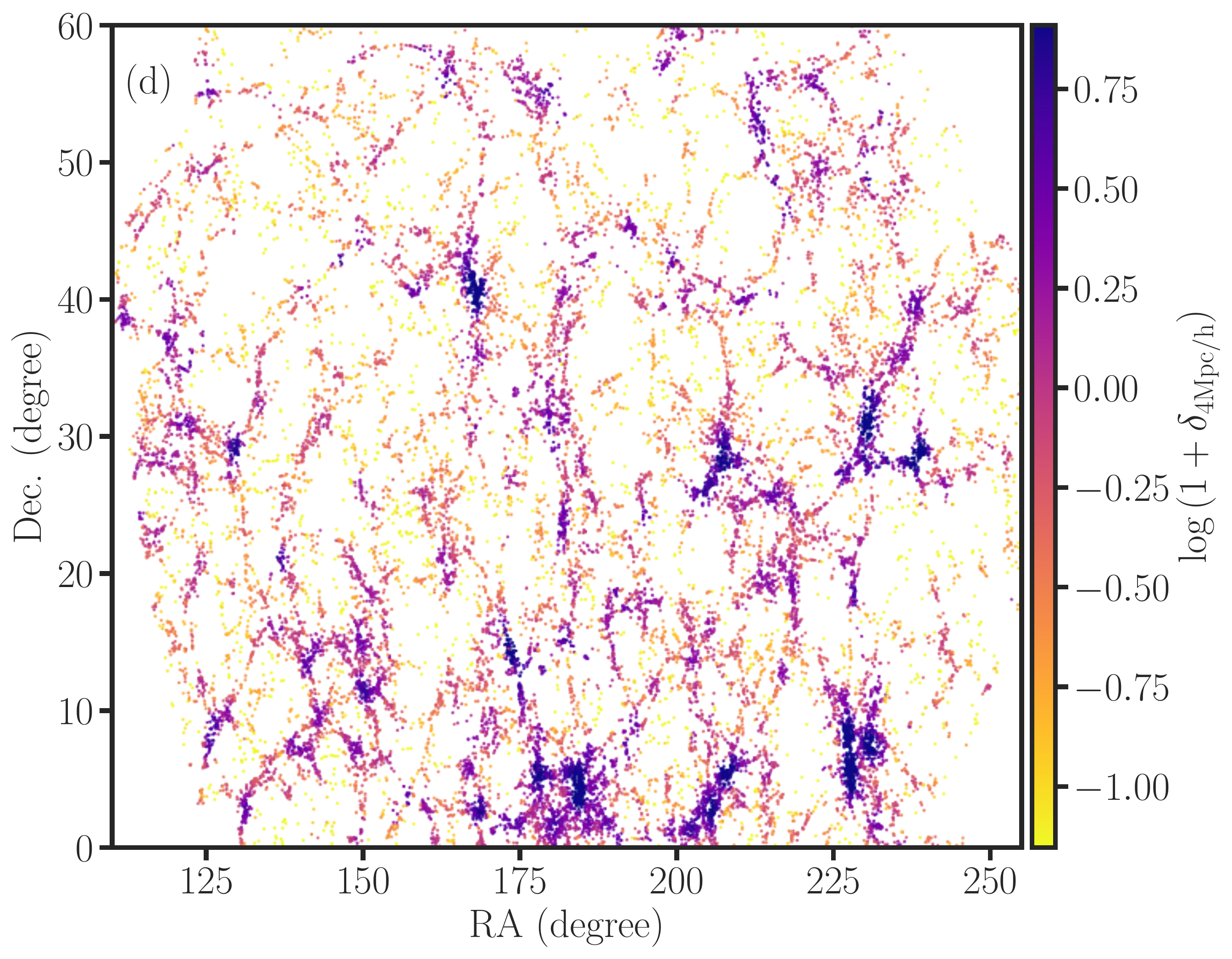}{0.45\textwidth}{}}
\gridline{\fig{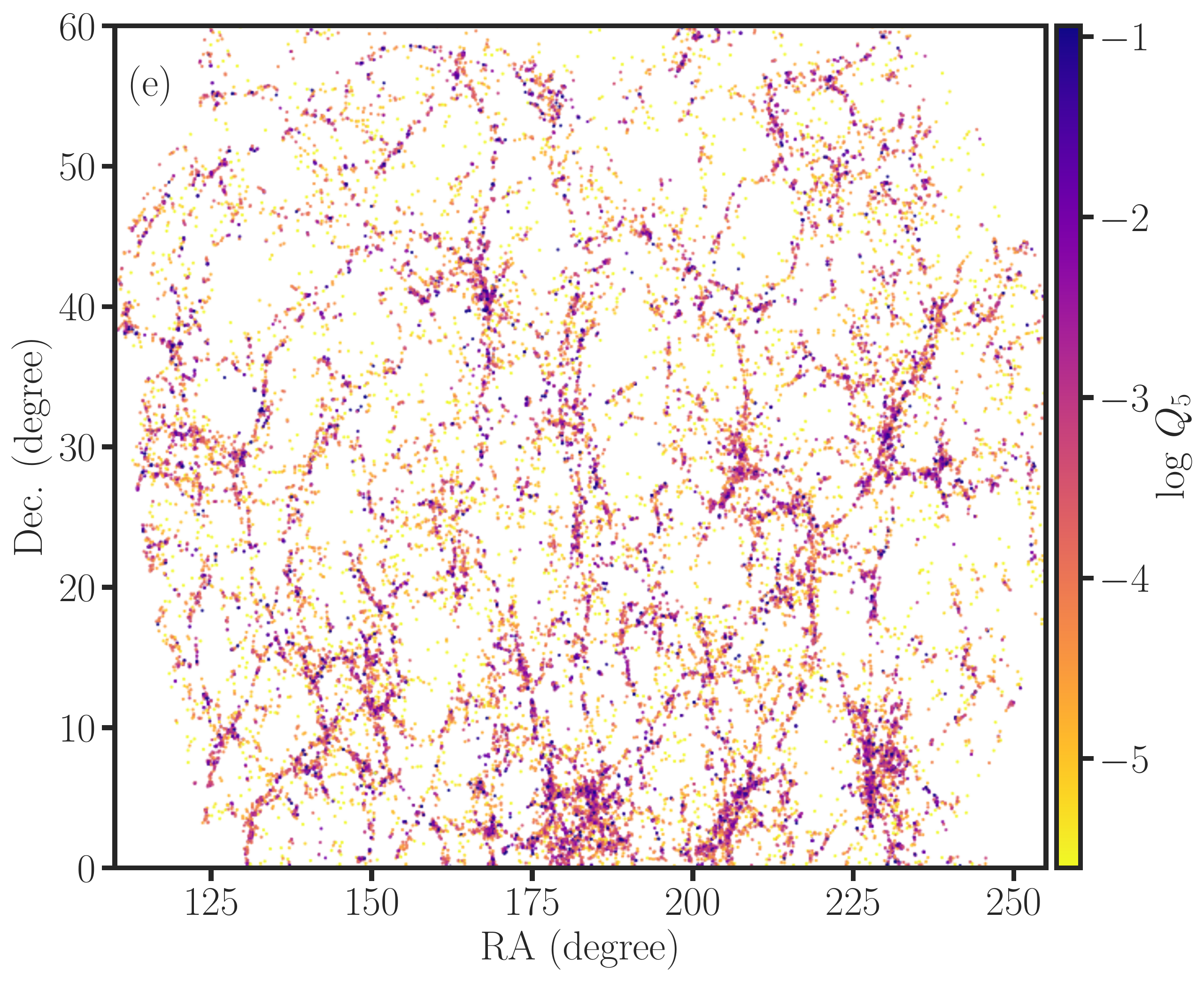}{0.45\textwidth}{}
         \fig{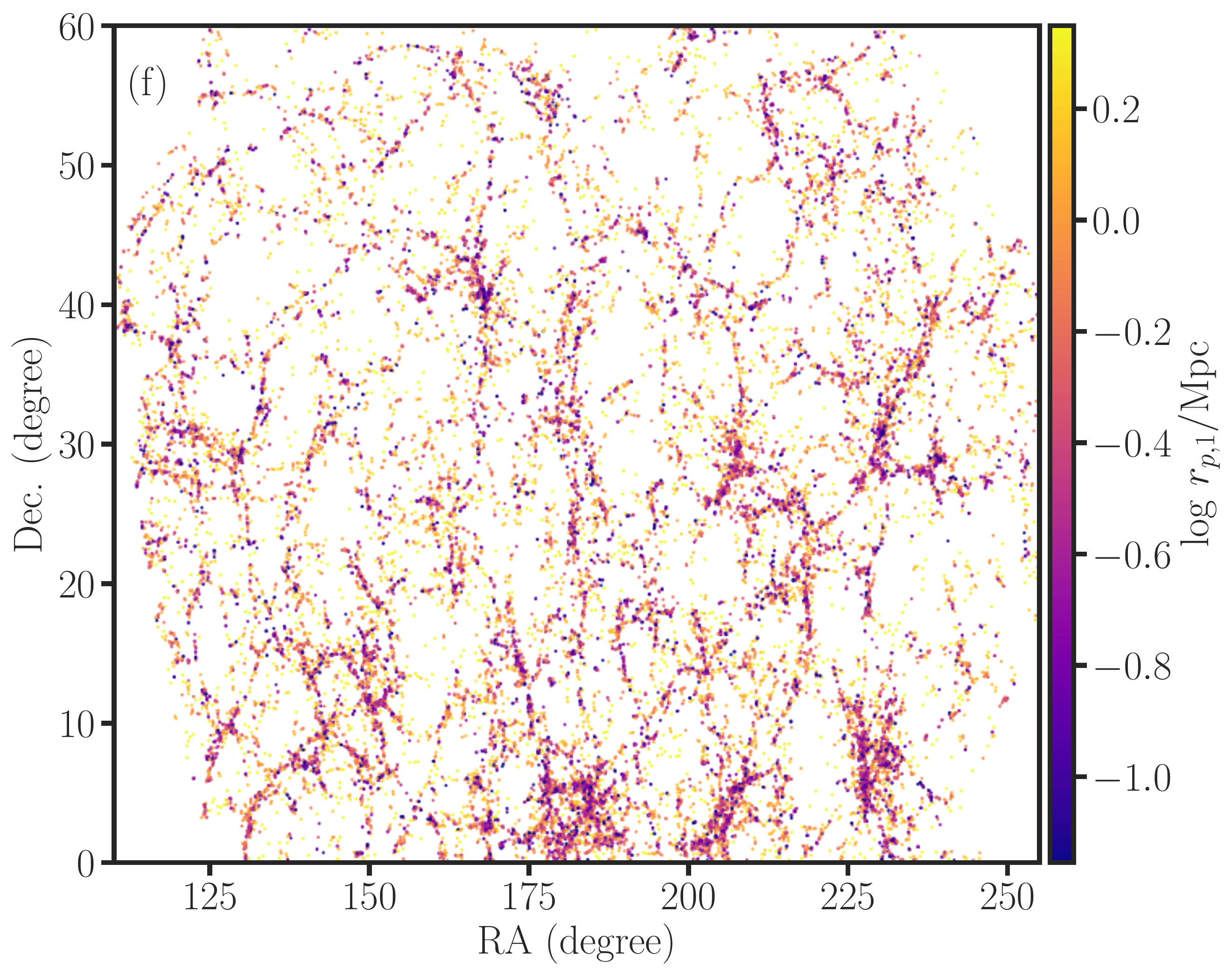}{0.45\textwidth}{}}
\caption{Visualizing the various environmental indicators. Panel (a): the SDSS galaxy distribution at $z= 0.075-0.08$; panel (b): the 5th nearest neighbor overdensity; panel (c): the mass overdensity within $1\,h^{-1}$\,Mpc;
panel (d): the mass overdensity within $4\,h^{-1}$\,Mpc; panel (e): the tidal parameter of the five nearest neighbors; panel (f): the nearest neighbor distance. The color bars are truncated outside $5\%-95\%$ range of the data. \label{fig:vis_sp}}
\end{figure*}

Furthermore, we quantify the tidal strength exerted by the five nearest neighbor galaxies using the tidal parameter \citep{Verley+07,Argudo-Fernandez+13} defined as follows:

\begin{equation}
\log \,Q_k = \log \left(\sum^k_{i=1} \frac{M_{\star, i}}{M_\star}\left(\frac{D}{r_{p,i}}\right)^3\right)
\end{equation}

\noindent where $M_\star$ and $D$ are the stellar mass and diameter of the main galaxy of interest, $M_{\star,i}$ is  the stellar mass and $r_{p,i}$ is the projected distance to the $i$th nearest neighbor. Following \citet{Argudo-Fernandez+13}, we set $D = 2\alpha\,R_{90}$, where $R_{90}$ is the $r$-band Petrosian radius containing 90\% of the total flux of the main galaxy and it is scaled by a factor $\alpha = 1.43$ to recover the  isophotal diameter $D_{25}$ at the surface brightness of 25 mag/arcsec$^2$. In our measurements of $Q_{k}$ and mass overdensities, we use redshift and $M_\star$ data from the SDSS DR17. The $M_\star$ estimates used here were derived using the methodology of \citet{Chen+12} and they agree well ($\rho = 0.97$ and $\sigma \approx 0.1$\,dex) with the estimates from \citet{Salim+18}. When $M_\star$ estimates of nearby neighbors are not available, we estimate them using their $i$-band luminosities and the mean relationship between $g-i$ color and the mass-to-light ratio ($\log\, M_\star/L_i = -0.5 +0.66 \times (g-i)$). If a neighbor of a galaxy is classified as QSO, we set its $\log\,M_\star = 0$. Our main conclusions do not change if remove galaxies with QSO neighbors. Figure~\ref{fig:vis_sp} visualizes the information content of some of the environmental indicators measured by the author. They are conspicuously correlated but are also discrepant.

\citet{Tempel+14} utilized a modified friends-of-friends (FoF) algorithm with a variable linking length ($\sim 0.3-0.7$\,Mpc) in the transverse and radial directions to identify groups of galaxies within certain neighborhood radius. Their flux-limited catalogue (based on SDSS DR10) includes 82,458 galaxy groups with two or more members. The flux-limited and volume-limited catalogs agree well, especially for large groups/clusters. About $80\%-90\%$ of groups are identified at the same locations in both catalogs, and $\sim 60\%-70\%$ of the matched groups have similar mass estimates within a factor of two. We use the flux-limited catalog, which includes measurements of several group characteristic (richness, size, radial velocity dispersion, etc) and normalized environmental luminosity densities estimated using different smoothing radii (1, 2, 4, and 8$\,h^{-1}$Mpc). Following \citet{Poggianti+09}, groups with more six members are classified as rich groups if the have group radial velocity $\sigma_g < 400$\,km\,s$^{-1}$ or as clusters if $\sigma_g > 400$\,km\,s$^{-1}$.

\citet{Tempel+14}'s measurements indicate that starbursts and upper SFMS galaxies have high 1$\,h^{-1}$Mpc luminosity densities but low 8$\,h^{-1}$Mpc luminosity overdensities relative to other galaxies (see Appendix~\ref{sec:appA}). Suspecting this could be due to a mass-to-light ratio bias, the author of this paper did the mass overdensity measurements. Starbursts and upper SFMS galaxies have relatively low mass overdensities at all scales. \citet{Tempel+14}'s luminosity overdensities are used only in the Appendix. Note that in all overdensity calculations, the mass or luminosity of the primary galaxy is included. Galaxies within 4$\,h^{-1}$Mpc from the survey edge are excluded using distance measurements in \citet{Tempel+14}'s catalog.

In \citet{Lim+17} catalog (based on SDSS DR13), the groups were identified with an updated version of the iterative halo-based group finder of \citet{YangX+05,YangX+07}. This version of the group finder uses $M_h$ proxies that scale with $M_\star$ of central galaxies and the the ratio of $M_\star$ between the central galaxy and the $k$th (typically $k=4$) most massive satellite to estimate preliminary $M_h$ for every galaxy. If the galaxy is isolated, the mean relation between the halo mass and stellar mass of isolated galaxies based on EAGLE simulations is used. The group finder then assigns galaxies into groups using halo properties such as halo radius and velocity dispersion. NFW density profile \citep{Navarro+97} and a Gaussian distribution for the redshifts of galaxies within a halo ($P(z-z_\mathrm{group})$) is assumed. An abundance matching between the mass function of the preliminary groups and a theoretical halo-mass function is used to update $M_h$ at each iteration. The update continues until group memberships converge. Mock galaxy samples constructed from EAGLE simulations were used to test and calibrate the group finder. The tests  showed that the group finder can find $\sim$ 95\% of the true member galaxies for $\sim 85$\% of the groups in the SDSS. The typical uncertainty of $M_h$ is $\sim 0.2$\,dex. The group finder also classifies galaxies into centrals and satellites.

Similarly, in the self-calibrated model of \citet{Tinker21}, which also attempts to improve the halo-based algorithm of \citet{YangX+05}, groups are rank-ordered by their weighted total group luminosity; the multiplicative weight factors are calibrated by comparing the predictions of the group catalog based $N$-body simulations to measurements of galaxy clustering and the total $r$-band luminosity of satellites, $L_\mathrm{sat}$, around spectroscopic central galaxies. The latter quantity is based on deep-imaging data from the DECALS Legacy Imaging Survey. Besides probing $M_h$, $L_\mathrm{sat}$ is also sensitive to the formation histories of halo, with younger haloes having more substructure, and thus more satellite galaxies. To break this degeneracy, the concentration of the central galaxy is used in the calibration. The concentration index of a galaxy is defined as the ratio between the radius that contains 90\% of the galaxy light to the half-light radius, $C = R_{90}/R_{50}$. \citet{Tinker21} gave several reasons for using $C$ in the halo mass estimate: (1) $C$ for central galaxies does not correlate with large-scale environment (proxy for halo age) at fixed $M_\star$.  (2) The correlation of $C$ with $L_\mathrm{sat}$ is roughly independent of $M_\star$. (3) $C$ varies minimally with galaxy luminosity/mass. Furthermore, \citet{Tinker21} showed that his halo masses estimates are in good agreement with weak-lensing estimates for star-forming and quiescent central galaxies. However, the results inferred from his group catalog also differ in several ways from previous catalogs. He found significantly different fractions of satellite in SFGs and QGs than other methods. He claimed that his method is more sensitive to low-mass halos, $M_h < 10^{12}\,M_\odot$.

\subsection{Statistical Methods}

To estimate the 95\% simultaneous confidence intervals of the fractions of galaxies of a given class (say starbursts), that are isolated, pairs, in groups, or in cluster, we assume a multinomial distribution and compute the confidence intervals of the multinomial proportions using the Goodman method\footnote{For a multinomial distribution with $k$ categories and a total sample size $N_\mathrm{tot}$, the Goodman method approximates the true proportions $p_1,p_2 \cdots p_k$ with the observed sample proportion for $i$th category $\hat{p}_i$ as  $(p_i - \hat{p}_i) \le \pm \chi(\alpha/k, 1) \sqrt{\hat{p}_i(1-\hat{p}_i)/N_\mathrm{tot}}$ for $i= 1, 2, \cdots k$, where $\alpha=0.05$ and $\chi^2(\alpha/k, 1)$ is the ($1-\alpha/k) \times 100$th percentile of the chi-square distribution with 1 degree of freedom.}\citep{Goodman65} as implemented in \texttt{statsmodels} python package. The Multinomial distribution is an extension of the binomial distribution for $k > 2$ categories.

For \citet{Lim+17}'s central/satellite classification, the mean central fraction in a given $M_\star$ range is estimated as $\hat{f}_\mathrm{cen}= N_\mathrm{cen}/N_\mathrm{tot}$, where $N_\mathrm{cen}$ and $N_\mathrm{tot}$ are number of centrals and the total number of galaxies in a given sample, respectively. The standard error of $\hat{f}_\mathrm{cen}$ is estimated by the Normal approximation of the Binomial distribution as $\sigma_{\hat{f}_\mathrm{cen}} = \sqrt{\frac{\hat{f}_\mathrm{cen} (1-\hat{f}_\mathrm{cen})}{N_\mathrm{tot}}}$. Since \citet{Tinker21} provides the probability of satellite ($p_\mathrm{sat} = 1-p_\mathrm{cen}$) for each galaxy, in his case, assuming the Poisson-binomial distribution for the number of centrals in a given sample $\hat{f}_\mathrm{cen} = \sum \limits _{{i=1}}^{N_\mathrm{tot}}p_{\mathrm{cen},i}/N_\mathrm{tot}$ and $\sigma_{\hat{f}_\mathrm{cen}}= \sqrt{\sum \limits _{{i=1}}^{N_\mathrm{tot}}p_{\mathrm{cen},i}(1-p_{\mathrm{cen},i})}/N_\mathrm{tot}$. The Binomial distribution is a special case of the Poisson-binomial distribution, when all probabilities are equal to each other.

The Anderson-Darling (AD) test is used to test the null hypothesis that two samples come from the same but unspecified distribution. This non-parametric test is more powerful than the Kolmogorov-Smirnov (KS) test; it is especially sensitive to the tail of a distribution and requires a small sample size. The KS statistics compares two empirical cumulative distribution functions (ECDFs) by looking only at their maximum absolute difference. In contrast, the AD statistics compares two ECDFs by looking at the weighted sum of all their squared differences, which are calculated at each point in the joint sample. The weights are determined by the inverse-variance of the joint ECDF at each point; points in the tail of the distribution receive more weights than points close to the median. The AD test is done using the \texttt{scipy.stats} python package.

To match two samples in a multi-dimensional space, the function \texttt{NearestNeighbors} in \texttt{sklearn} python package is used with a setting of euclidian distance and ball tree algorithm ($\mathrm{leaf\,size=30)}$;  each input measurement is transformed to the logarithmic scale and is standardized by subtracting its mean and dividing by its standard deviation.

\section{Results}\label{sec:res}

This section first compares the environments of galaxies classified by their $\delSSFR$, ranging from starbursts to QGs. It demonstrates the existence of a general trend with $\delSSFR$, in the sense that starbursts occupy the lowest density environments among SFGs and QGs occupy the highest densities. The section then compares the environments of QPSBs and (post-starburst) AGNs with those of starbursts and normal galaxies. It ends by assessing the consistency of the evolution from starbursts $\rightarrow$ AGNs $\rightarrow$ QPSBs $\rightarrow$ QGs in terms of their structures and multiscale environments.

\subsection{The Environments of Starbursts and Comparison Samples and their Trends with $\delSSFR$}

\begin{figure*}[hbt!]
\gridline{\fig{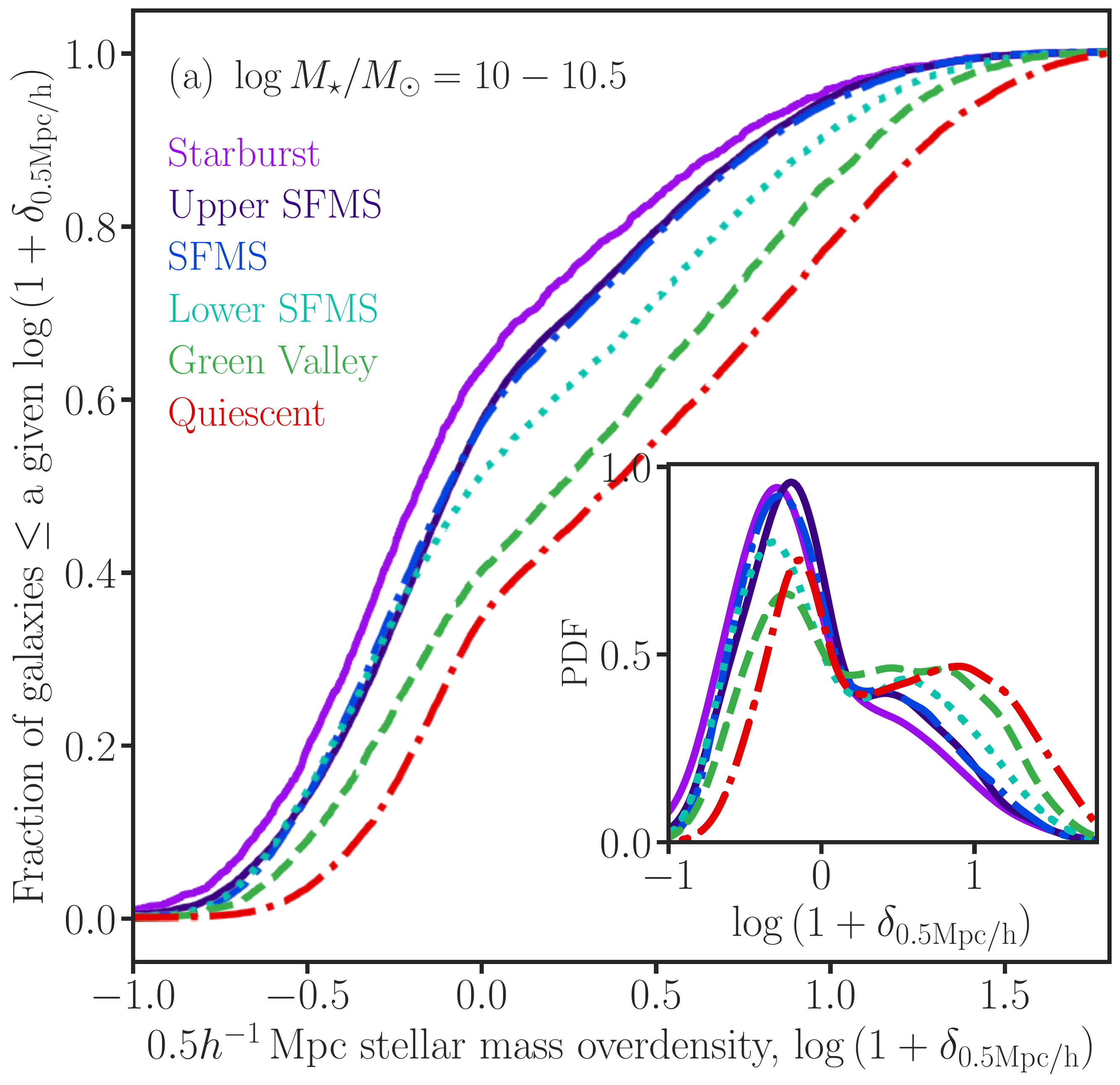}{0.47\textwidth}{}
          \fig{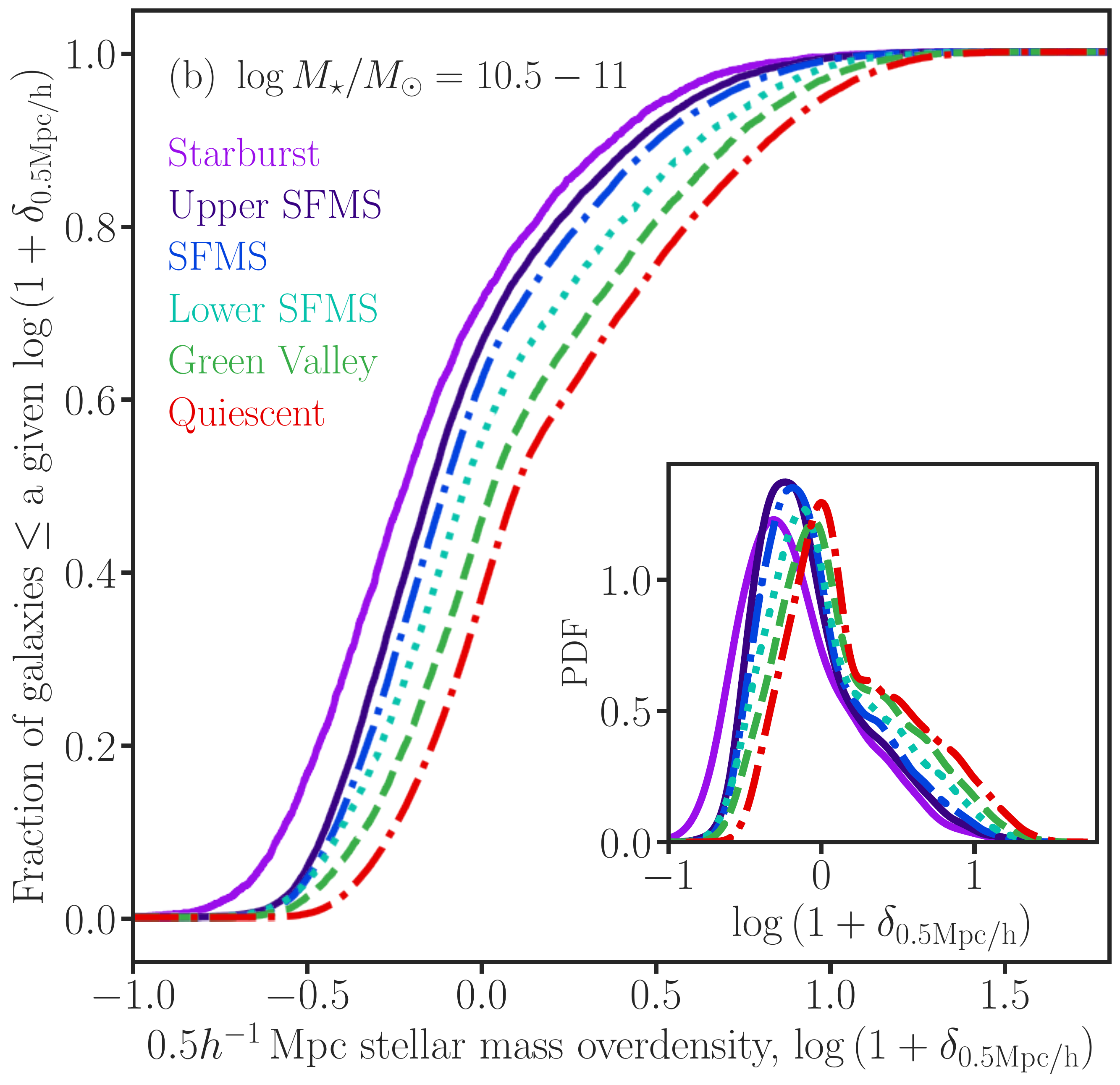}{0.47\textwidth}{}}
\gridline{\fig{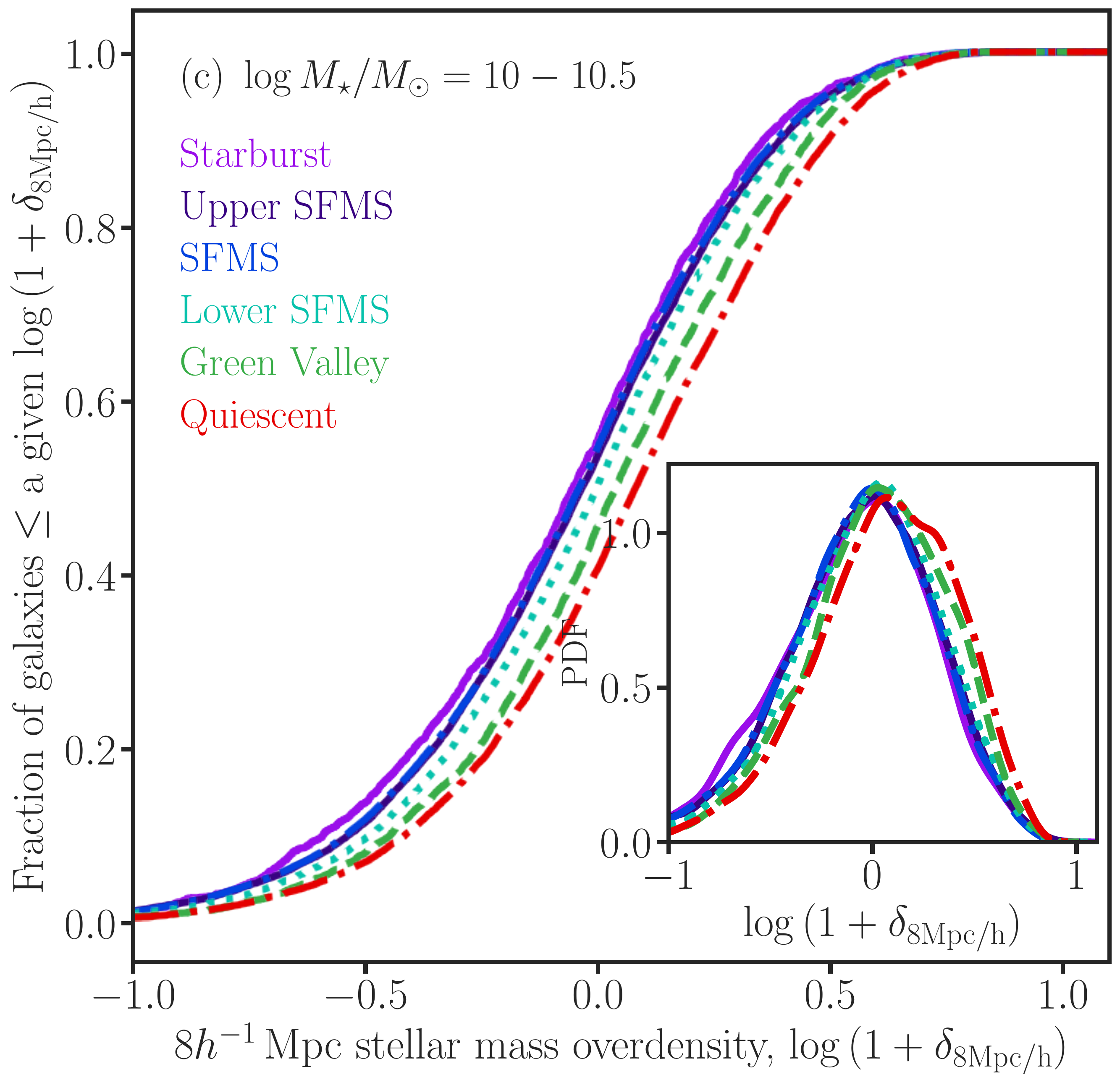}{0.47\textwidth}{}
          \fig{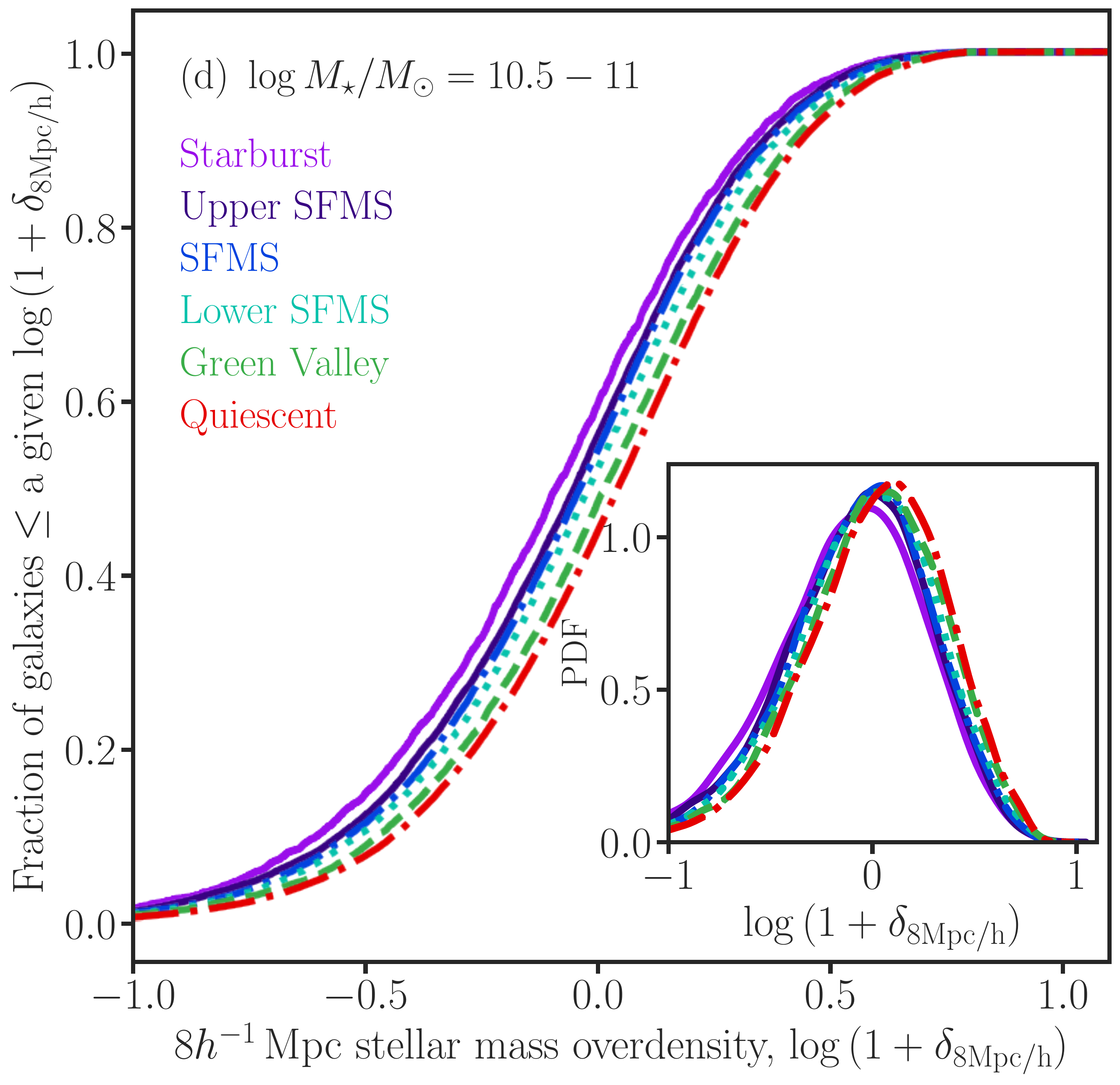}{0.47\textwidth}{}}
\caption{The cumulative distributions of the normalized environmental mass densities within $0.5\,h^{-1}$Mpc ($\deltahalf$) or $8\,h^{-1}$Mpc ($\deltaeight$) for galaxies grouped into two $M_\star$ ranges and $\delSSFR$. The $\deltahalf$ distributions of starbursts are different from those of normal SFGs, and they are narrow and peak at lower densities compared to those of QGs. In contrast, the $\deltaeight$ distributions of starbursts are slightly lower than those of normal SFGs, but they are much lower than those of green-valley galaxies and QGs.\label{fig:del18_sb}}
\end{figure*}

\begin{figure*}[hbt!]
\fig{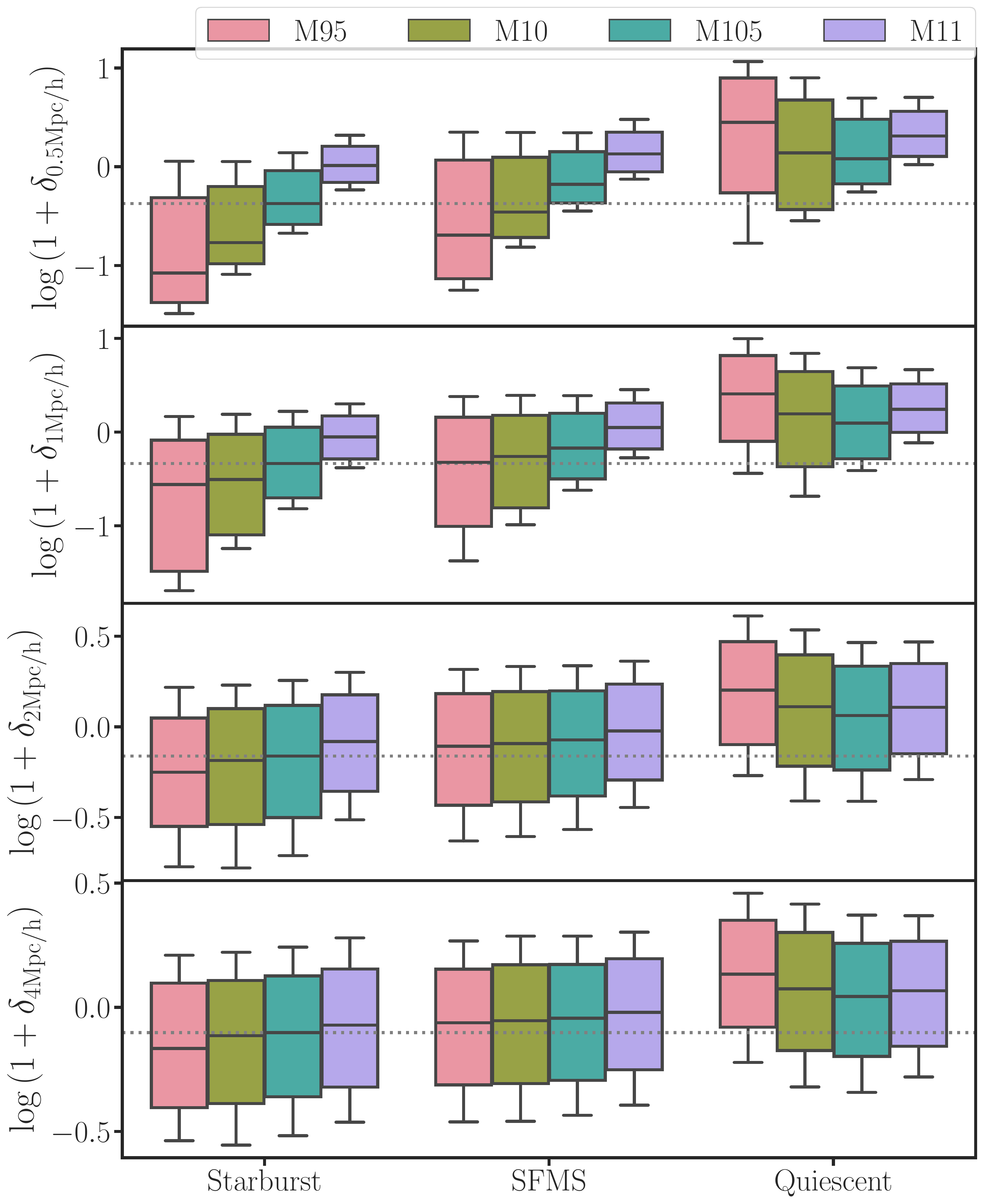}{0.9\textwidth}{}
\caption{Comparing the $0.5-4\,h^{-1}$\,Mpc mass overdensities of starbursts with those of SFMS galaxies and QGs at different $M_\star$ ranges. The names of the samples indicate the starting points of the $\log\, M_\star$ ranges, which span 0.5\,dex. For example, M11 denotes $\log M_\star/M_\odot = 11-11.5$. The box marks 15\%, 50\%, and 85\% of the mass overdensity distribution at a given scale and the error bar extends to show the rest of the distribution, ignoring outliers. The dashed line shows the median overdensity for M105: $\log M_\star/M_\odot = 10.5-11$ at a given scale. The small-scale mass overdensities strongly depend on $M_\star$. At all $M_\star$ ranges, the mass overdensity distributions of starbursts are different from those of SFMS galaxies and QGs at $\sim 0.5-2\,h^{-1}$\,Mpc scales. At $4\,h^{-1}$\,Mpc scale, they are more similar to those SFMS galaxies than they are to the distributions of QGs. \label{fig:massdel18}}
\end{figure*}

Starbursts inhabit lower density environments than galaxies with lower $\delSSFR$. For example, Figure~\ref{fig:del18_sb} shows the distributions of stellar mass overdensities within $0.5\,h^{-1}$\,Mpc and $8\,h^{-1}$\,Mpc in two narrow $M_\star$ ranges: $\Mlow$ and $\Mhigh$, and for subsets of galaxies ranging from starbursts to QGs. Notably, $\delSSFR$ is significantly anticorrelated with $\deltahalf$ ($\rho = -0.39$ and $p < .001$) and $\deltaeight$ ($\rho = -0.15$ and $p < .001$); the $\deltahalf$ and  $\deltaeight$ distributions of the different samples generally shift toward higher overdensity as $\delSSFR$ decreases, as the galaxies move toward quiescence. Table~\ref{tab:del18_sb} presents the summary statistics (median, 15\%, and 85\%) for the distributions of overdensities measured within apertures $r_{ap} \in \{0.5, 1, 2, 4, 8\}\,h^{-1}$Mpc for all $M_\star$ ranges. Furthermore, Figure~\ref{fig:massdel18} compares the summary statistics of the overdensity distributions of starbursts, SFMS galaxies, and QGs for all $M_\star$ ranges. Starbursts have lower overdensities than QGs at all scales and for all $M_\star$ ranges. The difference between the overdensties of starbursts and SFMS galaxies is more evident at the scales of $\sim 0.5-2h^{-1}$\,Mpc. The environmental difference of starbursts/SFGs and QGs is not sensitive to the adopted mass bins of 0.5\,dex. The results still hold if the $M_\star$ bins are made 0.25\,dex.

\begin{rotatetable*}
\movetableright=0.1in
\begin{deluxetable*}{llcccccc}
\tabletypesize{\footnotesize}
\tablecaption{The Multiscale Environments of Starbursts, QPSBs, and Comparison Samples \label{tab:del18_sb}}
\tablewidth{0pt}
\tablenum{1}
\tablehead{
\colhead{Sample} & \colhead{Measurement} & \colhead{Starburst} & \colhead{QPSB}  & \colhead{AGN H$\delta_A > 4$\,{\AA}} & \colhead{AGN H$\delta_A < 3$\,{\AA}} & \colhead{SFMS} & \colhead{QG}
}
\decimalcolnumbers
\startdata
 & $\log\,(1+\delta_\mathrm{0.5Mpc})$ & $-0.21\,(-0.42, 0.09)$ & $-0.31\,(-0.47, 0.03)$ & $-0.23\,(-0.41, 0.13)$ & $-0.02\,(-0.27, 0.32)$ & $-0.12\,(-0.33, 0.22)$ & $0.07\,(-0.19,  0.43)$\\
  & $\log\,(1+\delta_\mathrm{1Mpc})$ & $-0.23\,(-0.53, 0.15)$ & $-0.33\,(-0.59,  0.27)$ & $-0.27\,(-0.53, 0.23)$ & $-0.03\,(-0.36, 0.37)$ & $-0.13\,(-0.44,  0.28)$ & $0.07\,(-0.27, 0.49)$\\
& $\log\,(1+ \delta_\mathrm{2Mpc})$  & $-0.18\,(-0.62, 0.21)$ &  $-0.20\,(-0.61, 0.23)$ & $-0.28\,(-0.72, 0.21) $ &  $-0.03\,(-0.44,  0.38) $ & $-0.11\,(-0.54, 0.31)$ & $0.05\, (-0.35, 0.46) $\\
M11 & $\log\,(1+\delta_\mathrm{4Mpc})$  & $-0.14\,(-0.57, 0.24)$ & $-0.15\,(-0.47, 0.28)$ & $-0.19\,(-0.72, 0.25)$ & $-0.03\,(-0.43, 0.34)$ & $-0.08\,(-0.51,  0.30)$ & $0.04\,(-0.36,  0.40)$\\
& $\log\,(1+\delta_\mathrm{8Mpc})$  & $-0.11\,(-0.51, 0.24)$ & $-0.11\,(-0.41, 0.25)$ & $-0.13\,(-0.53, 0.29)$ & $-0.02\,(-0.38, 0.29)$ & $-0.06\,(-0.43, 0.27)$ & $0.03\,(-0.32, 0.33)$\\  
& $\log\,(1+\delta_5)$  & $-0.19\,(-0.69, 0.39)$ & $-0.26\,(-0.64, 0.42)$ & $-0.24\,(-0.77, 0.48)$ & $-0.05\,(-0.59, 0.67)$ & $-0.13\,(-0.64, 0.49)$ & $0.08\,(-0.48, 0.86)$\\  
& $\log\,Q_5$  & $-4.05\,(-5.30, -2.31)$ & $-4.24\,(-5.33, -2.78)$ & $-4.04\,(-5.39,-2.19)$ & $-3.59\,(-4.97, -1.83)$ & $-3.85\,(-5.09, -2.24)$ & $-3.41\,(-4.97, -1.77)$\\  
\hline
 & $\log\,(1+\delta_\mathrm{0.5Mpc})$ & $-0.22\,(-0.52, 0.26)$ & $-0.26\,(-0.54, 0.23)$ & $-0.16\,(-0.40, 0.30)$ & $-0.02\,(-0.31, 0.51)$ & $-0.11\,(-0.39, 0.39)$ & $0.11\,(-0.20,  0.72)$\\
  & $\log\,(1+\delta_\mathrm{1Mpc})$ & $-0.23\,(-0.67, 0.36)$ & $-0.26\,(-0.72,  0.33)$ & $-0.21\,(-0.56, 0.35)$ & $-0.04\,(-0.48, 0.51)$ & $-0.13\,(-0.57,  0.44)$ & $0.13\,(-0.37, 0.71)$\\
& $\log\,(1+ \delta_\mathrm{2Mpc})$  & $-0.18\,(-0.83, 0.32)$ &  $-0.21\,(-0.81, 0.31)$ & $-0.17\,(-0.75, 0.32) $ &  $-0.03\,(-0.58,  0.45) $ & $-0.10\,(-0.67, 0.40)$ & $0.10\, (-0.46, 0.59)$\\
M105 & $\log\,(1+\delta_\mathrm{4Mpc})$  & $-0.13\,(-0.68, 0.29)$ & $-0.15\,(-0.63, 0.28)$ & $-0.11\,(-0.70, 0.30)$ & $-0.03\,(-0.50, 0.38)$ & $-0.07\,(-0.56, 0.34)$ & $0.07\,(-0.41, 0.47)$\\
& $\log\,(1+\delta_\mathrm{8Mpc})$  & $-0.08\,(-0.50, 0.26)$ & $-0.10\,(-0.52, 0.25)$ & $-0.06\,(-0.52, 0.26)$ & $-0.02\,(-0.40, 0.31)$ & $-0.04\,(-0.43, 0.29)$ & $0.04\,(-0.34,  0.37)$\\  
& $\log\,(1+\delta_5)$  & $-0.15\,(-0.65, 0.42)$ & $-0.18\,(-0.64, 0.45)$ & $-0.14\,(-0.70,  0.46)$ & $-0.05\,(-0.60,  0.67)$ & $-0.11\,(-0.63, 0.53)$ & $0.12\,(-0.50, 0.96)$\\  
& $\log\,Q_5$  & $-4.14\,(-5.39, 2.40)$ & $-4.44\,(-5.57,-2.85)$ & $-4.25\,(-5.44, -2.61)$ & $-3.78\,(-5.17, -2.10)$ & $-3.86\,(-5.10, -2.28)$ & $-3.63\,(-5.18, -1.99)$\\  
\hline
 & $\log\,(1+\delta_\mathrm{0.5Mpc})$ & $-0.18\,(-0.59, 0.56)$ & $-0.05\,(-0.51, 0.66)$ & $-0.07\,(-0.46, 0.62)$ & $0.03\,(-0.44, 0.80)$ & $-0.11\,(-0.53, 0.66)$ & $0.39\,(-0.28, 1.15)$\\
 & $\log\,(1+\delta_\mathrm{1Mpc})$ & $-0.20\,(-0.92, 0.55)$ & $-0.11\,(-0.86,  0.55)$ & $-0.11\,(-0.73, 0.59)$ & $-0.02\,(-0.76, 0.61)$ & $-0.12\,(-0.84,  0.53)$ & $0.29\,(-0.58, 0.93)$\\
& $\log\,(1+ \delta_\mathrm{2Mpc})$  & $-0.13\,(-0.95, 0.42)$ &  $-0.10\,(-0.80, 0.38)$ & $-0.03\,(-0.84, 0.50) $ & $-0.02\,(-0.64, 0.53) $ & $-0.09\,(-0.76, 0.45)$ & $0.20\, (-0.48, 0.73)$\\
M10 & $\log\,(1+\delta_\mathrm{4Mpc})$  & $-0.09\,(-0.68, 0.33)$ & $-0.07\,(-0.62, 0.34)$ & $-0.05\,(-0.55, 0.37)$ & $-0.03\,(-0.54, 0.43)$ & $-0.06\,(-0.57, 0.37)$ & $0.13\,(-0.39, 0.55)$\\
& $\log\,(1+\delta_\mathrm{8Mpc})$  & $-0.06\,(-0.50, 0.28)$ & $-0.02\,(-0.50, 0.30)$ & $0.00\,(-0.39, 0.31)$ & $-0.02\,(-0.39, 0.35)$ & $-0.04\,(-0.44, 0.30)$ & $0.08\,(-0.31, 0.42)$\\  
& $\log\,(1+\delta_5)$  & $-0.11\,(-0.63, 0.47)$ & $-0.12\,(-0.66, 0.61)$ & $-0.05\,(-0.59, 0.68)$ & $0.00\,(-0.61, 0.78)$ & $-0.10\,(-0.64, 0.57)$ & $0.29\,(-0.47, 1.20)$\\  
& $\log\,Q_5$  & $-3.98\,(-5.18, -2.10)$ & $-4.21\,(-5.68, -2.46)$ & $-4.06\,(-5.39, -2.54)$ & $-3.85\,(-5.29, -2.04)$ & $-3.64\,(-4.92, -2.05)$ & $-3.40\,(-5.17, -1.77)$\\  
\hline
 & $\log\,(1+\delta_\mathrm{0.5Mpc})$ & $-0.19\,(-0.68, 0.91)$ & $0.20\,(-0.54, 1.27)$ & $0.22\,(-0.62, 0.74)$ & $0.01\,(-0.73, 0.80)$ & $-0.14\,(-0.76, 0.90)$ & $0.77\,(-0.38, 1.43)$\\
 & $\log\,(1+\delta_\mathrm{1Mpc})$ & $-0.20\,(-1.29, 0.58)$ & $0.07\,(-0.83,  0.84)$ & $-0.14\,(-1.03, 0.45)$ & $-0.11\,(-1.02, 0.60)$ & $-0.10\,(-1.13,  0.62)$ & $0.53\,(-0.33, 1.13)$\\
& $\log\,(1+ \delta_\mathrm{2Mpc})$  & $-0.18\,,(-0.92, 0.44)$ & $0.06\,(-0.39, 0.61)$ & $0.02\,(-0.63, 0.45) $ & $-0.09\,(-0.78, 0.44) $ & $-0.06\,(-0.79, 0.49)$ & $0.36\, (-0.27, 0.87)$\\
M95 & $\log\,(1+\delta_\mathrm{4Mpc})$  & $-0.13\,(-0.65, 0.35)$ & $0.04\,(-0.41, 0.42)$ & $0.01\,(-0.48, 0.32)$ & $-0.01\,(-0.60, 0.37)$ & $-0.04\,(-0.56, 0.39)$ & $0.23\,(-0.24, 0.64)$\\
& $\log\,(1+\delta_\mathrm{8Mpc})$  & $-0.10\,(-0.48, 0.27)$ & $0.02\,(-0.28, 0.35)$ & $0.02\,(-0.32, 0.35)$ & $-0.02\,(-0.36, 0.27)$ & $-0.02\,(-0.42, 0.31)$ & $0.15\,(-0.20, 0.48)$\\  
& $\log\,(1+\delta_5)$  & $-0.15\,(-0.65, 0.45)$ & $0.09\,(-0.47, 0.94)$ & $-0.06\,(-0.52  0.69)$ & $-0.02\,(-0.65, 0.78)$ & $-0.07\,(-0.64, 0.61)$ & $0.68\,(-0.26, 1.44)$\\  
& $\log\,Q_5$  & $-3.78\,(-5.12, -2.00)$ & $-3.44\,(-4.76, -1.70)$ & $-3.38\,(-4.74, -1.77)$ & $-3.21\,(-4.72, -1.61)$ & $-2.67\,(-4.57, -1.30)$ & $-2.64\,(-4.78, -1.18)$\\  
\enddata
\tablecomments{Column (1) : the $M_\star$ ranges of the samples binned by 0.5\,dex. The names of the samples indicate the minima of the $\log\, M_\star$ ranges. For example, M11 denotes $\log M_\star/M_\odot = 11-11.5$. Column (2) : gives the measurements of mass overdensities with radii of \{0.5,1, 2, 4, 8\}\,$h^{-1}$Mpc, the 5th nearest neighbor overdensity, and the tidal parameter due to the five nearest neighbors. We use the notation $X\,(Y, Z)$ to denote $X$ = median (50\%), $Y=$15\%, and $Z=85$\% of a distribution. The results for additional comparison samples are given in Appendix~\ref{sec:appB}.}
\end{deluxetable*}
\end{rotatetable*}

As expected from the clear trends in Figure~\ref{fig:del18_sb} and Figure~\ref{fig:massdel18}, in almost all cases the AD test rejects the null hypothesis that the multiscale overdensities of starbursts for each $M_\star$ range are the same as those of the other samples ($p < .001$). The exceptions being that the distributions of $\deltafour$ and $\deltaeight$ of starbursts and upper SFMS galaxies in $\log M_\star/M_\odot = 11-11.5$ are similar ($p=.01$ for $\deltaeight$ and $p=.1$ for $\deltafour$). Furthermore, the AD test indicates that the mass overdensity distributions of upper and lower SFMS galaxies are different at all scales and for all $M_\star$ ranges ($p < .001$). Therefore, SFGs do not scatter randomly above and below SFMS.

\begin{figure}[hbt]
\fig{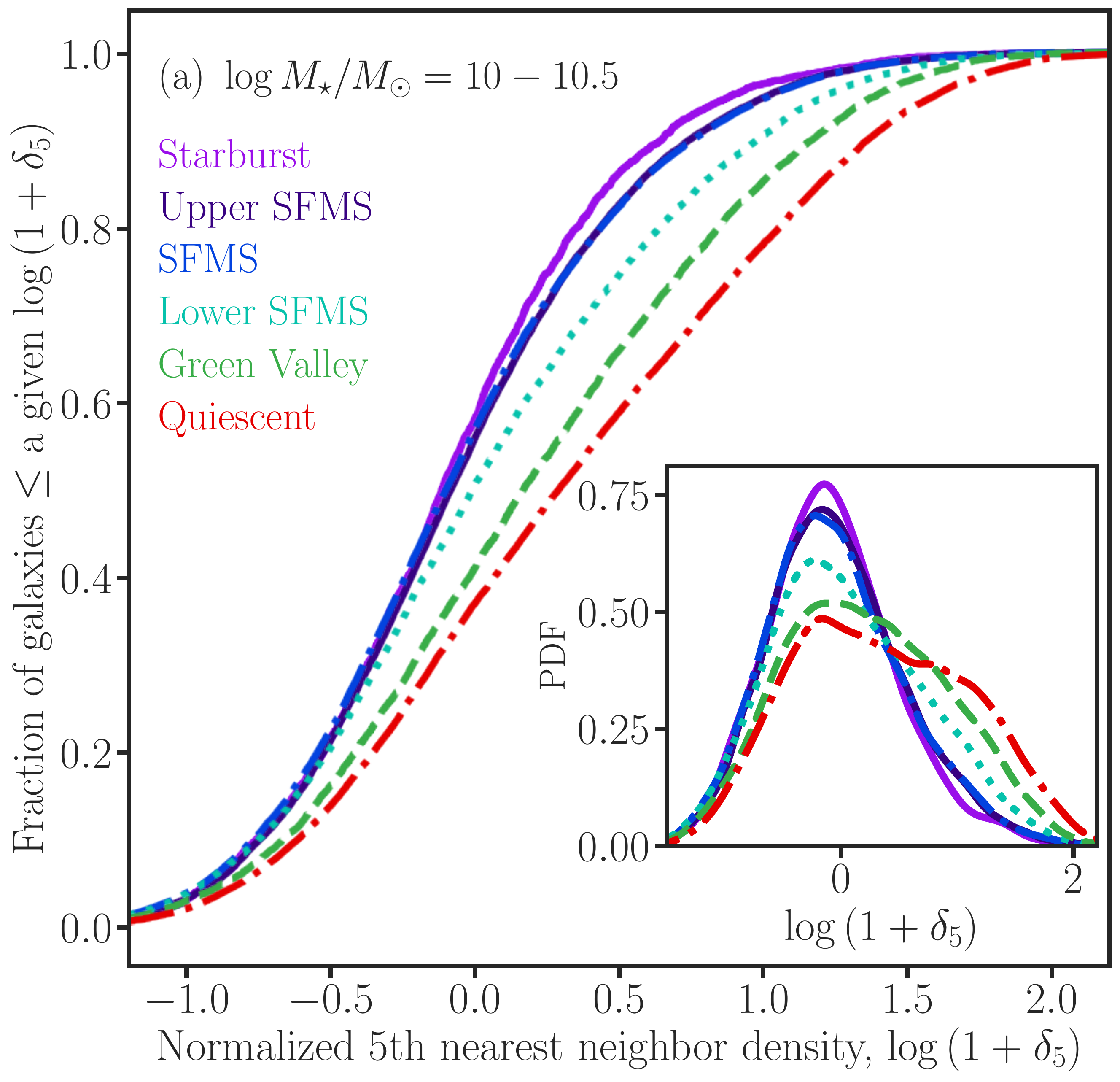}{0.5\textwidth}{}
\fig{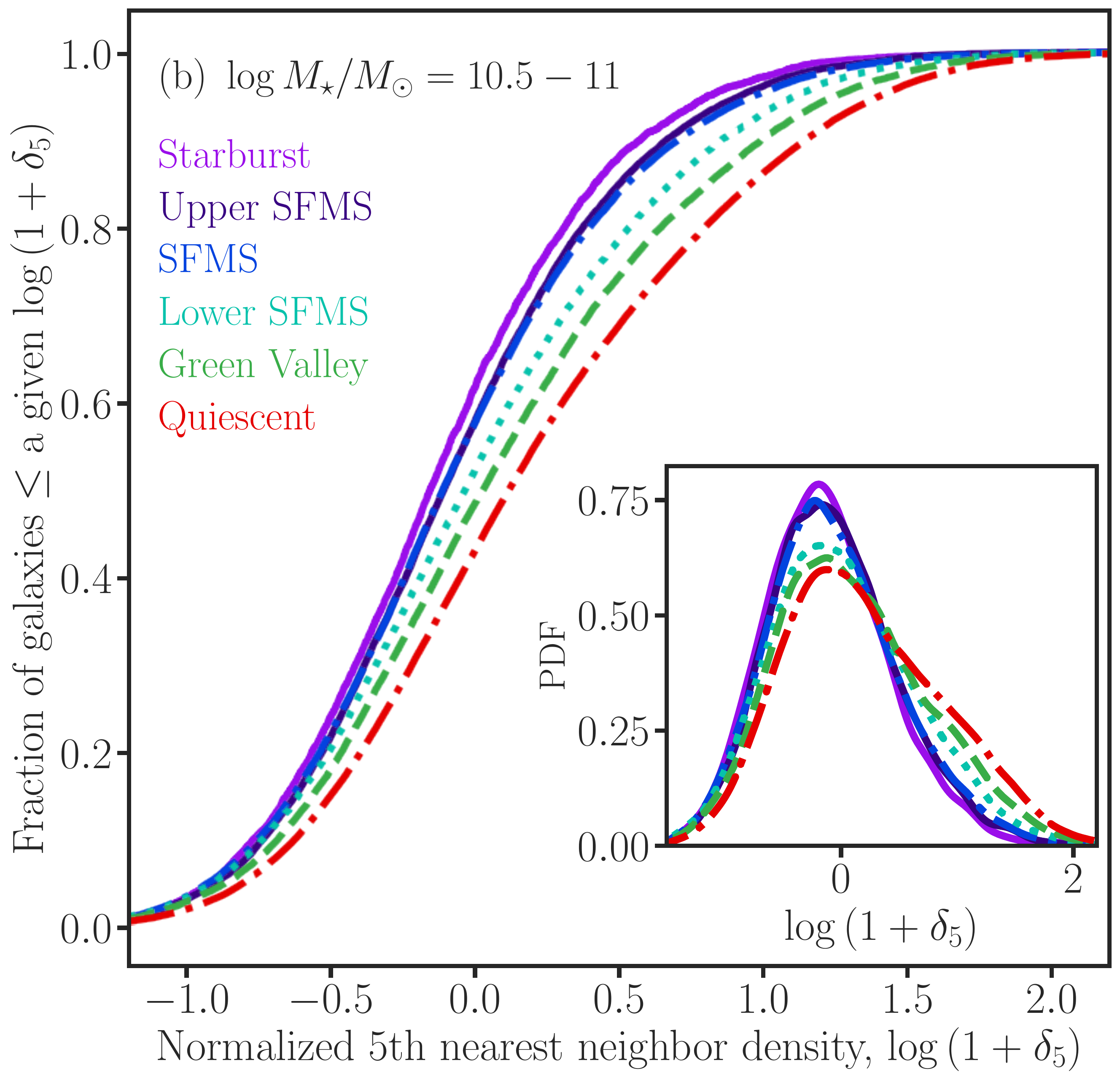}{0.5\textwidth}{}
\caption{Similar to Figure~\ref{fig:del18_sb}, but here the cumulative distributions of the normalized 5th nearest neighbor number densities ($\delta_5$) are shown. The $\delta_5$ of starbursts are slightly lower than those of normal SFGs, but they are much lower than those of green-valley galaxies and QGs.\label{fig:del5_sb}}
\end{figure}

The overdensity of galaxy counts within the fifth nearest neighbor distances ($\delta_5$) of the samples divided by $\delSSFR$ give similar results as fixed aperture mass overdensities (Figure~\ref{fig:del5_sb} and Table~\ref{tab:del18_sb}). The median and the dispersion of $\delta_5$ increase as $\delSSFR$ decreases. Starbursts have the lowest $\delta_5$, while QGs have the highest $\delta_5$. The AD test confirms that the differences in $\delta_5$ distributions shown in Figure~\ref{fig:del5_sb} are significant ($p < .001$). In general, the $\delta_5$ distributions of starbursts are significantly different from the distributions of the SFMS galaxies at all $M_\star$ ranges ($p \lesssim .01$). Likewise, the $\delta_5$ distributions of upper SFMS and lower SFMS galaxies are different for all $M_\star$ ranges ($p < .001$). The trends presented in Figure~\ref{fig:del5_sb} are similar if $\delta_3$ is used instead.

\begin{figure*}[hbt!]
\gridline{\fig{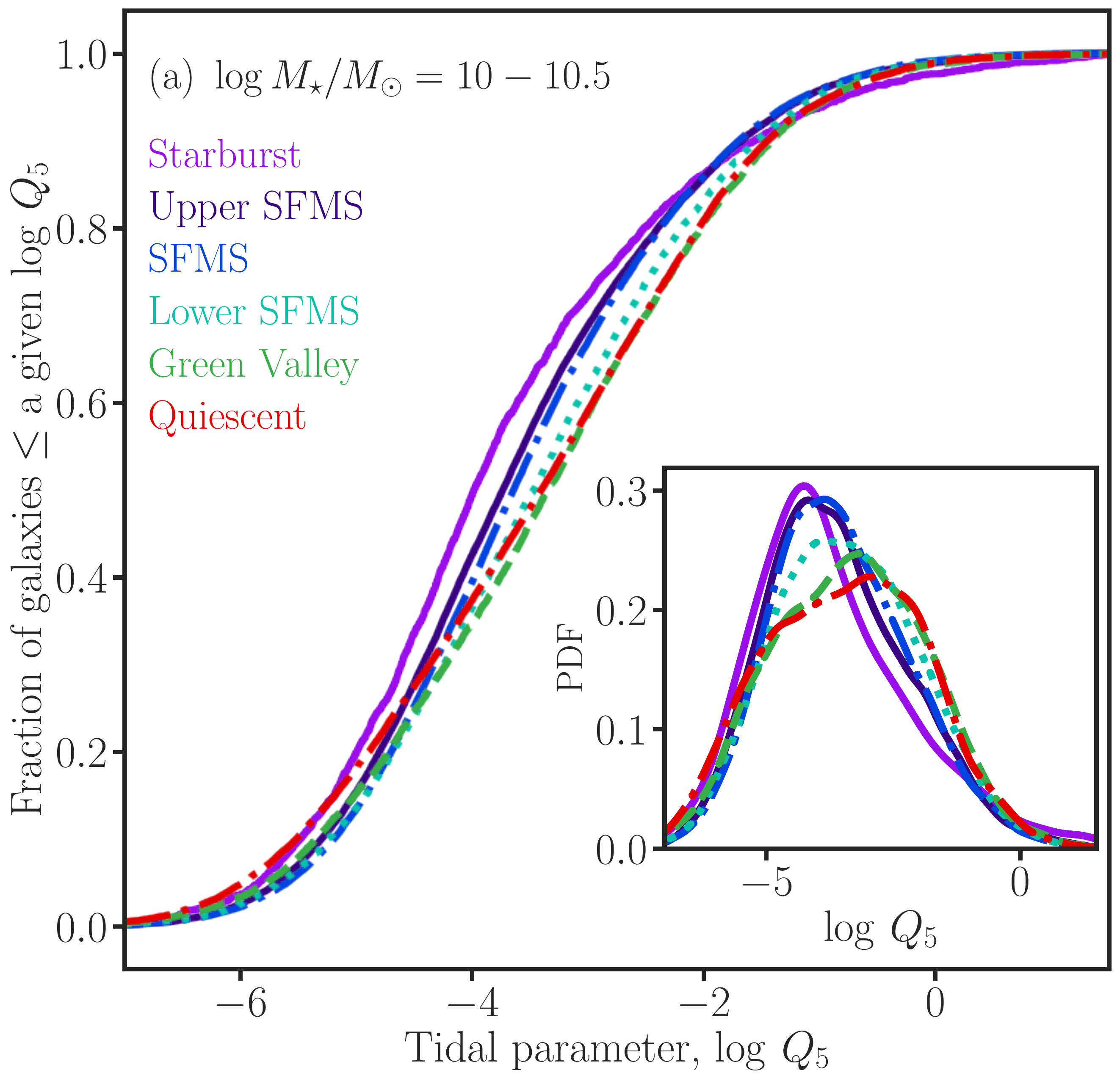}{0.49\textwidth}{}
          \fig{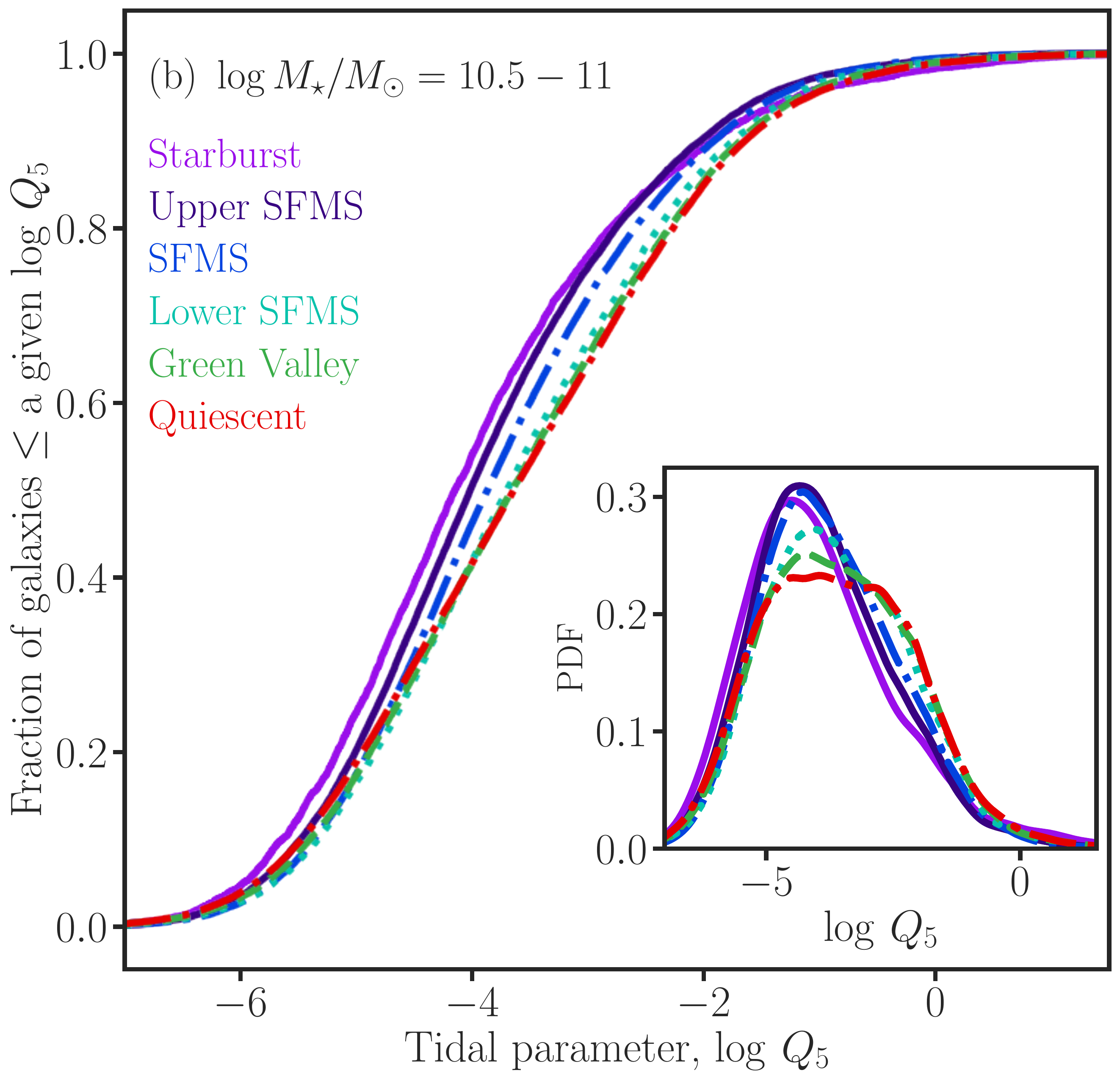}{0.49\textwidth}{}}
\gridline{\fig{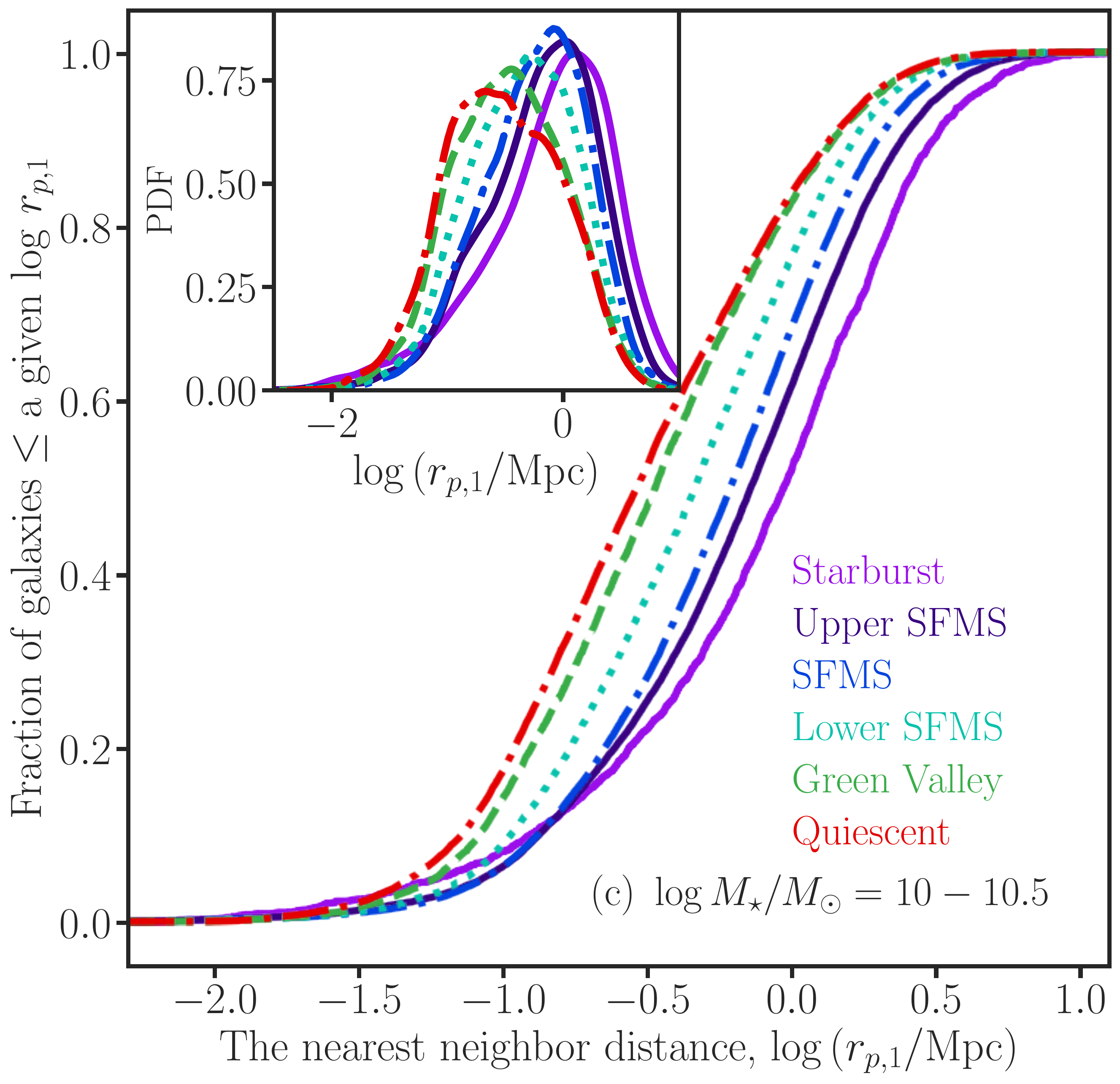}{0.49\textwidth}{}
          \fig{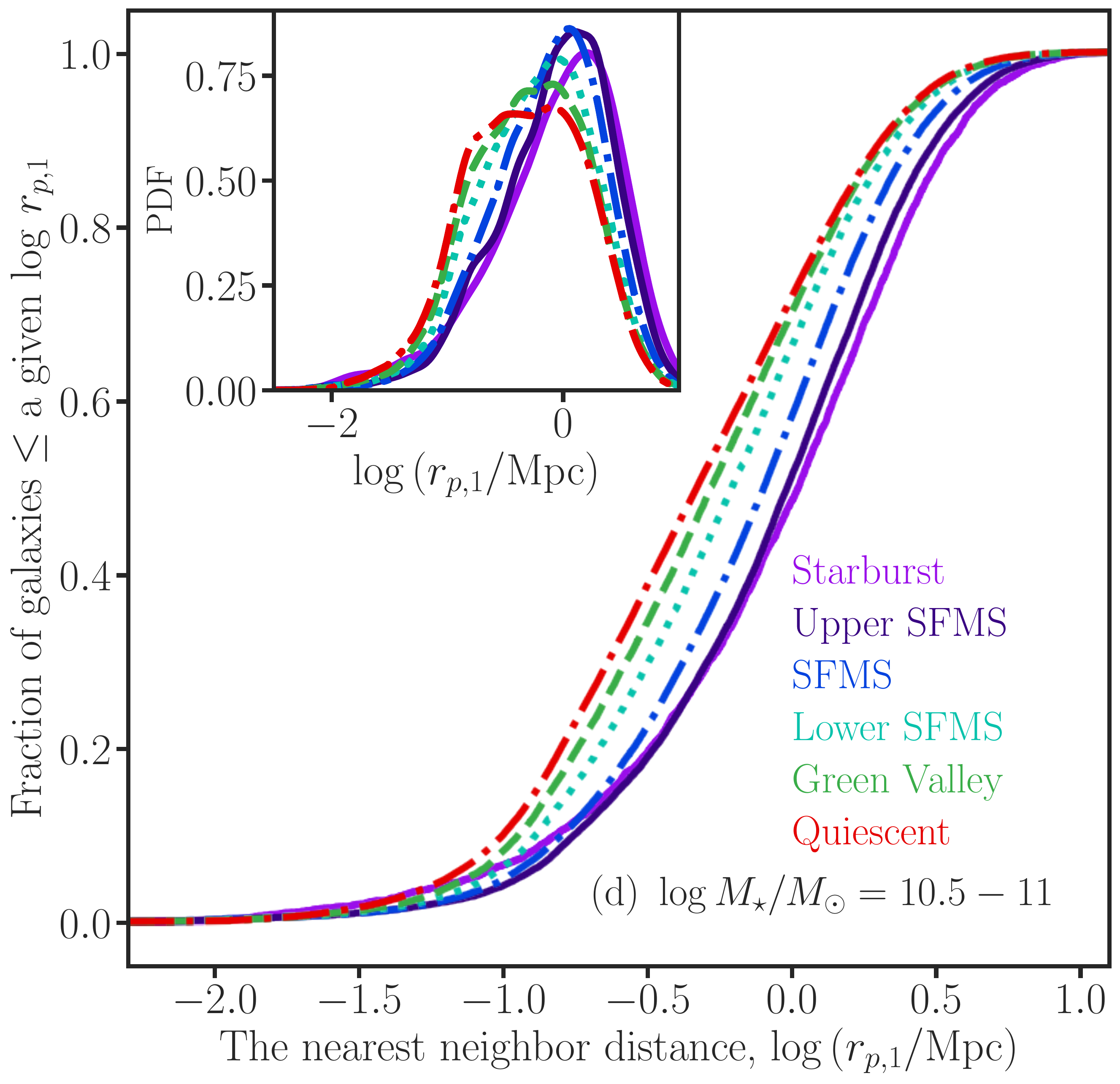}{0.49\textwidth}{}}
\caption{Similar to Figure~\ref{fig:del5_sb}, but here the cumulative distributions of the tidal parameter of the five nearest neighbors ($Q_5$) and the projected distance to the nearest neighbor ($r_{p,1}$) are shown. The $Q_5$ and $r_{p,1}$ of starbursts are slightly lower than to those of SFMS galaxies, and are much lower than those of green-valley galaxies and QGs. Note that $\sim 10\%-15\%$ of the starbursts have $Q_5 > -2$ or nearby neighbors within $r_{p,1} < 0.1-0.2$\,Mpc, with only $\sim 2\%-3\%$ enhancement compared to SFMS galaxies. About $55\%-60\%$ of $Q_5$ is contributed by the nearest neighbor. \label{fig:Q5_sb}}
\end{figure*}

Furthermore, Figure~\ref{fig:Q5_sb} compares the ECDFs of the tidal parameter $Q_5$ and the nearest neighbor distance $r_{p,1}$ of starbursts and the other samples grouped by $\delSSFR$ for two $M_\star$ ranges: $\Mlow$ and $\Mhigh$. The two environmental indicators are strongly anticorrelated ($\rho=-0.9$); the first nearest neighbors contributes about $55\%-60\%$ to the total $Q_5$ values. Again, starbursts have the lower $Q_5$ or the higher $r_{p,1}$ than those of the other samples (see also Table~\ref{tab:del18_sb}). The distributions of $r_{p,1}$  and $Q_5$ of starbursts are significantly different (AD test $p <.001$) from all of the other samples at all $M_\star$ ranges. In most cases, the ECDFs of $Q_5$ and $r_{p,1}$ of various samples are significantly different from each other. In other words, $\delSSFR$ and $r_{p,1}$ show a weak but significant trend ($\rho \approx 0.2$), in the sense that the median nearest neighbor distances decrease from starbursts to SFMS galaxies to QGs. Only $15$\% (8\%) of the starbursts with $\Mlow$ have nearby neighbors within $r_{p,1} < 0.2$\,Mpc ($< 0.1$\,Mpc). The difference in percentage between the nearby neighbors of starbursts and SFMS galaxies is only $\sim 2\%-3$\%. Therefore, we do not find compelling evidence that most starbursts are triggered by interactions with their nearest neighbors. 

\begin{figure*}[ht!]
\gridline{\fig{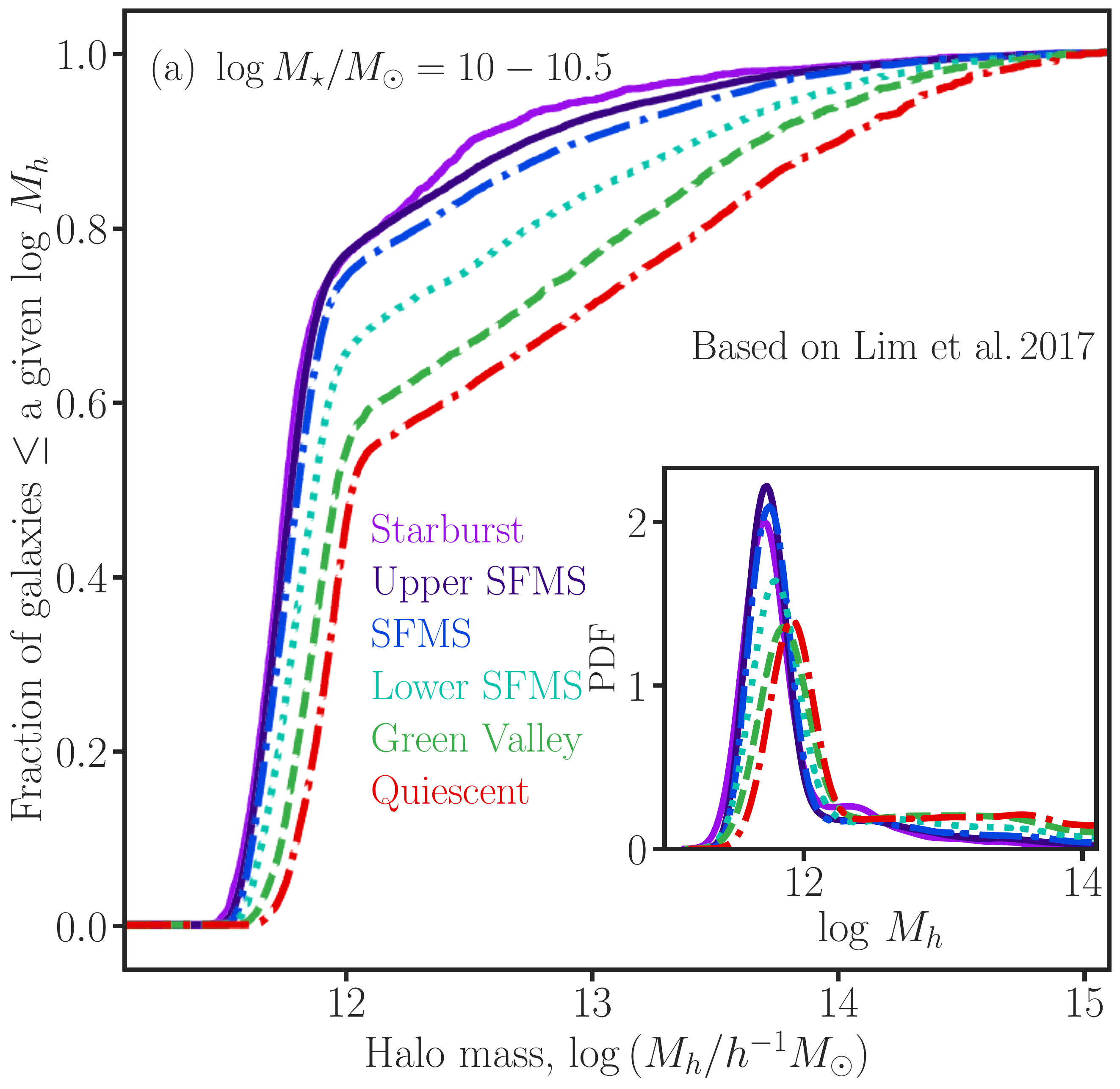}{0.48\textwidth}{}
          \fig{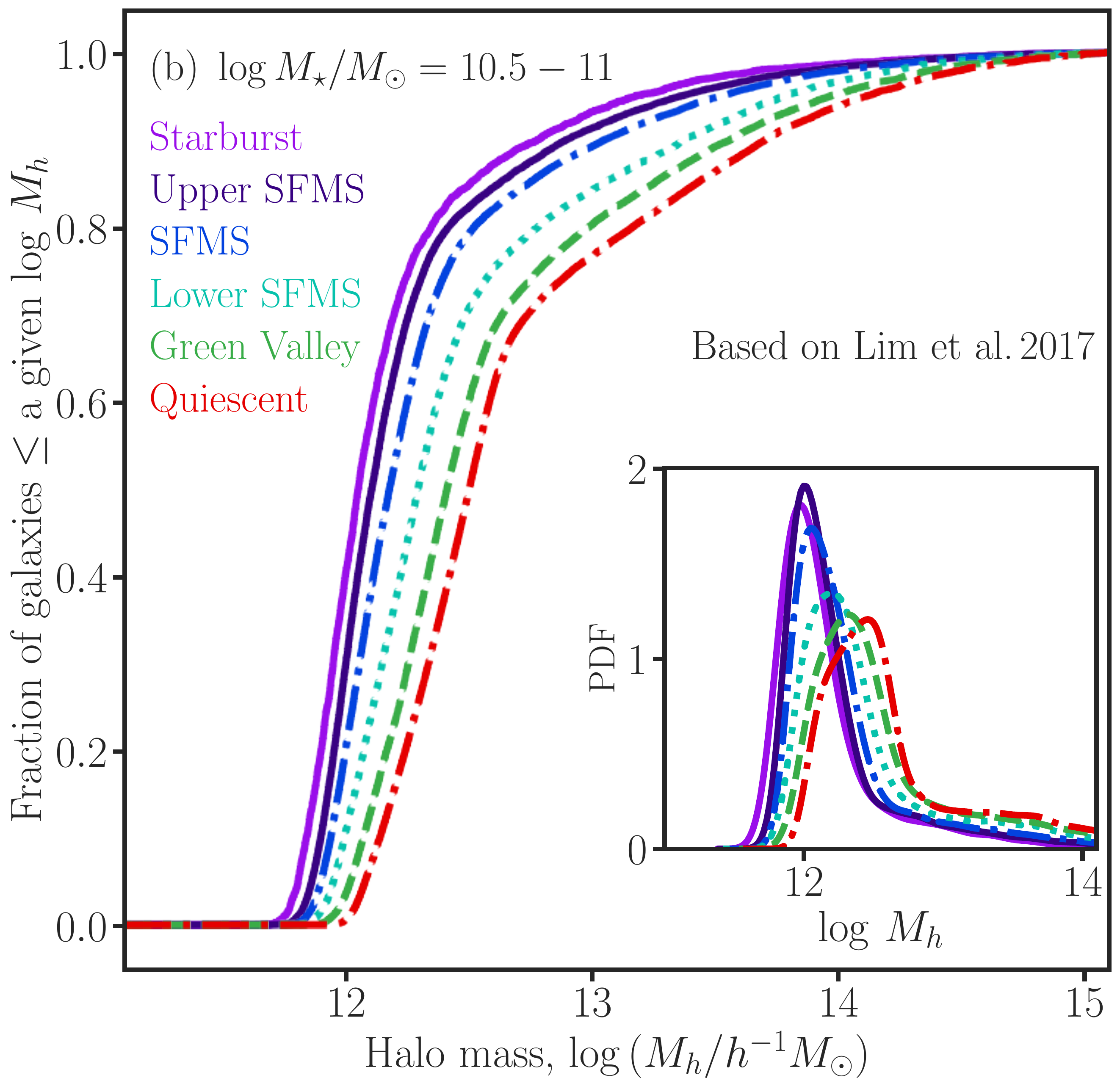}{0.48\textwidth}{}}
\gridline{\fig{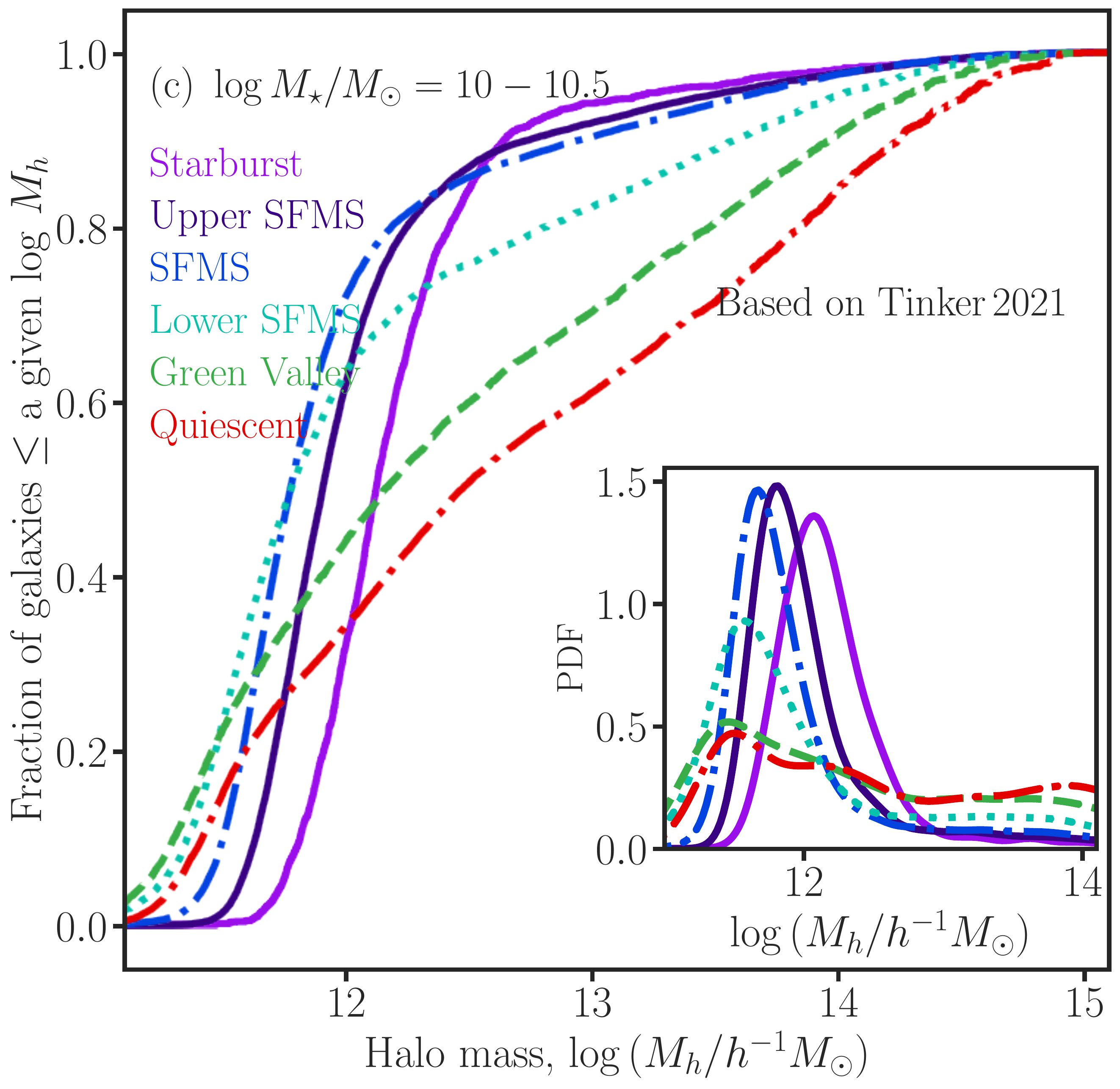}{0.48\textwidth}{}
          \fig{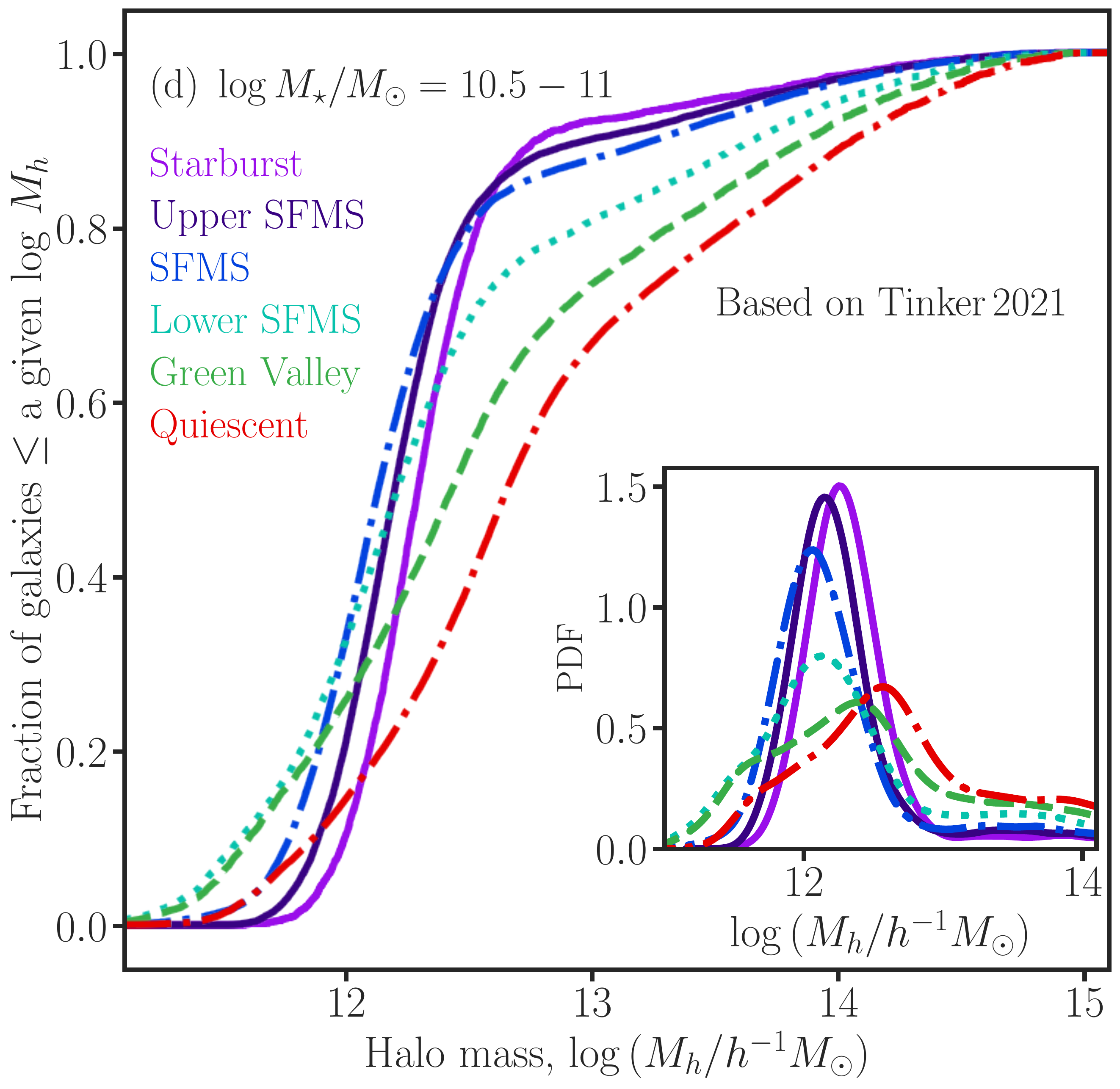}{0.48\textwidth}{}}
\caption{The cumulative distributions of halo mass ($M_h$) estimates from \citet[][panels (a) and (b)]{Lim+17} and \citet[][panels (c) and (d)]{Tinker21} for galaxies grouped by $M_\star$ and $\delSSFR$. Both $M_h$ estimates indicate that $\sim 90\%$ of starbursts have $M_h < 10^{13}\,h^{-1}M_\odot$, which is a significantly higher percentage than that of QGs below this halo mass. The two estimates however disagree on the trends of $M_h$ with $\delSSFR$ below $M_h \lesssim 3 \times 10^{12}\,h^{-1}M_\odot$. \label{fig:Mh_sb}}
\end{figure*}

Figure~\ref{fig:Mh_sb} compares the ECDFs of $M_h$ estimated by \citet{Lim+17} and \citet{Tinker21}, again dividing the samples by $M_\star$ and $\delSSFR$. Both estimates broadly agree that starbursts have lower $M_h$ than QGs. However, below few times $10^{12}\, M_\odot$, \citet{Tinker21}'s estimates indicate that starbursts have high $M_h$ among SFGs and that $M_h$ increases with $\delSSFR$, while \citet{Lim+17}'s estimates indicate the opposite. For the purpose of this work, the difference between the two $M_h$ estimates is not so important. It suggests that starbursts are useful in discriminating between different methods of connecting observed galaxies with simulated dark matter haloes. Simulators should also consider comparing their predictions with less model-dependent environmental indicators such as $\deltahalf$ and $\deltaone$.

\begin{figure*}[hbt!]
\includegraphics[width=0.5\textwidth]{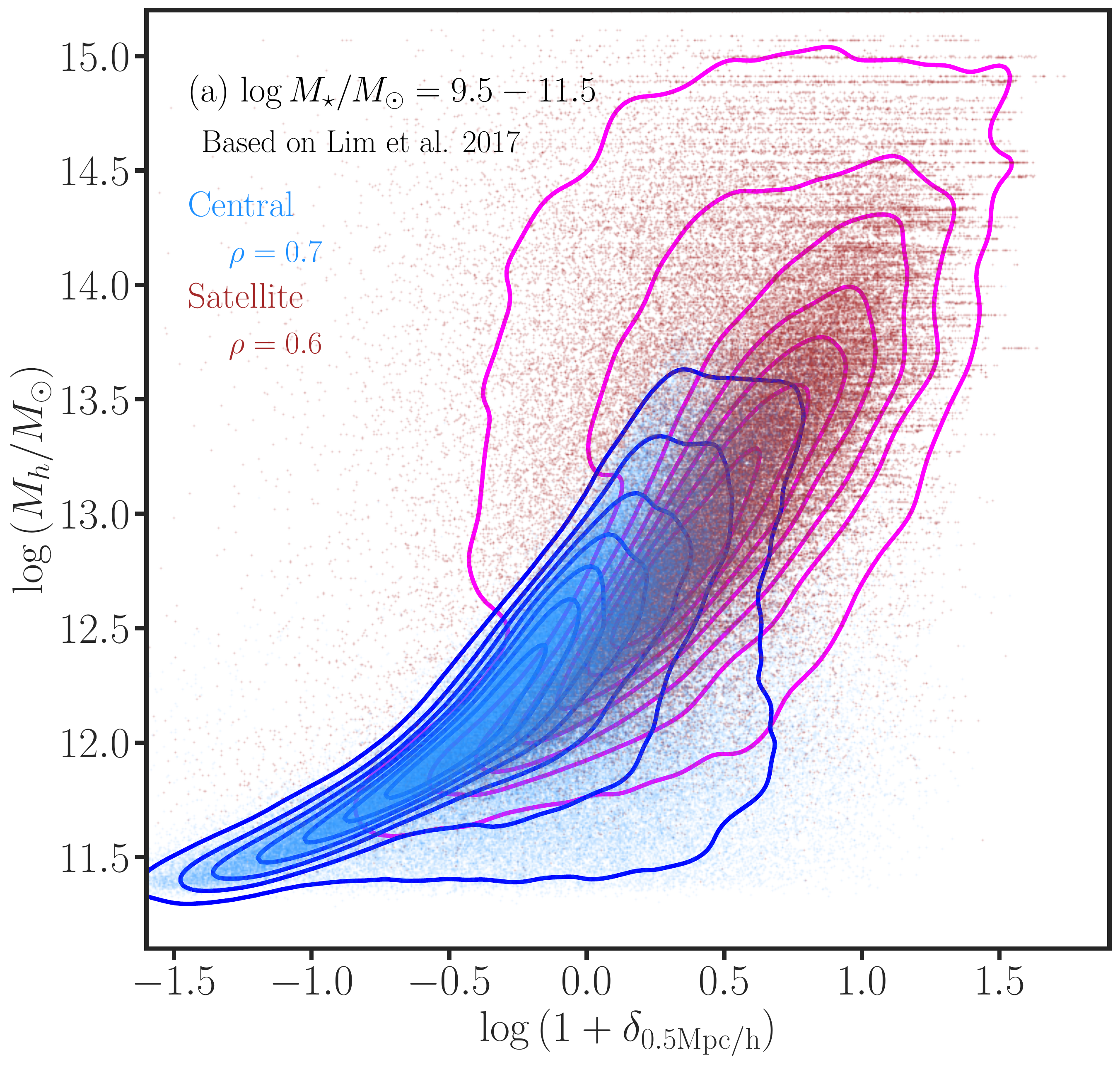}
\includegraphics[width=0.5\textwidth]{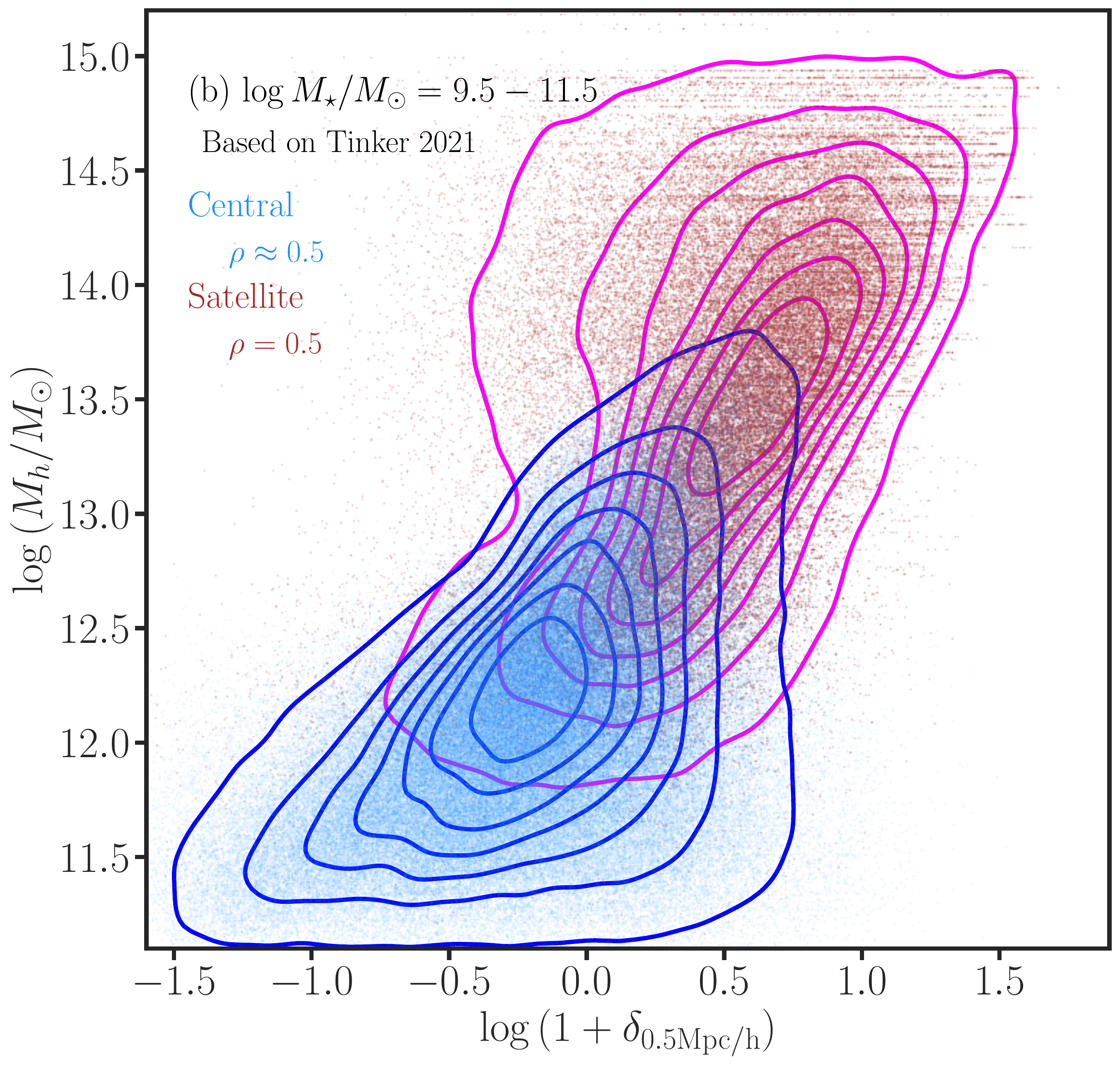}
\caption{The correlations between the stellar mass overdensity $\deltahalf$ and halo mass $M_h$ for satellites and centrals. Panel (a) uses \citet{Lim+17}'s measurements of $M_h$ and central/satellite classification, whereas panel (b) uses \citet{Tinker21}'s measurements. The $\rho$ values are the Spearman correlation coefficients. \label{fig:Sig05Mh_sat}}
\end{figure*}

\begin{figure*}[hbt!]
\gridline{\fig{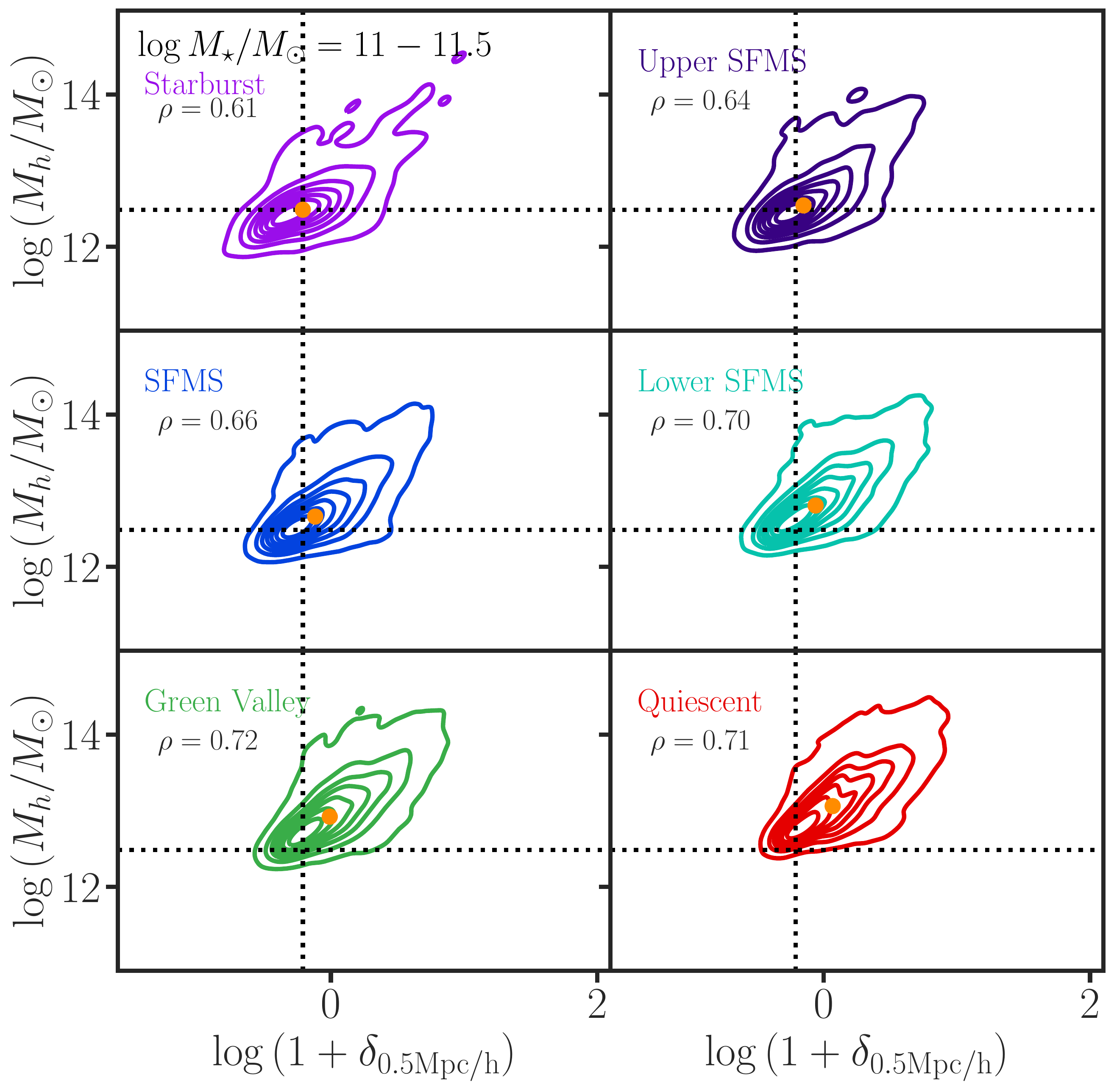}{0.48\textwidth}{}
\fig{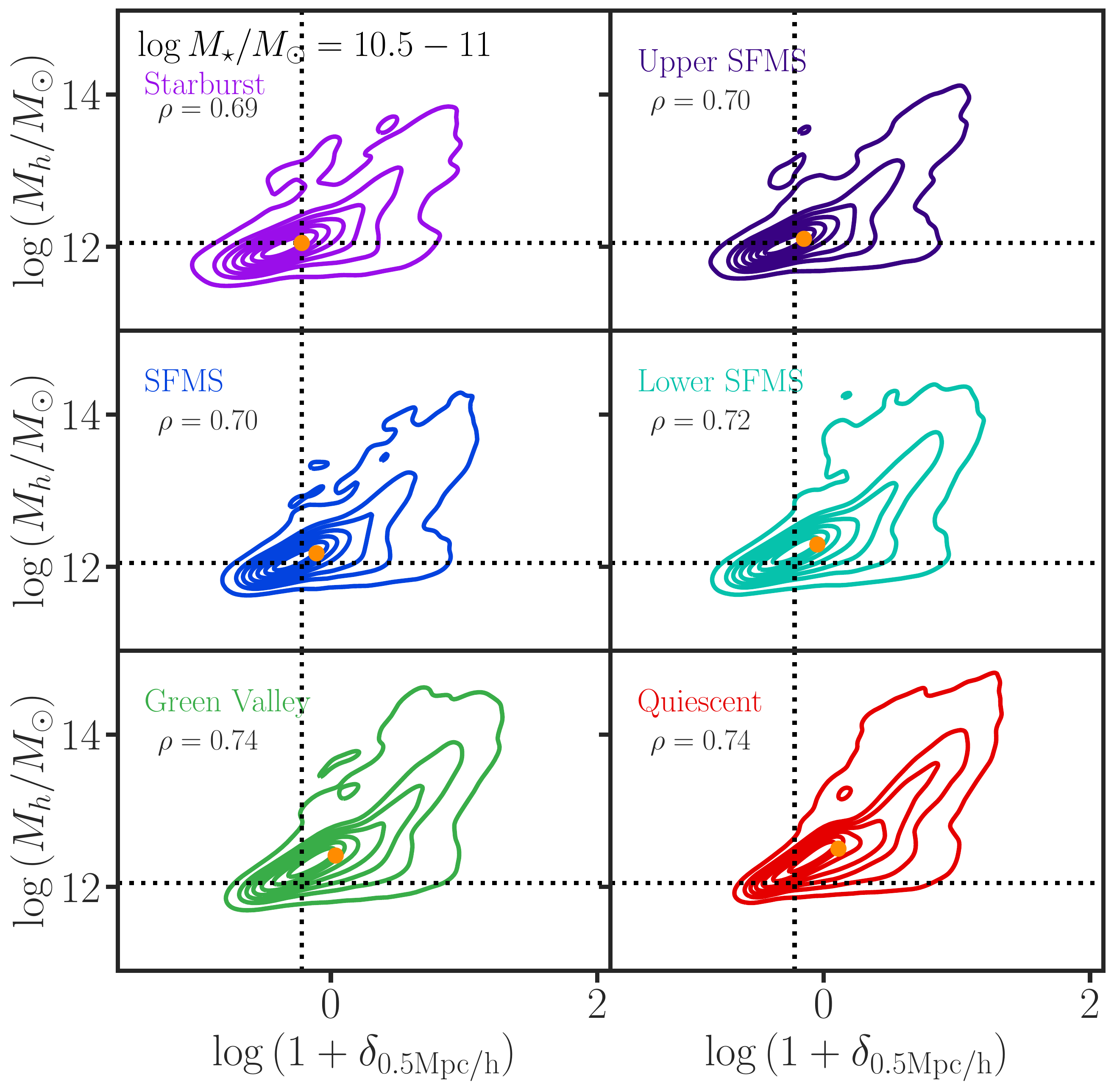}{0.48\textwidth}{}}
\gridline{\fig{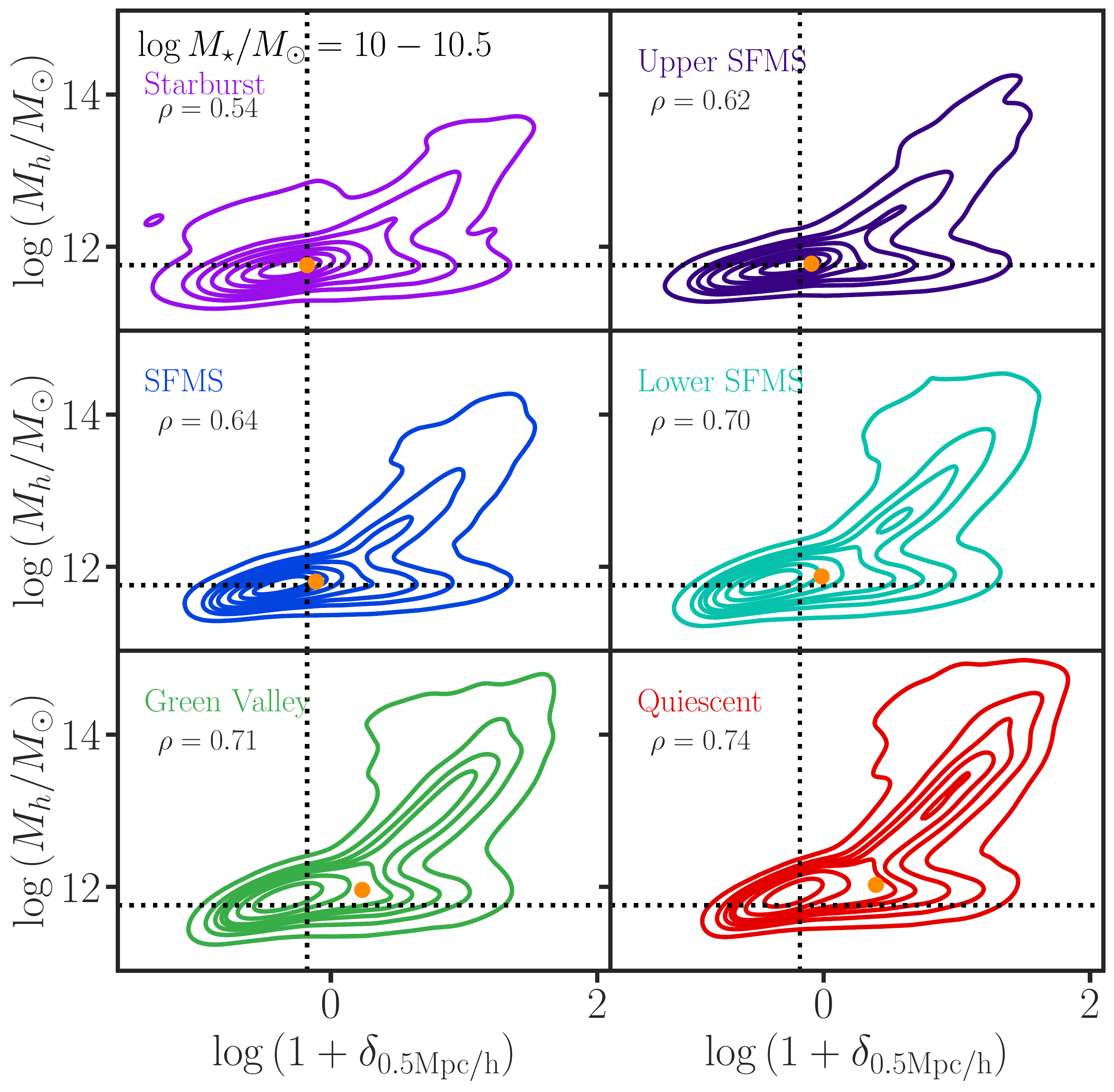}{0.48\textwidth}{}
\fig{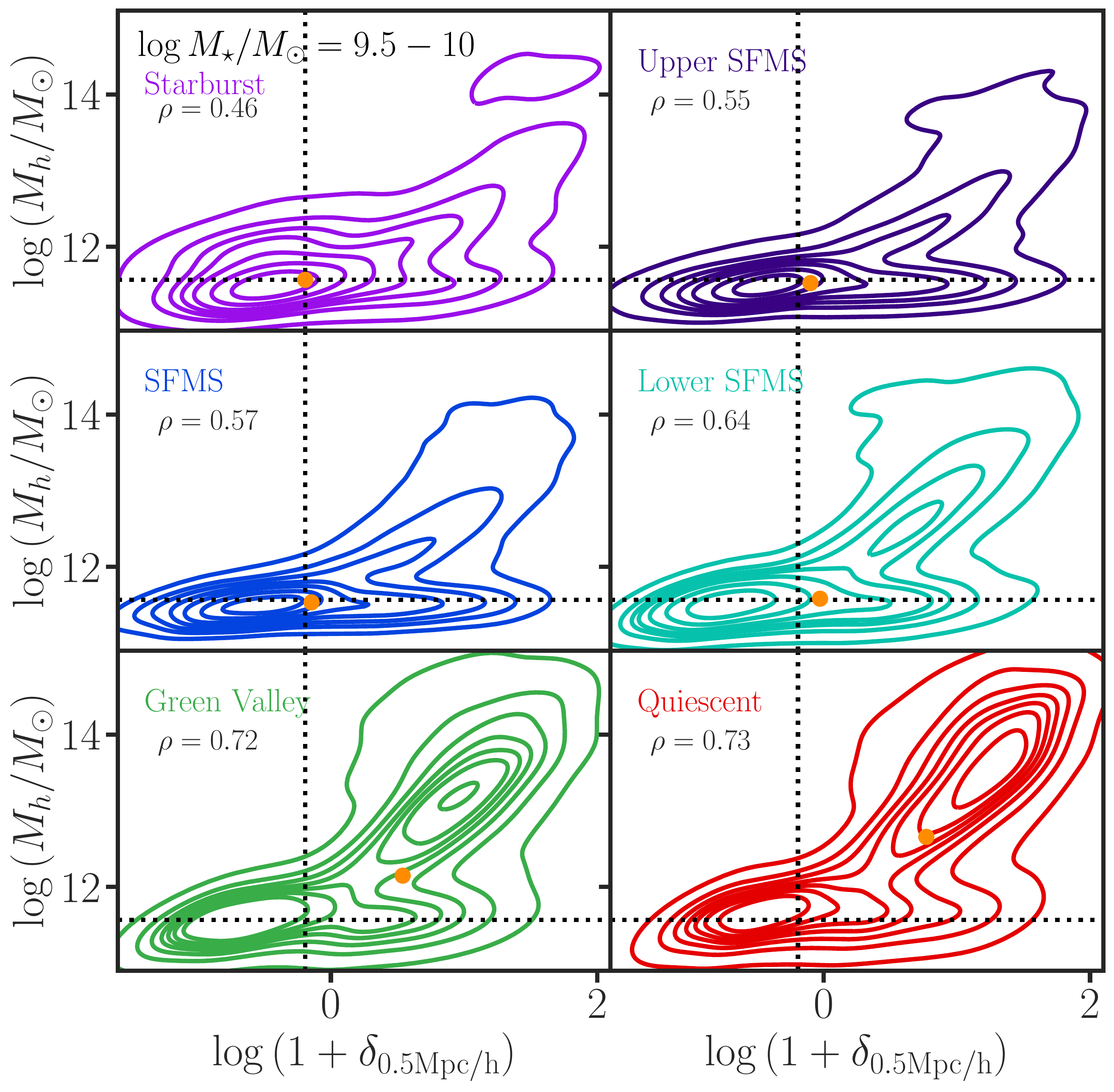}{0.48\textwidth}{}}
\caption{The correlations between the stellar mass overdensity $\deltahalf$ and halo mass $M_h$ \citep{Lim+17} as a function of $M_\star$ and $\delSSFR$. The $\rho$ value in each panel is the Spearman correlation coefficient for the indicated $M_\star$ range and $\delSSFR$ class. For guidance, the dashed lines mark the median values of $\deltahalf$ and $M_h$ of starbursts; they are same for all panels in a given $M_\star$ range. The orange point denotes the median values of $\deltahalf$ and $M_h$ for a given panel. Generally, $\deltahalf$ shows a moderate ($\rho \approx 0.4-0.7$) correlation with $M_h$. The strength of the correlation decreases as $M_\star$ increases or as $\delSSFR$ increases from QGs to starbursts. The contours represent number density of galaxies. The plume of points on the upper right corner of the panels are mainly satellites. Centrals occupy the horizontal contours at low $M_h$. Their distribution shifts leftward and downward as $M_\star$ decreases, as expected from $M_\star-M_h$ relation. \label{fig:Sig05Mh}}
\end{figure*}

The stellar mass overdensities significantly correlate with \citet{Lim+17}'s $M_h$; the smaller the scale the overdensities probe, the stronger the correlation (see Appendix~\ref{sec:appC}). For example, the correlation between $\deltahalf$ and $M_h$ for the whole sample has $\rho \approx 0.8$ and that of $\deltaeight$ and $M_h$ has $\rho \approx 0.3$. Furthermore, the nature of the relationship between $\deltahalf$ and $M_h$ (i.e., its shape, scatter, and $\rho$) also depend subtly on satellite/central classification, $\delSSFR$, and $M_\star$ (Figure~\ref{fig:Sig05Mh_sat} \&~\ref{fig:Sig05Mh}). Generally, $\rho$ decreases as $M_\star$ or $\delSSFR$ increases. For a given $M_\star$ range, the correlation for QGs has $\rho \approx 0.7$, while that of the SFGs has $\rho \approx 0.5-0.7$. In particular, starbursts with with $M_\star < 3 \times 10^{10}\,M_\odot$ have the lowest $\rho \approx 0.5$ among SFG subsamples. The bimodal distributions of galaxies in Figure~\ref{fig:Sig05Mh} change with $M_\star$ due to changing satellite fractions; the plume of points in the upper corner of the panels (for $M_\star < 10^{11}\,M_\odot$) are mainly satellites. Centrals occupy the almost horizontal cluster of points at low $M_h$, which marches downward and leftward in Figure~\ref{fig:Sig05Mh} as $M_\star$ decreases, as expected from the $M_\star-M_h$ relation. In short, $\deltahalf$ and $\deltaone$ are decent proxies for $M_h$.

\begin{deluxetable*}{ccccccc}
\tablenum{2}
\tablecaption{The Mean Central Fractions and Median (15\%, 85\%) Halo Masses of Starbursts, QPSBs, and Comparison Samples.  \label{tab:fcentMh_sb}}
\tablewidth{0pt}
\tablehead{
\colhead{Sample} & \colhead{Measurement} & \colhead{Starburst} & \colhead{QPSB} & \colhead{AGN H$\delta_A > 4$\,{\AA}} & \colhead{AGN H$\delta_A < 3$\,{\AA}} & \colhead{QG}
}
\decimalcolnumbers
\startdata    
& $f$ Central Tinker & $0.839 \pm 0.005$ & $0.983 \pm 0.007$ & $0.902 \pm 0.012$ & $0.788 \pm 0.002$ & $0.777 \pm 0.001$\\
& $f$ Central Lim et al. & $0.832 \pm 0.016$ & $0.907 \pm 0.030 $ & $0.845 \pm 0.034$ & $0.785 \pm 0.007$ & $0.767 \pm 0.002$ \\
M11 & $\log M_h$ Tinker & $12.3\,(11.8, 13.2)$ & $12.4\,(12.2, 12.6)$ &$12.4\,(12.2, 12.7)$  & $12.4\,(11.7, 13.6)$ & $12.9\,(11.8,13.8)$\\
& $\log M_h$ Lim et al. & $12.5\,(12.3, 12.8)$ & $12.7\,(12.6, 13.0)$ &$12.6\,(12.4, 13.1)$  & $12.9\,(12.6, 13.4)$ & $13.1\,(12.7,13.6)$\\
\hline
 & $f$ Central Tinker & $0.801 \pm 0.002$ & $0.880 \pm 0.008$ & $0.874 \pm 0.007$ & $0.770 \pm 0.003$ & $0.708 \pm 0.001$ \\
& $f$ Central Lim et al. & $0.799 \pm 0.006$ & $0.833 \pm 0.021$ & $0.786 \pm 0.017$ & $0.723 \pm 0.007$ & $0.647 \pm 0.002$ \\
M105 & $\log M_h$  Tinker & $12.4\,(11.7, 13.4)$ & $12.5\,(12.2, 12.7)$ & $12.4\,(12.1, 12.7)$ & $12.3\,(11.7, 13.4)$ & $12.5\,(11.7,13.8)$ \\
& $\log M_h$  Lim et al. & $12.1\,(11.9, 12.5)$ & $12.3\,(12.1, 12.6)$ & $12.2\,(12.0, 12.6)$ & $12.3\,(12.1, 13.0)$ & $12.5\,(12.2,13.4)$\\
\hline
  & $f$ Central Tinker & $0.816 \pm 0.003$ & $0.798 \pm 0.012$ & $0.828 \pm 0.013$ & $0.692 \pm 0.006$ & $0.606 \pm 0.001$ \\
  & $f$ Central Lim et al. & $0.787 \pm 0.008$ & $0.728 \pm 0.027$ & $0.733 \pm 0.027$ & $0.666 \pm 0.015$ & $0.526 \pm 0.003$ \\
M10 & $\log M_h$  Tinker & $12.1\,(11.8, 12.9)$ & $12.2\,(11.9, 12.6)$ &$12.2\,(11.8, 12.6)$ & $12.0\,(11.4, 13.1)$ & $12.4\,(11.5,13.9)$\\
& $\log M_h$ Lim et al. & $11.8\,(11.6, 12.3)$ & $12.0\,(11.8, 12.6)$ & $11.9\,(11.7, 12.5)$ & $11.9\,(11.8, 12.9)$ & $12.0\,(11.8,13.7)$\\
\hline
 & $f$ Central Tinker & $0.791 \pm 0.007$ & $0.664 \pm 0.024$ & $0.755 \pm 0.040$ & $0.718 \pm 0.017$ & $0.276 \pm 0.0016$\\
& $f$ Central Lim et al. & $0.766 \pm 0.016$ & $0.563 \pm 0.055$ & $ 0.642 \pm 0.066$ & $0.647 \pm 0.047$ & $0.362 \pm 0.009$\\
M95 & $\log M_h$  Tinker & $12.0\,(11.7, 12.4)$ & $12.0\,(11.5, 12.9)$ & $11.9\,(11.4, 12.3)$ & $11.6\,(11.2, 12.3)$ & $13.2\,(11.3,14.3)$ \\
& $\log M_h$  Lim et al. & $11.6\,(11.4, 12.2)$ & $11.7\,(11.5, 13.1)$ & $11.7\,(11.5, 12.6)$ & $11.6\,(11.5, 12.6)$  & $12.7\,(11.6,14.1)$\\
\enddata
\tablecomments{The halo mass are in units of $h^{-1}M_\odot$. The results for other comparison samples can be found in Appendix~\ref{sec:appB}.}
\end{deluxetable*}

We have checked that the group masses (with three or more members) estimated by \citet{Tempel+14} assuming NFW profile are also significantly correlated with our mass overdensity measurements. The strength of the correlations between the group mass and mass overdensities are similar to the corresponding results using \citet{Lim+17}'s $M_h$ estimates. In addition, the mass overdensities also significantly correlate with \citet{Tinker21}'s $M_h$ estimates. His estimates, however, give lower $\rho$ values. 

\begin{figure*}[ht!]
\gridline{\fig{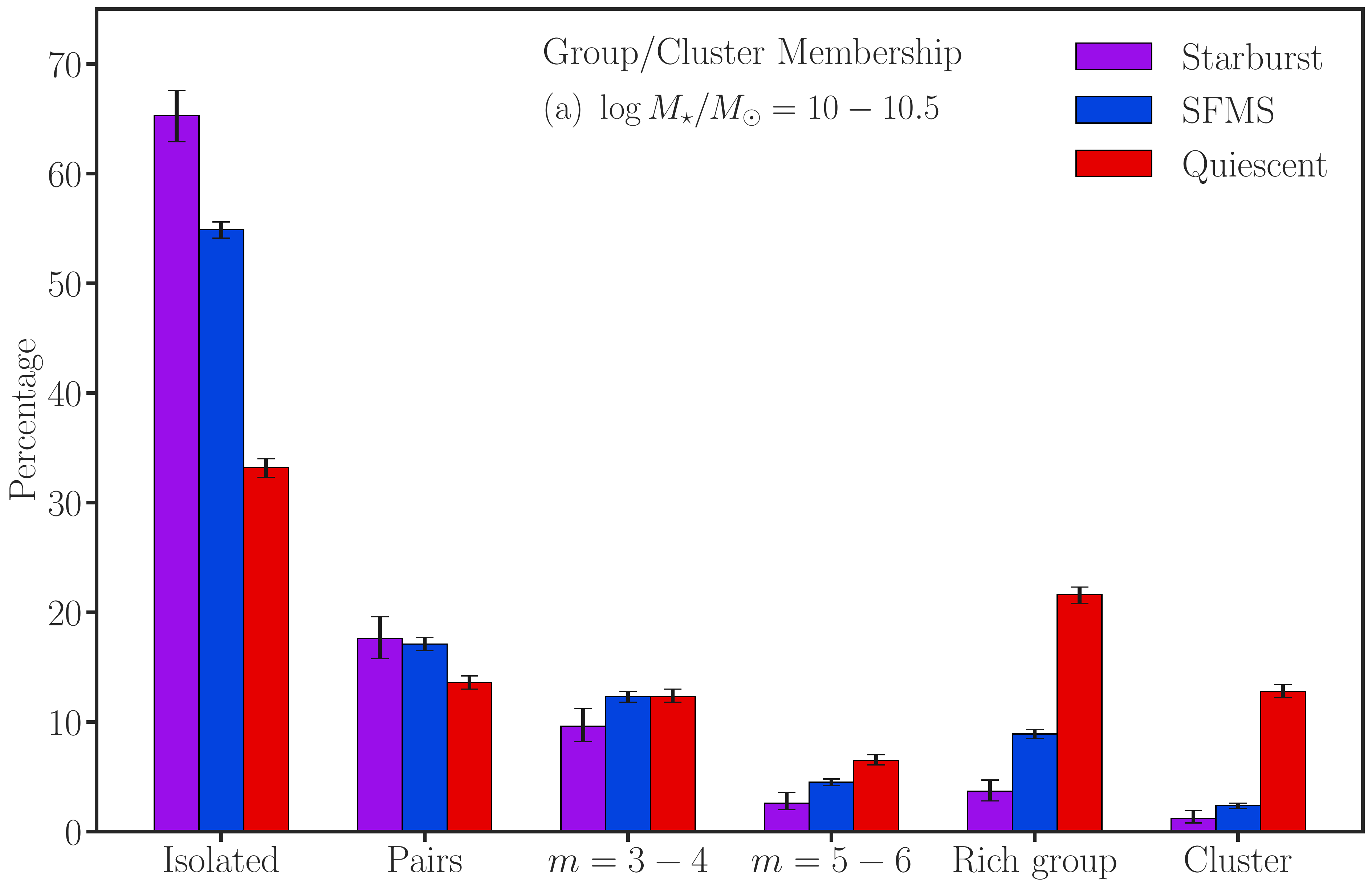}{0.48\textwidth}{}
\fig{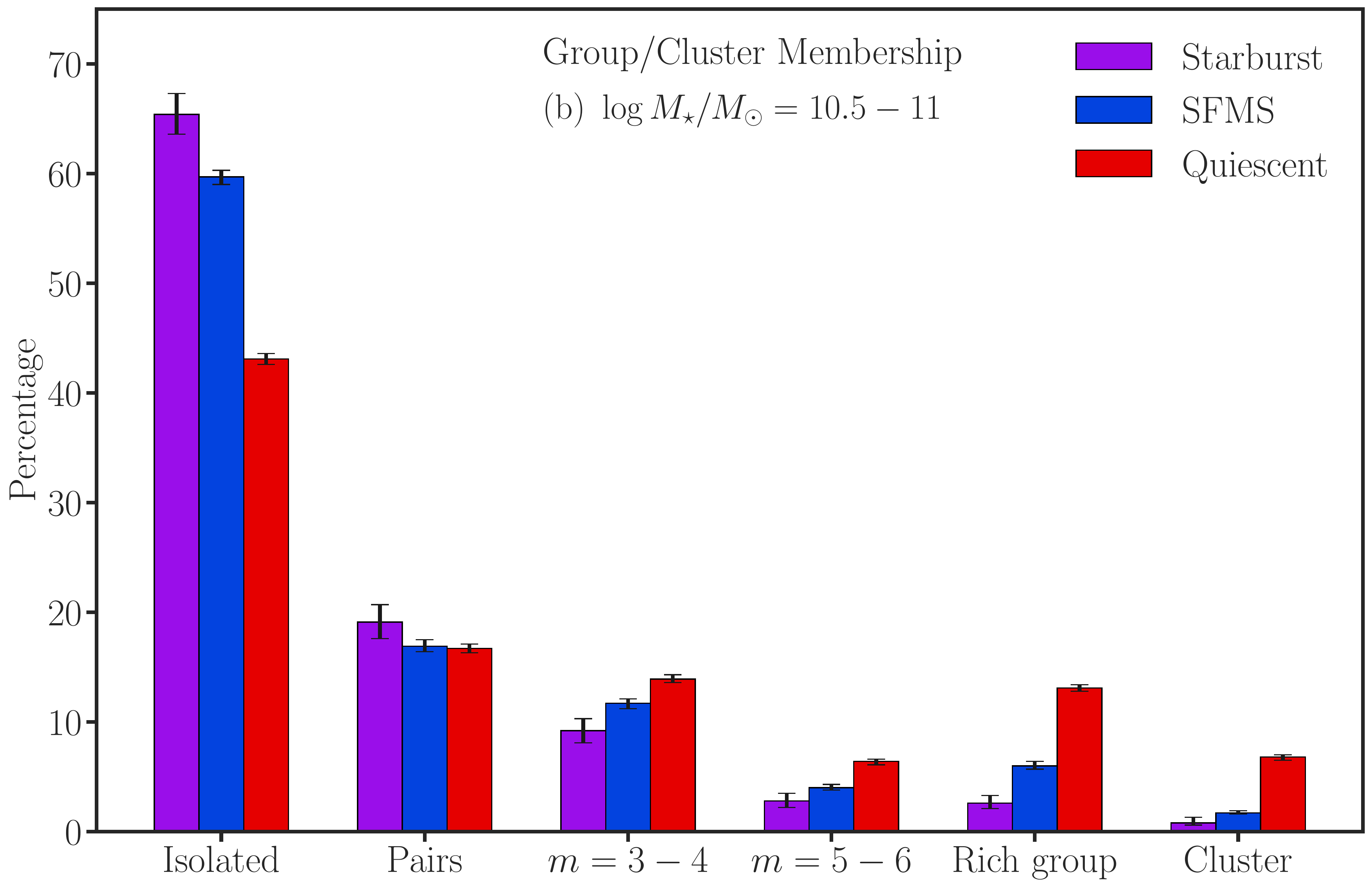}{0.48\textwidth}{}}
\caption{Comparing the mean percentages of starbursts that are isolated, in pairs, in groups with $m=3-6$ members, in rich groups ($m > 6$ and $\sigma < 400\,$km\,s$^{-1}$), or in clusters ($m > 6$ and $\sigma > 400$\,km\,s$^{-1}$) with their counterparts in SFMS galaxies and QGs. The error bars denote the 95\% multinomial confidence intervals.\label{fig:grpercent_sb}}
\end{figure*}

Furthermore, $\sim 80\%-85\%$ starbursts are isolated or pairs. The fraction of starbursts in rich groups or clusters is significantly smaller than those SFGs and QGs (Figure~\ref{fig:grpercent_sb}, Table~\ref{tab:richness2_sb}, and Appendix~\ref{sec:appB}). Noting that the number of galaxies in groups increases with decreasing $M_\star$, let us discuss the results for $\Mhigh$ sample as an example. In this $M_\star$ range, $65\%$ of starbursts are isolated (have no neighbors within the FoF linking length), $19\%$ are pairs, $3\%$ are in rich groups, and $1\%$ are in clusters. In contrast, $43\%$ QGs of are isolated, $17\%$ are pairs, $13\%$ are in rich groups, and $7\%$ are in clusters. Likewise, $60\%$ of SFMS galaxies are isolated, $17\%$ are pairs, $6\%$ are in rich groups, and $2\%$ are in clusters. In addition, the fraction of starbursts that are satellites is only $\sim 15\%-20\%$, and it is significantly lower than those of QGs of similar $M_\star$ (Table ~\ref{tab:fcentMh_sb}).

Having shown that starbursts live in distinctly low density environments, by all measures, and are unlike most non-bursty SFGs and QGs, next we check if the environments of PSBs conform with those of starbursts. 

\subsection{The Environments of QPSBs and Strong-H$\delta_A$ AGNs }

\begin{figure*}[ht!]
\gridline{\fig{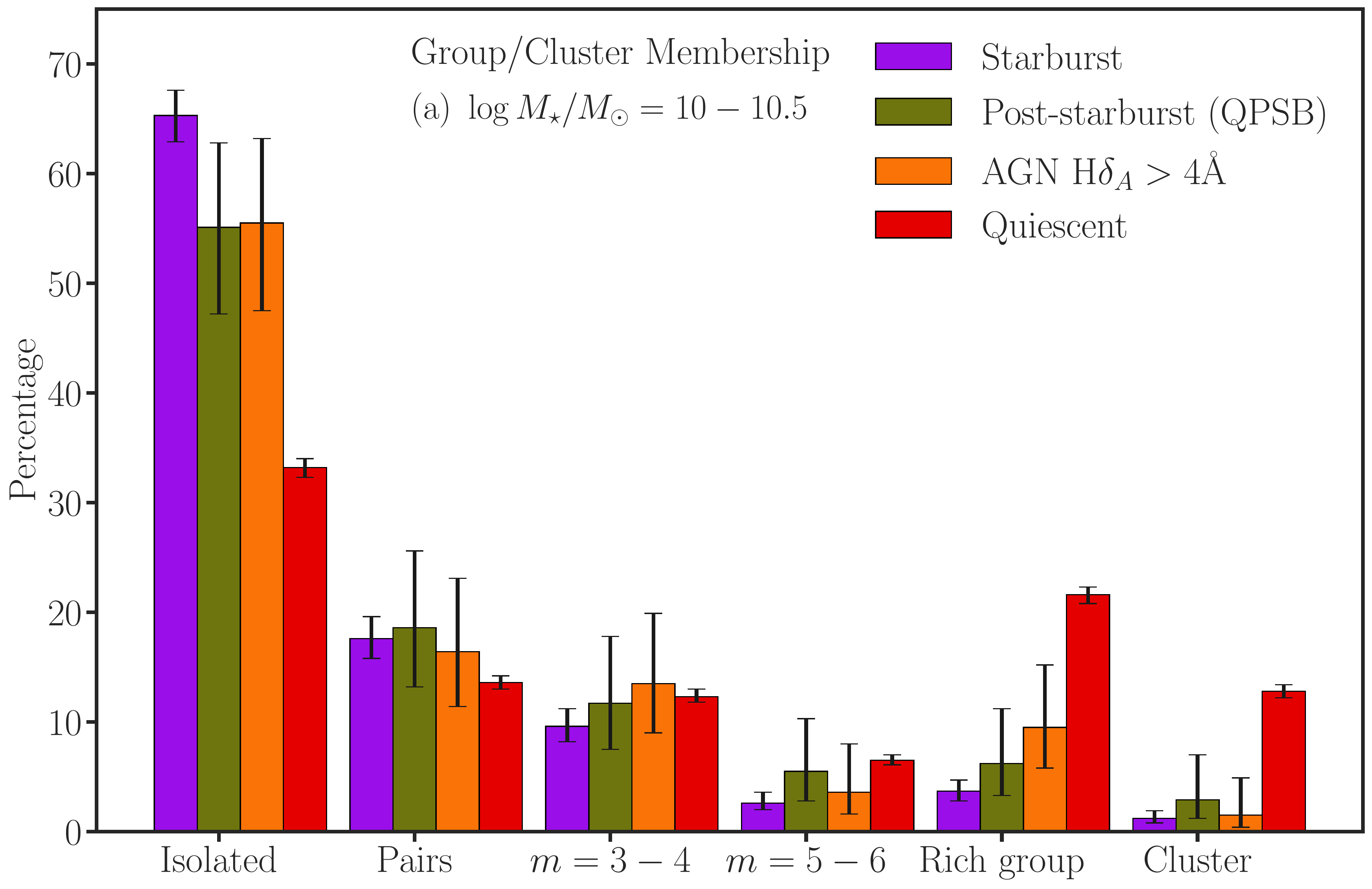}{0.48\textwidth}{}
\fig{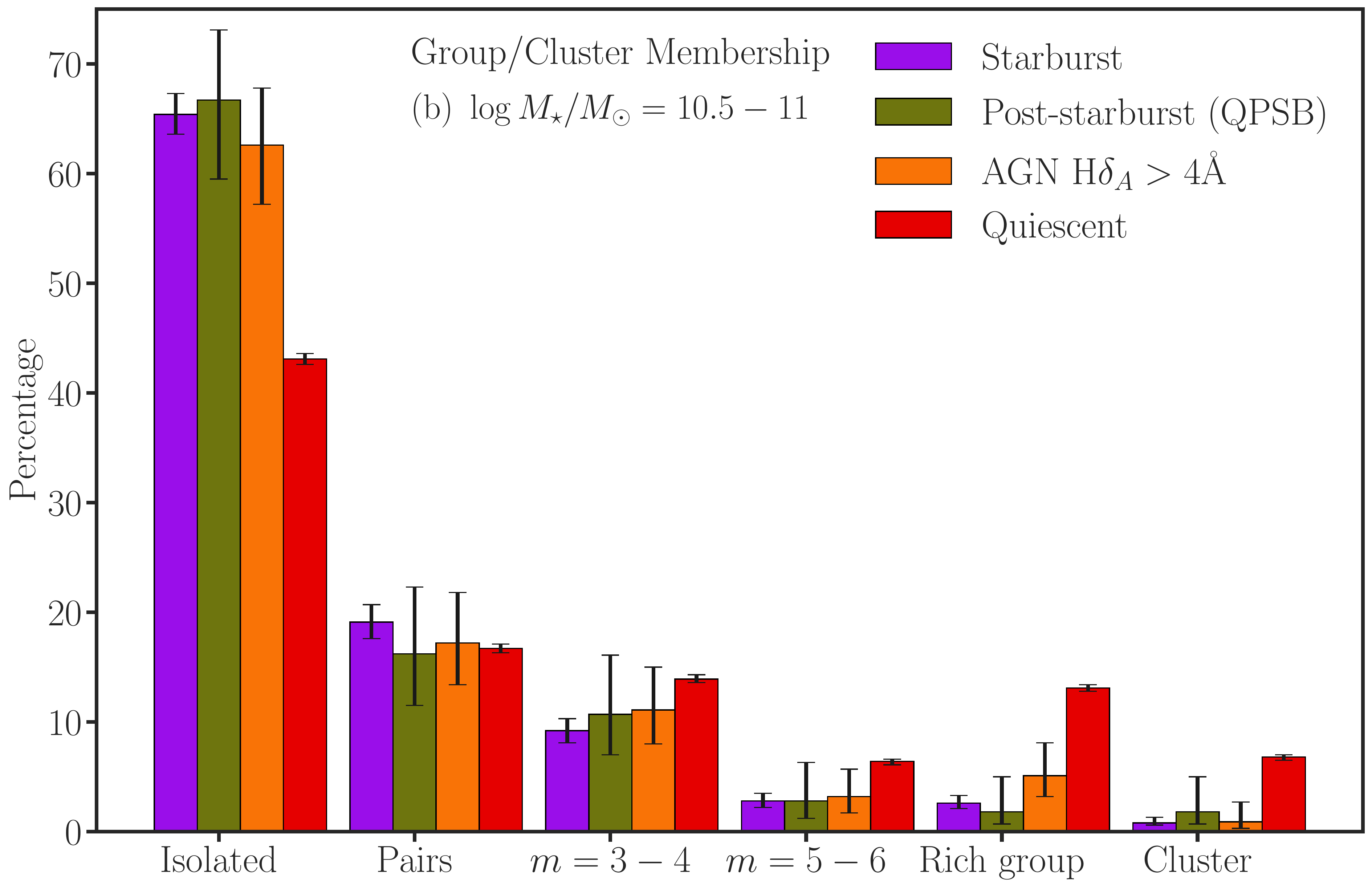}{0.48\textwidth}{}}
\caption{Similar to Figure~\ref{fig:grpercent_sb}, here the fractions of QPSBs and H$\delta_A > 4$\,{\AA} AGNs that are isolated, pairs, in groups or clusters are shown to be similar to the corresponding fractions of starbursts. \label{fig:grpercent_psb}}
\end{figure*}

Like starbursts, most QPSBs and AGNs with H$\delta_A> 4$\,{\AA} are isolated or centrals residing in dark matter halos of $M_h  < 10^{13}\, M_\odot$. The fractions of QPSBs and strong-H$\delta_A$ AGNs that are isolated, pairs, in groups, or in clusters are similar to those of starbursts (Figure~\ref{fig:grpercent_psb} and Table~\ref{tab:richness2_sb}). So are their fractions of centrals/satellites (Table~\ref{tab:fcentMh_sb}).

\begin{rotatetable*}
\movetableright=0.1in
\begin{deluxetable*}{cccccccc}
\tablenum{3}
\tablecaption{The Mean Fractions (and 95\% Confidence Intervals) of Galaxies that are Isolated, Pairs, in Groups, or in Clusters. \label{tab:richness2_sb}}
\tablewidth{0pt}
\tablehead{
\colhead{Sample} & \colhead{Measurement} &\colhead{Starburst} & \colhead{QPSB}  & \colhead{AGNs H$\delta_A > 4$\,{\AA}} & \colhead{AGN H$\delta_A < 3$\,{\AA}} &\colhead{SFMS} & \colhead{QG}}
\decimalcolnumbers
\startdata
M11 & Isolated &$0.632\,(0.576, 0.684)$ & $0.659\,(0.519, 0.776)$ & $0.612\,(0.490,  0.722)$  & $0.498\,(0.476, 0.52 )$ & $0.605\,(0.593, 0.616)$  & $0.447\,(0.441, 0.453) $ \\
      & Pairs & $0.200\,(0.158, 0.248)$  & $0.205\,(0.115, 0.338)$ & $0.190\,(0.112, 0.302)$ & $0.194\,(0.177, 0.212)$ & $0.184\,(0.175, 0.194)$ & $0.188\,(0.183, 0.193)$ \\
      & $2-3$ neighbors & $0.108\,(0.078, 0.148)$ & $0.080\,(0.031, 0.190)$  & $0.112\,(0.056, 0.212)$ & $0.139\,(0.124, 0.154)$ & $ 0.116\,(0.108, 0.124)$ & $0.156\,(0.152, 0.161)$\\ 
      & $4-5$ neighbors & $0.033\,(0.018, 0.060)$ & $0.023\,(0.004, 0.111)$  & $0.034\,(0.010,  0.112)$ & $0.057\,(0.048, 0.068)$ & $0.035\,(0.031, 0.040)$ & $0.066\,(0.062, 0.069)$ \\
      & Rich group & $0.015\,(0.006, 0.036)$  & $0.023\,(0.004, 0.111)$ & $0.052\,(0.019, 0.136])$ & $0.083\,(0.072, 0.960)$ & $0.045\,(0.040, 0.050)$  & $0.100\,(0.096, 0.104)$ \\
      & Cluster & $0.013\,(0.005, 0.033)$ & $0.011\,(0.001, 0.093)$ & $0.000\,(0.000, 0.057)$ & $0.029\,(0.023, 0.038)$ & $0.015\,(0.012, 0.018)$ & $0.043\,(0.041, 0.046)$ \\
\hline
M105 & Isolated & $0.654\,(0.636, 0.673)$  & $0.667\,(0.595, 0.731)$ & $0.626\,(0.572, 0.678)$  & $0.501\,(0.482, 0.521)$ & $0.597\,(0.590, 0.603)$ & $0.431\,(0.426, 0.436) $  \\
      & Pairs & $0.191\,(0.176, 0.207)$  &  $0.162\,(0.115, 0.223)$  & $0.172\,(0.134, 0.218)$ &  $0.175\,(0.161, 0.191)$ & $0.169\,(0.164, 0.175)$ & $0.167\,(0.163, 0.171) $ \\
      & $2-3$ neighbors & $0.092\,(0.081, 0.103)$  & $0.107\,(0.070, 0.161)$  & $0.111\,(0.080, 0.150)$ & $0.134\,(0.122, 0.148)$ & $ 0.117\,(0.112, 0.121)$ &  $0.139\,(0.136, 0.143)$ \\ 
      & $4-5$ neighbors & $0.028\,(0.022, 0.035)$  & $0.028\,(0.012, 0.063)$  & $0.032\,(0.017, 0.057)$ & $0.058\,(0.049, 0.067)$ & $0.040\,(0.038, 0.043)$ &  $0.064\,(0.061, 0.066)$ \\
      & Rich group & $0.026\,(0.021, 0.033)$  & $0.018\,(0.007, 0.050)$ & $0.051\,(0.032, 0.081)$ & $0.102\,(0.091, 0.115)$ & $0.060\,(0.057, 0.064)$ &  $0.131\,(0.128, 0.134)$ \\
      & Cluster & $0.008\,(0.006, 0.013)$ & $0.018\,(0.007, 0.050)$ & $0.009\,(0.003, 0.027)$ & $0.029\,(0.023, 0.036)$ & $0.017\,(0.016, 0.019)$  & $0.068\,(0.065, 0.070)$ \\
\hline
M10 & Isolated & $0.653\,(0.629, 0.676)$ &  $0.551\,(0.472, 0.628)$ & $0.555\,(0.475, 0.632)$  & $0.447\,(0.407, 0.488)$  & $0.549\,(0.541, 0.556)$  & $0.332\,(0.323, 0.340) $ \\
      & Pairs & $0.176\,(0.158, 0.196)$  & $0.186\,(0.132, 0.256)$  & $0.164\,(0.114, 0.231)$ & $0.175\,(0.146, 0.208)$  & $0.171\,(0.165, 0.177)$ & $0.136\,(0.130, 0.142)$ \\
      & $2-3$ neighbors & $0.096\,(0.082, 0.112)$ & $0.117\,(0.075, 0.178)$  & $0.135\,( 0.090, 0.199)$ & $0.136\,(0.111, 0.167)$ & $0.123\,(0.118, 0.128)$ & $0.123\,(0.118, 0.130)$  \\ 
      & $4-5$ neighbors & $0.026\,(0.020, 0.036)$  & $0.055\,(0.028, 0.103)$  & $0.036\,(0.016, 0.080)$ & $0.057\,(0.041, 0.079)$ & $0.045\,(0.042, 0.048)$ & $0.065\,(0.061, 0.070)$ \\
      & Rich group & $0.037\,(0.028, 0.047)$ & $0.062\,(0.033, 0.112)$ & $0.095\,(0.058, 0.152)$ & $0.139\,(0.113, 0.170)$ & $0.089\,(0.085, 0.093)$ & $0.216\,(0.208, 0.223)$ \\
      & Cluster &$0.012\,(0.008, 0.019)$ & $0.029\,(0.012, 0.070)$ & $0.015\,(0.004, 0.049)$ & $0.046\,(0.032, 0.066)$ &  $0.024\,(0.021, 0.026)$  & $0.128\,(0.122, 0.134)$ \\
\hline
M95 & Isolated & $0.623\,(0.576, 0.668)$  & $0.439\,(0.305, 0.583)$ & $0.500\,(0.331, 0.669)$  & $0.471\,(0.348, 0.598)$ &  $0.529\,(0.519, 0.539)$ & $0.216\,(0.196, 0.238)$  \\
      & Pairs & $0.179\,(0.145, 0.219)$  & $0.085\,(0.033, 0.202)$  & $0.148\,(0.062, 0.315)$ & $0.135\,(0.069, 0.246)$ & $0.175\,(0.167, 0.183)$ & $0.090\,(0.077, 0.106)$\\\
      & $2-3$ neighbors &  $0.103\,(0.077, 0.135)$ & $0.232\,(0.133, 0.373)$  & $0.111\,(0.040, 0.271)$ & $0.144\,(0.076, 0.257)$ & $0.123\,(0.116, 0.129)$ & $0.091\,(0.077, 0.107)$  \\ 
      & $4-5$ neighbors &$0.027\,(0.016, 0.047)$  & $0.012\,(0.001, 0.099)$  & $0.019\,(0.002, 0.145)$ & $0.038\,(0.011, 0.124)$ & $0.050\,(0.046, 0.055)$ & $0.063\,(0.052, 0.077)$  \\
      & Rich group & $0.045\,(0.029, 0.070)$  & $0.146\,(0.071, 0.277)$ & $0.185\,(0.085, 0.357)$ & $0.183\,(0.104, 0.301)$ & $0.099\,(0.094, 0.106)$ &  $0.336\,(0.312, 0.361)$ \\
      & Cluster & $0.022\,(0.012, 0.041)$ & $0.085\,(0.033, 0.202)$ & $0.037\,(0.007, 0.173)$ & $0.024\,(0.007, 0.110)$ &$0.024\,(0.021, 0.027)$ & $0.203\,(0.184, 0.225)$\\
\enddata
\tablecomments{The results for the upper SFMS, lower SFMS, and green-valley galaxies are given in the Appendix~\ref{sec:appB}.}
\end{deluxetable*}
\end{rotatetable*}

\begin{figure*}[ht!]
\gridline{\fig{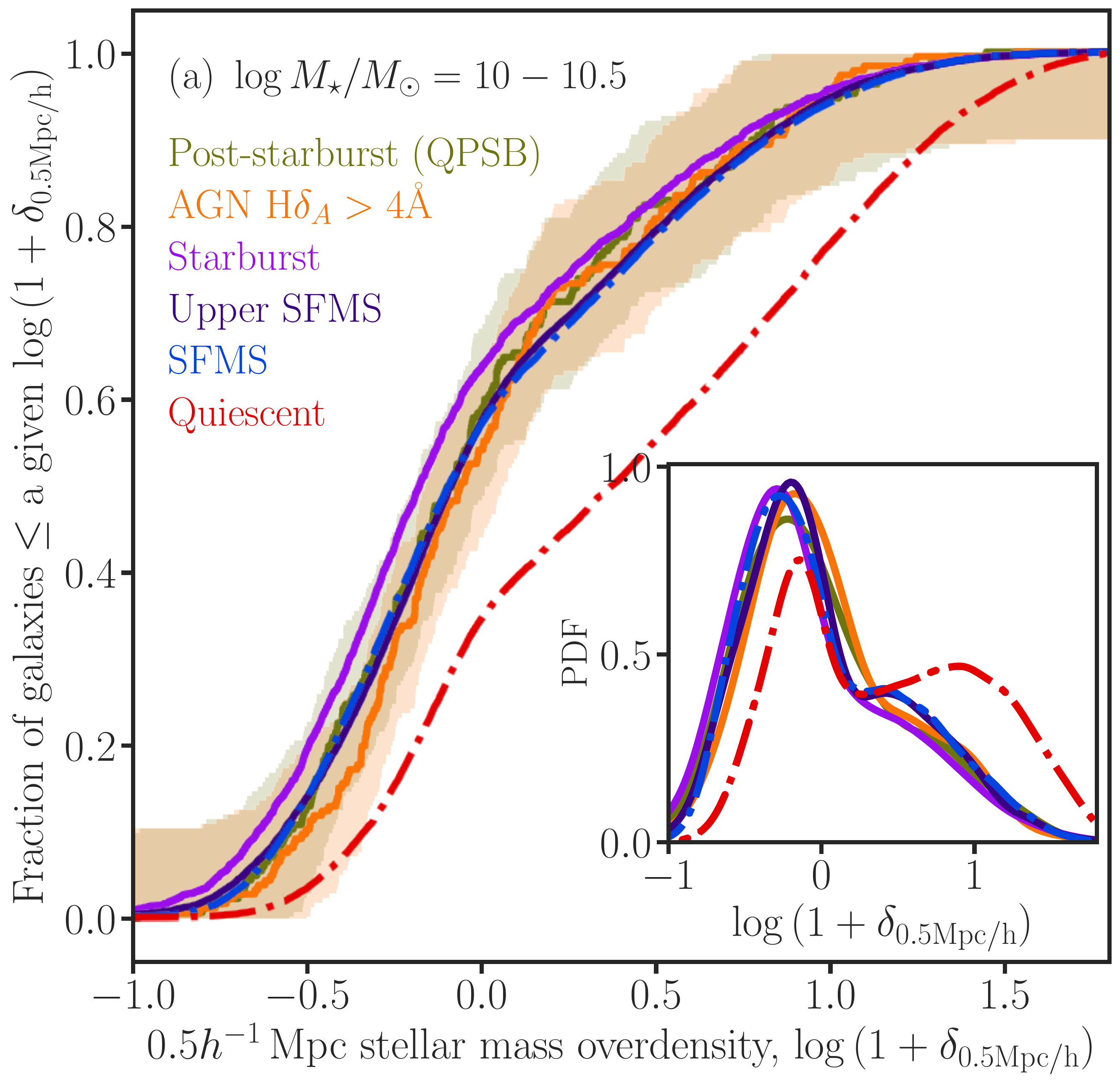}{0.48\textwidth}{}
          \fig{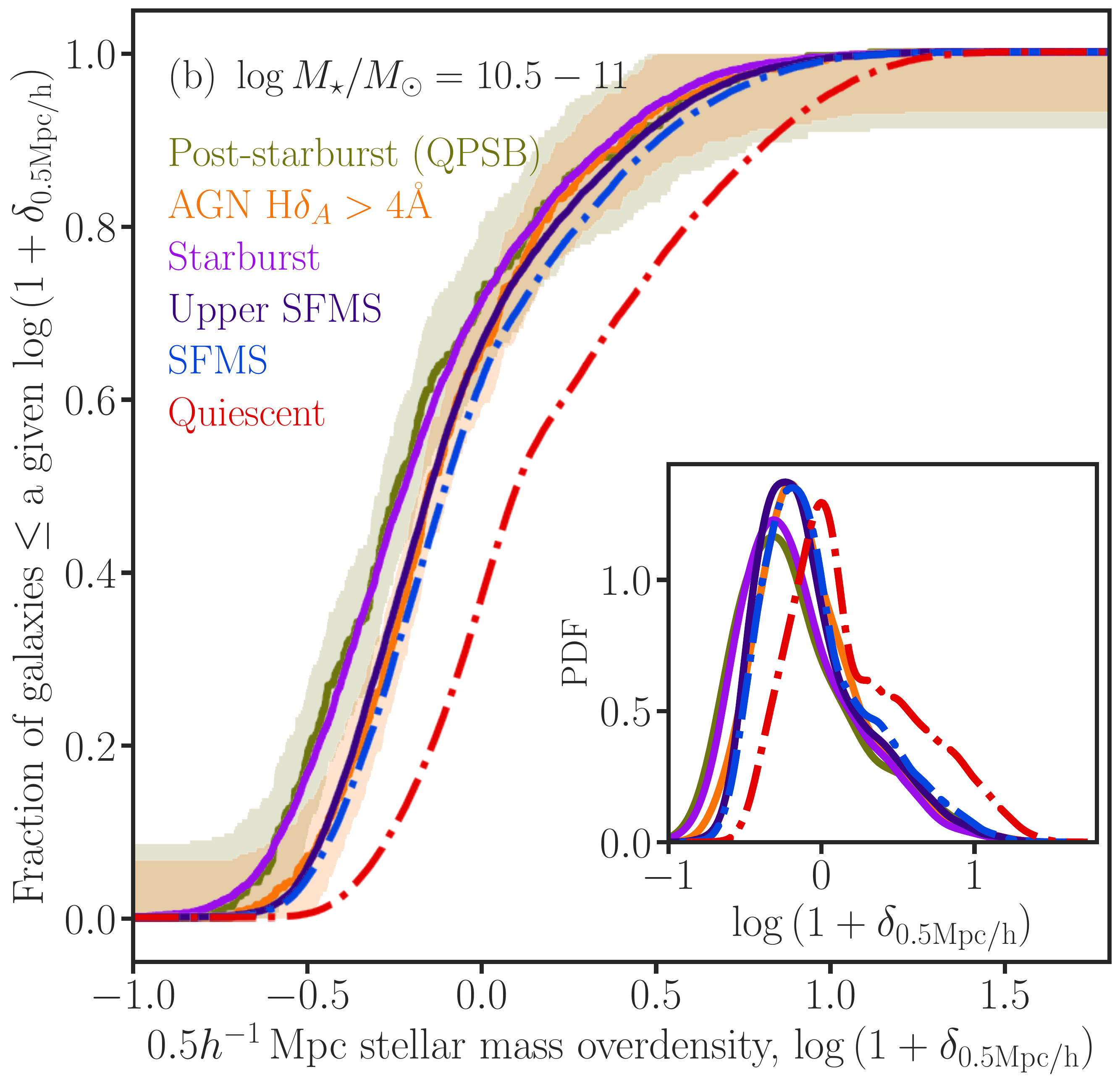}{0.48\textwidth}{}}
\gridline{\fig{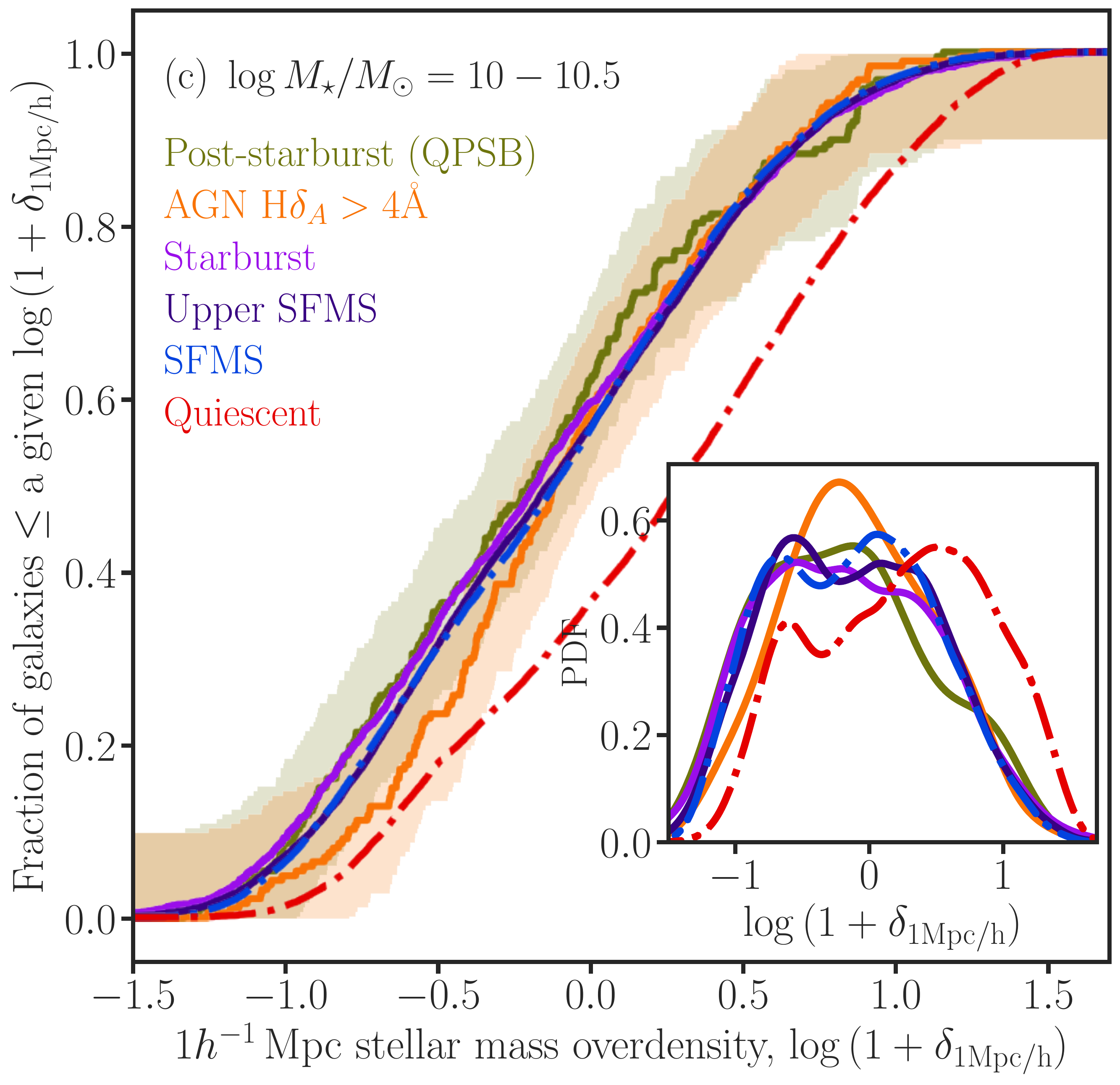}{0.48\textwidth}{}
          \fig{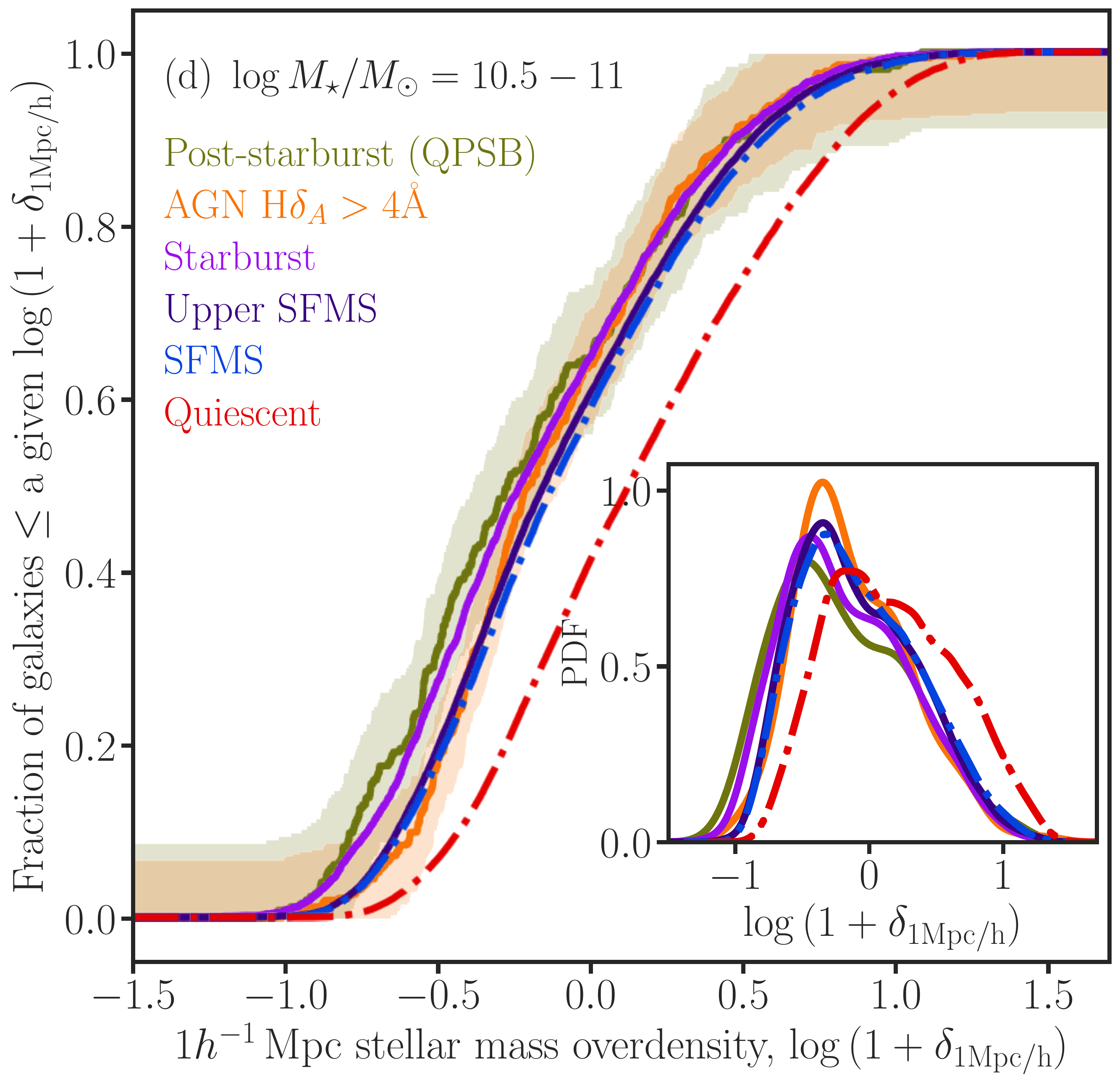}{0.48\textwidth}{}}                   
\caption{Comparing the cumulative distributions of mass densities $\deltahalf$ and $\deltaone$ of QPSBs to those of starbursts, SFMS galaxies, and QGs. 
Depending on $M_\star$, the $0.5-1\,h^{-1}$Mpc scale environments of QPSBs similar to or slightly lower than those of starbursts, but they are quite different from the environments of QGs. As shown in panel (a), QPSBs and starbursts do not necessarily have similar $\deltahalf$ distributions. This can be understood if only a subset of starbursts become QPSBs; the number of starbursts is much higher than QPSBs. The  environments of AGNs with H$\delta_A > 4$\,{\AA}  are similar to upper SFMS galaxies. \label{fig:del18_psb}}
\end{figure*}

The multiscale environments of QPSBs are generally similar to those of starbursts, especially at the scale of $>1h^{-1}$\,Mpc. The AD test, however, sometimes indicates that the $\deltahalf$ or $\deltaone$ distributions of the two populations are significantly different. Figure~\ref{fig:del18_psb} shows the distributions of $\deltahalf$ and $\deltaone$ in two $M_\star$ ranges for QPSBs and for the comparison samples of starbursts, AGNs with H$\delta_A > 4$\,{\AA}, SFGs, and QGs. Due to the small sample size, the Dvoretzky-Kiefer-Wolfowitz confidence interval\footnote{The true CDF, $F(\delta)$, is bounded by the CDF estimated from a sample of size $n$ as {$\displaystyle F_{n}(\delta)-\varepsilon \leq F(\delta)\leq F_{n}(\delta)+\varepsilon \;{\text{, where}}\;\varepsilon ={\sqrt {\frac {\ln {\frac {2}{\alpha }}}{2n}}}\; \text{and $\alpha$ is the significance level}$}. The CDFs of QPSBs in a given mass range have $\epsilon \approx 0.1$, using $\alpha = 0.05$. The CDFs of starbursts have $\epsilon \approx 0.02$ and those of the other samples have $\epsilon < 0.01$.} of the ECDFs of QPSBs and AGNs with H$\delta_A > 4$\,{\AA} are quite large. Nevertheless, $\deltahalf$ and $\deltaone$ of distributions of QPSBs and AGNs with H$\delta_A > 4$\,{\AA} are clearly different from those of QGs. For $\Mhigh$, the $\deltahalf$ and $\deltaone$ of QPSBs are similar those of starbursts ($p = .2$ for $\deltahalf$ and $p > .2$ for $\deltaone$) but are dissimilar to those of SFMS and upper SFMS galaxies ($p < .001$). In contrast, for $\Mlow$, $\deltahalf$ of QPSBs are more similar ($p > .2$) to upper SFMS and SFMS galaxies than they are to starbursts ($p=.001$). The multiscale environments of AGNs with H$\delta_A > 4$\,{\AA} are generally similar those of the upper SFMS galaxies. Except for galaxies with $M_\star > 10^{11}\,M_\odot$, the AD test does not indicate significant differences between the overdensity distributions of the two populations at all scales. In some cases, the distributions of $\deltahalf$ and $\deltaone$ of AGNs with H$\delta_A > 4$\,{\AA} are significantly different from those of starbursts and QPSBs (as also shown in Figure~\ref{fig:del18_psb}).

We see later that not all starbursts are equally likely to become QPSBs and the subtle discrepancies of the small-scale overdensities can be understood with a selective draw of QPSBs and strong-H$\delta_A$ AGNs from the parent samples of starbursts and/or upper SFMS galaxies. Starbursts and upper SFMS galaxies outnumber QPSBs and strong-H$\delta_A$ AGNs by more than ten times. Thus, the fact that their environments are different sometimes, does not necessarily imply that they are not evolutionary connected. Because the selection from the parent sample is not random, it is not necessary to match the overdensity distributions of the parent samples; there is no inconsistency as long as there are enough galaxies in the parent samples that have similar environments (and structures) as QPSBs and strong-H$\delta_A$ AGNs. 

\begin{figure*}[ht!]
\gridline{\fig{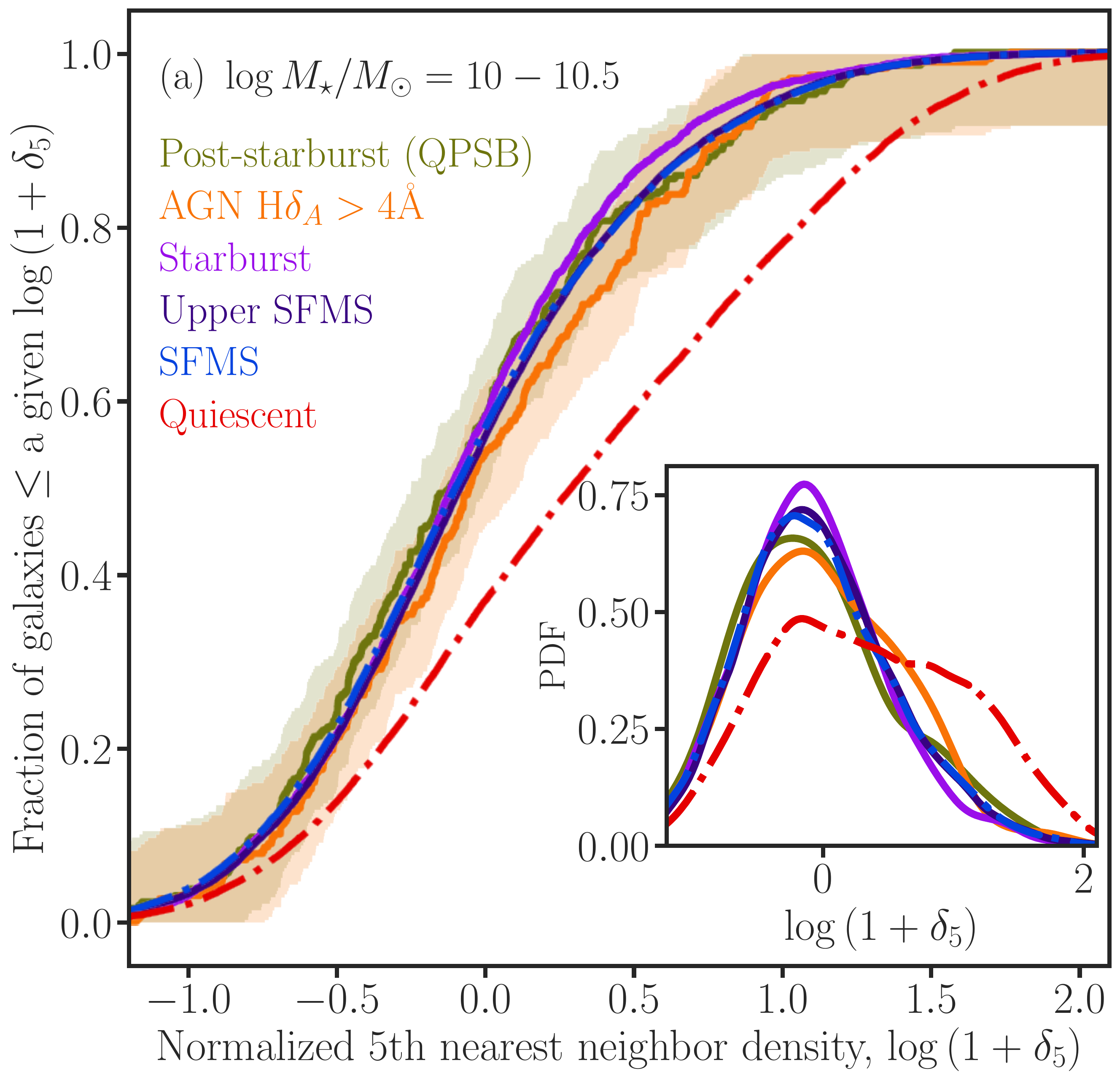}{0.48\textwidth}{}
          \fig{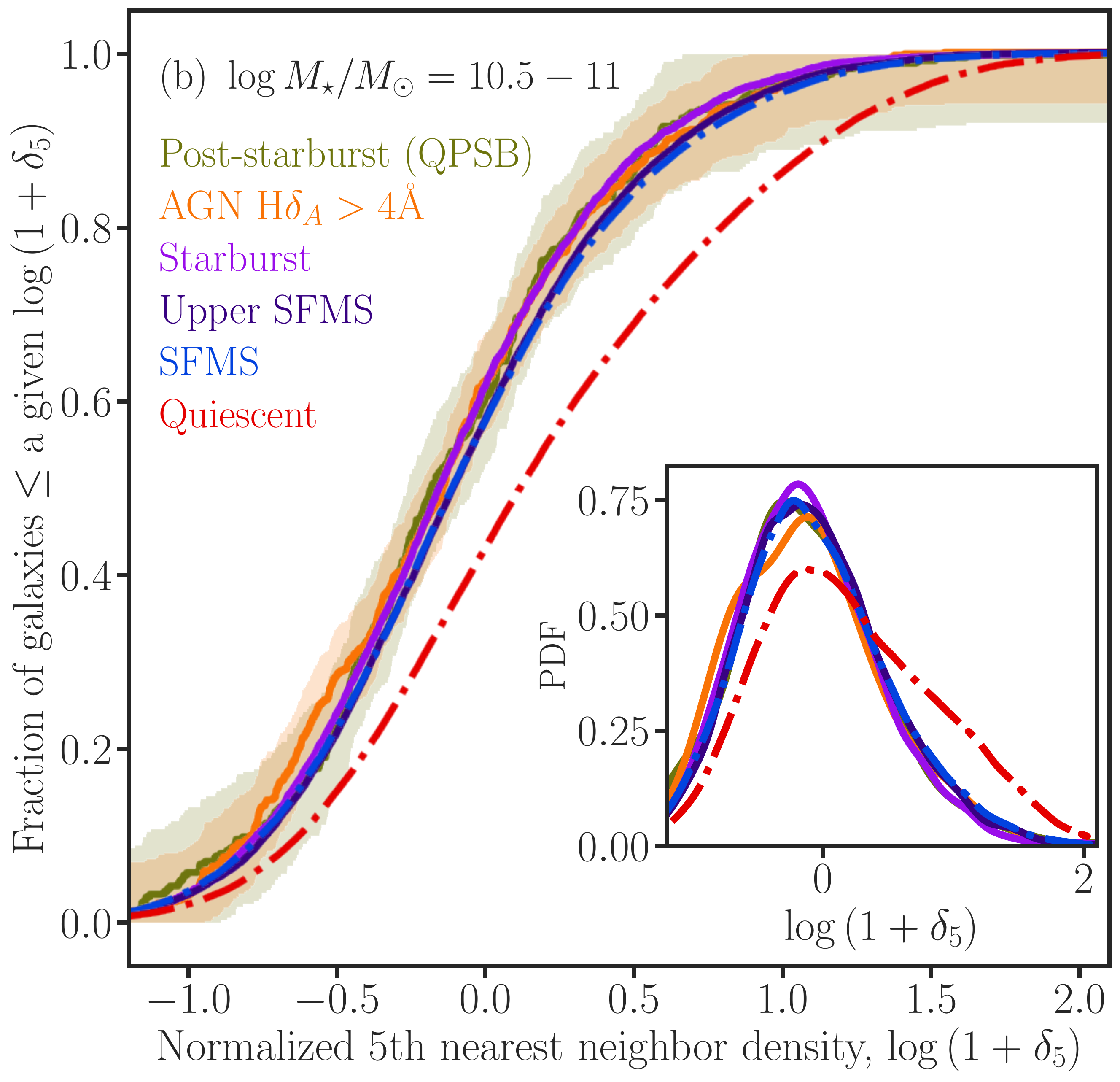}{0.48\textwidth}{}}
          \gridline{\fig{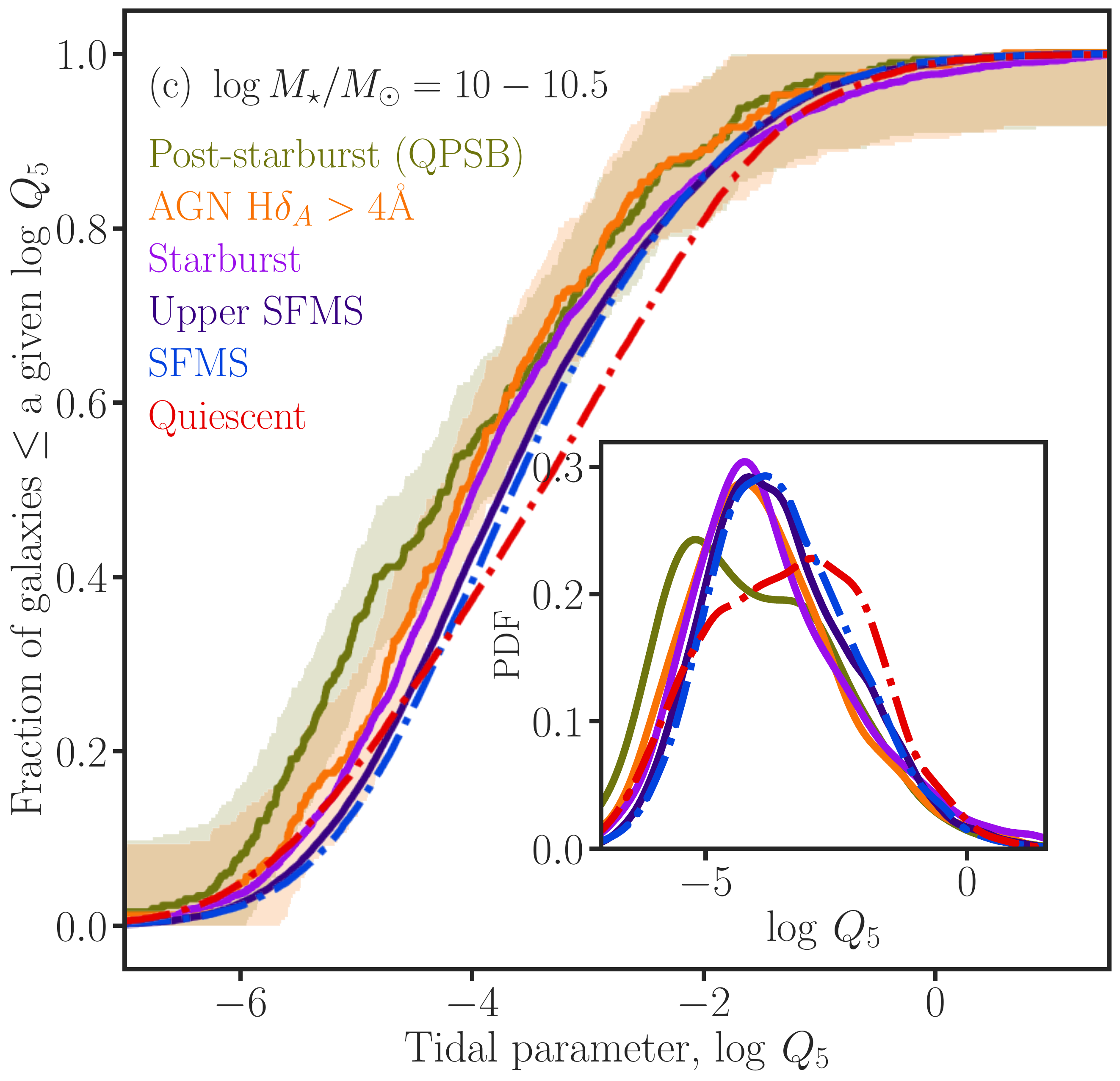}{0.48\textwidth}{}
          \fig{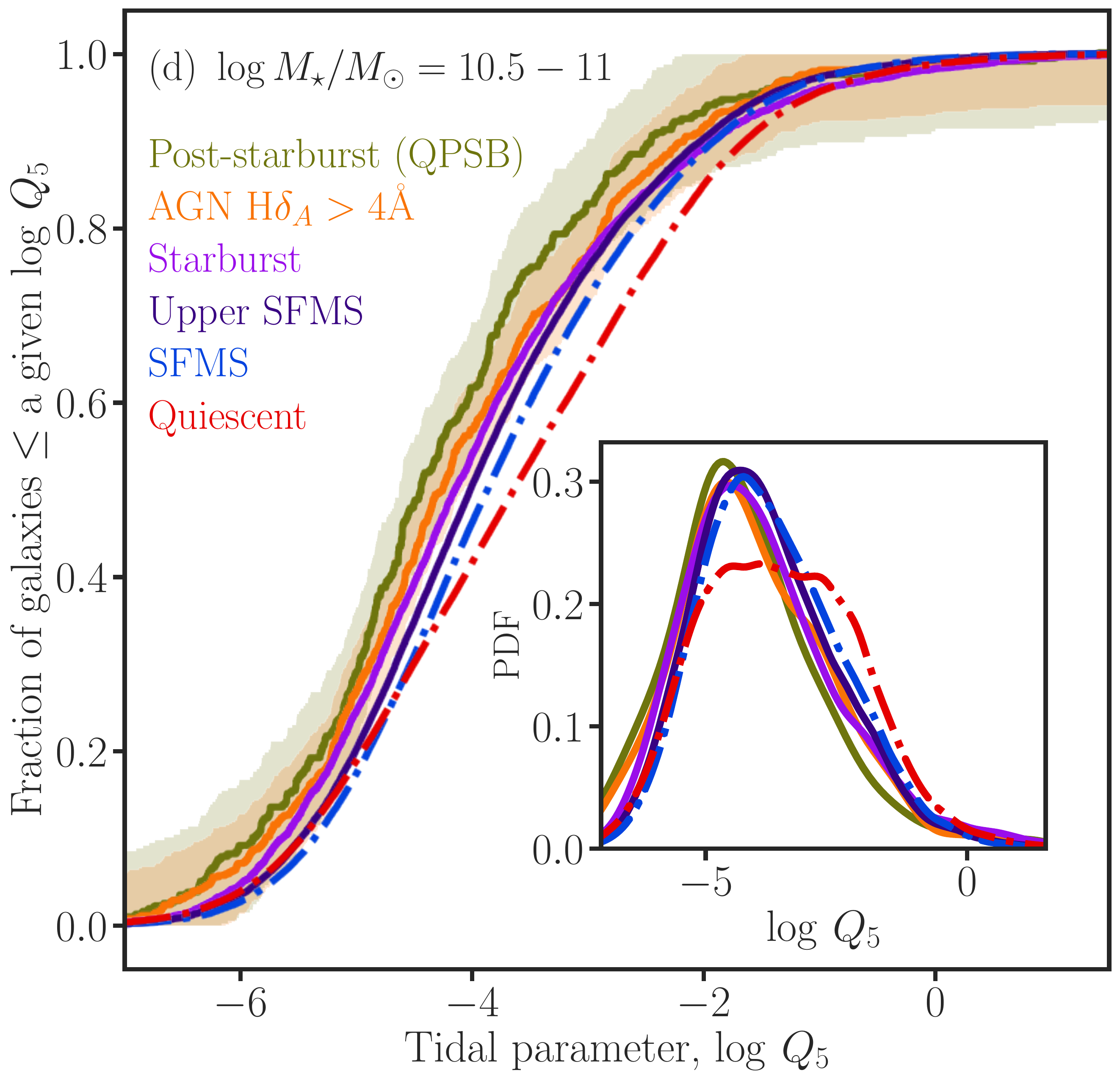}{0.48\textwidth}{}}
\caption{The ECDFs of the 5th nearest neighbor overdensity ($\delta_5$) and the tidal parameter($Q_5$) of QPSBs and AGNs with H$\delta_A > 4$\,{\AA} are compared with starbursts, other SFGs, and QGs. 
The distributions of $\delta_5$ and $Q_5$ of QPSBs and AGNs with H$\delta_A > 4$\,{\AA} are more similar to those of SFGs than to those of QGs. \label{fig:del5_psb}}
\end{figure*}

The $\delta_5$, $r_{p,1}$, and $Q_5$ distributions of QPSBs and strong-H$\delta_A$ AGNs are also clearly different from those of QGs (Figure~\ref{fig:del5_psb}). The $\delta_5$ distributions of QPSBs and strong-H$\delta_A$ AGNs are generally similar to the $\delta_5$ distributions starbursts and/or upper SFMS galaxies. The $r_{p,1}$ and $Q_5$ distributions of QPSBs are significantly different from starbursts and upper SFMS galaxies ($p < .001$). The $\delta_5$, $r_{p,1}$, and $Q_5$ distributions of QPSBs and strong-H$\delta_A$ AGNs are not inconsistent.

\begin{figure*}
\gridline{\fig{M10511_delM05Mpc_AGN_dist_R1.pdf}{0.47\textwidth}{}
\fig{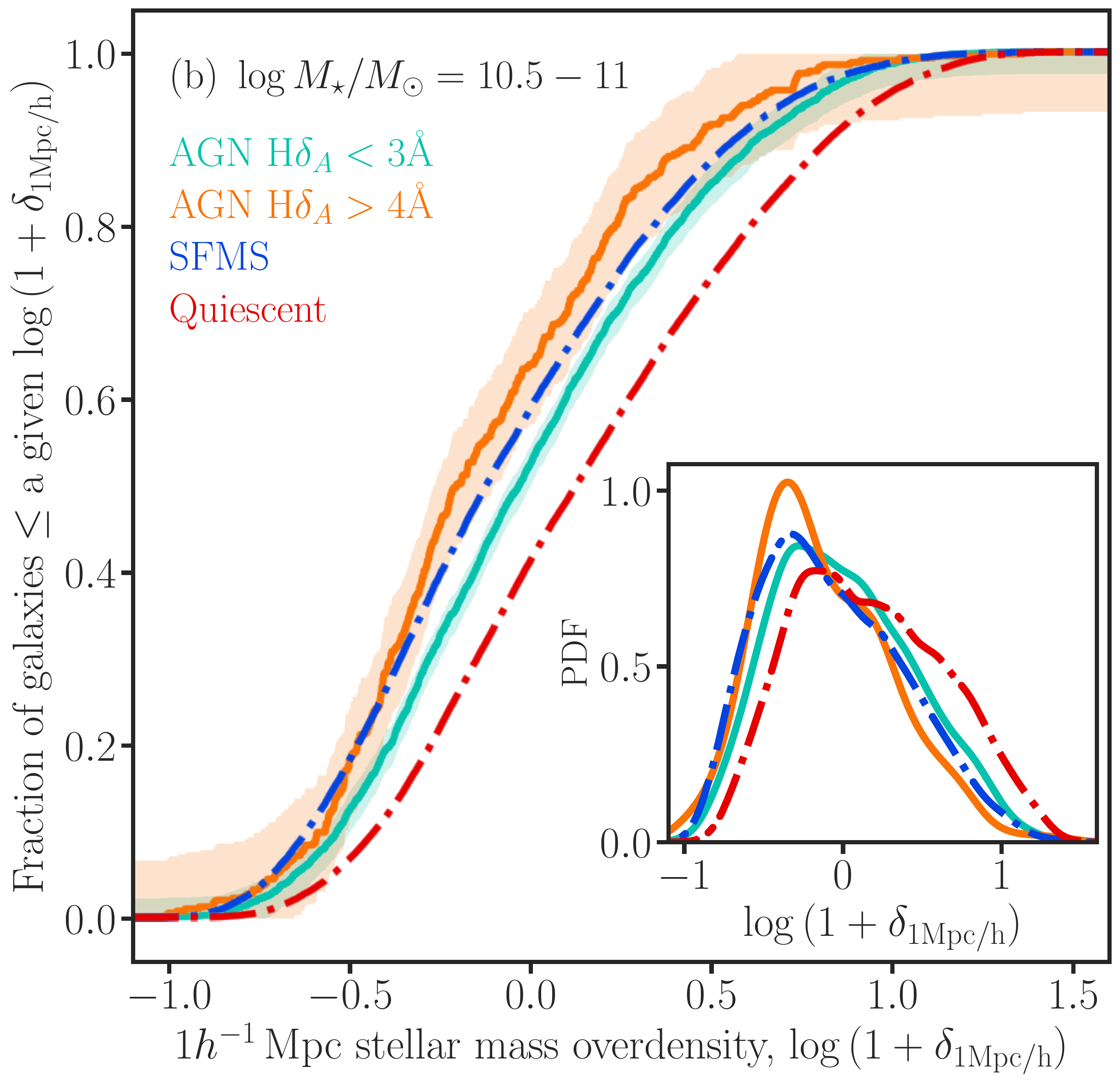}{0.47\textwidth}{}}
\gridline{\fig{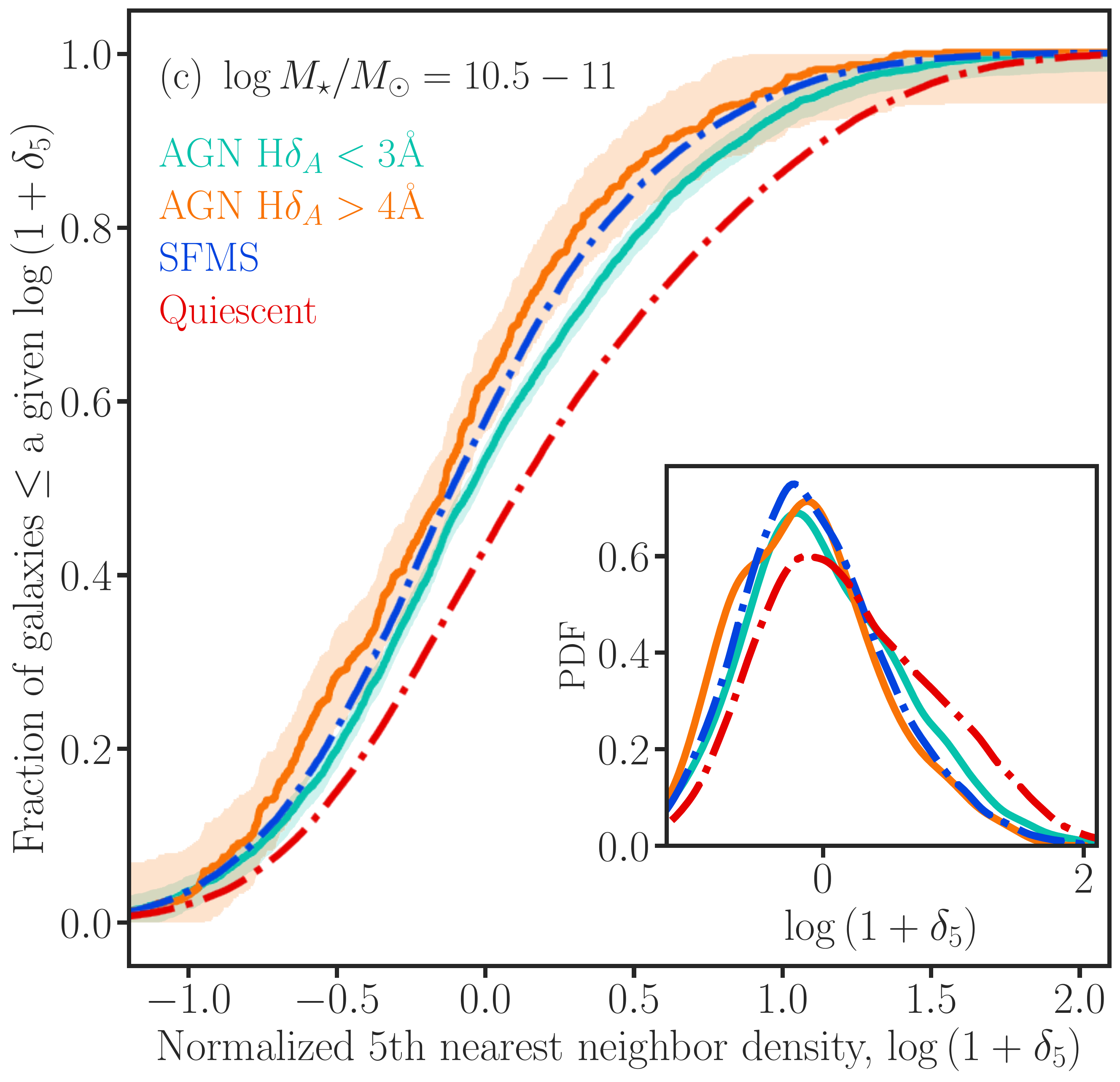}{0.47\textwidth}{}
        \fig{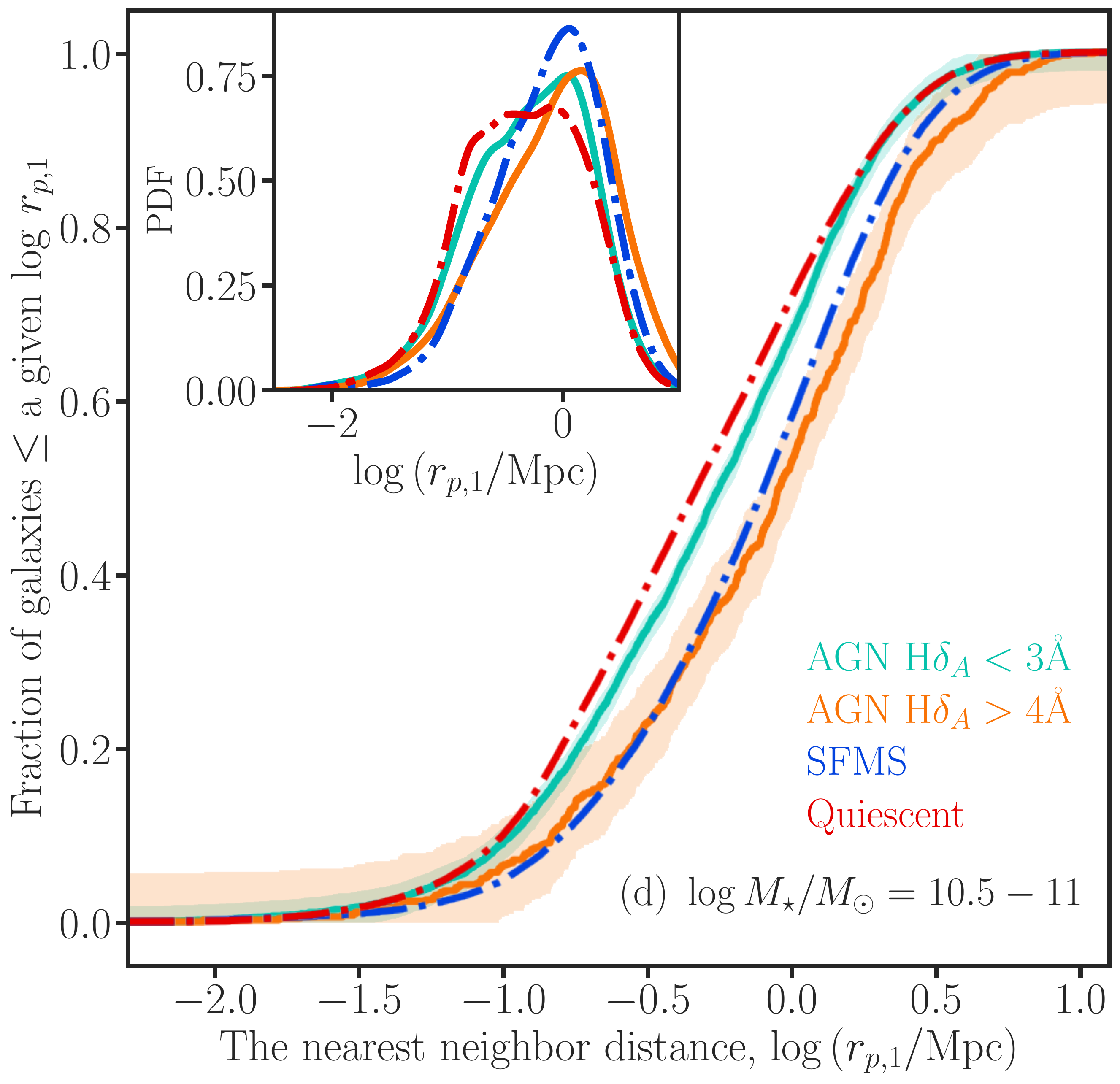}{0.47\textwidth}{}}          
          \caption{Comparing the environments of AGNs with those of the other samples. The environments of AGNs have lower densities than those of QGs. The overdensities (nearest neighbor distances) of AGNs with H$\delta_A > 4$\,{\AA} are significantly lower (longer) than those of SFMS galaxies while the opposite is true for AGNs with H$\delta_A < 3$\,{\AA} .\label{fig:del18_agn}}
\end{figure*}

Lastly, the environments of AGNs with H$\delta_A > 4$\,{\AA} and of those with H$\delta_A  < 3$\,{\AA} are very different. Figure~\ref{fig:del18_agn} compares, for example, the distributions of $\deltahalf$, $\deltaone$, $\delta_5$, and $r_{p,1}$ of AGNs with SFMS galaxies and QGs in $\Mhigh$. Strong-H$\delta_A$ AGNs live in environments that are slightly sparser than those SFMS galaxies while weak-H$\delta_A$ AGNs live in environments denser than those SFMS galaxies (more consistent with those of lower SFMS galaxies). Both weak and strong-H$\delta_A$ AGNs reside in distinctly lower density environments than QGs do.

\subsection{Structural and Environmental Consistency of the Evolution from Starbursts $\rightarrow$ AGNs $\rightarrow$ QPSBs $\rightarrow$ QGs}

\begin{figure*}[ht!]
\gridline{\fig{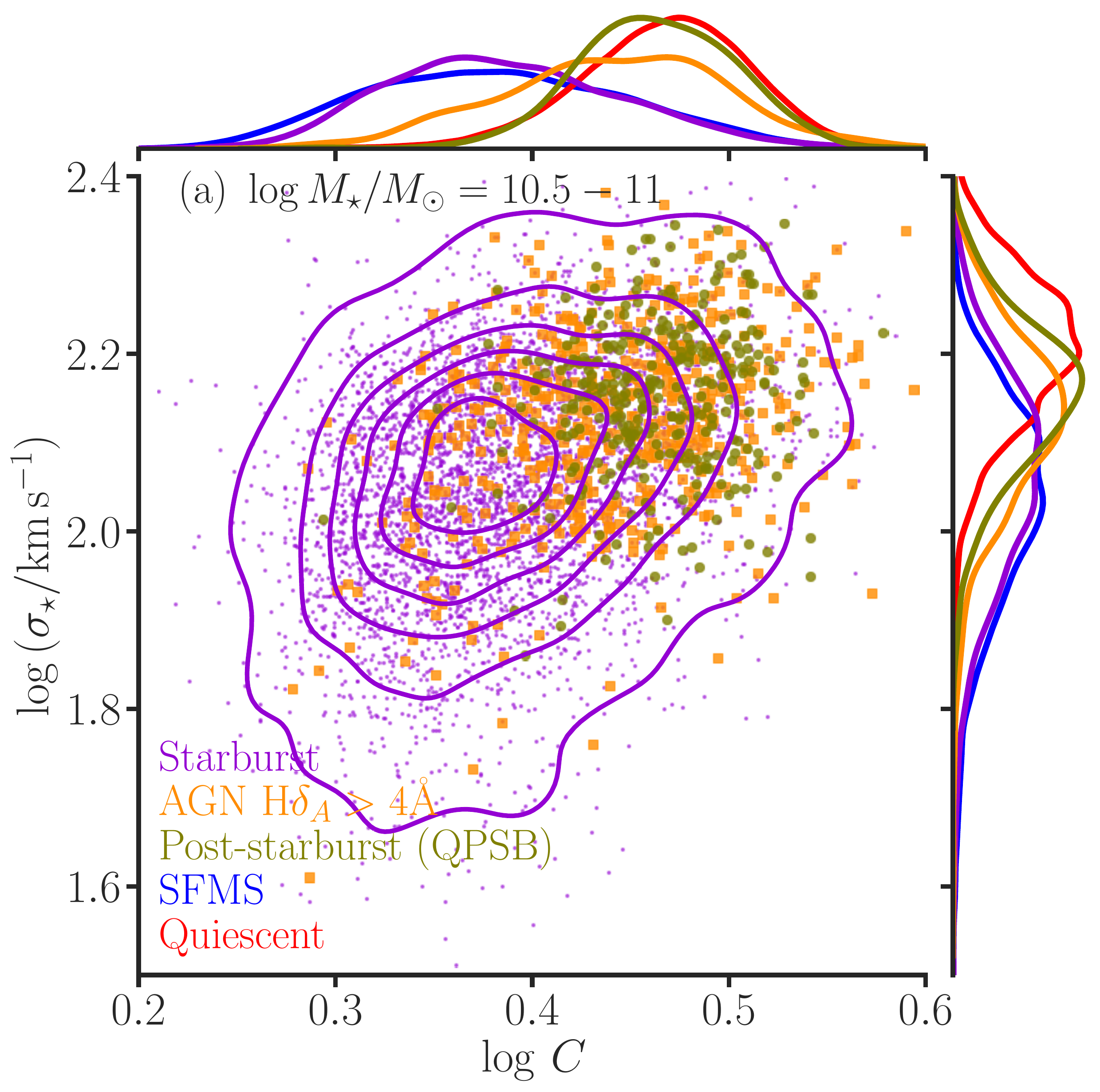}{0.48\textwidth}{}
\fig{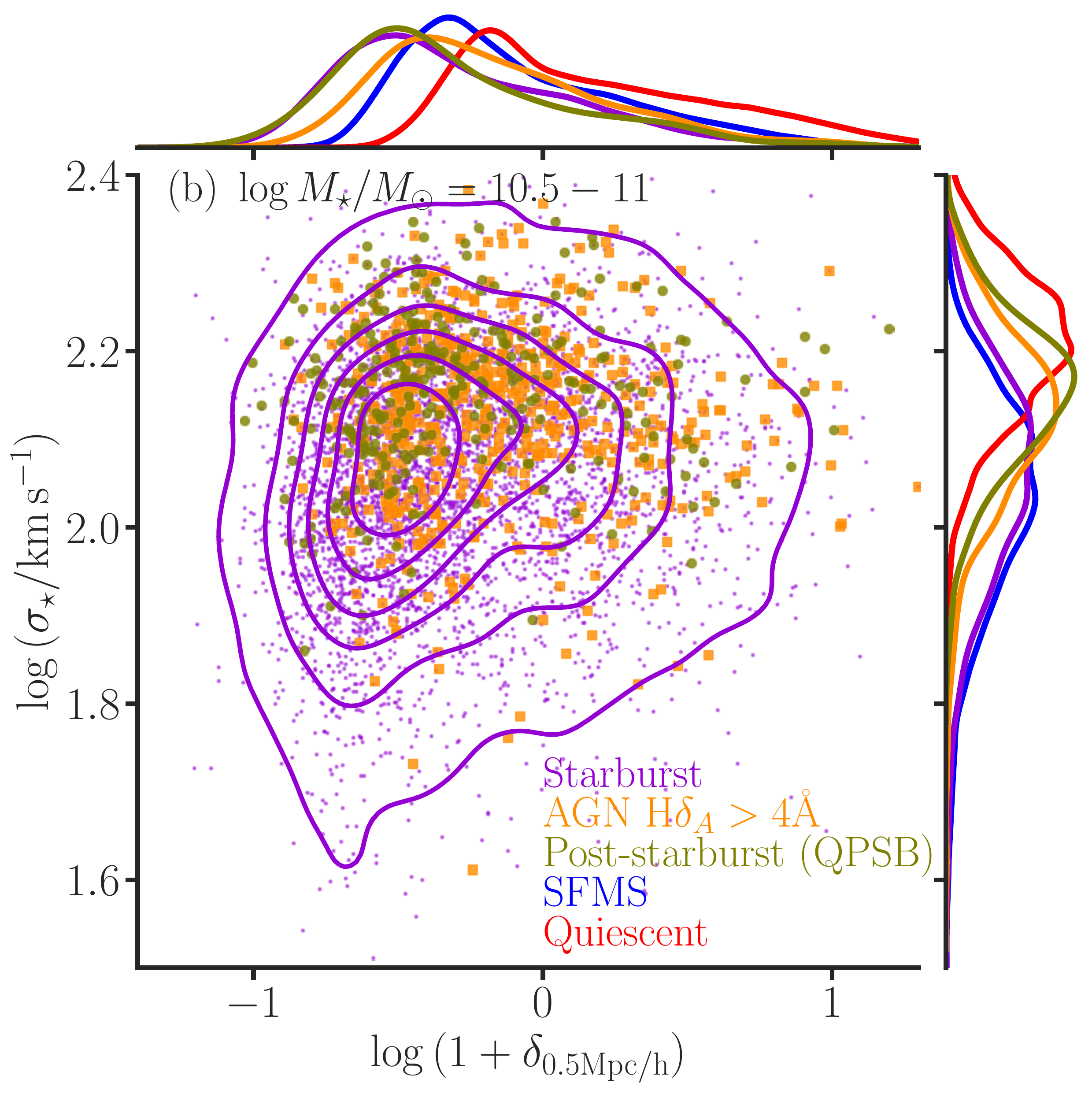}{0.48\textwidth}{}}
\gridline{\fig{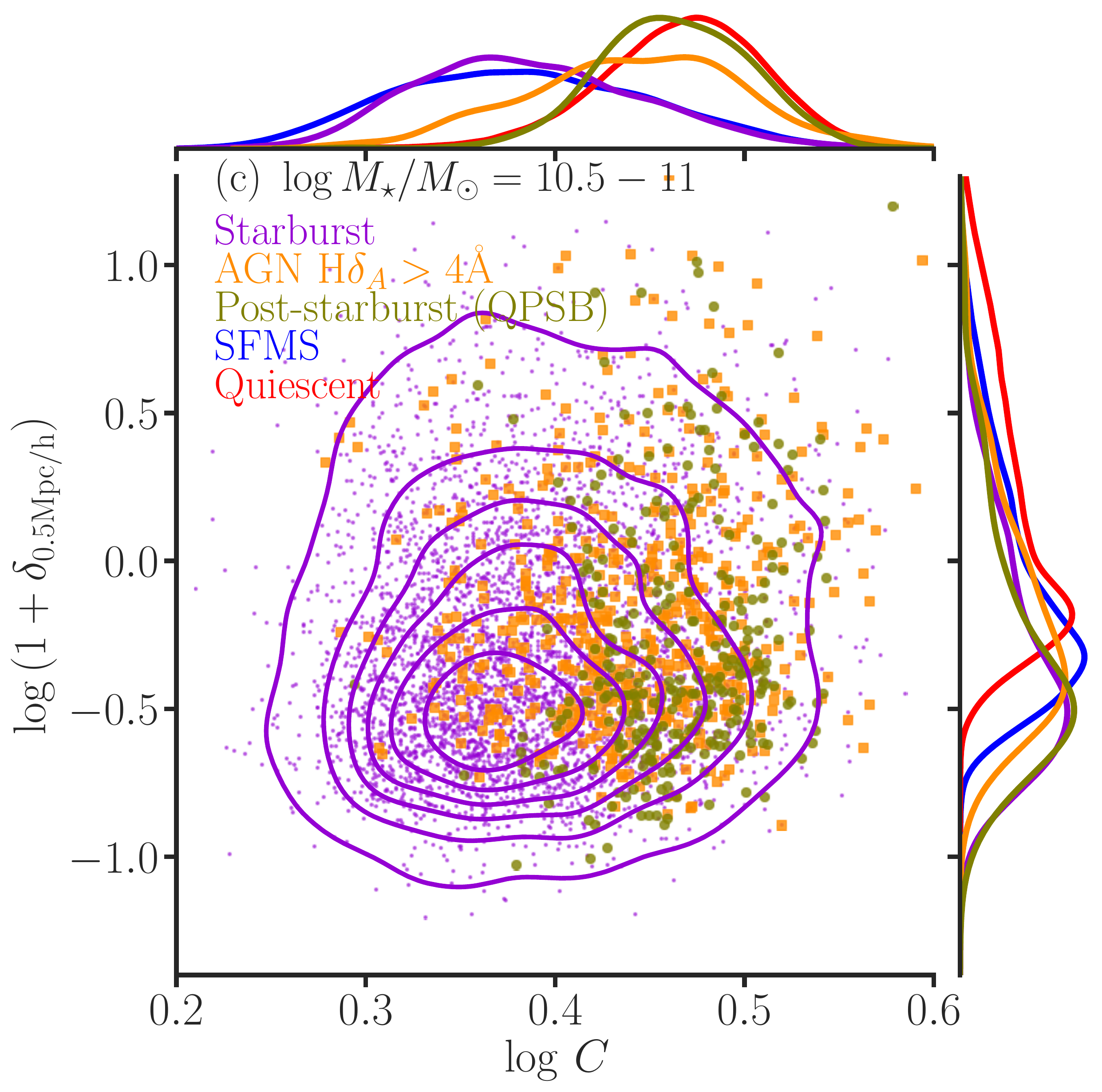}{0.48\textwidth}{}
\fig{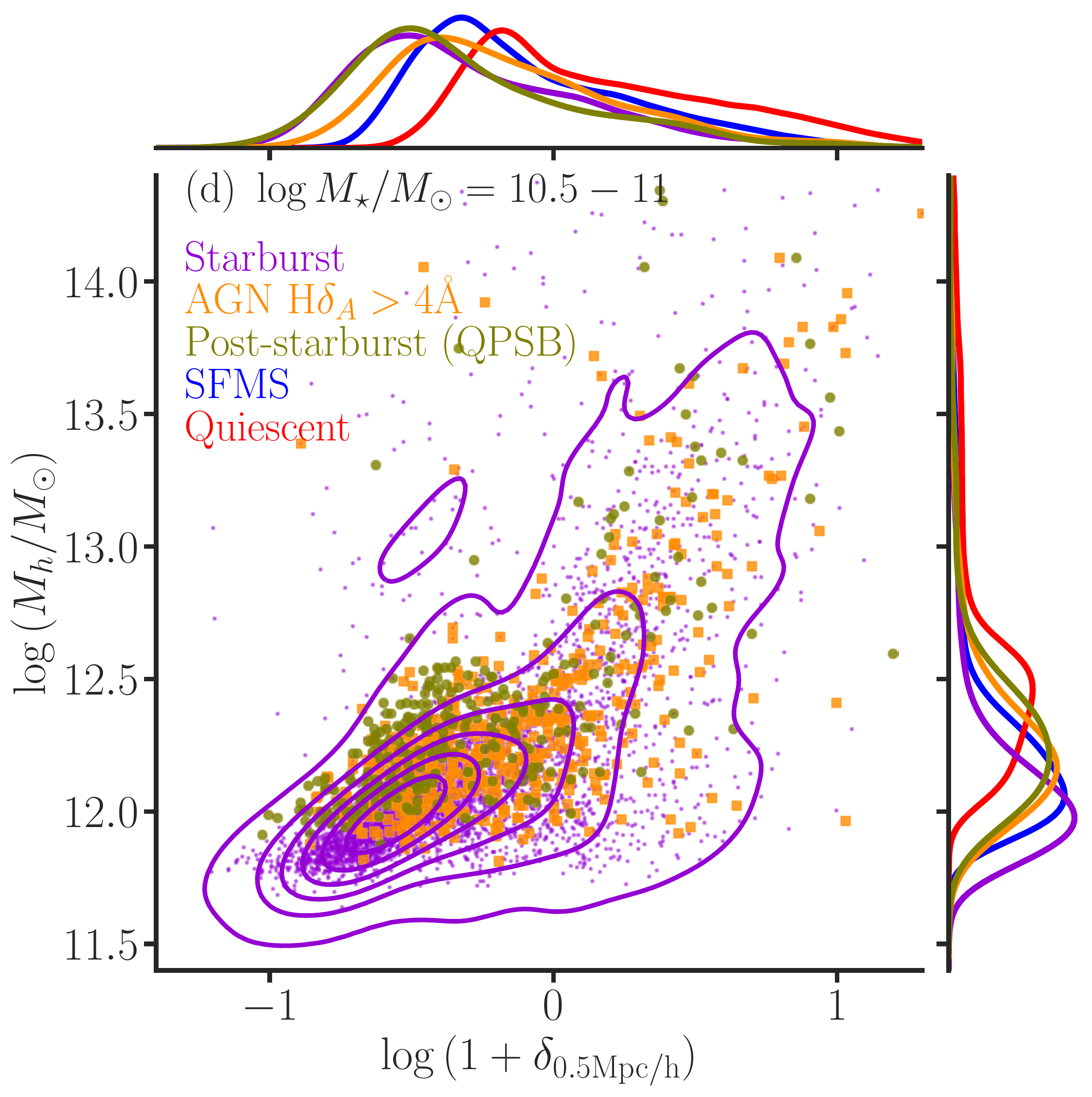}{0.48\textwidth}{}}
\caption{Comparing the stellar velocity dispersions ($\sigma_\star$), $z$-band concentration indices ($C$), $\deltahalf$, and $M_h$ from \citet{Lim+17} of starbursts, strong-H$\delta_A$ AGNs, QPSBs, SFMS galaxies, and QGs. Although the first three samples overlap in the plotted parameters, starbursts have similar $\sigma_\star$ and $C$ as SFMS galaxies (some are disc-dominated/less centrally concentrated), while QPSBs and strong-H$\delta_A$ AGNs have similar structures as QGs (i.e., most are bulge-dominated). Starbursts also have lower $M_h$ and $\deltahalf$ than QGs. Their structures correlate weakly ($\rho \approx 0.2$) with $M_h$ and $\deltahalf$. \label{fig:morphSB}}
\end{figure*}

Because the starburst to QPSB evolution is rapid ($\lesssim 1$\,Gyr), galaxies that are on this evolution track are expected to have similar $M_\star$, environment, and structure. We saw in the previous subsection that starbursts, strong-H$\delta_A$ AGNs, and QPSBs live in broadly similar environments, albeit exhibiting some subtle differences. Comparing their structure together with their environments (Figure~\ref{fig:morphSB}) shows that (1) starbursts have a wide range of $\sigma_\star$ and $C$ and the typical starburst is less centrally concentrated than QPSBs. (2) The distribution of $\sigma_\star$ and $C$ of QPSBs and strong-H$\delta_A$ AGNs are more similar to those of QGs than to those of SFMS galaxies, although their environments are more similar to the latter. (3) Most AGNs with H$\delta_A > 4$\,{\AA} have similar environments and structures as QPSBs, while only few are structurally similar to starbursts. (4) The correlation between environments and structures of these three populations is weak (e.g., $\rho \approx 0.2-0.3$ for the $\deltahalf$ vs. $\sigma_\star$). The implications of the trends in Figure~\ref{fig:morphSB} is that only some starbursts become QPSBs, and the mechanism that produces QPSBs is associated with structural transformations and AGN activities of galaxies in low-density environments. 

\begin{figure*}[ht!]
\gridline{\fig{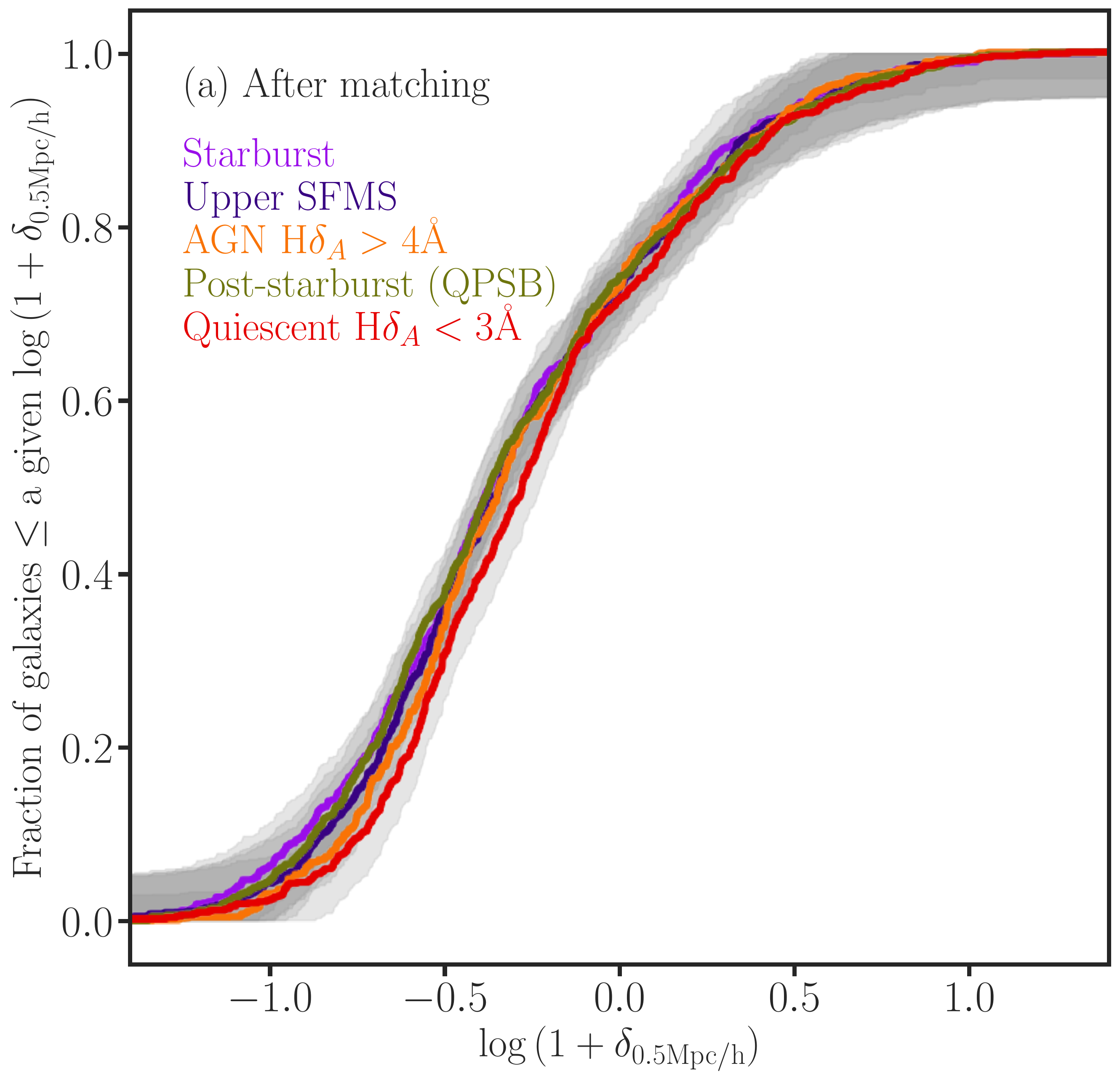}{0.32\textwidth}{}
        \fig{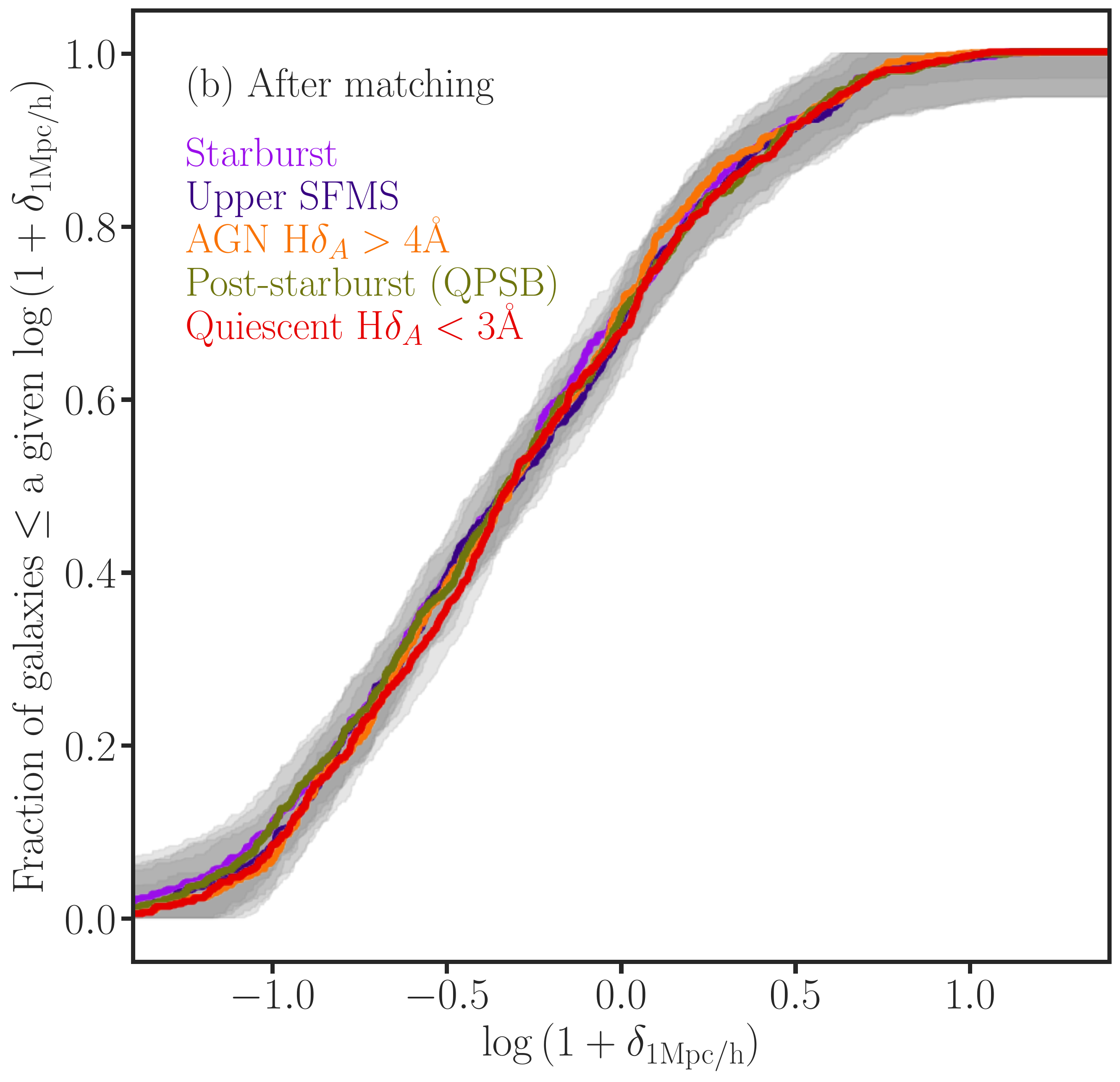}{0.32\textwidth}{}
        \fig{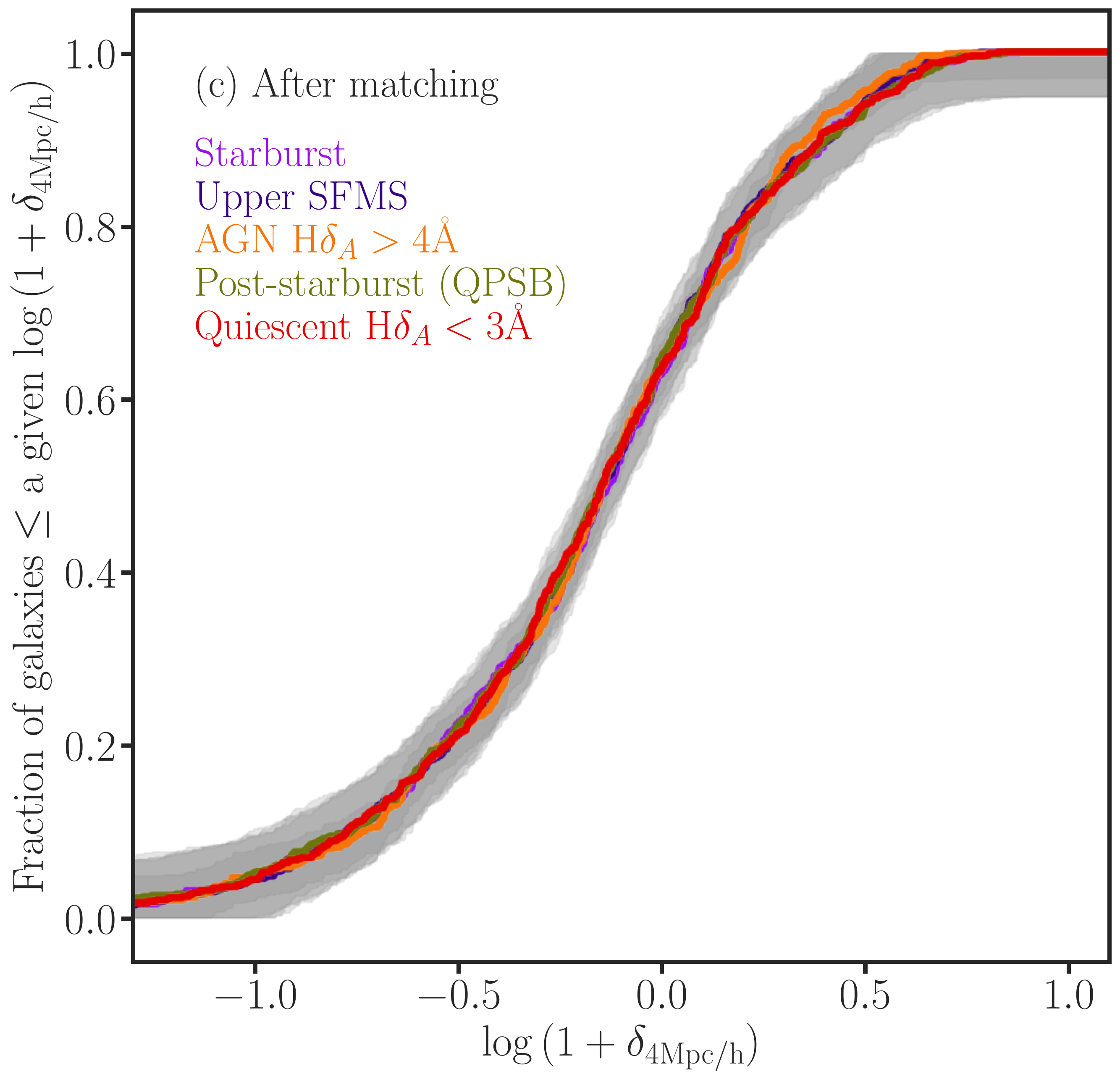}{0.32\textwidth}{}}
\gridline{\fig{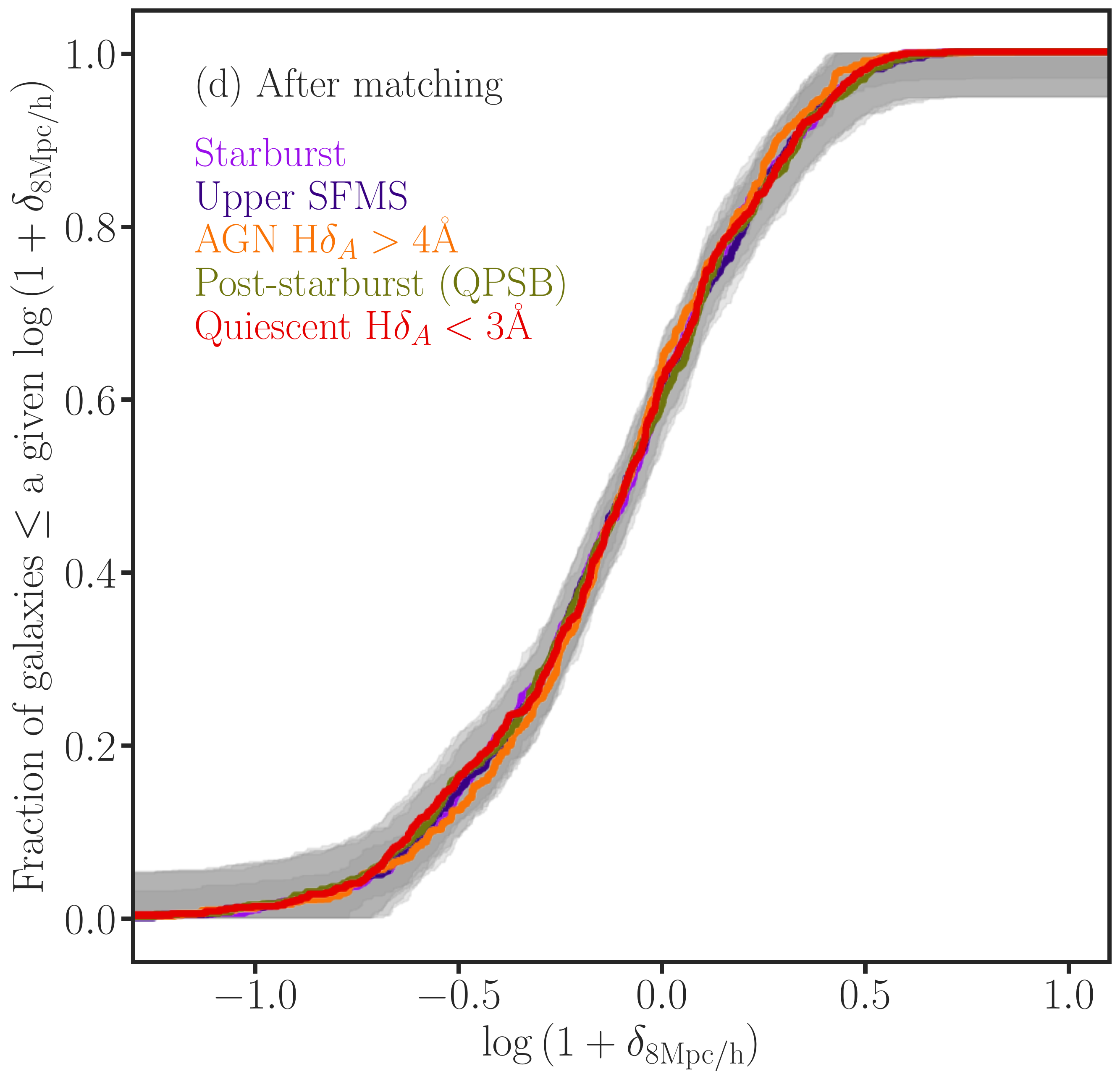}{0.32\textwidth}{}
        \fig{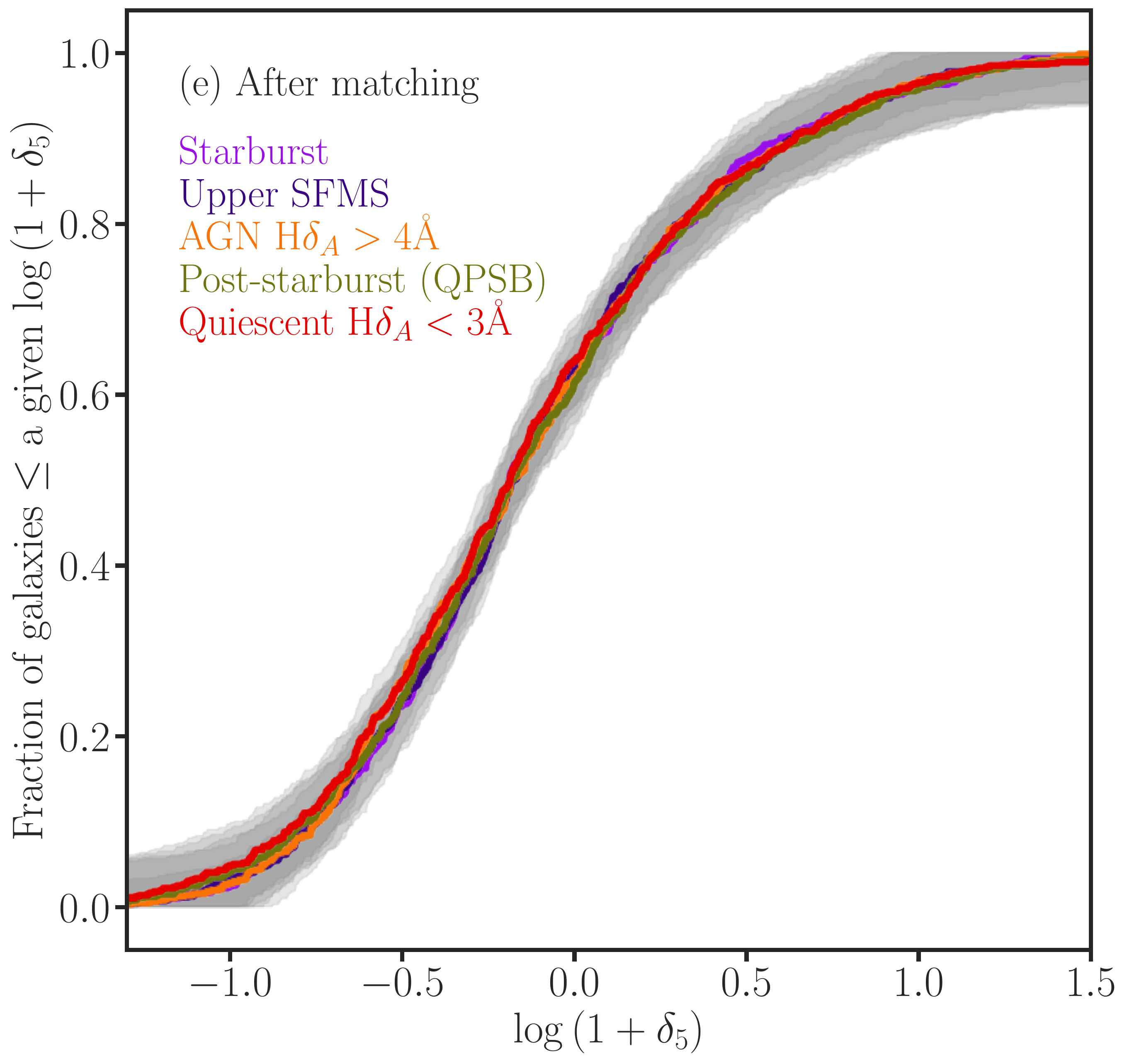}{0.32\textwidth}{} 
         \fig{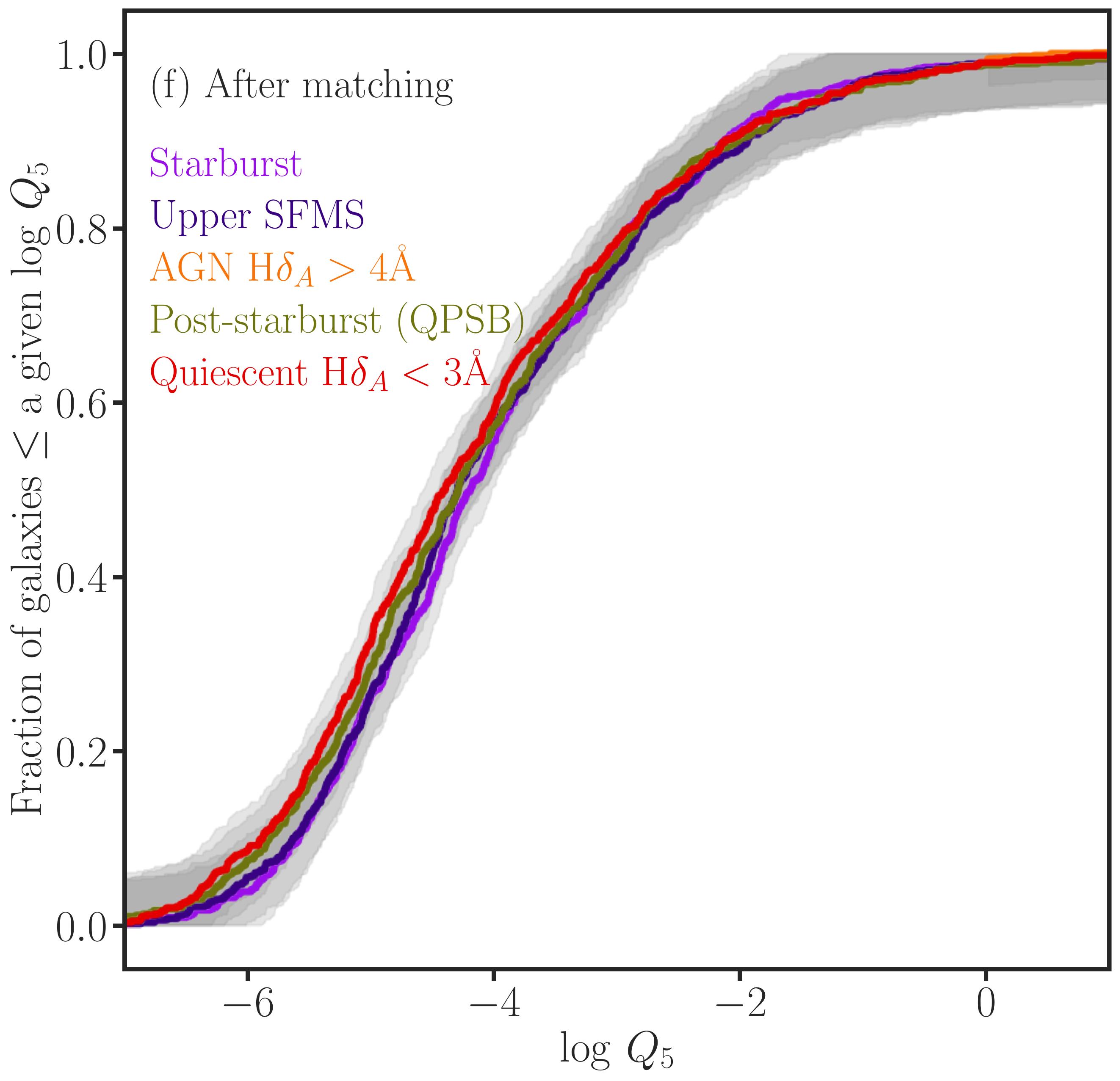}{0.32\textwidth}{}}      
\gridline{\fig{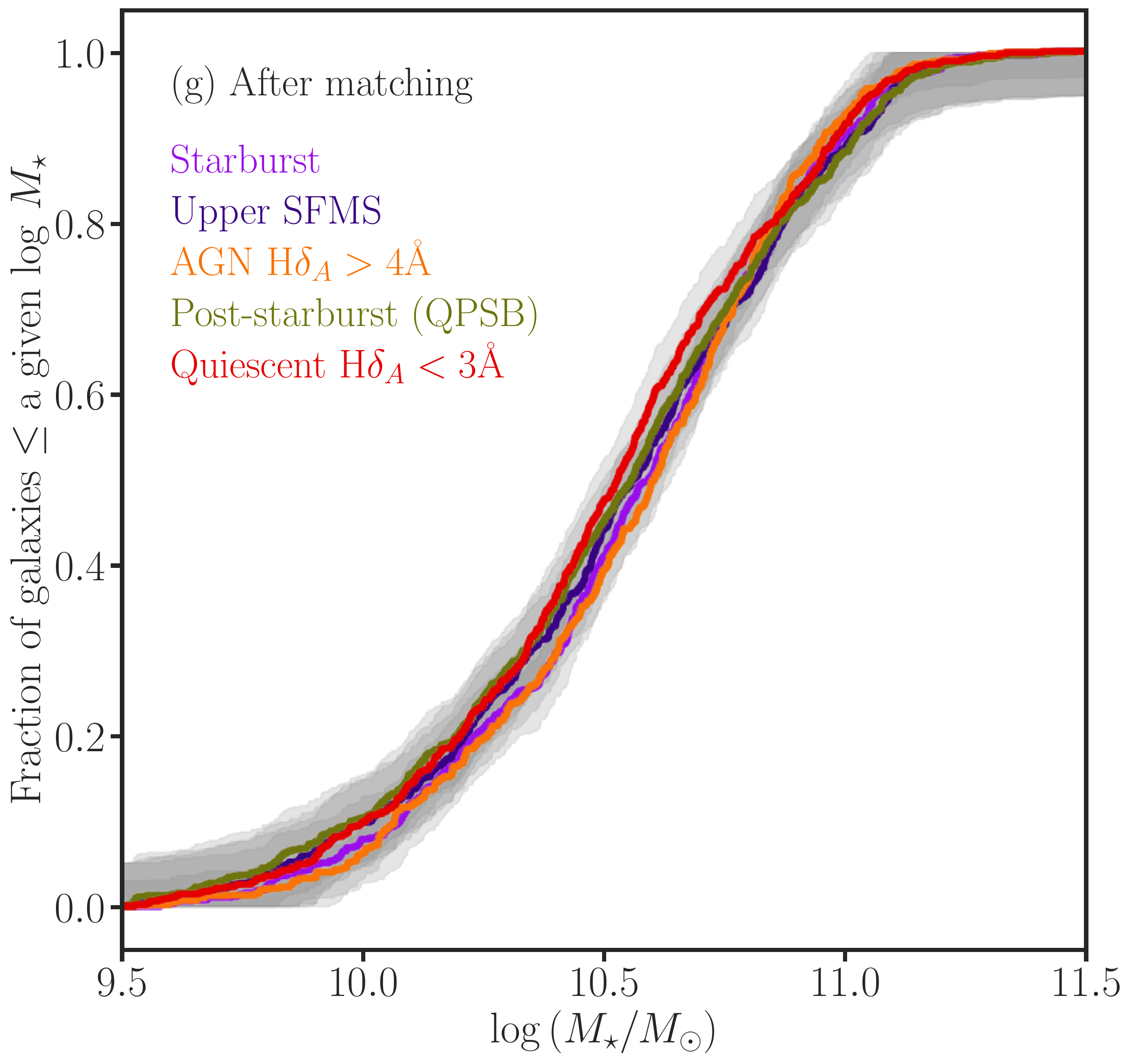}{0.32\textwidth}{}
        \fig{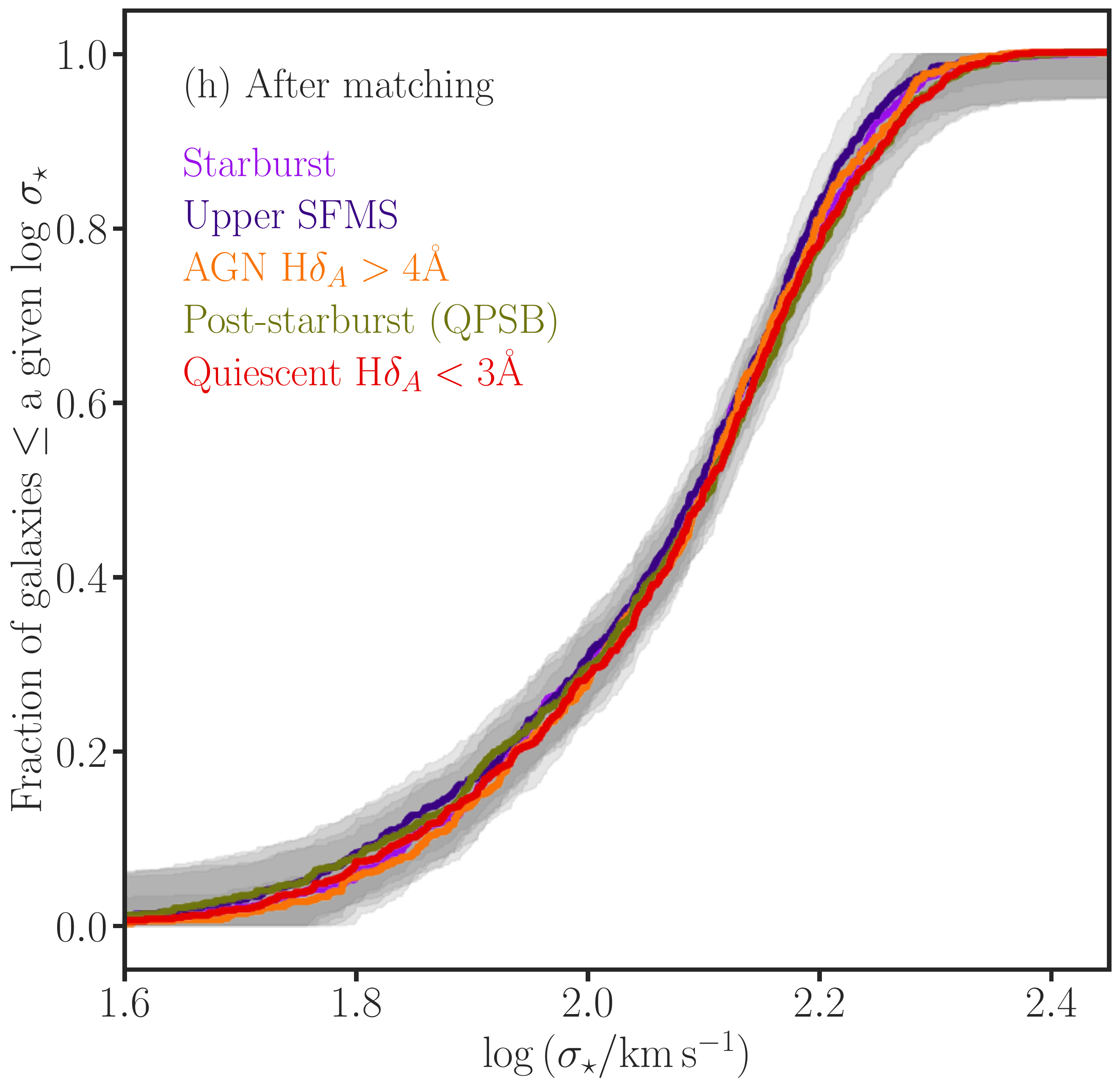}{0.32\textwidth}{} 
         \fig{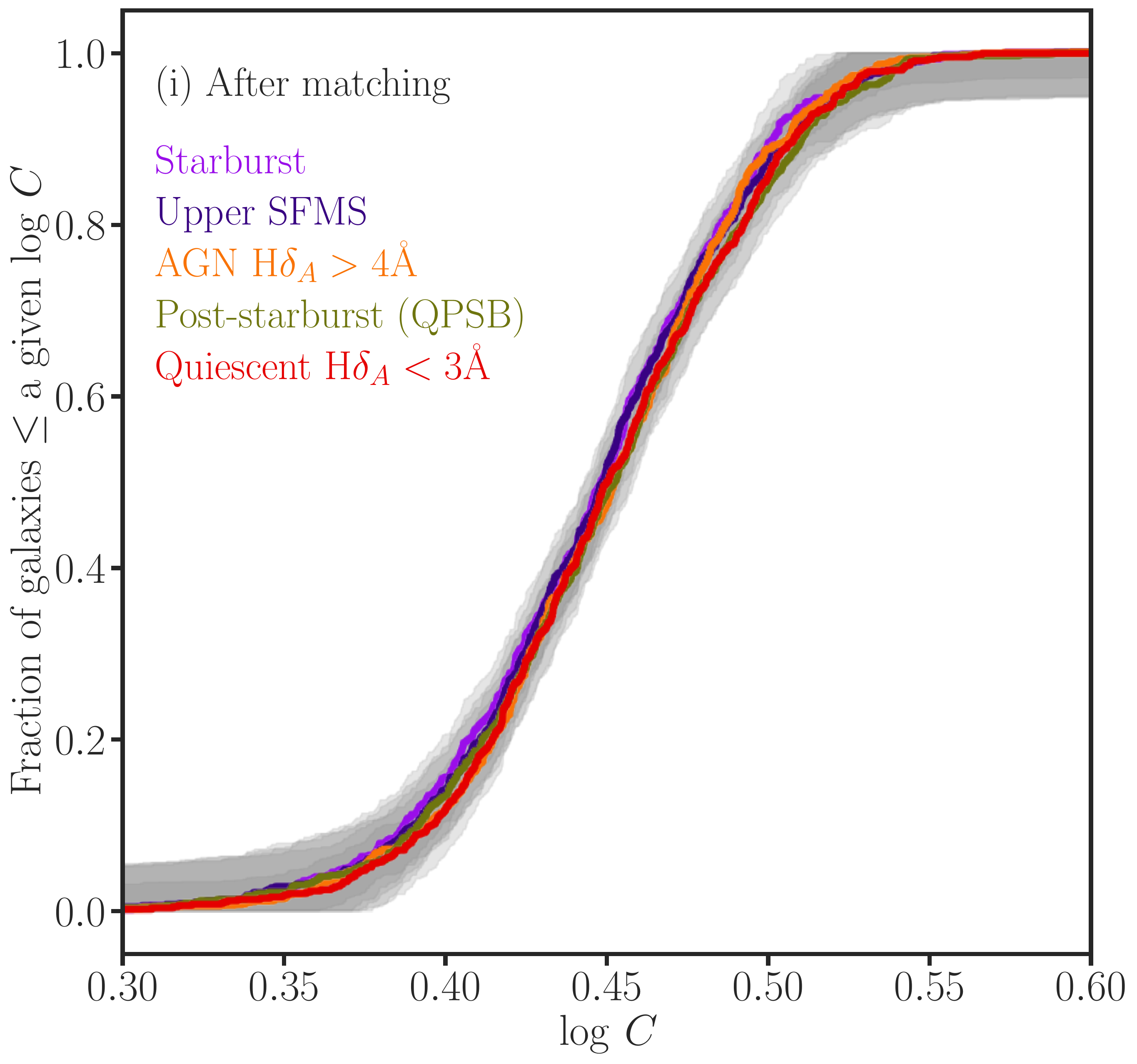}{0.32\textwidth}{}}                       
\caption{Simultaneously matching all QPSBs to starbursts, upper SFMS galaxies, strong-H$\delta_A$ AGNs, and QGs of similar multiscale ($0.5-8\,h^{-1}$Mpc) environmental indicators (panels (a) to (f)), $M_\star$, $\sigma_\star$, and $C$ (panels (g) to (i), respectively). The figure demonstrates that some starbursts, upper SFMS galaxies, and AGNs are progenitors of QPSBs, which in turn are progenitors of some QGs.\label{fig:matched_psb}}
\end{figure*}

Next, we match each QPSB with its closet starburst, upper SFMS galaxy, strong-H$\delta_A$ AGN, and QG simultaneously in multi-dimensional space of $M_\star$, $\sigma_\star$, $C$, and environments (Figure~\ref{fig:matched_psb}), thereby demonstrating that subsamples of SFGs, AGNs, and QGs that are similar to QPSBs in many aspects exist. In other words, we confirm using structure and multiscale environments  that QPSBs are descendants of some starbursts, as their spectral indices suggest, and in turn QPSBs are progenitors of some QGs. Moreover, most AGNs with H$\delta_A > 4$\,{\AA} are associated with the evolution from starbursts to QPSBs. They too are PSBs. All AGNs with H$\delta_A > 4$\,{\AA} can also be matched to some starbursts and upper SFMS galaxies of similar structures and environments. A significant fraction ($\sim 20\%-30\%$) of starbursts cannot be matched to QPSBs or QGs of similar $M_\star$, structures, and environments. Note that we only compare face-on galaxies with axis ratio $b/a > 0.3$ in this subsection. The $M_\star$ distributions before matching are given in Appendix~\ref{sec:appD}.

The matching procedure does not impose a limit on the maximum acceptable distance to remove cases where the first nearest neighbor may be too distant. This is not consequential in our case (Figure~\ref{fig:matched_psb}) as the samples are well-matched. Moreover, the nearest neighbor may be a match to one or more QPSBs. Excluding duplicate matches does not change the matched distributions of the different samples shown in Figure~\ref{fig:matched_psb}. Majority ($\sim 85\%$) of QPSBs have at least one match within a distance of $\sim 1 - 1.5$ in each matched sample; the 85\% (or 50\%) of the absolute difference between the variables used to match the samples is $\lesssim 0.2$\,dex (or $\lesssim 0.1$\,dex). Matching to the closest $k=3$ neighbors also gives similar results.

Given the considerable uncertainty of estimating the total $M_\star$ and the burst mass fraction in starbursts and their descendants, it is reasonable to match by the stellar mass at the time of observation and ignore the evolution of $M_\star$. The amount of stellar mass formed in a recent burst in massive local galaxies may be $5\%-30\%$. The molecular gas fractions of starbursts are $\sim 15\%-30\%$ \citep[e.g.,][]{Saintonge+17}. A significant fraction of PSBs may still have molecular gas fractions of $\sim 5\%-10\%$ \citep[e.g.,][]{French+15,Yesuf+17b}. Thus, the stellar mass added by a recent burst is likely below 30\% for most PSBs. The median (or 85\%) difference $\Delta \log\,M_\star$ between PSBs and the matched sample is $\sim 0.1$\,dex (or $\sim 0.2$\,dex).

\section{Discussion}\label{sec:disc}

Environment, galaxy mergers/interactions, rapid gas consumption during the starburst phase, stellar and AGN feedback have all been invoked to explain the origin of starbursts and/or QPSBs \citep{Mihos+96, Hopkins+08, Snyder+11,Davis+19, Lotz+21}. In this section, we discuss how the results presented in this paper may help discriminate among competing mechanisms. We also compare our results with those of previous studies. Future analyses of the multiscale environments and structures of starbursts and their descendent in cosmological simulations may provide more insightful interpretations.

\subsection{Starbursts and QPSBs are Rare in High-density Environments}

Majorities of starbursts and QPSBs are isolated/centrals and do not live in massive groups and clusters (Table~\ref{tab:fcentMh_sb} and Table~\ref{tab:richness2_sb}). Environmental effects such as ram pressure stripping cannot explain the origin of typical starbursts and QPSBs. In particular, for galaxies with $M_\star > 3 \times 10^{10} \, M_\odot$, $\sim 80\%-85\%$ starbursts and $\sim 85\%-95\%$ strong-H$\delta_A$ AGNs and QPSBs are centrals. Half of them have $M_h \approx 3 \times 10^{12} \, M_\odot$, $\sim 85\%-90$\% have $M_h \lesssim 10^{13} \,M_\odot$, and only $\sim 2\%-4\%$ have $M_h > 10^{14} \,M_\odot$. In contrast, $\sim 50\%-55\%$ of QGs have $M_h \lesssim 10^{13} \,M_\odot$ and  $\sim 8\%-14\%$ of QGs have $M_h > 10^{14} \, M_\odot$. These results broadly agree with expectations from cosmological simulations \citep{Davis+19,Lotz+21,Wilkinson+18}.

\citet{Davis+19} studied the cold interstellar medium of 1244 simulated QPSBs selected from the EAGLE cosmological simulations \citep{Schaye+15}. The simulated QPSBs have $M_\star > 3 \times 10^{9}\,M_\odot$, only 19\% of them are satellites or become satellites within $\pm 0.5$\,Gyr of the time of burst, and only $ \sim 10$\% show gas streams consistent with ram pressure stripping (compared to 5\% for the control sample). The satellite fraction of QPSBs in EAGLE agrees reasonably with our results. Satellite-related environmental effects are not the dominant formation channel of QPSBs in EAGLE simulations.
 
Similarly, \citet{Lotz+21} studied 647 QPSBs with $M_\star \ge 5 \times 10^{10}\,M_\odot$ in the Magneticum Pathfinder cosmological hydrodynamical simulations. The authors found that their simulated QPSBs at $z \approx 0$ live in low-mass halos, similar to stellar mass-matched SFGs, but unlike mass-matched QGs. More quantitatively, 89\% of QPSBs, 85\% SFGs, and 79\% of QGs in the simulation have $M_h < 10^{13}\,M_\odot$. In contrast, only 2\% of QPSBs, 3\% of SFGs, and 7\% QGs have $M_h > 10^{14}\,M_\odot$. 

\citet{Wilkinson+18} selected 196 starbursts at $z = 0.15$ with $M_\star > 10^{9}\,M_\odot$ in Illustris simulations using similar definition ($\delSSFR > 0.6$\,dex) to ours. After dividing their sample based on the merger history within 2\, Gyr prior to the starburst, the authors found that the mean values of $M_h$ for the merger starbursts and non-merger starbursts are $M_h = 3.4 \pm 2.0 \times 10^{12} \,M_\odot$ and $M_h = 6.1 \pm 2.1 \times 10^{12} \,M_\odot$, respectively. These values are not significantly discrepant with the observational estimates. Specifically, noting that most of the starbursts in Illustris have $M_\star <  10^{10}\,M_\odot$, we find, for the same $M_\star$ threshold, a median (15\%, 85\%) $M_h$ value of $\log M_h \approx  1\,(0.6, 3) \times 10^{12} M_\odot$ based on \citet{Tinker21}'s catalog. Similarly, the median values for \citet{Lim+17} is catalog is $M_h \approx  0.5\,(0.4, 2) \times 10^{12} M_\odot$. In addition, \citet{Wilkinson+18} found that their starbursts are located in the lower-density environments (under density within 0.5\,Mpc or 1\,Mpc as shown in their Figure 4) compared to random samples from Illustris, consistent with our results. Given our comprehensive results on multiscale environments of starbursts, a future study can definitely do better comparisons, preferably by also classifying the observed starbursts into mergers and non-mergers to confirm or refute the predicted trend.

Using EAGLE cosmological simulations, \citet{Matthee+19} found that the SFMS residuals at $z = 0.1$ arise from fluctuations on short ($<0.2-2$\,Gyr) and long ($\sim10$\,Gyr) timescales. The long-timescale fluctuations are related to $M_h$ and halo formation time. At $M_\star < 10^{10}\,M_\odot$, EAGLE predicted a clear trend that halos that form later have galaxies with higher SSFRs in them. The simulated starbursts (SSFR $\gtrsim 0.1$\,Gyr) at $M_\star < 10^{10}\,M_\odot$ have lower halo formation time and higher $M_h$ than SFMS galaxies at similar $M_\star$. At least qualitatively, this agrees with our result that the starbursts with $M_\star < 10^{10}\,M_\odot$ have lower environmental densities than SFMS galaxies do. It is expected that haloes that assembled more recently are substantially less clustered than those that assembled earlier \citep[][]{Gao+05,Wechsler+18}. \citet{Berti+21} also found that galaxies above the SFMS are less clustered than those below, at fixed $M_\star$. Our results are consistent with short and long-term changes in $\delSSFR$ as indicated by the existence of galaxies evolving from starbursts to QPSBs and by the clear difference in environments of upper SFMS and lower SFMS galaxies, quantified in several ways.

\subsection{Mergers and Interactions may Trigger Some Starbursts}

As discussed in the introduction, the merger hypothesis is a viable pathway for some galaxies to evolve from SFGs $\rightarrow$ starbursts $\rightarrow$ AGNs $\rightarrow$ QPSBs $\rightarrow$ QGs. Furthermore, it is a common view that classical bulges and ellipticals form mainly by dissipative, gas-rich mergers, while pseudobulges form by disc-related, secular processes \citep{Kormendy+04,Brooks+16}. Gas-rich mergers lead to central mass concentration and may destroy discs \citep{Bournaud+05, Martin+18}. The starbursts, QPSBs, and AGNs with high $C$ and $\sigma_\star$ (Figure~\ref{fig:morphSB}) are classical bulges or ellipticals \citep{Yesuf+20a}. Thus, demonstrating the existence of a population of galaxies along the aforementioned evolutionary sequence of similar $M_\star$, environments, $C$, and $\sigma_\star$ provides indirect support for the merger hypothesis. On the other hand, a large fraction of nearby starbursts have low $C$ and $\sigma_\star$, unlike most QPSBs and QGs. These pseudobulge starbursts likely were not mergers. Likewise, \citet{Dressler+13} and \citet{Abramson+13} found that most starbursts in their sample $z \approx 0.4$ are not centrally concentrated, just like the SFGs in their sample. These starbursts were identified by their anomalously strong H$\delta$ absorption or \ion{O}{2} emission. Most of them, however, do not lie significantly above SFMS\footnote{The SFRs in \citet{Dressler+13} and \citet{Abramson+13} were estimated using \emph{Spitzer}/MIPS 24\,$\mu$m and/or emission lines (\ion{O}{2}, H$\beta$ or/and H$\alpha$).}. Their structures were quantified using S\'{e}rsic indices in $J$- and $K$-bands. Using $C$ and S\'{e}rsic index based on $K$-band images from the GAMA survey \citep{Kelvin+12}, we have confirmed that the structures of $z < 0.2$ starbursts (and upper SFMS galaxies) are very different from QGs. Therefore, the low-concentration starbursts in our sample do not simply fade and become QPSBs and QGs in a short period of time -- a major structural rearrangement is required. Unlike \citet{Dressler+13}'s work, this study does not find that QPSBs live in similar environments as QGs at $z < 0.2$.

Well-separated galaxy interactions do not account for most starbursts, although they may trigger some starbursts. As shown in Figure~\ref{fig:Q5_sb},  less than 15\% (8\%) of massive galaxies above $10^{10}\,M_\odot$ have first nearest neighbors within $r_{p,1} < 200$\,kpc (100\,kpc). The fraction of starbursts with $r_{p,1} < 200$\,kpc is marginally different ($\lesssim 2\%-3\%$) from that of SFMS galaxies of similar $M_\star$. Likewise, \citet{Yamauchi+08} found that only $\sim 8\%$ of QPSBs have nearby companion within 50\,kpc compared to 5\% for normal galaxies. Therefore, tidal interactions are not the dominant mechanism for triggering starbursts although they may trigger some starbursts \citep[e.g., ][]{ Li+08,Patton+20}. 

In summary, mergers may drive the evolution of some galaxies from SFGs $\rightarrow$ starbursts $\rightarrow$ AGNs $\rightarrow$ QPSBs. The existence of a significant number of isolated and diffuse/pseudobulge starbursts indicates that not all starbursts are triggered by mergers/interactions. What else can be a triggering mechanism? Next, we discuss the possible role of gas accretion modulated by environment and feedback in triggering starbursts. This hypothesis will be tested rigorously in the future.

\subsection{Relationship Between Environment and Gas Accretion or AGN Activity}

Observational estimates indicate that gas accretion rates from minor mergers are not enough ($\sim 0.1-0.3\,M_\odot\,\mathrm{yr}^{-1}$) to sustain the observed SFR in local galaxies \citep{Sancisi+08, DiTeodoro+14}. Moreover, halo gas estimates made using background quasars of 17 low-$z$ starbursts or PSBs indicates a total gas mass of $\sim 3-6 \times 10^{9}\,M_\odot$ in their the circum-galactic medium \citep{Heckman+17}. These galaxies are mostly normal late-type galaxies. ``A few appear to be interacting with companions, but few (if any) appear to be recent or on-going mergers." Galaxy simulations also indicate that gas infall rates onto dark matter halos are dominated by the diffuse component over the merger contribution, at least for galaxies with $M_h < 10^{13}\,M_\odot$ \citep{Fakhouri+10,Wright+21}. 

Gas accretions onto halos or galaxies are regulated by the environment and feedback \citep[e.g.,][]{vandeVoort+17, Brennan+18, Correa+18}. Based on predictions of cosmological simulations, plenty of cold gas is expected in the halos of starbursts like ours. Most starbursts in our sample have $M_h \approx \mathrm{few} \times 10^{12}\,M_\odot$ and live in the low-density environments, which are conducive for gas accretion onto their halos. As discussed in the introduction, starbursts are also overwhelmingly asymmetric/disturbed. Diffuse gas accretion rates of a few $M_\odot\,\mathrm{yr^{-1}}$ may plausibly explain the ubiquity of asymmetries in strongly star-forming galaxies \citep{Keres+05,Bournaud+05,Jog+09, Yesuf+21}. According to cosmological simulations, gas is accreted onto halos in hot and cold modes. The contribution of the hot mode increases with $M_h$ as haloes become massive enough to efficiently shock heat the accreting gas \citep{Keres+05,Dekel+06}. The median hot fraction increases sharply around $M_h \approx 10^{12}\, M_\odot$ and almost all the gas is accreted in the hot mode for halos above $10^{13}\,M_\odot$ \citep{Wright+21}. Starbursts and upper SFMS galaxies in EAGLE simulations have higher gas accretion rates than galaxies below SFMS. Furthermore, gas accretion rates in massive halos and in dense environments ($\delta_{10}$) in EAGLE are strongly suppressed, especially for satellite galaxies at smaller halo-centric distances \citep{vandeVoort+17}. However, the accretion rates of simulated, massive, central galaxies (in similar $M_\star$ range as ours) are not appreciably impacted by their environments. They have gas accretion rates comparable to their SFRs. EAGLE and other simulations also predict that stellar and AGN feedback can curtail gas accretion rates onto galaxies significantly \citep[e.g.,][]{Brennan+18, Correa+18}. We have shown that strong-H$\delta_A$ AGNs have similar large and small-scale environments, $C$, and $\sigma_\star$ (black hole mass) to those of starbursts and QPSBs of similar $M_\star$. This is expected if AGN feedback acts as a maintenance mode, preventing cooling flows onto these galaxies. 

Apart from the maintenance role, however, AGNs may not primarily impact the evolution from starbursts to QPSBs; it is rather driven by gas consumption. First, a significant time delay is observed between starbursts and AGNs that show PSB spectral signatures \citep{Wild+10, Yesuf+14}. Second, PSBs are a mixture of gas-poor and gas-rich systems, and about half of QPSBs have large amounts of cold gas and dust \citep{Zwaan+13, French+15, Rowlands+15, Alatalo+16, Yesuf+17b, LiZhihui+19, YesufHo20b,Bezanson+22}, indicating that AGN feedback did not completely remove or destroy them. Strong AGNs in normal SFGs are also not necessarily gas-deficient \citep{Ho+08, Husemann+17, Rosario+18, Shangguan+20, YesufHo20a}. Furthermore, no consensus has been reached on the effects of AGN-driven outflows in galaxies in general or in PSBs in particular \citep[][and references therein]{Yesuf+17a, Baron+22}. Some simulations also found that AGNs do not clear out gas in PSBs \citep[][but see \citet{Lotz+21}]{Snyder+11,Davis+19}, and not all simulated PSBs are gas-deficient. 

Bulge-dominated central/isolated galaxies of similar $M_\star$ and structures can be evolutionarily linked through a life cycle wherein the molecular gas amount already inside galaxies regulates their SFRs and levels of AGN activity \citep{YesufHo20a}. These galaxies first consume their gas mostly through bursty star formation, then pass through a transition phase of intermediate gas richness in which star formation and AGNs coexist, before  retiring as gas-poor QGs. Strong AGNs are gas-rich; neither their gas reservoirs nor their abilities to form stars seem to be impacted instantaneously by AGN feedback. The molecular gas masses of these galaxies were estimated using an empirical relation among molecular gas, dust absorption, and metallicity \citep{YesufHo19}. Incidentally, galaxies in low-density environments show more dust absorption than those at high-density environments \citep{Kauffmann+04}.  \citet{YesufHo20a}'s sample includes starbursts and PSBs, although most galaxies in their sample are ordinary. In addition, using the same empirical calibration, \citet{YesufHo20b} found median (15\%, 85\%) molecular gas mass and fraction of $\log\,M_\mathrm{H_2}/M_\odot = 9.3\,(9.1, 9.6)$ and $f_{H2} = 4\,(2.5, 8)$\%, respectively, for strong-H$\delta_A$ Seyferts with $\Mhigh$. Therefore, strong AGNs are gas-rich regardless of their SFHs. Simulations also indicate that gas content of galaxies do not change within $\sim 1\,$Gyr due to halo gas accretion \citep{Scholz-Diaz+21}. Gas amount already in the galaxies determines the level of star formation and AGN activity. Environment and feedback regulate its long term replenishments and the cooling rate of halo gas.

\citet{Sabater+15} inferred from SDSS observations that the large-scale environments ($\delta_{10}$) and galaxy interactions only affect AGN activity indirectly by influencing the central gas supply in the nuclear region. After controlling for the effects of $M_\star$ and central SFR (which is correlated with cold gas), the authors found that the effects of the environment on AGN fraction and AGN luminosity are minimal.

Lastly, previous estimates of $M_h$ based on AGN clustering, weak lensing, or galaxy groups found that optically selected AGNs reside preferentially in halos of $M_h \approx \mathrm{few} \times 10^{12}\,h^{-1} M_\odot$ \citep{Mandelbaum+09,ZhangZiwen+21}, consistent with our finding. Obviously, the discussion above does not do justice to the rich literature on AGN environments, with a plethora of differing results \citep[see references in][]{Sabater+15,Man+19,ZhangZiwen+21}. We add a new result that AGNs with $H\delta_A  > 4${\AA} live in lower-density environments than AGNs with $H\delta_A  < 3${\AA} do. The strong-$H\delta_A$ AGNs are only $\sim 10\%$ of narrow-line AGNs.

\subsection{Comparison with Previous Studies of Starbursts and QPSBs at $z < 0.3$}

The advent of large redshift surveys such as the SDSS led to the discovery that QPSBs are not particularly prevalent in clusters or high-density environments, but rather live in a wide range of environments \citep{Zabludoff+96,Quintero+04, Blake+04, Goto05b, Hogg+06, Helmboldt+08, Yan+09,Paccagnella+19}. With few exceptions, most these studies reached a similar conclusion as ours. Namely, QPSBs live mainly in lower density environments than QGs. Although the environments of QPSBs are widely studied, only few studies \citep{Owers+07, Pawlik+18} attempted to directly compare their environments with those of starbursts and/or AGNs using the same datasets and methodology. Next, we give an overview of these studies and compare our results.

\citet{Quintero+04} studied the environments of 1143 k+a galaxies in SDSS at $z=0.05-0.2$, which were identified as outliers in the diagram of H$\alpha$ emission EW and the A/K stellar population fraction. The authors found that k+a galaxies in their sample do not primarily live in the high-density environments or clusters typical of bulge-dominated galaxies. In particular, \citet{Quintero+04} found that the mean overdensity of galaxies around k+a galaxies, measured within $1\,h^{-1}$ Mpc and $8\,h^{-1}$ scales are similar to the mean overdensities of disk-dominated galaxies (S\'{e}rsic $n < 2$). However, the comparison samples were not mass-matched, and the average values were taken over a wide range in $M_\star$, potentially washing out the mass trends (Figure~\ref{fig:massdel18} and Figure~\ref{fig:del18_psb}). 

In the follow-up study of \citet{Quintero+04}'s k+a sample, \citet{Hogg+06} studied the fractional abundance of the k+a galaxies and comparison samples as function of environmental density. The authors quantified environment by measuring: (1) number densities of galaxies in 8\,$h^{-1}$Mpc radius, (2) transverse distances to nearest Virgo-like galaxy clusters, and (3) the distance to the nearest neighbor galaxy, $r_{p,1}$. They found that the fractions of k+a galaxies depend on the environmental indicators above in similar way to those of SFGs. \citet{Hogg+06} concluded that: (1) starbursts/QPSBs are not triggered by external tidal impulses from close passages of massive galaxies (because they did not find strong dependance of k+a fraction with $r_{p,1}$). (2) A small fraction of QPSBs are created by ICM interactions on infall into clusters (because the k+a fraction does not vary strongly with the distance from the cluster center, particularly outside of the virial radius). Although \citet{Hogg+06} did not study starbursts directly, we reach similar conclusions. Similarly, in 521 SDSS clusters at $z < 0.1$ studied by \citet{vonderLinden+10}, the fraction of galaxies that have strong Balmer absorption lines is independent of the distance from the cluster center; it is only marginally significant within $0.2R_{200}$ and for $M _\star > 3 \times 10^{10}M_\odot$.

\citet{Owers+07} studied the environment and spatial clustering of 418 starbursts at $z < 0.16$ selected from the 2dFGRS and compared them with random control samples matched in redshift and $r$-band apparent magnitude. The environmental indicators they used include $r_{p,1}$, the tidal parameter, and $\delta_{5}$. The authors found that the distribution the tidal parameter and $r_{p,1}$ of starbursts differ significantly ($p=.0017$ and $p=.015$, respectively) from those of the random control samples; a higher proportion of starbursts have $\log\,Q_1 > -1$ than the control sample and 38\% of starbursts have bright neighbors within $\sim 20$\,kpc. The authors concluded that galaxy-galaxy mergers/interactions are important in triggering $\sim 30\%-40\%$ starbursts in their samples, and that the rest of starbursts, in which environmental influences are not obvious, must be internally driven. On the other hand, the authors found that starbursts are less clustered on $\sim 1-15\,h^{-1}$\,Mpc scales compared to control samples, in agreement with our results. Their starbursts were selected using H$\alpha$ EW. After showing that the 56 QPSBs in 2dFGRS studied by \citet{Blake+04} have broadly similar environmental properties as starbursts, \citet{Owers+07} surmised that only a small fraction of their starbursts evolve through the k+a phase. In comparison, our analysis is done using samples of starbursts and QPSBs that are more than ten times larger than those of the two studies. It also controls for effects of $M_\star$ and structure better. We find only marginal excess of neighbors around starbursts in our data (Figure~\ref{fig:Q5_sb}). In the future, we plan to improve our analysis by combining spectroscopic and photometric redshifts of nearby neighbors of starbursts. From cursory visual inspections our sample, we do not think that 38\% of our starbursts have close but well-separated neighbors within $\sim 20$\,kpc. 

\citet{Helmboldt+08} showed that $M_\star$, $\sigma_\star$, and $r_{p,4}$ of early-type SFGs are similar to those of QPSBs ($N= 435$, $z = 0.05-0.08$, and H$\delta_A > 2\,${\AA}). To visually identify E/S0 morphology, the authors only selected bright ($m_r < 16$) SFGs that are nearby enough that their spiral arms or prominent discs would be apparent in their $g$-band images. The authors found the fractions of early-type SFGs and QPSBs are similar, and they increase in parallel with $r_{n,4}$ (i.e., with decreasing environmental density). The environments of these galaxies are more similar to those of typical SFGs than to those of most early-type QGs, in agreement with our results. Unlike \citet{Helmboldt+08}, we do not restrict the comparison to bright SFGs and do not exclude AGNs from our analysis. Our selection is not based on morphology, but we agree that QPSBs have E/S0 morphology as indicated by their $C$ and $\sigma_\star$.

Furthermore, \citet{Pawlik+18} studied 189 strong Balmer absorption-line galaxies, including those that are QPSBs or AGNs. This sample is taken from the SDSS and has $z < 0.05$. About 80\% of galaxies in this sample have $M_\odot < 3 \times 10^{10}\, M_\odot$. The authors studied the SFHs, structures, and environments (as quantified by $\delta_{5}$) of different classes of galaxies with strong Balmer absorption lines to ascertain whether or not they are evolutionarily connected. The goals of the authors were similar to ours except that the authors were also interested in measuring faint merger signatures (asymmetries), which necessitated restricting their sample to $z < 0.05$. Consequently, their sample is much smaller and different from ours. Nevertheless, the authors showed that low-mass ($9.5 < \log\,(M/M_\odot) < 10.5$), QPSBs have similar distribution of $\delta_{5}$ as SFGs. At high-mass ($\log\,(M/M_\odot) > 10.5$), the authors tentatively found QPSBs are preferentially found in lower density environments than the control sample of SFGs. We conclusively confirm these results with a larger sample. More specifically, having split our sample into two $M_\star$ ranges as in \citet{Pawlik+18} and also combining upper SFMS and lower SFMS galaxies together, we find that low-mass QPSBs have a similar $\delta_{5}$ distribution to that of low-mass SFGs of (AD test $p > .25$), whereas high-mass QPSBs have a different $\delta_{5}$ distribution from that of high-mass SFGs (AD test $p = .005$). A similar result is also shown in Figure~\ref{fig:del5_psb}. 

\citet{Pawlik+18} also found that QPSBs and strong Balmer-line AGNs have similar $r$-band Sersic indices, SFHs, and $\delta_{5}$. This also agrees with our more compressive results on the structures ($\sigma_\star$ and $C$) and environments of the two populations quantified several ways, not only using $\delta_{5}$. Therefore, \citet{Pawlik+18}'s and our results confirm that these two populations have the same physical origin and are the same class of galaxies, i.e., PSBs. Our work goes further to demonstrate that these two populations can be evolutionary linked to a subset of starbursts, (upper) SFMS galaxies, and QGs that have similar structures and multiscale environments. \citet{Pawlik+18} did not study starbursts; their randomly selected SFG control sample did not make distinctions between SFGs of different $\delSSFR$ as this study does. 

Furthermore, \citet{Pawlik+18} proposed $M_\star$ dependent evolutionary scenarios that lead to QPSBs (see their Figure 12 and \cite{Dressler+13}). Our results do not fully support these scenarios but let us explain what they are. In the low-mass range, (a) after a violent triggering event (e.g., galaxy mergers), a SFG experiences a starburst and a morphological transformation, which are likely followed by a strong AGN activity, and then the galaxy becomes a bulge-dominated QG; (b) a less violent mechanism leads to a starburst that fades more gradually through a star-forming PSB phase, and it may or may not lead to a morphological transformation. Depending on the remaining gas reservoir, the galaxy either returns to the SFMS or becomes a QPSB or PSB AGN and subsequently joins the red sequence. In high-mass range, the aforementioned scenario (a) also happens. Besides, (c) an already quiescent galaxy experiences a relatively weak burst due to a minor merger, after which it passes through a brief PSB phase only to return back to the red sequence. In other words, to explain some PSBs, the authors proposed a cycle of rejuvenation starting from QGs but not passing through SFMS at high $M_\star$, and a cycle of minor burst originating from SFMS but not passing through the red sequence at low $M_\star$. Our analysis is not inconsistent with the proposed cycle of burst at low-mass; some low-mass starbursts have low environmental density and low central concentration, unlike QPSBs. However, the cycle of burst without morphological transformation may also happen in the high-mass range; some high-mass starbursts also do not evolve to QPSBs because they have low central concentration, although the environments of the two populations are similar at the high-mass range. Moreover, the our analysis does not support the proposed cycle of rejuvenation. First, this paper shows that QPSBs in all $M_\star$ ranges can be matched to starbursts, SFMS galaxies, strong-H$\delta_A$ AGNs, and QGs. Second, it shows that the environments of QGs, as a whole, are very different from those starbursts and PSBs. Therefore, a rejuvenation of QGs is not a dominant mechanism in originating QPSBs.

\subsection{Caveats}

The SFR used in this work is based on UV-optical-mid IR SED fitting, and it is sensitive to changes on $\sim 100$ Myr timescale. Although H$\alpha$ emission is more sensitive, this study does not use it to define starbursts. \citet{Baron+22} recently showed that H$\alpha$-based SFR can be orders of magnitude smaller than the FIR-based SFR, suggesting that some highly obscured starbursts can be misclassified as transition PSBs if H$\alpha$ is used \citep[see also][]{Poggianti+00}. In addition, spatial gradients in star formation could be such that the H$\alpha$ measured in SDSS fiber may not account for the star formation in outskirts of galaxies. Although the SED-based SFRs are sufficient for our purpose, future work may investigate their sensitivity to burst amplitude, mass fraction, and burst duration. It is already shown that these SFRs may overestimate the instantaneous SFRs in PSBs \citep{Hayward+14, Salim+16, YesufHo20b}. Differences in star formation timescales can cause H$\alpha$ to be absent during a PSB phase, while UV and mid-IR emission still persist after the O stars have died off. Thus, our QPSBs and strong-H$\delta_A$ AGNs are selected independent of their SFR estimates from their SEDs. The upper SFMS galaxies that are matched to QPSBs are consistent with being either low-amplitude starbursts or fading starbursts. We do not distinguish between the two cases at the moment.

It should be noted that the SDSS fiber spectroscopy only cover $\sim$ 3\,kpc radius at $z = 0.1$. Our work and the aforementioned studies could only select QPSBs based on their central stellar populations. Using MaNGA integral field survey data, \citet{Chen+19} found 31 galaxies with central post-starburst (CPSBs) regions, 37 galaxies with off-centre ring-like post-starburst regions (RPSB), and 292 galaxies with irregular PSBs regions. The authors found that CPSBs have suppressed star formation throughout their bulge and disc, while RPSBs have recently suppressed star formation in their outer regions and active nuclear star formation. Most RPSBs are located in lower SFMS (median $D_n(4000)$ = 1.4), whereas CPSBs are mainly located in the green valley. RPSBs have similar kinematics ($v_\mathrm{rot}/\sigma_\star$) as a control sample of non-PSBs matched in $M_\star$ and global $D_n(4000)$ index. Most of them have S\'{e}rsic index $n < 2$ (median $n=1.6$). \citet{Chen+19} provided two evidence that RPSBs and CPSBs are produced by different physical mechanisms. First, the two populations have very different SFHs (mass-weighted ages and  spectral indices) at all radii. Second, the CPSBs have lower $v_\mathrm{rot}/\sigma_\star$ at all radii. We checked that the environments of RPSBs are consistent with those of starbursts and QPSBs ($\log\,(1+\delta_\mathrm{0.5Mpc}) = 0.04\,(-0.7, 0.4)$, $\log\,(1+\delta_\mathrm{8Mpc}) = -0.1\,(-0.4, 0.2)$, $\log\, M_h/M_\odot = 12.0\,(11.3, 12.3)$, 75\% are centrals, and $\log\, M_\star/M_\odot = 10.1\,(9.5, 10.6)$).

The SDSS spectroscopic sample is incomplete because it is flux-limited and because of the fiber collision -- the 55$^{\prime\prime}$ minimum separation requirement between spectroscopic fibers. The former affects galaxies in our sample with $M_\star \lesssim 10^{10}\, M_\odot$ at $z \gtrsim 0.1$, while the latter affects galaxies that reside in groups or are pairs. Restricting the redshifts of galaxies in our two lowest mass ranges to $z < 0.1$ does not change the main conclusions. The fraction of missing pairs in \citet{Tempel+14}'s catalog is about 8\%. \citet{Lim+17}'s group catalog incorporates redshifts for the missing galaxies from different sources to achieve about 98\% completeness. \citet{Tinker21} assigned the spectroscopic redshift and other properties of the nearest-neighbor galaxy when they are not measured. The environmental indicators measured by this author use the SDSS DR17, which has improved redshift completeness. Furthermore, there are good agreements between the measurements based on SDSS and similar measurements based on the Galaxy And Mass Assembly (GAMA) survey \citep{Brough+13}; the GAMA survey is more complete and deeper than SDSS. Although it is important to ameliorate the effects of survey incompleteness in the future (e.g., using photometric redshifts and deeper redshift surveys), they are not large enough to change the conclusions of this study \citep[see also][]{Hogg+06, Luo+14}. Lastly, halo mass estimates in the group catalogs clearly need improvements and their discrepancies need a resolution.

\section{Summary and Conclusions}\label{sec:conc}

Using SDSS data, we study how star formation and black hole activity depend on environments in galaxies at $z=0.02-0.16$ and with $M_\star = 3\times 10^9 - 3\times 10^{11}\,M_\odot$. In particular, we check the consistency of the evolution from SFGs $\rightarrow$ starbursts $\rightarrow$ AGNs $\rightarrow$ QPSBs $\rightarrow$ QGs using multiple environmental indicators, $C$, and $\sigma_\star$ for galaxies in four $M_\star$ bins of 0.5\,dex. The following are our conclusions: 

\begin{itemize}

\item All QPSBs can be matched to some SFGs, starbursts, AGNs, and QGs that have similar $M_\star$, $C$, $\sigma_\star$, and multiscale environments. The environments of QPSBs are quite different from those of QGs. But, they are broadly similar to those SFGs at the scale of $\gtrsim 2\,$Mpc.

\item The environments of starbursts are broadly similar to or slightly lower density than those of SFGs. Starbursts clearly do not reside in high density environments populated by most QGs. Their distributions of small-scale environments ($r_{p,\,1}$, $\deltahalf$, and $M_h$) are also significantly different from those of non-bursty SFGs; larger fractions of starbursts have fewer (massive) neighbors within $\lesssim 1\,h^{-1}$Mpc than normal SFGs.

\item The environments of AGNs with H$\delta_A > 4$\,{\AA} are similar to QPSBs. Moreover, their distributions of $C$ and $\sigma_\star$ are also similar. Therefore, AGNs with H$\delta_A > 4$\,{\AA} are also PSBs. They all can be matched to some starbursts that have similar $M_\star$, $C$, $\sigma_\star$, and multiscale environments. AGNs with H$\delta_A > 4$\,{\AA} live in lower density environments than those of AGNs with H$\delta_A < 3$\,{\AA}. The latter live in environments similar to those of lower SFMS galaxies.

\item Depending on the $M_\star$ range, $\sim 70\%-90$\% of starbursts, QPSBs, and AGNs with H$\delta_A > 4$\,{\AA} are isolated or central galaxies (are not satellites), and $\sim 85\%$ of them have $M_h < 10^{13}\,M_\odot$ and only $\sim 2\%-4\%$ have $M_h < 10^{14}\,M_\odot$. The distributions $M_h$ and central fractions of starbursts, QPSBs, and AGNs with H$\delta_A > 4$\,{\AA} are significantly different from those of QGs.

\item Unlike QGs, only a small fraction starbursts, QPSBs, and AGNs with H$\delta_A > 4$\,{\AA} are found in clusters and rich groups. For example, in $\Mhigh$ range, $65\%$ of starbursts are isolated, $19\%$ are pairs, $9\%$ have 2 or 3 neighbors, $3\%$ are in rich groups, and  $\sim 1\%$ in clusters.

\item The distributions of environmental indicators of upper SFMS and lower SFMS are significantly different; consistent with previous finding  that the scatter of SFMS is not random \citep{Berti+21}. Furthermore, some upper SFMS and lower SFMS galaxies may be progenitors of QPSBs because they have similar structures and environments. These SFGs are either fading/weak starbursts or some QPSBs may have originated from a truncation of normal SFGs.

\item A significant fraction ($\sim 20\%-30\%$) of starbursts cannot be matched to QPSBs or QGs of similar $M_\star$, structures, and environments. Some starbursts have low $C$ and $\sigma_\star$, similar to late-type SFGs and/or their environments can be inconsistently lower density than those of QPSBs. Thus, a significant fraction of starbursts may not quench rapidly.

\item Most starbursts are not triggered by tidal interactions. About $80\%-90\%$ of starbursts do not have nearby neighbors within 200\,kpc and their tidal parameters are in fact slightly lower than SFMS galaxies. 

\item The mass overdensity within 0.5\,$h^{-1}$\,Mpc, $\deltahalf$, is significantly correlated with halo mass, $M_h$. It may be provide useful information when $M_h$ is hard to measure accurately.

\end{itemize}

Our work implies that the evolution from SFGs $\rightarrow$ starbursts $\rightarrow$ AGNs $\rightarrow$QPSBs $\rightarrow$ QGs or rejuvenation of QGs in the opposite direction is not a common path taken by typical galaxies at $z < 0.2$. This is mainly because the environments of typical QGs are very different from those of QPSB and starbursts. The importance of rapid quenching of starbursts to the build-up of red sequence need to be reassessed in the future taking their structures and environments into account, in addition to their inferred SFHs. Although we tried to be thorough in our analysis of multiscale environments, further checks should be done using upcoming and future data to confirm the intriguing dependance of $\delSSFR$ of SFGs on the small-scale environmental indicators. It would also be useful to improve our analysis with a selection of more complete samples of weak starbursts or/and PSBs with ongoing star formation or/and AGN activity. The merger hypothesis for the origin of PSBs is obviously incomplete. Future investigations need to quantify the prevalence of both major and minor mergers in starbursts and constrain alternative explanations. For example, a future study on the role of gas accretion in triggering starbursts in low-density environments will be useful since mergers cannot fully account for prevalence of disturbances/asymmetries and low-$C$ discs in starbursts \citep[e.g.,][]{Yesuf+21}. Likewise, a multivariate comparison of observed properties of starbursts and PSBs with similar measurements based on mock observations of cosmological simulations will significantly improve our understanding of the rapid quenching of star formation.

\emph{Software}: astropy \citep{AstropyI,AstropyII}, scipy \citep{Scipy}, scikit-learn \citep{scikit-learn}, and statmodels \citep{Statsmodels}

\begin{acknowledgments}

We thank the anonymous referee and John Silverman very much for their suggestions and comments that significantly improved the paper. H. Yesuf was supported by The Research Fund for International Young Scientists of NSFC (11950410492). Kavli IPMU is supported by World Premier International Research Center Initiative (WPI), MEXT, Japan 

Funding for SDSS-III has been provided by the Alfred P. Sloan Foundation, the Participating Institutions, the National Science Foundation, and the U.S. Department of Energy Office of Science. The SDSS-III web site is http://www.sdss3.org/.
SDSS-III is managed by the Astrophysical Research Consortium for the Participating Institutions of the SDSS-III Collaboration including the University of Arizona, the Brazilian Participation Group, Brookhaven National Laboratory, Carnegie Mellon University, University of Florida, the French Participation Group, the German Participation Group, Harvard University, the Instituto de Astrofisica de Canarias, the Michigan State/Notre Dame/JINA Participation Group, Johns Hopkins University, Lawrence Berkeley National Laboratory, Max Planck Institute for Astrophysics, Max Planck Institute for Extraterrestrial Physics, New Mexico State University, New York University, Ohio State University, Pennsylvania State University, University of Portsmouth, Princeton University, the Spanish Participation Group, University of Tokyo, University of Utah, Vanderbilt University, University of Virginia, University of Washington, and Yale University.
\end{acknowledgments}


\begin{thebibliography}{}
\expandafter\ifx\csname natexlab\endcsname\relax\def\natexlab#1{#1}\fi
\providecommand{\url}[1]{\href{#1}{#1}}
\providecommand{\dodoi}[1]{doi:~\href{http://doi.org/#1}{\nolinkurl{#1}}}
\providecommand{\doeprint}[1]{\href{http://ascl.net/#1}{\nolinkurl{http://ascl.net/#1}}}
\providecommand{\doarXiv}[1]{\href{https://arxiv.org/abs/#1}{\nolinkurl{https://arxiv.org/abs/#1}}}

\bibitem[{{Abramson} {et~al.}(2013){Abramson}, {Dressler}, {Gladders},
  {Oemler}, {Poggianti}, {Monson}, {Persson}, \& {Vulcani}}]{Abramson+13}
{Abramson}, L.~E., {Dressler}, A., {Gladders}, M.~D., {et~al.} 2013, \apj, 777,
  124, \dodoi{10.1088/0004-637X/777/2/124}

\bibitem[{{Aihara} {et~al.}(2011){Aihara}, {Allende Prieto}, {An}, {Anderson},
  {Aubourg}, {Balbinot}, {Beers}, {Berlind}, {Bickerton}, {Bizyaev}, {Blanton},
  {Bochanski}, {Bolton}, {Bovy}, {Brandt}, {Brinkmann}, {Brown}, {Brownstein},
  {Busca}, {Campbell}, {Carr}, {Chen}, {Chiappini}, {Comparat}, {Connolly},
  {Cortes}, {Croft}, {Cuesta}, {da Costa}, {Davenport}, {Dawson}, {Dhital},
  {Ealet}, {Ebelke}, {Edmondson}, {Eisenstein}, {Escoffier}, {Esposito},
  {Evans}, {Fan}, {Femen{\'\i}a Castell{\'a}}, {Font-Ribera}, {Frinchaboy},
  {Ge}, {Gillespie}, {Gilmore}, {Gonz{\'a}lez Hern{\'a}ndez}, {Gott}, {Gould},
  {Grebel}, {Gunn}, {Hamilton}, {Harding}, {Harris}, {Hawley}, {Hearty}, {Ho},
  {Hogg}, {Holtzman}, {Honscheid}, {Inada}, {Ivans}, {Jiang}, {Johnson},
  {Jordan}, {Jordan}, {Kazin}, {Kirkby}, {Klaene}, {Knapp}, {Kneib},
  {Kochanek}, {Koesterke}, {Kollmeier}, {Kron}, {Lampeitl}, {Lang}, {Le Goff},
  {Lee}, {Lin}, {Long}, {Loomis}, {Lucatello}, {Lundgren}, {Lupton}, {Ma},
  {MacDonald}, {Mahadevan}, {Maia}, {Makler}, {Malanushenko}, {Malanushenko},
  {Mandelbaum}, {Maraston}, {Margala}, {Masters}, {McBride}, {McGehee},
  {McGreer}, {M{\'e}nard}, {Miralda-Escud{\'e}}, {Morrison}, {Mullally},
  {Muna}, {Munn}, {Murayama}, {Myers}, {Naugle}, {Neto}, {Nguyen}, {Nichol},
  {O'Connell}, {Ogando}, {Olmstead}, {Oravetz}, {Padmanabhan},
  {Palanque-Delabrouille}, {Pan}, {Pandey}, {P{\^a}ris}, {Percival},
  {Petitjean}, {Pfaffenberger}, {Pforr}, {Phleps}, {Pichon}, {Pieri}, {Prada},
  {Price-Whelan}, {Raddick}, {Ramos}, {Reyl{\'e}}, {Rich}, {Richards}, {Rix},
  {Robin}, {Rocha-Pinto}, {Rockosi}, {Roe}, {Rollinde}, {Ross}, {Ross},
  {Rossetto}, {S{\'a}nchez}, {Sayres}, {Schlegel}, {Schlesinger}, {Schmidt},
  {Schneider}, {Sheldon}, {Shu}, {Simmerer}, {Simmons}, {Sivarani}, {Snedden},
  {Sobeck}, {Steinmetz}, {Strauss}, {Szalay}, {Tanaka}, {Thakar}, {Thomas},
  {Tinker}, {Tofflemire}, {Tojeiro}, {Tremonti}, {Vandenberg}, {Vargas
  Maga{\~n}a}, {Verde}, {Vogt}, {Wake}, {Wang}, {Weaver}, {Weinberg}, {White},
  {White}, {Yanny}, {Yasuda}, {Yeche}, \& {Zehavi}}]{Aihara+11}
{Aihara}, H., {Allende Prieto}, C., {An}, D., {et~al.} 2011, \apjs, 193, 29,
  \dodoi{10.1088/0067-0049/193/2/29}

\bibitem[{{Alatalo} {et~al.}(2016){Alatalo}, {Lisenfeld}, {Lanz}, {Appleton},
  {Ardila}, {Cales}, {Kewley}, {Lacy}, {Medling}, {Nyland}, {Rich}, \&
  {Urry}}]{Alatalo+16}
{Alatalo}, K., {Lisenfeld}, U., {Lanz}, L., {et~al.} 2016, \apj, 827, 106,
  \dodoi{10.3847/0004-637X/827/2/106}

\bibitem[{{Almaini} {et~al.}(2017){Almaini}, {Wild}, {Maltby}, {Hartley},
  {Simpson}, {Hatch}, {McLure}, {Dunlop}, \& {Rowlands}}]{Almaini+17}
{Almaini}, O., {Wild}, V., {Maltby}, D.~T., {et~al.} 2017, \mnras, 472, 1401,
  \dodoi{10.1093/mnras/stx1957}

\bibitem[{{Argudo-Fern{\'a}ndez} {et~al.}(2013){Argudo-Fern{\'a}ndez},
  {Verley}, {Bergond}, {Sulentic}, {Sabater}, {Fern{\'a}ndez Lorenzo}, {Leon},
  {Espada}, {Verdes-Montenegro}, {Santander-Vela}, {Ruiz}, \&
  {S{\'a}nchez-Exp{\'o}sito}}]{Argudo-Fernandez+13}
{Argudo-Fern{\'a}ndez}, M., {Verley}, S., {Bergond}, G., {et~al.} 2013, \aap,
  560, A9, \dodoi{10.1051/0004-6361/201321326}

\bibitem[{{Astropy Collaboration} {et~al.}(2013){Astropy Collaboration},
  {Robitaille}, {Tollerud}, {Greenfield}, {Droettboom}, {Bray}, {Aldcroft},
  {Davis}, {Ginsburg}, {Price-Whelan}, {Kerzendorf}, {Conley}, {Crighton},
  {Barbary}, {Muna}, {Ferguson}, {Grollier}, {Parikh}, {Nair}, {Unther},
  {Deil}, {Woillez}, {Conseil}, {Kramer}, {Turner}, {Singer}, {Fox}, {Weaver},
  {Zabalza}, {Edwards}, {Azalee Bostroem}, {Burke}, {Casey}, {Crawford},
  {Dencheva}, {Ely}, {Jenness}, {Labrie}, {Lim}, {Pierfederici}, {Pontzen},
  {Ptak}, {Refsdal}, {Servillat}, \& {Streicher}}]{AstropyI}
{Astropy Collaboration}, {Robitaille}, T.~P., {Tollerud}, E.~J., {et~al.} 2013,
  \aap, 558, A33, \dodoi{10.1051/0004-6361/201322068}

\bibitem[{{Astropy Collaboration} {et~al.}(2018){Astropy Collaboration},
  {Price-Whelan}, {Sip{\H{o}}cz}, {G{\"u}nther}, {Lim}, {Crawford}, {Conseil},
  {Shupe}, {Craig}, {Dencheva}, {Ginsburg}, {VanderPlas}, {Bradley},
  {P{\'e}rez-Su{\'a}rez}, {de Val-Borro}, {Aldcroft}, {Cruz}, {Robitaille},
  {Tollerud}, {Ardelean}, {Babej}, {Bach}, {Bachetti}, {Bakanov}, {Bamford},
  {Barentsen}, {Barmby}, {Baumbach}, {Berry}, {Biscani}, {Boquien}, {Bostroem},
  {Bouma}, {Brammer}, {Bray}, {Breytenbach}, {Buddelmeijer}, {Burke},
  {Calderone}, {Cano Rodr{\'\i}guez}, {Cara}, {Cardoso}, {Cheedella}, {Copin},
  {Corrales}, {Crichton}, {D'Avella}, {Deil}, {Depagne}, {Dietrich}, {Donath},
  {Droettboom}, {Earl}, {Erben}, {Fabbro}, {Ferreira}, {Finethy}, {Fox},
  {Garrison}, {Gibbons}, {Goldstein}, {Gommers}, {Greco}, {Greenfield},
  {Groener}, {Grollier}, {Hagen}, {Hirst}, {Homeier}, {Horton}, {Hosseinzadeh},
  {Hu}, {Hunkeler}, {Ivezi{\'c}}, {Jain}, {Jenness}, {Kanarek}, {Kendrew},
  {Kern}, {Kerzendorf}, {Khvalko}, {King}, {Kirkby}, {Kulkarni}, {Kumar},
  {Lee}, {Lenz}, {Littlefair}, {Ma}, {Macleod}, {Mastropietro}, {McCully},
  {Montagnac}, {Morris}, {Mueller}, {Mumford}, {Muna}, {Murphy}, {Nelson},
  {Nguyen}, {Ninan}, {N{\"o}the}, {Ogaz}, {Oh}, {Parejko}, {Parley}, {Pascual},
  {Patil}, {Patil}, {Plunkett}, {Prochaska}, {Rastogi}, {Reddy Janga},
  {Sabater}, {Sakurikar}, {Seifert}, {Sherbert}, {Sherwood-Taylor}, {Shih},
  {Sick}, {Silbiger}, {Singanamalla}, {Singer}, {Sladen}, {Sooley},
  {Sornarajah}, {Streicher}, {Teuben}, {Thomas}, {Tremblay}, {Turner},
  {Terr{\'o}n}, {van Kerkwijk}, {de la Vega}, {Watkins}, {Weaver}, {Whitmore},
  {Woillez}, {Zabalza}, \& {Astropy Contributors}}]{AstropyII}
{Astropy Collaboration}, {Price-Whelan}, A.~M., {Sip{\H{o}}cz}, B.~M., {et~al.}
  2018, \aj, 156, 123, \dodoi{10.3847/1538-3881/aabc4f}

\bibitem[{{Baldry} {et~al.}(2006){Baldry}, {Balogh}, {Bower}, {Glazebrook},
  {Nichol}, {Bamford}, \& {Budavari}}]{Baldry+06}
{Baldry}, I.~K., {Balogh}, M.~L., {Bower}, R.~G., {et~al.} 2006, \mnras, 373,
  469, \dodoi{10.1111/j.1365-2966.2006.11081.x}

\bibitem[{{Ball} {et~al.}(2008){Ball}, {Loveday}, \& {Brunner}}]{Ball+08}
{Ball}, N.~M., {Loveday}, J., \& {Brunner}, R.~J. 2008, \mnras, 383, 907,
  \dodoi{10.1111/j.1365-2966.2007.12627.x}

\bibitem[{{Balogh} {et~al.}(2004){Balogh}, {Eke}, {Miller}, {Lewis}, {Bower},
  {Couch}, {Nichol}, {Bland-Hawthorn}, {Baldry}, {Baugh}, {Bridges}, {Cannon},
  {Cole}, {Colless}, {Collins}, {Cross}, {Dalton}, {de Propris}, {Driver},
  {Efstathiou}, {Ellis}, {Frenk}, {Glazebrook}, {Gomez}, {Gray}, {Hawkins},
  {Jackson}, {Lahav}, {Lumsden}, {Maddox}, {Madgwick}, {Norberg}, {Peacock},
  {Percival}, {Peterson}, {Sutherland}, \& {Taylor}}]{Balogh+04}
{Balogh}, M., {Eke}, V., {Miller}, C., {et~al.} 2004, \mnras, 348, 1355,
  \dodoi{10.1111/j.1365-2966.2004.07453.x}

\bibitem[{{Balogh} {et~al.}(1999){Balogh}, {Morris}, {Yee}, {Carlberg}, \&
  {Ellingson}}]{Balogh+99}
{Balogh}, M.~L., {Morris}, S.~L., {Yee}, H.~K.~C., {Carlberg}, R.~G., \&
  {Ellingson}, E. 1999, \apj, 527, 54, \dodoi{10.1086/308056}

\bibitem[{{Bamford} {et~al.}(2009){Bamford}, {Nichol}, {Baldry}, {Land},
  {Lintott}, {Schawinski}, {Slosar}, {Szalay}, {Thomas}, {Torki}, {Andreescu},
  {Edmondson}, {Miller}, {Murray}, {Raddick}, \& {Vandenberg}}]{Bamford+09}
{Bamford}, S.~P., {Nichol}, R.~C., {Baldry}, I.~K., {et~al.} 2009, \mnras, 393,
  1324, \dodoi{10.1111/j.1365-2966.2008.14252.x}

\bibitem[{{Baron} {et~al.}(2022){Baron}, {Netzer}, {Lutz}, {Prochaska}, \&
  {Davies}}]{Baron+22}
{Baron}, D., {Netzer}, H., {Lutz}, D., {Prochaska}, J.~X., \& {Davies}, R.~I.
  2022, \mnras, 509, 4457, \dodoi{10.1093/mnras/stab3232}

\bibitem[{{Behroozi} {et~al.}(2022){Behroozi}, {Hearin}, \&
  {Moster}}]{Behroozi+22}
{Behroozi}, P., {Hearin}, A., \& {Moster}, B.~P. 2022, \mnras, 509, 2800,
  \dodoi{10.1093/mnras/stab3193}

\bibitem[{{Berti} {et~al.}(2021){Berti}, {Coil}, {Hearin}, \&
  {Behroozi}}]{Berti+21}
{Berti}, A.~M., {Coil}, A.~L., {Hearin}, A.~P., \& {Behroozi}, P.~S. 2021, \aj,
  161, 49, \dodoi{10.3847/1538-3881/abcc6a}

\bibitem[{{Bezanson} {et~al.}(2022){Bezanson}, {Spilker}, {Suess}, {Setton},
  {Feldmann}, {Greene}, {Kriek}, {Narayanan}, \& {Verrico}}]{Bezanson+22}
{Bezanson}, R., {Spilker}, J.~S., {Suess}, K.~A., {et~al.} 2022, \apj, 925,
  153, \dodoi{10.3847/1538-4357/ac3dfa}

\bibitem[{{Blake} {et~al.}(2004){Blake}, {Pracy}, {Couch}, {Bekki}, {Lewis},
  {Glazebrook}, {Baldry}, {Baugh}, {Bland-Hawthorn}, {Bridges}, {Cannon},
  {Cole}, {Colless}, {Collins}, {Dalton}, {De Propris}, {Driver}, {Efstathiou},
  {Ellis}, {Frenk}, {Jackson}, {Lahav}, {Lumsden}, {Maddox}, {Madgwick},
  {Norberg}, {Peacock}, {Peterson}, {Sutherland}, \& {Taylor}}]{Blake+04}
{Blake}, C., {Pracy}, M.~B., {Couch}, W.~J., {et~al.} 2004, \mnras, 355, 713,
  \dodoi{10.1111/j.1365-2966.2004.08351.x}

\bibitem[{{Blanton} \& {Berlind}(2007)}]{Blanton+07}
{Blanton}, M.~R., \& {Berlind}, A.~A. 2007, \apj, 664, 791,
  \dodoi{10.1086/512478}

\bibitem[{{Blanton} \& {Moustakas}(2009)}]{Blanton+09}
{Blanton}, M.~R., \& {Moustakas}, J. 2009, \araa, 47, 159,
  \dodoi{10.1146/annurev-astro-082708-101734}

\bibitem[{{Boselli} \& {Gavazzi}(2006)}]{Boselli+06}
{Boselli}, A., \& {Gavazzi}, G. 2006, \pasp, 118, 517, \dodoi{10.1086/500691}

\bibitem[{{Bournaud} {et~al.}(2005){Bournaud}, {Jog}, \&
  {Combes}}]{Bournaud+05}
{Bournaud}, F., {Jog}, C.~J., \& {Combes}, F. 2005, \aap, 437, 69,
  \dodoi{10.1051/0004-6361:20042036}

\bibitem[{{Brennan} {et~al.}(2018){Brennan}, {Choi}, {Somerville},
  {Hirschmann}, {Naab}, \& {Ostriker}}]{Brennan+18}
{Brennan}, R., {Choi}, E., {Somerville}, R.~S., {et~al.} 2018, \apj, 860, 14,
  \dodoi{10.3847/1538-4357/aac2c4}

\bibitem[{{Brinchmann} {et~al.}(2004){Brinchmann}, {Charlot}, {White},
  {Tremonti}, {Kauffmann}, {Heckman}, \& {Brinkmann}}]{Brinchmann+04}
{Brinchmann}, J., {Charlot}, S., {White}, S.~D.~M., {et~al.} 2004, \mnras, 351,
  1151, \dodoi{10.1111/j.1365-2966.2004.07881.x}

\bibitem[{{Brooks} \& {Christensen}(2016)}]{Brooks+16}
{Brooks}, A., \& {Christensen}, C. 2016, in Astrophysics and Space Science
  Library, Vol. 418, Galactic Bulges, ed. E.~{Laurikainen}, R.~{Peletier}, \&
  D.~{Gadotti}, 317, \dodoi{10.1007/978-3-319-19378-6\_12}

\bibitem[{{Brough} {et~al.}(2013){Brough}, {Croom}, {Sharp}, {Hopkins},
  {Taylor}, {Baldry}, {Gunawardhana}, {Liske}, {Norberg}, {Robotham}, {Bauer},
  {Bland-Hawthorn}, {Colless}, {Foster}, {Kelvin}, {Lara-Lopez},
  {L{\'o}pez-S{\'a}nchez}, {Loveday}, {Owers}, {Pimbblet}, \&
  {Prescott}}]{Brough+13}
{Brough}, S., {Croom}, S., {Sharp}, R., {et~al.} 2013, \mnras, 435, 2903,
  \dodoi{10.1093/mnras/stt1489}

\bibitem[{{Bruzual} \& {Charlot}(2003)}]{Bruzual+03}
{Bruzual}, G., \& {Charlot}, S. 2003, \mnras, 344, 1000,
  \dodoi{10.1046/j.1365-8711.2003.06897.x}

\bibitem[{{Burton} {et~al.}(2013){Burton}, {Jarvis}, {Smith}, {Bonfield},
  {Hardcastle}, {Stevens}, {Bourne}, {Baes}, {Brough}, {Cava}, {Cooray},
  {Dariush}, {De Zotti}, {Dunne}, {Eales}, {Hopwood}, {Ibar}, {Ivison},
  {Liske}, {Loveday}, {Maddox}, {Negrello}, {Smith}, \& {Valiante}}]{Burton+13}
{Burton}, C.~S., {Jarvis}, M.~J., {Smith}, D.~J.~B., {et~al.} 2013, \mnras,
  433, 771, \dodoi{10.1093/mnras/stt770}

\bibitem[{{Carollo} {et~al.}(2016){Carollo}, {Cibinel}, {Lilly}, {Pipino},
  {Bonoli}, {Finoguenov}, {Miniati}, {Norberg}, \& {Silverman}}]{Carollo+16}
{Carollo}, C.~M., {Cibinel}, A., {Lilly}, S.~J., {et~al.} 2016, \apj, 818, 180,
  \dodoi{10.3847/0004-637X/818/2/180}

\bibitem[{{Catinella} {et~al.}(2018){Catinella}, {Saintonge}, {Janowiecki},
  {Cortese}, {Dav{\'e}}, {Lemonias}, {Cooper}, {Schiminovich}, {Hummels},
  {Fabello}, {Ger{\'e}b}, {Kilborn}, \& {Wang}}]{Catinella+18}
{Catinella}, B., {Saintonge}, A., {Janowiecki}, S., {et~al.} 2018, \mnras, 476,
  875, \dodoi{10.1093/mnras/sty089}

\bibitem[{{Chabrier}(2003)}]{Chabrier03}
{Chabrier}, G. 2003, \pasp, 115, 763, \dodoi{10.1086/376392}

\bibitem[{{Chary} \& {Elbaz}(2001)}]{Chary+01}
{Chary}, R., \& {Elbaz}, D. 2001, \apj, 556, 562, \dodoi{10.1086/321609}

\bibitem[{{Chen} {et~al.}(2012){Chen}, {Kauffmann}, {Tremonti}, {White},
  {Heckman}, {Kova{\v{c}}}, {Bundy}, {Chisholm}, {Maraston}, {Schneider},
  {Bolton}, {Weaver}, \& {Brinkmann}}]{Chen+12}
{Chen}, Y.-M., {Kauffmann}, G., {Tremonti}, C.~A., {et~al.} 2012, \mnras, 421,
  314, \dodoi{10.1111/j.1365-2966.2011.20306.x}

\bibitem[{{Chen} {et~al.}(2019){Chen}, {Shi}, {Wild}, {Tremonti}, {Rowlands},
  {Bizyaev}, {Yan}, {Lin}, \& {Riffel}}]{Chen+19}
{Chen}, Y.-M., {Shi}, Y., {Wild}, V., {et~al.} 2019, \mnras, 489, 5709,
  \dodoi{10.1093/mnras/stz2494}

\bibitem[{{Cibinel} {et~al.}(2019){Cibinel}, {Daddi}, {Sargent}, {Le Floc'h},
  {Liu}, {Bournaud}, {Oesch}, {Amram}, {Calabr{\`o}}, {Duc}, {Pannella},
  {Puglisi}, {Perret}, {Elbaz}, \& {Kokorev}}]{Cibinel+19}
{Cibinel}, A., {Daddi}, E., {Sargent}, M.~T., {et~al.} 2019, \mnras, 485, 5631,
  \dodoi{10.1093/mnras/stz690}

\bibitem[{{Cooper} {et~al.}(2009){Cooper}, {Newman}, \& {Yan}}]{Cooper+09}
{Cooper}, M.~C., {Newman}, J.~A., \& {Yan}, R. 2009, \apj, 704, 687,
  \dodoi{10.1088/0004-637X/704/1/687}

\bibitem[{{Cooper} {et~al.}(2008){Cooper}, {Newman}, {Weiner}, {Yan},
  {Willmer}, {Bundy}, {Coil}, {Conselice}, {Davis}, {Faber}, {Gerke},
  {Guhathakurta}, {Koo}, \& {Noeske}}]{Cooper+08}
{Cooper}, M.~C., {Newman}, J.~A., {Weiner}, B.~J., {et~al.} 2008, \mnras, 383,
  1058, \dodoi{10.1111/j.1365-2966.2007.12613.x}

\bibitem[{{Correa} {et~al.}(2018){Correa}, {Schaye}, {van de Voort}, {Duffy},
  \& {Wyithe}}]{Correa+18}
{Correa}, C.~A., {Schaye}, J., {van de Voort}, F., {Duffy}, A.~R., \& {Wyithe},
  J. S.~B. 2018, \mnras, 478, 255, \dodoi{10.1093/mnras/sty871}

\bibitem[{{Cortese} {et~al.}(2021){Cortese}, {Catinella}, \&
  {Smith}}]{Cortese+21}
{Cortese}, L., {Catinella}, B., \& {Smith}, R. 2021, \pasa, 38, e035,
  \dodoi{10.1017/pasa.2021.18}

\bibitem[{{Croton} {et~al.}(2006){Croton}, {Springel}, {White}, {De Lucia},
  {Frenk}, {Gao}, {Jenkins}, {Kauffmann}, {Navarro}, \& {Yoshida}}]{Croton+06}
{Croton}, D.~J., {Springel}, V., {White}, S. D.~M., {et~al.} 2006, \mnras, 365,
  11, \dodoi{10.1111/j.1365-2966.2005.09675.x}

\bibitem[{{Darvish} {et~al.}(2016){Darvish}, {Mobasher}, {Sobral}, {Rettura},
  {Scoville}, {Faisst}, \& {Capak}}]{Darvish+16}
{Darvish}, B., {Mobasher}, B., {Sobral}, D., {et~al.} 2016, \apj, 825, 113,
  \dodoi{10.3847/0004-637X/825/2/113}

\bibitem[{{Davis} {et~al.}(2019){Davis}, {van de Voort}, {Rowlands},
  {McAlpine}, {Wild}, \& {Crain}}]{Davis+19}
{Davis}, T.~A., {van de Voort}, F., {Rowlands}, K., {et~al.} 2019, \mnras, 484,
  2447, \dodoi{10.1093/mnras/stz180}

\bibitem[{{Dekel} \& {Birnboim}(2006)}]{Dekel+06}
{Dekel}, A., \& {Birnboim}, Y. 2006, \mnras, 368, 2,
  \dodoi{10.1111/j.1365-2966.2006.10145.x}

\bibitem[{{Dekel} \& {Burkert}(2014)}]{Dekel+14}
{Dekel}, A., \& {Burkert}, A. 2014, \mnras, 438, 1870,
  \dodoi{10.1093/mnras/stt2331}

\bibitem[{{Di Teodoro} \& {Fraternali}(2014)}]{DiTeodoro+14}
{Di Teodoro}, E.~M., \& {Fraternali}, F. 2014, \aap, 567, A68,
  \dodoi{10.1051/0004-6361/201423596}

\bibitem[{{Donnari} {et~al.}(2021){Donnari}, {Pillepich}, {Joshi}, {Nelson},
  {Genel}, {Marinacci}, {Rodriguez-Gomez}, {Pakmor}, {Torrey}, {Vogelsberger},
  \& {Hernquist}}]{Donnari+21}
{Donnari}, M., {Pillepich}, A., {Joshi}, G.~D., {et~al.} 2021, \mnras, 500,
  4004, \dodoi{10.1093/mnras/staa3006}

\bibitem[{{Dressler}(1980)}]{Dressler+80}
{Dressler}, A. 1980, \apj, 236, 351, \dodoi{10.1086/157753}

\bibitem[{{Dressler} \& {Gunn}(1983)}]{Dressler+83}
{Dressler}, A., \& {Gunn}, J.~E. 1983, \apj, 270, 7, \dodoi{10.1086/161093}

\bibitem[{{Dressler} {et~al.}(2013){Dressler}, {Oemler}, {Poggianti},
  {Gladders}, {Abramson}, \& {Vulcani}}]{Dressler+13}
{Dressler}, A., {Oemler}, Augustus, J., {Poggianti}, B.~M., {et~al.} 2013,
  \apj, 770, 62, \dodoi{10.1088/0004-637X/770/1/62}

\bibitem[{{Elbaz} {et~al.}(2007){Elbaz}, {Daddi}, {Le Borgne}, {Dickinson},
  {Alexander}, {Chary}, {Starck}, {Brandt}, {Kitzbichler}, {MacDonald},
  {Nonino}, {Popesso}, {Stern}, \& {Vanzella}}]{Elbaz+07}
{Elbaz}, D., {Daddi}, E., {Le Borgne}, D., {et~al.} 2007, \aap, 468, 33,
  \dodoi{10.1051/0004-6361:20077525}

\bibitem[{{Ellison} {et~al.}(2013){Ellison}, {Mendel}, {Scudder}, {Patton}, \&
  {Palmer}}]{Ellison+13}
{Ellison}, S.~L., {Mendel}, J.~T., {Scudder}, J.~M., {Patton}, D.~R., \&
  {Palmer}, M. J.~D. 2013, \mnras, 430, 3128, \dodoi{10.1093/mnras/sts546}

\bibitem[{{Ellison} {et~al.}(2010){Ellison}, {Patton}, {Simard}, {McConnachie},
  {Baldry}, \& {Mendel}}]{Ellison+10}
{Ellison}, S.~L., {Patton}, D.~R., {Simard}, L., {et~al.} 2010, \mnras, 407,
  1514, \dodoi{10.1111/j.1365-2966.2010.17076.x}

\bibitem[{{Fabian}(2012)}]{Fabian+12}
{Fabian}, A.~C. 2012, \araa, 50, 455,
  \dodoi{10.1146/annurev-astro-081811-125521}

\bibitem[{{Fakhouri} \& {Ma}(2010)}]{Fakhouri+10}
{Fakhouri}, O., \& {Ma}, C.-P. 2010, \mnras, 401, 2245,
  \dodoi{10.1111/j.1365-2966.2009.15844.x}

\bibitem[{{French}(2021)}]{French+21}
{French}, K.~D. 2021, \pasp, 133, 072001, \dodoi{10.1088/1538-3873/ac0a59}

\bibitem[{{French} {et~al.}(2015){French}, {Yang}, {Zabludoff}, {Narayanan},
  {Shirley}, {Walter}, {Smith}, \& {Tremonti}}]{French+15}
{French}, K.~D., {Yang}, Y., {Zabludoff}, A., {et~al.} 2015, \apj, 801, 1,
  \dodoi{10.1088/0004-637X/801/1/1}

\bibitem[{{Gao} {et~al.}(2005){Gao}, {Springel}, \& {White}}]{Gao+05}
{Gao}, L., {Springel}, V., \& {White}, S. D.~M. 2005, \mnras, 363, L66,
  \dodoi{10.1111/j.1745-3933.2005.00084.x}

\bibitem[{{Genzel} {et~al.}(2015){Genzel}, {Tacconi}, {Lutz}, {Saintonge},
  {Berta}, {Magnelli}, {Combes}, {Garc{\'\i}a-Burillo}, {Neri}, {Bolatto},
  {Contini}, {Lilly}, {Boissier}, {Boone}, {Bouch{\'e}}, {Bournaud}, {Burkert},
  {Carollo}, {Colina}, {Cooper}, {Cox}, {Feruglio}, {F{\"o}rster Schreiber},
  {Freundlich}, {Gracia-Carpio}, {Juneau}, {Kovac}, {Lippa}, {Naab}, {Salome},
  {Renzini}, {Sternberg}, {Walter}, {Weiner}, {Weiss}, \& {Wuyts}}]{Genzel+15}
{Genzel}, R., {Tacconi}, L.~J., {Lutz}, D., {et~al.} 2015, \apj, 800, 20,
  \dodoi{10.1088/0004-637X/800/1/20}

\bibitem[{{G{\'o}mez} {et~al.}(2003){G{\'o}mez}, {Nichol}, {Miller}, {Balogh},
  {Goto}, {Zabludoff}, {Romer}, {Bernardi}, {Sheth}, {Hopkins}, {Castander},
  {Connolly}, {Schneider}, {Brinkmann}, {Lamb}, {SubbaRao}, \&
  {York}}]{Gomez+03}
{G{\'o}mez}, P.~L., {Nichol}, R.~C., {Miller}, C.~J., {et~al.} 2003, \apj, 584,
  210, \dodoi{10.1086/345593}

\bibitem[{Goodman(1965)}]{Goodman65}
Goodman, L.~A. 1965, Technometrics, 7, 247

\bibitem[{{Goto}(2005{\natexlab{a}})}]{Goto05}
{Goto}, T. 2005{\natexlab{a}}, \mnras, 357, 937,
  \dodoi{10.1111/j.1365-2966.2005.08701.x}

\bibitem[{{Goto}(2005{\natexlab{b}})}]{Goto05b}
{Goto}, T. 2005{\natexlab{b}}, \mnras, 360, 322,
  \dodoi{10.1111/j.1365-2966.2005.09036.x}

\bibitem[{{Haas} {et~al.}(2012){Haas}, {Schaye}, \& {Jeeson-Daniel}}]{Haas+12}
{Haas}, M.~R., {Schaye}, J., \& {Jeeson-Daniel}, A. 2012, \mnras, 419, 2133,
  \dodoi{10.1111/j.1365-2966.2011.19863.x}

\bibitem[{{Hayward} {et~al.}(2014){Hayward}, {Lanz}, {Ashby}, {Fazio},
  {Hernquist}, {Mart{\'\i}nez-Galarza}, {Noeske}, {Smith}, {Wuyts}, \&
  {Zezas}}]{Hayward+14}
{Hayward}, C.~C., {Lanz}, L., {Ashby}, M. L.~N., {et~al.} 2014, \mnras, 445,
  1598, \dodoi{10.1093/mnras/stu1843}

\bibitem[{{Heckman} {et~al.}(2017){Heckman}, {Borthakur}, {Wild},
  {Schiminovich}, \& {Bordoloi}}]{Heckman+17}
{Heckman}, T., {Borthakur}, S., {Wild}, V., {Schiminovich}, D., \& {Bordoloi},
  R. 2017, \apj, 846, 151, \dodoi{10.3847/1538-4357/aa80dc}

\bibitem[{{Helmboldt} {et~al.}(2008){Helmboldt}, {Walterbos}, \&
  {Goto}}]{Helmboldt+08}
{Helmboldt}, J.~F., {Walterbos}, R.~A.~M., \& {Goto}, T. 2008, \mnras, 387,
  1537, \dodoi{10.1111/j.1365-2966.2008.13229.x}

\bibitem[{{Henriques} {et~al.}(2017){Henriques}, {White}, {Thomas}, {Angulo},
  {Guo}, {Lemson}, \& {Wang}}]{Henriques+17}
{Henriques}, B. M.~B., {White}, S. D.~M., {Thomas}, P.~A., {et~al.} 2017,
  \mnras, 469, 2626, \dodoi{10.1093/mnras/stx1010}

\bibitem[{{Ho} {et~al.}(2008){Ho}, {Darling}, \& {Greene}}]{Ho+08}
{Ho}, L.~C., {Darling}, J., \& {Greene}, J.~E. 2008, \apj, 681, 128,
  \dodoi{10.1086/588207}

\bibitem[{{Hogg} {et~al.}(2006){Hogg}, {Masjedi}, {Berlind}, {Blanton},
  {Quintero}, \& {Brinkmann}}]{Hogg+06}
{Hogg}, D.~W., {Masjedi}, M., {Berlind}, A.~A., {et~al.} 2006, \apj, 650, 763,
  \dodoi{10.1086/507172}

\bibitem[{{Hopkins} {et~al.}(2008){Hopkins}, {Hernquist}, {Cox}, \&
  {Kere{\v{s}}}}]{Hopkins+08}
{Hopkins}, P.~F., {Hernquist}, L., {Cox}, T.~J., \& {Kere{\v{s}}}, D. 2008,
  \apjs, 175, 356, \dodoi{10.1086/524362}

\bibitem[{{Husemann} {et~al.}(2017){Husemann}, {Davis}, {Jahnke},
  {Dannerbauer}, {Urrutia}, \& {Hodge}}]{Husemann+17}
{Husemann}, B., {Davis}, T.~A., {Jahnke}, K., {et~al.} 2017, \mnras, 470, 1570,
  \dodoi{10.1093/mnras/stx1123}

\bibitem[{{Hwang} {et~al.}(2010){Hwang}, {Elbaz}, {Lee}, {Jeong}, {Park},
  {Lee}, \& {Lee}}]{HwangHS+10}
{Hwang}, H.~S., {Elbaz}, D., {Lee}, J.~C., {et~al.} 2010, \aap, 522, A33,
  \dodoi{10.1051/0004-6361/201014807}

\bibitem[{{Jog} \& {Combes}(2009)}]{Jog+09}
{Jog}, C.~J., \& {Combes}, F. 2009, \physrep, 471, 75,
  \dodoi{10.1016/j.physrep.2008.12.002}

\bibitem[{{Kampczyk} {et~al.}(2013){Kampczyk}, {Lilly}, {de Ravel}, {Le
  F{\`e}vre}, {Bolzonella}, {Carollo}, {Diener}, {Knobel}, {Kova{\v{c}}},
  {Maier}, {Renzini}, {Sargent}, {Vergani}, {Abbas}, {Bardelli}, {Bongiorno},
  {Bordoloi}, {Caputi}, {Contini}, {Coppa}, {Cucciati}, {de la Torre},
  {Franzetti}, {Garilli}, {Iovino}, {Kneib}, {Koekemoer}, {Lamareille}, {Le
  Borgne}, {Le Brun}, {Leauthaud}, {Mainieri}, {Mignoli}, {Pello}, {Peng},
  {Perez Montero}, {Ricciardelli}, {Scodeggio}, {Silverman}, {Tanaka}, {Tasca},
  {Tresse}, {Zamorani}, {Zucca}, {Bottini}, {Cappi}, {Cassata}, {Cimatti},
  {Fumana}, {Guzzo}, {Kartaltepe}, {Marinoni}, {McCracken}, {Memeo}, {Meneux},
  {Oesch}, {Porciani}, {Pozzetti}, \& {Scaramella}}]{Kampczyk+13}
{Kampczyk}, P., {Lilly}, S.~J., {de Ravel}, L., {et~al.} 2013, \apj, 762, 43,
  \dodoi{10.1088/0004-637X/762/1/43}

\bibitem[{{Kauffmann} {et~al.}(2004){Kauffmann}, {White}, {Heckman},
  {M{\'e}nard}, {Brinchmann}, {Charlot}, {Tremonti}, \&
  {Brinkmann}}]{Kauffmann+04}
{Kauffmann}, G., {White}, S. D.~M., {Heckman}, T.~M., {et~al.} 2004, \mnras,
  353, 713, \dodoi{10.1111/j.1365-2966.2004.08117.x}

\bibitem[{{Kawinwanichakij} {et~al.}(2017){Kawinwanichakij}, {Papovich},
  {Quadri}, {Glazebrook}, {Kacprzak}, {Allen}, {Bell}, {Croton}, {Dekel},
  {Ferguson}, {Forrest}, {Grogin}, {Guo}, {Kocevski}, {Koekemoer}, {Labb{\'e}},
  {Lucas}, {Nanayakkara}, {Spitler}, {Straatman}, {Tran}, {Tomczak}, \& {van
  Dokkum}}]{Kawinwanichakij+17}
{Kawinwanichakij}, L., {Papovich}, C., {Quadri}, R.~F., {et~al.} 2017, \apj,
  847, 134, \dodoi{10.3847/1538-4357/aa8b75}

\bibitem[{{Kelvin} {et~al.}(2012){Kelvin}, {Driver}, {Robotham}, {Hill},
  {Alpaslan}, {Baldry}, {Bamford}, {Bland-Hawthorn}, {Brough}, {Graham},
  {H{\"a}ussler}, {Hopkins}, {Liske}, {Loveday}, {Norberg}, {Phillipps},
  {Popescu}, {Prescott}, {Taylor}, \& {Tuffs}}]{Kelvin+12}
{Kelvin}, L.~S., {Driver}, S.~P., {Robotham}, A. S.~G., {et~al.} 2012, \mnras,
  421, 1007, \dodoi{10.1111/j.1365-2966.2012.20355.x}

\bibitem[{{Kennicutt} \& {Evans}(2012)}]{Kennicutt+12}
{Kennicutt}, R.~C., \& {Evans}, N.~J. 2012, \araa, 50, 531,
  \dodoi{10.1146/annurev-astro-081811-125610}

\bibitem[{{Kere{\v{s}}} {et~al.}(2005){Kere{\v{s}}}, {Katz}, {Weinberg}, \&
  {Dav{\'e}}}]{Keres+05}
{Kere{\v{s}}}, D., {Katz}, N., {Weinberg}, D.~H., \& {Dav{\'e}}, R. 2005,
  \mnras, 363, 2, \dodoi{10.1111/j.1365-2966.2005.09451.x}

\bibitem[{{Kewley} {et~al.}(2006){Kewley}, {Groves}, {Kauffmann}, \&
  {Heckman}}]{Kewley+06}
{Kewley}, L.~J., {Groves}, B., {Kauffmann}, G., \& {Heckman}, T. 2006, \mnras,
  372, 961, \dodoi{10.1111/j.1365-2966.2006.10859.x}

\bibitem[{{Kormendy} \& {Ho}(2013)}]{Kormendy+13}
{Kormendy}, J., \& {Ho}, L.~C. 2013, \araa, 51, 511,
  \dodoi{10.1146/annurev-astro-082708-101811}

\bibitem[{{Kormendy} \& {Kennicutt}(2004)}]{Kormendy+04}
{Kormendy}, J., \& {Kennicutt}, Robert~C., J. 2004, \araa, 42, 603,
  \dodoi{10.1146/annurev.astro.42.053102.134024}

\bibitem[{{Koulouridis} {et~al.}(2006){Koulouridis}, {Chavushyan}, {Plionis},
  {Krongold}, \& {Dultzin-Hacyan}}]{Koulouridis+06}
{Koulouridis}, E., {Chavushyan}, V., {Plionis}, M., {Krongold}, Y., \&
  {Dultzin-Hacyan}, D. 2006, \apj, 651, 93, \dodoi{10.1086/507070}

\bibitem[{{Larson} {et~al.}(2016){Larson}, {Sanders}, {Barnes}, {Ishida},
  {Evans}, {U}, {Mazzarella}, {Kim}, {Privon}, {Mirabel}, \&
  {Flewelling}}]{Larson+16}
{Larson}, K.~L., {Sanders}, D.~B., {Barnes}, J.~E., {et~al.} 2016, \apj, 825,
  128, \dodoi{10.3847/0004-637X/825/2/128}

\bibitem[{{Lemaux} {et~al.}(2017){Lemaux}, {Tomczak}, {Lubin}, {Wu}, {Gal},
  {Rumbaugh}, {Kocevski}, \& {Squires}}]{Lemaux+17}
{Lemaux}, B.~C., {Tomczak}, A.~R., {Lubin}, L.~M., {et~al.} 2017, \mnras, 472,
  419, \dodoi{10.1093/mnras/stx1579}

\bibitem[{{Li} {et~al.}(2008){Li}, {Kauffmann}, {Heckman}, {Jing}, \&
  {White}}]{Li+08}
{Li}, C., {Kauffmann}, G., {Heckman}, T.~M., {Jing}, Y.~P., \& {White}, S.
  D.~M. 2008, \mnras, 385, 1903, \dodoi{10.1111/j.1365-2966.2008.13000.x}

\bibitem[{{Li} {et~al.}(2019){Li}, {French}, {Zabludoff}, \&
  {Ho}}]{LiZhihui+19}
{Li}, Z., {French}, K.~D., {Zabludoff}, A.~I., \& {Ho}, L.~C. 2019, \apj, 879,
  131, \dodoi{10.3847/1538-4357/ab1f68}

\bibitem[{{Lim} {et~al.}(2017){Lim}, {Mo}, {Lu}, {Wang}, \& {Yang}}]{Lim+17}
{Lim}, S.~H., {Mo}, H.~J., {Lu}, Y., {Wang}, H., \& {Yang}, X. 2017, \mnras,
  470, 2982, \dodoi{10.1093/mnras/stx1462}

\bibitem[{{Lin} {et~al.}(2010){Lin}, {Cooper}, {Jian}, {Koo}, {Patton}, {Yan},
  {Willmer}, {Coil}, {Chiueh}, {Croton}, {Gerke}, {Lotz}, {Guhathakurta}, \&
  {Newman}}]{LinLihwai+10}
{Lin}, L., {Cooper}, M.~C., {Jian}, H.-Y., {et~al.} 2010, \apj, 718, 1158,
  \dodoi{10.1088/0004-637X/718/2/1158}

\bibitem[{{Lotz} {et~al.}(2008){Lotz}, {Jonsson}, {Cox}, \&
  {Primack}}]{Lotz+08}
{Lotz}, J.~M., {Jonsson}, P., {Cox}, T.~J., \& {Primack}, J.~R. 2008, \mnras,
  391, 1137, \dodoi{10.1111/j.1365-2966.2008.14004.x}

\bibitem[{{Lotz} {et~al.}(2021){Lotz}, {Dolag}, {Remus}, \&
  {Burkert}}]{Lotz+21}
{Lotz}, M., {Dolag}, K., {Remus}, R.-S., \& {Burkert}, A. 2021, \mnras, 506,
  4516, \dodoi{10.1093/mnras/stab2037}

\bibitem[{{Luo} {et~al.}(2014){Luo}, {Yang}, \& {Zhang}}]{Luo+14}
{Luo}, W., {Yang}, X., \& {Zhang}, Y. 2014, \apjl, 789, L16,
  \dodoi{10.1088/2041-8205/789/1/L16}

\bibitem[{{Mahajan}(2013)}]{Mahajan13}
{Mahajan}, S. 2013, \mnras, 431, L117, \dodoi{10.1093/mnrasl/slt021}

\bibitem[{{Maltby} {et~al.}(2018){Maltby}, {Almaini}, {Wild}, {Hatch},
  {Hartley}, {Simpson}, {Rowlands}, \& {Socolovsky}}]{Maltby+18}
{Maltby}, D.~T., {Almaini}, O., {Wild}, V., {et~al.} 2018, \mnras, 480, 381,
  \dodoi{10.1093/mnras/sty1794}

\bibitem[{{Man} {et~al.}(2019){Man}, {Peng}, {Kong}, {Guo}, {Zhang}, \&
  {Dou}}]{Man+19}
{Man}, Z.-y., {Peng}, Y.-j., {Kong}, X., {et~al.} 2019, \mnras, 488, 89,
  \dodoi{10.1093/mnras/stz1706}

\bibitem[{{Mandelbaum} {et~al.}(2009){Mandelbaum}, {Li}, {Kauffmann}, \&
  {White}}]{Mandelbaum+09}
{Mandelbaum}, R., {Li}, C., {Kauffmann}, G., \& {White}, S. D.~M. 2009, \mnras,
  393, 377, \dodoi{10.1111/j.1365-2966.2008.14235.x}

\bibitem[{{Martin} {et~al.}(2018){Martin}, {Kaviraj}, {Devriendt}, {Dubois}, \&
  {Pichon}}]{Martin+18}
{Martin}, G., {Kaviraj}, S., {Devriendt}, J.~E.~G., {Dubois}, Y., \& {Pichon},
  C. 2018, \mnras, 480, 2266, \dodoi{10.1093/mnras/sty1936}

\bibitem[{{Matthee} \& {Schaye}(2019)}]{Matthee+19}
{Matthee}, J., \& {Schaye}, J. 2019, \mnras, 484, 915,
  \dodoi{10.1093/mnras/stz030}

\bibitem[{{McNab} {et~al.}(2021){McNab}, {Balogh}, {van der Burg}, {Forestell},
  {Webb}, {Vulcani}, {Rudnick}, {Muzzin}, {Cooper}, {McGee}, {Biviano},
  {Cerulo}, {Chan}, {De Lucia}, {Demarco}, {Finoguenov}, {Forrest}, {Golledge},
  {Jablonka}, {Lidman}, {Nantais}, {Old}, {Pintos-Castro}, {Poggianti},
  {Reeves}, {Wilson}, {Yee}, \& {Zaritsky}}]{McNab+21}
{McNab}, K., {Balogh}, M.~L., {van der Burg}, R. F.~J., {et~al.} 2021, \mnras,
  508, 157, \dodoi{10.1093/mnras/stab2558}

\bibitem[{{Mihos} \& {Hernquist}(1996)}]{Mihos+96}
{Mihos}, J.~C., \& {Hernquist}, L. 1996, \apj, 464, 641, \dodoi{10.1086/177353}

\bibitem[{{Muldrew} {et~al.}(2012){Muldrew}, {Croton}, {Skibba}, {Pearce},
  {Ann}, {Baldry}, {Brough}, {Choi}, {Conselice}, {Cowan}, {Gallazzi}, {Gray},
  {Gr{\"u}tzbauch}, {Li}, {Park}, {Pilipenko}, {Podgorzec}, {Robotham},
  {Wilman}, {Yang}, {Zhang}, \& {Zibetti}}]{Muldrew+12}
{Muldrew}, S.~I., {Croton}, D.~J., {Skibba}, R.~A., {et~al.} 2012, \mnras, 419,
  2670, \dodoi{10.1111/j.1365-2966.2011.19922.x}

\bibitem[{{Muzzin} {et~al.}(2012){Muzzin}, {Wilson}, {Yee}, {Gilbank},
  {Hoekstra}, {Demarco}, {Balogh}, {van Dokkum}, {Franx}, {Ellingson}, {Hicks},
  {Nantais}, {Noble}, {Lacy}, {Lidman}, {Rettura}, {Surace}, \&
  {Webb}}]{Muzzin+12}
{Muzzin}, A., {Wilson}, G., {Yee}, H.~K.~C., {et~al.} 2012, \apj, 746, 188,
  \dodoi{10.1088/0004-637X/746/2/188}

\bibitem[{{Navarro} {et~al.}(1997){Navarro}, {Frenk}, \& {White}}]{Navarro+97}
{Navarro}, J.~F., {Frenk}, C.~S., \& {White}, S. D.~M. 1997, \apj, 490, 493,
  \dodoi{10.1086/304888}

\bibitem[{{Noeske} {et~al.}(2007){Noeske}, {Weiner}, {Faber}, {Papovich},
  {Koo}, {Somerville}, {Bundy}, {Conselice}, {Newman}, {Schiminovich}, {Le
  Floc'h}, {Coil}, {Rieke}, {Lotz}, {Primack}, {Barmby}, {Cooper}, {Davis},
  {Ellis}, {Fazio}, {Guhathakurta}, {Huang}, {Kassin}, {Martin}, {Phillips},
  {Rich}, {Small}, {Willmer}, \& {Wilson}}]{Noeske+07}
{Noeske}, K.~G., {Weiner}, B.~J., {Faber}, S.~M., {et~al.} 2007, \apjl, 660,
  L43, \dodoi{10.1086/517926}

\bibitem[{{Nolan} {et~al.}(2007){Nolan}, {Raychaudhury}, \&
  {Kab{\'a}n}}]{Nolan+07}
{Nolan}, L.~A., {Raychaudhury}, S., \& {Kab{\'a}n}, A. 2007, \mnras, 375, 381,
  \dodoi{10.1111/j.1365-2966.2006.11326.x}

\bibitem[{{Noll} {et~al.}(2009){Noll}, {Burgarella}, {Giovannoli}, {Buat},
  {Marcillac}, \& {Mu{\~n}oz-Mateos}}]{Noll+09}
{Noll}, S., {Burgarella}, D., {Giovannoli}, E., {et~al.} 2009, \aap, 507, 1793,
  \dodoi{10.1051/0004-6361/200912497}

\bibitem[{{Owers} {et~al.}(2007){Owers}, {Blake}, {Couch}, {Pracy}, \&
  {Bekki}}]{Owers+07}
{Owers}, M.~S., {Blake}, C., {Couch}, W.~J., {Pracy}, M.~B., \& {Bekki}, K.
  2007, \mnras, 381, 494, \dodoi{10.1111/j.1365-2966.2007.12239.x}

\bibitem[{{Owers} {et~al.}(2019){Owers}, {Hudson}, {Oman}, {Bland-Hawthorn},
  {Brough}, {Bryant}, {Cortese}, {Couch}, {Croom}, {van de Sande}, {Federrath},
  {Groves}, {Hopkins}, {Lawrence}, {Lorente}, {McDermid}, {Medling},
  {Richards}, {Scott}, {Taranu}, {Welker}, \& {Yi}}]{Owers+19}
{Owers}, M.~S., {Hudson}, M.~J., {Oman}, K.~A., {et~al.} 2019, \apj, 873, 52,
  \dodoi{10.3847/1538-4357/ab0201}

\bibitem[{{Paccagnella} {et~al.}(2019){Paccagnella}, {Vulcani}, {Poggianti},
  {Moretti}, {Fritz}, {Gullieuszik}, \& {Fasano}}]{Paccagnella+19}
{Paccagnella}, A., {Vulcani}, B., {Poggianti}, B.~M., {et~al.} 2019, \mnras,
  482, 881, \dodoi{10.1093/mnras/sty2728}

\bibitem[{{Paccagnella} {et~al.}(2017){Paccagnella}, {Vulcani}, {Poggianti},
  {Fritz}, {Fasano}, {Moretti}, {Jaff{\'e}}, {Biviano}, {Gullieuszik},
  {Bettoni}, {Cava}, {Couch}, \& {D'Onofrio}}]{Paccagnella+17}
{Paccagnella}, A., {Vulcani}, B., {Poggianti}, B.~M., {et~al.} 2017, \apj, 838, 148, \dodoi{10.3847/1538-4357/aa64d7}

\bibitem[{{Park} {et~al.}(2007){Park}, {Choi}, {Vogeley}, {Gott}, {Blanton}, \&
  {SDSS Collaboration}}]{Park+07}
{Park}, C., {Choi}, Y.-Y., {Vogeley}, M.~S., {et~al.} 2007, \apj, 658, 898,
  \dodoi{10.1086/511059}

\bibitem[{{Patel} {et~al.}(2011){Patel}, {Kelson}, {Holden}, {Franx}, \&
  {Illingworth}}]{Patel+11}
{Patel}, S.~G., {Kelson}, D.~D., {Holden}, B.~P., {Franx}, M., \&
  {Illingworth}, G.~D. 2011, \apj, 735, 53, \dodoi{10.1088/0004-637X/735/1/53}

\bibitem[{{Patton} {et~al.}(2020){Patton}, {Wilson}, {Metrow}, {Ellison},
  {Torrey}, {Brown}, {Hani}, {McAlpine}, {Moreno}, \& {Woo}}]{Patton+20}
{Patton}, D.~R., {Wilson}, K.~D., {Metrow}, C.~J., {et~al.} 2020, \mnras, 494,
  4969, \dodoi{10.1093/mnras/staa913}

\bibitem[{{Pawlik} {et~al.}(2016){Pawlik}, {Wild}, {Walcher}, {Johansson},
  {Villforth}, {Rowlands}, {Mendez-Abreu}, \& {Hewlett}}]{Pawlik+16}
{Pawlik}, M.~M., {Wild}, V., {Walcher}, C.~J., {et~al.} 2016, \mnras, 456,
  3032, \dodoi{10.1093/mnras/stv2878}

\bibitem[{{Pawlik} {et~al.}(2018){Pawlik}, {Taj Aldeen}, {Wild},
  {Mendez-Abreu}, {Lah{\'e}n}, {Johansson}, {Jimenez}, {Lucas}, {Zheng},
  {Walcher}, \& {Rowlands}}]{Pawlik+18}
{Pawlik}, M.~M., {Taj Aldeen}, L., {Wild}, V., {et~al.} 2018, \mnras, 477,
  1708, \dodoi{10.1093/mnras/sty589}

\bibitem[{Pedregosa {et~al.}(2011)Pedregosa, Varoquaux, Gramfort, Michel,
  Thirion, Grisel, Blondel, Prettenhofer, Weiss, Dubourg, Vanderplas, Passos,
  Cournapeau, Brucher, Perrot, \& Duchesnay}]{scikit-learn}
Pedregosa, F., Varoquaux, G., Gramfort, A., {et~al.} 2011, Journal of Machine
  Learning Research, 12, 2825

\bibitem[{{Peng} {et~al.}(2010){Peng}, {Lilly}, {Kova{\v{c}}}, {Bolzonella},
  {Pozzetti}, {Renzini}, {Zamorani}, {Ilbert}, {Knobel}, {Iovino}, {Maier},
  {Cucciati}, {Tasca}, {Carollo}, {Silverman}, {Kampczyk}, {de Ravel},
  {Sanders}, {Scoville}, {Contini}, {Mainieri}, {Scodeggio}, {Kneib}, {Le
  F{\`e}vre}, {Bardelli}, {Bongiorno}, {Caputi}, {Coppa}, {de la Torre},
  {Franzetti}, {Garilli}, {Lamareille}, {Le Borgne}, {Le Brun}, {Mignoli},
  {Perez Montero}, {Pello}, {Ricciardelli}, {Tanaka}, {Tresse}, {Vergani},
  {Welikala}, {Zucca}, {Oesch}, {Abbas}, {Barnes}, {Bordoloi}, {Bottini},
  {Cappi}, {Cassata}, {Cimatti}, {Fumana}, {Hasinger}, {Koekemoer},
  {Leauthaud}, {Maccagni}, {Marinoni}, {McCracken}, {Memeo}, {Meneux}, {Nair},
  {Porciani}, {Presotto}, \& {Scaramella}}]{Peng+10}
{Peng}, Y.-j., {Lilly}, S.~J., {Kova{\v{c}}}, K., {et~al.} 2010, \apj, 721,
  193, \dodoi{10.1088/0004-637X/721/1/193}

\bibitem[{{Pillepich} {et~al.}(2018){Pillepich}, {Springel}, {Nelson}, {Genel},
  {Naiman}, {Pakmor}, {Hernquist}, {Torrey}, {Vogelsberger}, {Weinberger}, \&
  {Marinacci}}]{Pillepich+18}
{Pillepich}, A., {Springel}, V., {Nelson}, D., {et~al.} 2018, \mnras, 473,
  4077, \dodoi{10.1093/mnras/stx2656}

\bibitem[{{Poggianti} \& {Wu}(2000)}]{Poggianti+00}
{Poggianti}, B.~M., \& {Wu}, H. 2000, \apj, 529, 157, \dodoi{10.1086/308243}

\bibitem[{{Poggianti} {et~al.}(2009){Poggianti}, {Arag{\'o}n-Salamanca},
  {Zaritsky}, {De Lucia}, {Milvang-Jensen}, {Desai}, {Jablonka}, {Halliday},
  {Rudnick}, {Varela}, {Bamford}, {Best}, {Clowe}, {Noll}, {Saglia},
  {Pell{\'o}}, {Simard}, {von der Linden}, \& {White}}]{Poggianti+09}
{Poggianti}, B.~M., {Arag{\'o}n-Salamanca}, A., {Zaritsky}, D., {et~al.} 2009,
  \apj, 693, 112, \dodoi{10.1088/0004-637X/693/1/112}

\bibitem[{{Pracy} {et~al.}(2009){Pracy}, {Kuntschner}, {Couch}, {Blake},
  {Bekki}, \& {Briggs}}]{Pracy+09}
{Pracy}, M.~B., {Kuntschner}, H., {Couch}, W.~J., {et~al.} 2009, \mnras, 396,
  1349, \dodoi{10.1111/j.1365-2966.2009.14836.x}

\bibitem[{{Quintero} {et~al.}(2004){Quintero}, {Hogg}, {Blanton}, {Schlegel},
  {Eisenstein}, {Gunn}, {Brinkmann}, {Fukugita}, {Glazebrook}, \&
  {Goto}}]{Quintero+04}
{Quintero}, A.~D., {Hogg}, D.~W., {Blanton}, M.~R., {et~al.} 2004, \apj, 602,
  190, \dodoi{10.1086/380601}

\bibitem[{{Rosario} {et~al.}(2018){Rosario}, {Burtscher}, {Davies}, {Koss},
  {Ricci}, {Lutz}, {Riffel}, {Alexander}, {Genzel}, {Hicks}, {Lin},
  {Maciejewski}, {M{\"u}ller-S{\'a}nchez}, {Orban de Xivry}, {Riffel},
  {Schartmann}, {Schawinski}, {Schnorr-M{\"u}ller}, {Saintonge}, {Shimizu},
  {Sternberg}, {Storchi-Bergmann}, {Sturm}, {Tacconi}, {Treister}, \&
  {Veilleux}}]{Rosario+18}
{Rosario}, D.~J., {Burtscher}, L., {Davies}, R.~I., {et~al.} 2018, \mnras, 473,
  5658, \dodoi{10.1093/mnras/stx2670}

\bibitem[{{Rowlands} {et~al.}(2015){Rowlands}, {Wild}, {Nesvadba}, {Sibthorpe},
  {Mortier}, {Lehnert}, \& {da Cunha}}]{Rowlands+15}
{Rowlands}, K., {Wild}, V., {Nesvadba}, N., {et~al.} 2015, \mnras, 448, 258,
  \dodoi{10.1093/mnras/stu2714}

\bibitem[{{Rowlands} {et~al.}(2018){Rowlands}, {Wild}, {Bourne}, {Bremer},
  {Brough}, {Driver}, {Hopkins}, {Owers}, {Phillipps}, {Pimbblet}, {Sansom},
  {Wang}, {Alpaslan}, {Bland-Hawthorn}, {Colless}, {Holwerda}, \&
  {Taylor}}]{Rowlands+18}
{Rowlands}, K., {Wild}, V., {Bourne}, N., {et~al.} 2018, \mnras, 473, 1168,
  \dodoi{10.1093/mnras/stx1903}

\bibitem[{{Sabater} {et~al.}(2015){Sabater}, {Best}, \& {Heckman}}]{Sabater+15}
{Sabater}, J., {Best}, P.~N., \& {Heckman}, T.~M. 2015, \mnras, 447, 110,
  \dodoi{10.1093/mnras/stu2429}

\bibitem[{{Saintonge} {et~al.}(2017){Saintonge}, {Catinella}, {Tacconi},
  {Kauffmann}, {Genzel}, {Cortese}, {Dav{\'e}}, {Fletcher},
  {Graci{\'a}-Carpio}, {Kramer}, {Heckman}, {Janowiecki}, {Lutz}, {Rosario},
  {Schiminovich}, {Schuster}, {Wang}, {Wuyts}, {Borthakur}, {Lamperti}, \&
  {Roberts-Borsani}}]{Saintonge+17}
{Saintonge}, A., {Catinella}, B., {Tacconi}, L.~J., {et~al.} 2017, \apjs, 233,
  22, \dodoi{10.3847/1538-4365/aa97e0}

\bibitem[{{Salim} {et~al.}(2018){Salim}, {Boquien}, \& {Lee}}]{Salim+18}
{Salim}, S., {Boquien}, M., \& {Lee}, J.~C. 2018, \apj, 859, 11,
  \dodoi{10.3847/1538-4357/aabf3c}

\bibitem[{{Salim} {et~al.}(2016){Salim}, {Lee}, {Janowiecki}, {da Cunha},
  {Dickinson}, {Boquien}, {Burgarella}, {Salzer}, \& {Charlot}}]{Salim+16}
{Salim}, S., {Lee}, J.~C., {Janowiecki}, S., {et~al.} 2016, \apjs, 227, 2,
  \dodoi{10.3847/0067-0049/227/1/2}

\bibitem[{{Sancisi} {et~al.}(2008){Sancisi}, {Fraternali}, {Oosterloo}, \& {van
  der Hulst}}]{Sancisi+08}
{Sancisi}, R., {Fraternali}, F., {Oosterloo}, T., \& {van der Hulst}, T. 2008,
  \aapr, 15, 189, \dodoi{10.1007/s00159-008-0010-0}

\bibitem[{{Sanders} \& {Mirabel}(1996)}]{Sanders+96}
{Sanders}, D.~B., \& {Mirabel}, I.~F. 1996, \araa, 34, 749,
  \dodoi{10.1146/annurev.astro.34.1.749}

\bibitem[{{Sanders} {et~al.}(1988){Sanders}, {Soifer}, {Elias}, {Madore},
  {Matthews}, {Neugebauer}, \& {Scoville}}]{Sanders+88}
{Sanders}, D.~B., {Soifer}, B.~T., {Elias}, J.~H., {et~al.} 1988, \apj, 325,
  74, \dodoi{10.1086/165983}

\bibitem[{{Sazonova} {et~al.}(2021){Sazonova}, {Alatalo}, {Rowlands},
  {Deustua}, {French}, {Heckman}, {Lanz}, {Lisenfeld}, {Luo}, {Medling},
  {Nyland}, {Otter}, {Petric}, {Snyder}, \& {Urry}}]{Sazonova+21}
{Sazonova}, E., {Alatalo}, K., {Rowlands}, K., {et~al.} 2021, \apj, 919, 134,
  \dodoi{10.3847/1538-4357/ac0f7f}

\bibitem[{{Schaye} {et~al.}(2015){Schaye}, {Crain}, {Bower}, {Furlong},
  {Schaller}, {Theuns}, {Dalla Vecchia}, {Frenk}, {McCarthy}, {Helly},
  {Jenkins}, {Rosas-Guevara}, {White}, {Baes}, {Booth}, {Camps}, {Navarro},
  {Qu}, {Rahmati}, {Sawala}, {Thomas}, \& {Trayford}}]{Schaye+15}
{Schaye}, J., {Crain}, R.~A., {Bower}, R.~G., {et~al.} 2015, \mnras, 446, 521,
  \dodoi{10.1093/mnras/stu2058}

\bibitem[{{Scholz-D{\'\i}az} {et~al.}(2021){Scholz-D{\'\i}az}, {S{\'a}nchez
  Almeida}, \& {Dalla Vecchia}}]{Scholz-Diaz+21}
{Scholz-D{\'\i}az}, L., {S{\'a}nchez Almeida}, J., \& {Dalla Vecchia}, C. 2021,
  \mnras, 505, 4655, \dodoi{10.1093/mnras/stab1629}

\bibitem[{Seabold \& Perktold(2010)}]{Statsmodels}
Seabold, S., \& Perktold, J. 2010, in 9th Python in Science Conference

\bibitem[{{Setton} {et~al.}(2022){Setton}, {Verrico}, {Bezanson}, {Greene},
  {Suess}, {Goulding}, {Spilker}, {Kriek}, {Feldmann}, {Narayanan},
  {Hall-Hooper}, \& {Kado-Fong}}]{Setton+22}
{Setton}, D.~J., {Verrico}, M., {Bezanson}, R., {et~al.} 2022, arXiv e-prints,
  arXiv:2203.08835.
\newblock \doarXiv{2203.08835}

\bibitem[{{Shangguan} {et~al.}(2020){Shangguan}, {Ho}, {Bauer}, {Wang}, \&
  {Treister}}]{Shangguan+20}
{Shangguan}, J., {Ho}, L.~C., {Bauer}, F.~E., {Wang}, R., \& {Treister}, E.
  2020, \apj, 899, 112, \dodoi{10.3847/1538-4357/aba8a1}

\bibitem[{{Shangguan} {et~al.}(2019){Shangguan}, {Ho}, {Li}, {Zhuang}, {Xie},
  \& {Li}}]{Shangguan+19}
{Shangguan}, J., {Ho}, L.~C., {Li}, R., {et~al.} 2019, \apj, 870, 104,
  \dodoi{10.3847/1538-4357/aaf21a}

\bibitem[{{Skibba} {et~al.}(2009){Skibba}, {Bamford}, {Nichol}, {Lintott},
  {Andreescu}, {Edmondson}, {Murray}, {Raddick}, {Schawinski}, {Slosar},
  {Szalay}, {Thomas}, \& {Vandenberg}}]{Skibba+09}
{Skibba}, R.~A., {Bamford}, S.~P., {Nichol}, R.~C., {et~al.} 2009, \mnras, 399,
  966, \dodoi{10.1111/j.1365-2966.2009.15334.x}

\bibitem[{{Snyder} {et~al.}(2011){Snyder}, {Cox}, {Hayward}, {Hernquist}, \&
  {Jonsson}}]{Snyder+11}
{Snyder}, G.~F., {Cox}, T.~J., {Hayward}, C.~C., {Hernquist}, L., \& {Jonsson},
  P. 2011, \apj, 741, 77, \dodoi{10.1088/0004-637X/741/2/77}

\bibitem[{{Sobral} {et~al.}(2011){Sobral}, {Best}, {Smail}, {Geach},
  {Cirasuolo}, {Garn}, \& {Dalton}}]{Sobral+11}
{Sobral}, D., {Best}, P.~N., {Smail}, I., {et~al.} 2011, \mnras, 411, 675,
  \dodoi{10.1111/j.1365-2966.2010.17707.x}

\bibitem[{{Socolovsky} {et~al.}(2018){Socolovsky}, {Almaini}, {Hatch}, {Wild},
  {Maltby}, {Hartley}, \& {Simpson}}]{Socolovsky+18}
{Socolovsky}, M., {Almaini}, O., {Hatch}, N.~A., {et~al.} 2018, \mnras, 476,
  1242, \dodoi{10.1093/mnras/sty312}

\bibitem[{{Speagle} {et~al.}(2014){Speagle}, {Steinhardt}, {Capak}, \&
  {Silverman}}]{Speagle+14}
{Speagle}, J.~S., {Steinhardt}, C.~L., {Capak}, P.~L., \& {Silverman}, J.~D.
  2014, \apjs, 214, 15, \dodoi{10.1088/0067-0049/214/2/15}

\bibitem[{{Suess} {et~al.}(2021){Suess}, {Kriek}, {Price}, \&
  {Barro}}]{Suess+21}
{Suess}, K.~A., {Kriek}, M., {Price}, S.~H., \& {Barro}, G. 2021, \apj, 915,
  87, \dodoi{10.3847/1538-4357/abf1e4}

\bibitem[{{Tacchella} {et~al.}(2016){Tacchella}, {Dekel}, {Carollo},
  {Ceverino}, {DeGraf}, {Lapiner}, {Mandelker}, \& {Primack
  Joel}}]{Tacchella+16}
{Tacchella}, S., {Dekel}, A., {Carollo}, C.~M., {et~al.} 2016, \mnras, 457,
  2790, \dodoi{10.1093/mnras/stw131}

\bibitem[{{Tekola} {et~al.}(2014){Tekola}, {Berlind}, \&
  {V{\"a}is{\"a}nen}}]{Tekola+14}
{Tekola}, A.~G., {Berlind}, A.~A., \& {V{\"a}is{\"a}nen}, P. 2014, \mnras, 439,
  3033, \dodoi{10.1093/mnras/stu168}

\bibitem[{{Tekola} {et~al.}(2012){Tekola}, {V{\"a}is{\"a}nen}, \&
  {Berlind}}]{Tekola+12}
{Tekola}, A.~G., {V{\"a}is{\"a}nen}, P., \& {Berlind}, A. 2012, \mnras, 419,
  1176, \dodoi{10.1111/j.1365-2966.2011.19773.x}

\bibitem[{{Tempel} {et~al.}(2014){Tempel}, {Tamm}, {Gramann}, {Tuvikene},
  {Liivam{\"a}gi}, {Suhhonenko}, {Kipper}, {Einasto}, \& {Saar}}]{Tempel+14}
{Tempel}, E., {Tamm}, A., {Gramann}, M., {et~al.} 2014, \aap, 566, A1,
  \dodoi{10.1051/0004-6361/201423585}

\bibitem[{{Tinker}(2021)}]{Tinker21}
{Tinker}, J.~L. 2021, \apj, 923, 154, \dodoi{10.3847/1538-4357/ac2aaa}

\bibitem[{{Tran} {et~al.}(2004){Tran}, {Franx}, {Illingworth}, {van Dokkum},
  {Kelson}, \& {Magee}}]{Tran+04}
{Tran}, K.-V.~H., {Franx}, M., {Illingworth}, G.~D., {et~al.} 2004, \apj, 609,
  683, \dodoi{10.1086/421237}

\bibitem[{{Tumlinson} {et~al.}(2017){Tumlinson}, {Peeples}, \&
  {Werk}}]{Tumlinson+17}
{Tumlinson}, J., {Peeples}, M.~S., \& {Werk}, J.~K. 2017, \araa, 55, 389,
  \dodoi{10.1146/annurev-astro-091916-055240}

\bibitem[{{van de Voort} {et~al.}(2017){van de Voort}, {Bah{\'e}}, {Bower},
  {Correa}, {Crain}, {Schaye}, \& {Theuns}}]{vandeVoort+17}
{van de Voort}, F., {Bah{\'e}}, Y.~M., {Bower}, R.~G., {et~al.} 2017, \mnras,
  466, 3460, \dodoi{10.1093/mnras/stw3356}

\bibitem[{{Veilleux} {et~al.}(2002){Veilleux}, {Kim}, \&
  {Sanders}}]{Veilleux+02}
{Veilleux}, S., {Kim}, D.~C., \& {Sanders}, D.~B. 2002, \apjs, 143, 315,
  \dodoi{10.1086/343844}

\bibitem[{{Vergani} {et~al.}(2010){Vergani}, {Zamorani}, {Lilly}, {Lamareille},
  {Halliday}, {Scodeggio}, {Vignali}, {Ciliegi}, {Bolzonella}, {Bondi},
  {Kova{\v{c}}}, {Knobel}, {Zucca}, {Caputi}, {Pozzetti}, {Bardelli},
  {Mignoli}, {Iovino}, {Carollo}, {Contini}, {Kneib}, {Le F{\`e}vre},
  {Mainieri}, {Renzini}, {Bongiorno}, {Coppa}, {Cucciati}, {de la Torre}, {de
  Ravel}, {Franzetti}, {Garilli}, {Kampczyk}, {Le Borgne}, {Le Brun}, {Maier},
  {Pello}, {Peng}, {Perez Montero}, {Ricciardelli}, {Silverman}, {Tanaka},
  {Tasca}, {Tresse}, {Abbas}, {Bottini}, {Cappi}, {Cassata}, {Cimatti},
  {Guzzo}, {Koekemoer}, {Leauthaud}, {Maccagni}, {Marinoni}, {McCracken},
  {Memeo}, {Meneux}, {Oesch}, {Porciani}, {Scaramella}, {Capak}, {Sanders},
  {Scoville}, \& {Taniguchi}}]{Vergani+10}
{Vergani}, D., {Zamorani}, G., {Lilly}, S., {et~al.} 2010, \aap, 509, A42,
  \dodoi{10.1051/0004-6361/200912802}

\bibitem[{{Verley} {et~al.}(2007){Verley}, {Leon}, {Verdes-Montenegro},
  {Combes}, {Sabater}, {Sulentic}, {Bergond}, {Espada}, {Garc{\'\i}a},
  {Lisenfeld}, \& {Odewahn}}]{Verley+07}
{Verley}, S., {Leon}, S., {Verdes-Montenegro}, L., {et~al.} 2007, \aap, 472,
  121, \dodoi{10.1051/0004-6361:20077481}

\bibitem[{Virtanen {et~al.}(2020)Virtanen, Gommers, Oliphant, Haberland, Reddy,
  Cournapeau, Burovski, Peterson, Weckesser, Bright, {van der Walt}, Brett,
  Wilson, Millman, Mayorov, Nelson, Jones, Kern, Larson, Carey, Polat, Feng,
  Moore, {VanderPlas}, Laxalde, Perktold, Cimrman, Henriksen, Quintero, Harris,
  Archibald, Ribeiro, Pedregosa, {van Mulbregt}, \& {SciPy 1.0
  Contributors}}]{Scipy}
Virtanen, P., Gommers, R., Oliphant, T.~E., {et~al.} 2020, Nature Methods, 17,
  261, \dodoi{10.1038/s41592-019-0686-2}

\bibitem[{{von der Linden} {et~al.}(2010){von der Linden}, {Wild}, {Kauffmann},
  {White}, \& {Weinmann}}]{vonderLinden+10}
{von der Linden}, A., {Wild}, V., {Kauffmann}, G., {White}, S. D.~M., \&
  {Weinmann}, S. 2010, \mnras, 404, 1231,
  \dodoi{10.1111/j.1365-2966.2010.16375.x}

\bibitem[{{Vulcani} {et~al.}(2020){Vulcani}, {Fritz}, {Poggianti}, {Bettoni},
  {Franchetto}, {Moretti}, {Gullieuszik}, {Jaff{\'e}}, {Biviano}, {Radovich},
  \& {Mingozzi}}]{Vulcani+20}
{Vulcani}, B., {Fritz}, J., {Poggianti}, B.~M., {et~al.} 2020, \apj, 892, 146,
  \dodoi{10.3847/1538-4357/ab7bdd}

\bibitem[{{Wang} {et~al.}(2018){Wang}, {Mo}, {Chen}, {Yang}, {Yang}, {Wang},
  {van den Bosch}, {Jing}, {Kang}, {Lin}, {Lim}, {Huang}, {Lu}, {Li}, {Cui},
  {Zhang}, {Tweed}, {Wei}, {Li}, \& {Shi}}]{WangHuiyuan+18}
{Wang}, H., {Mo}, H.~J., {Chen}, S., {et~al.} 2018, \apj, 852, 31,
  \dodoi{10.3847/1538-4357/aa9e01}

\bibitem[{{Wechsler} \& {Tinker}(2018)}]{Wechsler+18}
{Wechsler}, R.~H., \& {Tinker}, J.~L. 2018, \araa, 56, 435,
  \dodoi{10.1146/annurev-astro-081817-051756}

\bibitem[{{Wetzel} {et~al.}(2012){Wetzel}, {Tinker}, \& {Conroy}}]{Wetzel+12}
{Wetzel}, A.~R., {Tinker}, J.~L., \& {Conroy}, C. 2012, \mnras, 424, 232,
  \dodoi{10.1111/j.1365-2966.2012.21188.x}

\bibitem[{{Wild} {et~al.}(2010){Wild}, {Heckman}, \& {Charlot}}]{Wild+10}
{Wild}, V., {Heckman}, T., \& {Charlot}, S. 2010, \mnras, 405, 933,
  \dodoi{10.1111/j.1365-2966.2010.16536.x}

\bibitem[{{Wilkinson} {et~al.}(2017){Wilkinson}, {Pimbblet}, \&
  {Stott}}]{Wilkinson+17}
{Wilkinson}, C.~L., {Pimbblet}, K.~A., \& {Stott}, J.~P. 2017, \mnras, 472,
  1447, \dodoi{10.1093/mnras/stx2034}

\bibitem[{{Wilkinson} {et~al.}(2018){Wilkinson}, {Pimbblet}, {Stott}, {Few}, \&
  {Gibson}}]{Wilkinson+18}
{Wilkinson}, C.~L., {Pimbblet}, K.~A., {Stott}, J.~P., {Few}, C.~G., \&
  {Gibson}, B.~K. 2018, \mnras, 479, 758, \dodoi{10.1093/mnras/sty1493}

\bibitem[{{Wilman} {et~al.}(2010){Wilman}, {Zibetti}, \&
  {Budav{\'a}ri}}]{Wilman+10}
{Wilman}, D.~J., {Zibetti}, S., \& {Budav{\'a}ri}, T. 2010, \mnras, 406, 1701,
  \dodoi{10.1111/j.1365-2966.2010.16845.x}

\bibitem[{{Woo} {et~al.}(2013){Woo}, {Dekel}, {Faber}, {Noeske}, {Koo},
  {Gerke}, {Cooper}, {Salim}, {Dutton}, {Newman}, {Weiner}, {Bundy}, {Willmer},
  {Davis}, \& {Yan}}]{WooJoanna+13}
{Woo}, J., {Dekel}, A., {Faber}, S.~M., {et~al.} 2013, \mnras, 428, 3306,
  \dodoi{10.1093/mnras/sts274}

\bibitem[{{Wright} {et~al.}(2021){Wright}, {Lagos}, {Power}, \&
  {Correa}}]{Wright+21}
{Wright}, R.~J., {Lagos}, C. d.~P., {Power}, C., \& {Correa}, C.~A. 2021,
  \mnras, 504, 5702, \dodoi{10.1093/mnras/stab1057}

\bibitem[{{Wu} {et~al.}(2020){Wu}, {van der Wel}, {Bezanson}, {Gallazzi},
  {Pacifici}, {Straatman}, {Bari{\v{s}}i{\'c}}, {Bell}, {Chauke}, {D'Eugenio},
  {Franx}, {Muzzin}, {Sobral}, \& {van Houdt}}]{WuPo-Feng+20}
{Wu}, P.-F., {van der Wel}, A., {Bezanson}, R., {et~al.} 2020, \apj, 888, 77,
  \dodoi{10.3847/1538-4357/ab5fd9}

\bibitem[{{Yamauchi} {et~al.}(2008){Yamauchi}, {Yagi}, \& {Goto}}]{Yamauchi+08}
{Yamauchi}, C., {Yagi}, M., \& {Goto}, T. 2008, \mnras, 390, 383,
  \dodoi{10.1111/j.1365-2966.2008.13756.x}

\bibitem[{{Yan} {et~al.}(2009){Yan}, {Newman}, {Faber}, {Coil}, {Cooper},
  {Davis}, {Weiner}, {Gerke}, \& {Koo}}]{Yan+09}
{Yan}, R., {Newman}, J.~A., {Faber}, S.~M., {et~al.} 2009, \mnras, 398, 735,
  \dodoi{10.1111/j.1365-2966.2009.15192.x}

\bibitem[{{Yang} {et~al.}(2005){Yang}, {Mo}, {van den Bosch}, \&
  {Jing}}]{YangX+05}
{Yang}, X., {Mo}, H.~J., {van den Bosch}, F.~C., \& {Jing}, Y.~P. 2005, \mnras,
  356, 1293, \dodoi{10.1111/j.1365-2966.2005.08560.x}

\bibitem[{{Yang} {et~al.}(2007){Yang}, {Mo}, {van den Bosch}, {Pasquali}, {Li},
  \& {Barden}}]{YangX+07}
{Yang}, X., {Mo}, H.~J., {van den Bosch}, F.~C., {et~al.} 2007, \apj, 671, 153,
  \dodoi{10.1086/522027}

\bibitem[{{Yang} {et~al.}(2008){Yang}, {Zabludoff}, {Zaritsky}, \&
  {Mihos}}]{YangYujin+08}
{Yang}, Y., {Zabludoff}, A.~I., {Zaritsky}, D., \& {Mihos}, J.~C. 2008, \apj,
  688, 945, \dodoi{10.1086/591656}

\bibitem[{{Yano} {et~al.}(2016){Yano}, {Kriek}, {van der Wel}, \&
  {Whitaker}}]{Yano+16}
{Yano}, M., {Kriek}, M., {van der Wel}, A., \& {Whitaker}, K.~E. 2016, \apjl,
  817, L21, \dodoi{10.3847/2041-8205/817/2/L21}

\bibitem[{{Yesuf} {et~al.}(2020){Yesuf}, {Faber}, {Koo}, {Woo}, {Primack}, \&
  {Luo}}]{Yesuf+20a}
{Yesuf}, H.~M., {Faber}, S.~M., {Koo}, D.~C., {et~al.} 2020, \apj, 889, 14,
  \dodoi{10.3847/1538-4357/ab5fe1}

\bibitem[{{Yesuf} {et~al.}(2014){Yesuf}, {Faber}, {Trump}, {Koo}, {Fang},
  {Liu}, {Wild}, \& {Hayward}}]{Yesuf+14}
{Yesuf}, H.~M., {Faber}, S.~M., {Trump}, J.~R., {et~al.} 2014, \apj, 792, 84,
  \dodoi{10.1088/0004-637X/792/2/84}

\bibitem[{{Yesuf} {et~al.}(2017{\natexlab{a}}){Yesuf}, {French}, {Faber}, \&
  {Koo}}]{Yesuf+17b}
{Yesuf}, H.~M., {French}, K.~D., {Faber}, S.~M., \& {Koo}, D.~C.
  2017{\natexlab{a}}, \mnras, 469, 3015, \dodoi{10.1093/mnras/stx1046}

\bibitem[{{Yesuf} \& {Ho}(2019)}]{YesufHo19}
{Yesuf}, H.~M., \& {Ho}, L.~C. 2019, \apj, 884, 177,
  \dodoi{10.3847/1538-4357/ab4202}

\bibitem[{{Yesuf} \& {Ho}(2020{\natexlab{a}})}]{YesufHo20b}
{Yesuf}, H.~M., \& {Ho}, L.~C. 2020{\natexlab{a}}, \apj, 900, 107, \dodoi{10.3847/1538-4357/abaa43}

\bibitem[{{Yesuf} \& {Ho}(2020{\natexlab{b}})}]{YesufHo20a}
{Yesuf}, H.~M., \& {Ho}, L.~C. 2020{\natexlab{b}}, \apj, 901, 42, \dodoi{10.3847/1538-4357/aba961}

\bibitem[{{Yesuf} {et~al.}(2021){Yesuf}, {Ho}, \& {Faber}}]{Yesuf+21}
{Yesuf}, H.~M., {Ho}, L.~C., \& {Faber}, S.~M. 2021, \apj, 923, 205,
  \dodoi{10.3847/1538-4357/ac27a7}

\bibitem[{{Yesuf} {et~al.}(2017{\natexlab{b}}){Yesuf}, {Koo}, {Faber},
  {Prochaska}, {Guo}, {Liu}, {Cunningham}, {Coil}, \&
  {Guhathakurta}}]{Yesuf+17a}
{Yesuf}, H.~M., {Koo}, D.~C., {Faber}, S.~M., {et~al.} 2017{\natexlab{b}},
  \apj, 841, 83, \dodoi{10.3847/1538-4357/aa6fae}

\bibitem[{{Zabludoff} {et~al.}(1996){Zabludoff}, {Zaritsky}, {Lin}, {Tucker},
  {Hashimoto}, {Shectman}, {Oemler}, \& {Kirshner}}]{Zabludoff+96}
{Zabludoff}, A.~I., {Zaritsky}, D., {Lin}, H., {et~al.} 1996, \apj, 466, 104,
  \dodoi{10.1086/177495}

\bibitem[{{Zhang} {et~al.}(2021){Zhang}, {Wang}, {Luo}, {Mo}, {Liang}, {Li},
  {Yang}, {Wang}, {Zhang}, {Hong}, {Wang}, {Wang}, {Li}, \&
  {Shi}}]{ZhangZiwen+21}
{Zhang}, Z., {Wang}, H., {Luo}, W., {et~al.} 2021, \aap, 650, A155,
  \dodoi{10.1051/0004-6361/202040150}

\bibitem[{{Ziparo} {et~al.}(2014){Ziparo}, {Popesso}, {Finoguenov}, {Biviano},
  {Wuyts}, {Wilman}, {Salvato}, {Tanaka}, {Nandra}, {Lutz}, {Elbaz},
  {Dickinson}, {Altieri}, {Aussel}, {Berta}, {Cimatti}, {Fadda}, {Genzel}, {Le
  Floc'h}, {Magnelli}, {Nordon}, {Poglitsch}, {Pozzi}, {Portal}, {Tacconi},
  {Bauer}, {Brandt}, {Cappelluti}, {Cooper}, \& {Mulchaey}}]{Ziparo+14}
{Ziparo}, F., {Popesso}, P., {Finoguenov}, A., {et~al.} 2014, \mnras, 437, 458,
  \dodoi{10.1093/mnras/stt1901}

\bibitem[{{Zolotov} {et~al.}(2015){Zolotov}, {Dekel}, {Mandelker}, {Tweed},
  {Inoue}, {DeGraf}, {Ceverino}, {Primack}, {Barro}, \& {Faber}}]{Zolotov+15}
{Zolotov}, A., {Dekel}, A., {Mandelker}, N., {et~al.} 2015, \mnras, 450, 2327,
  \dodoi{10.1093/mnras/stv740}

\bibitem[{{Zwaan} {et~al.}(2013){Zwaan}, {Kuntschner}, {Pracy}, \&
  {Couch}}]{Zwaan+13}
{Zwaan}, M.~A., {Kuntschner}, H., {Pracy}, M.~B., \& {Couch}, W.~J. 2013,
  \mnras, 432, 492, \dodoi{10.1093/mnras/stt496}

\end{thebibliography}

\appendix

\section{Mass Versus Luminosity Overdensities}\label{sec:appA}

\begin{figure*}[ht!]
\gridline{\fig{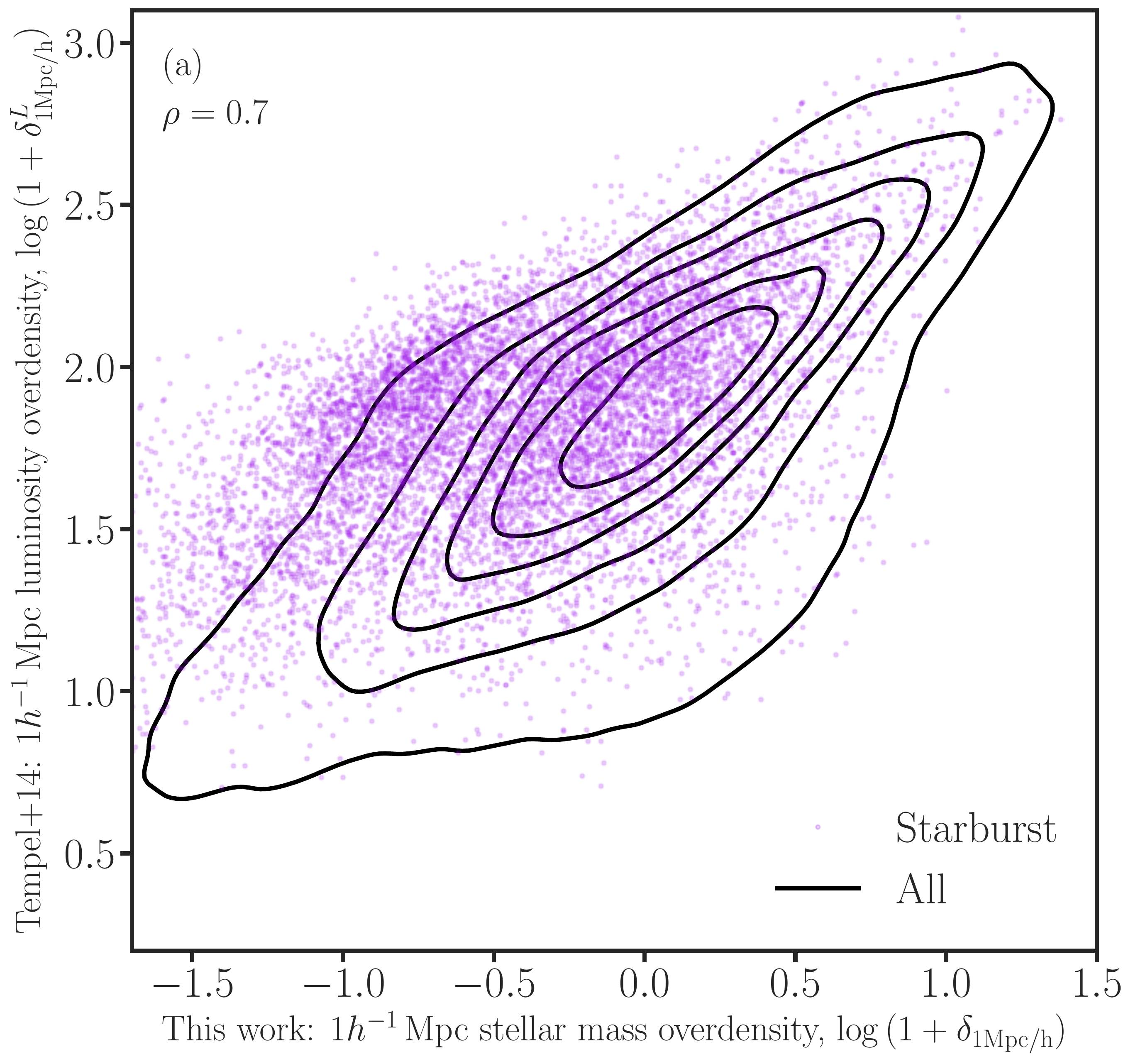}{0.48\textwidth}{}
          \fig{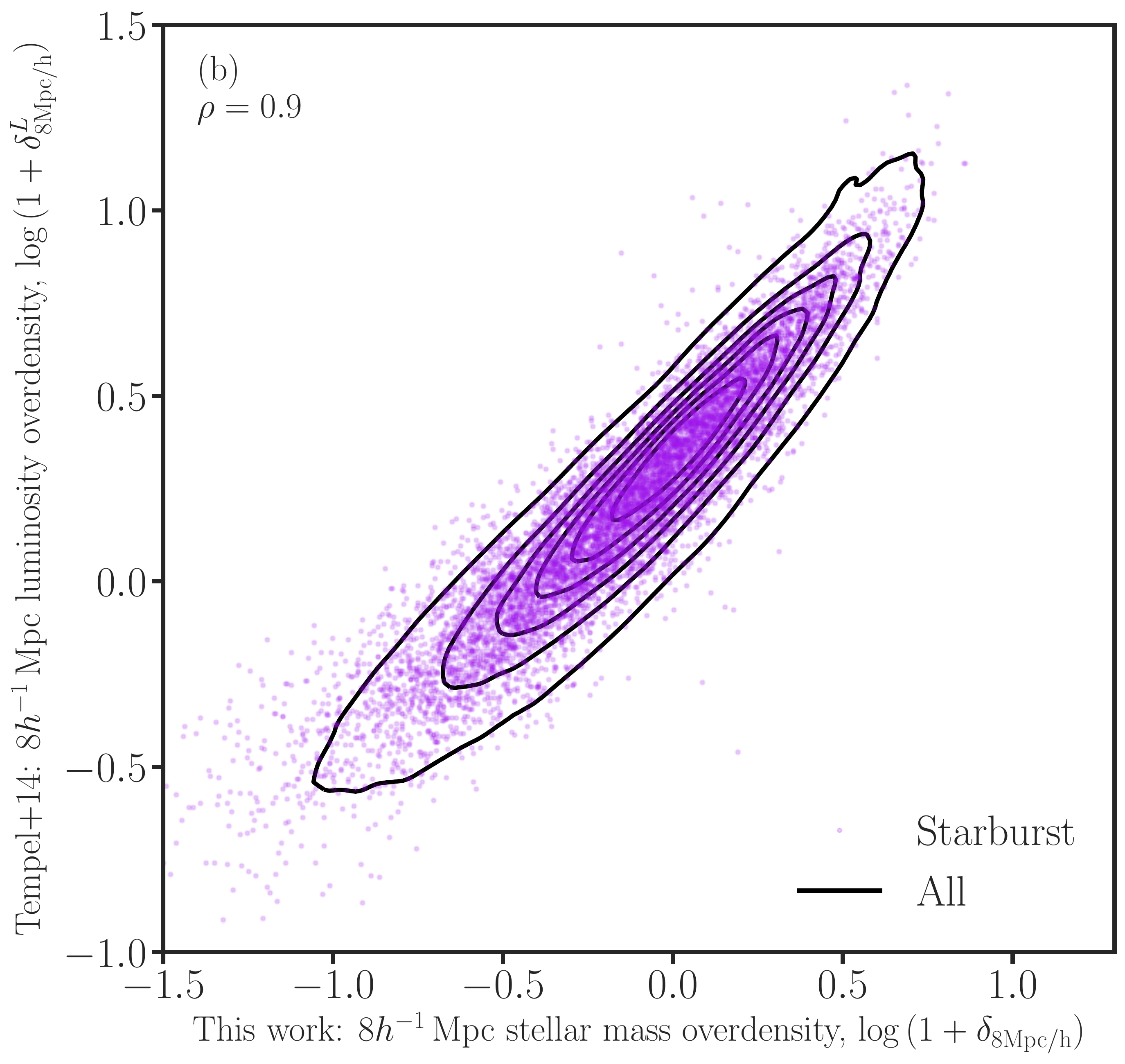}{0.48\textwidth}{}}
 \gridline{\fig{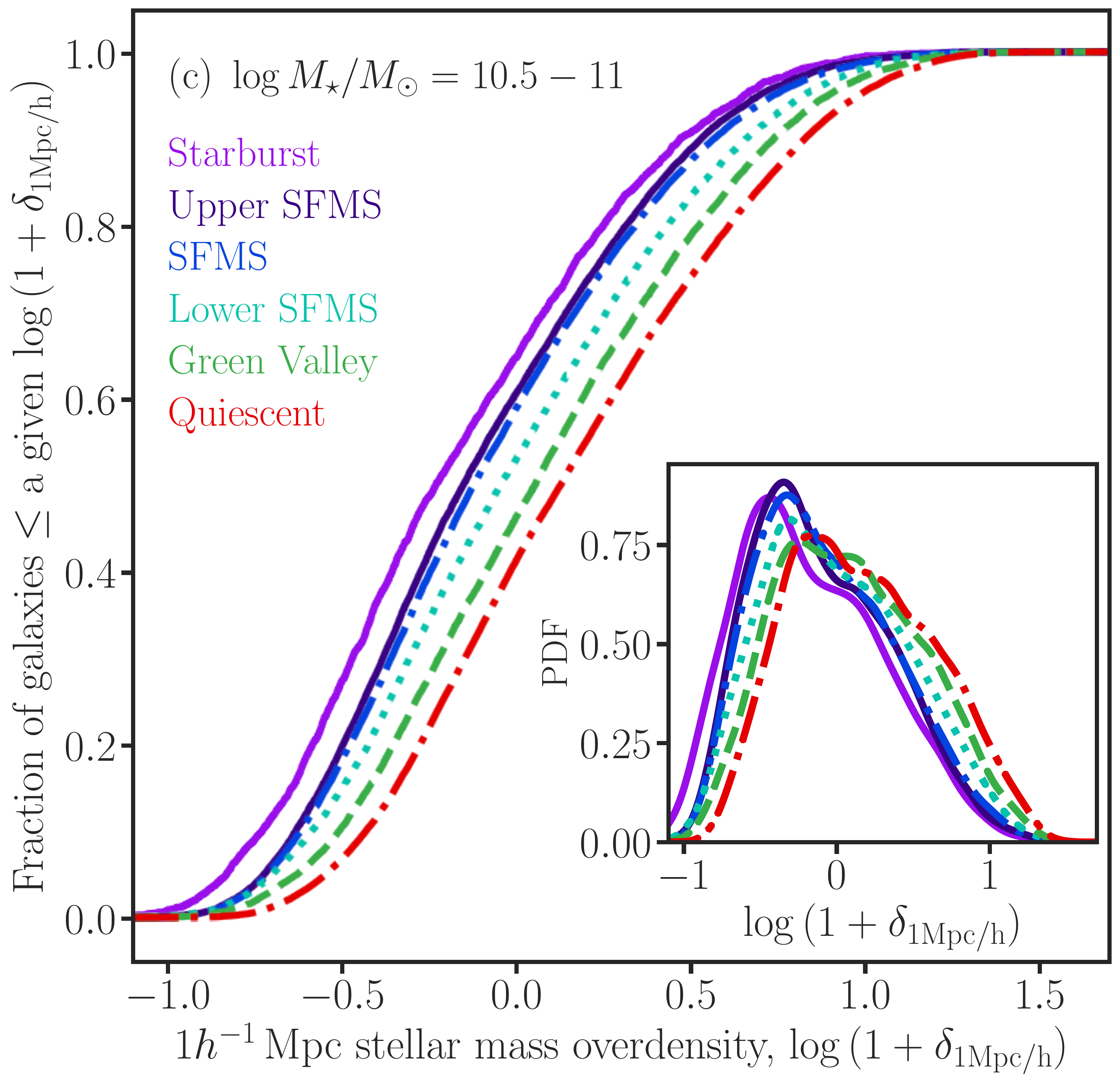}{0.48\textwidth}{}
          \fig{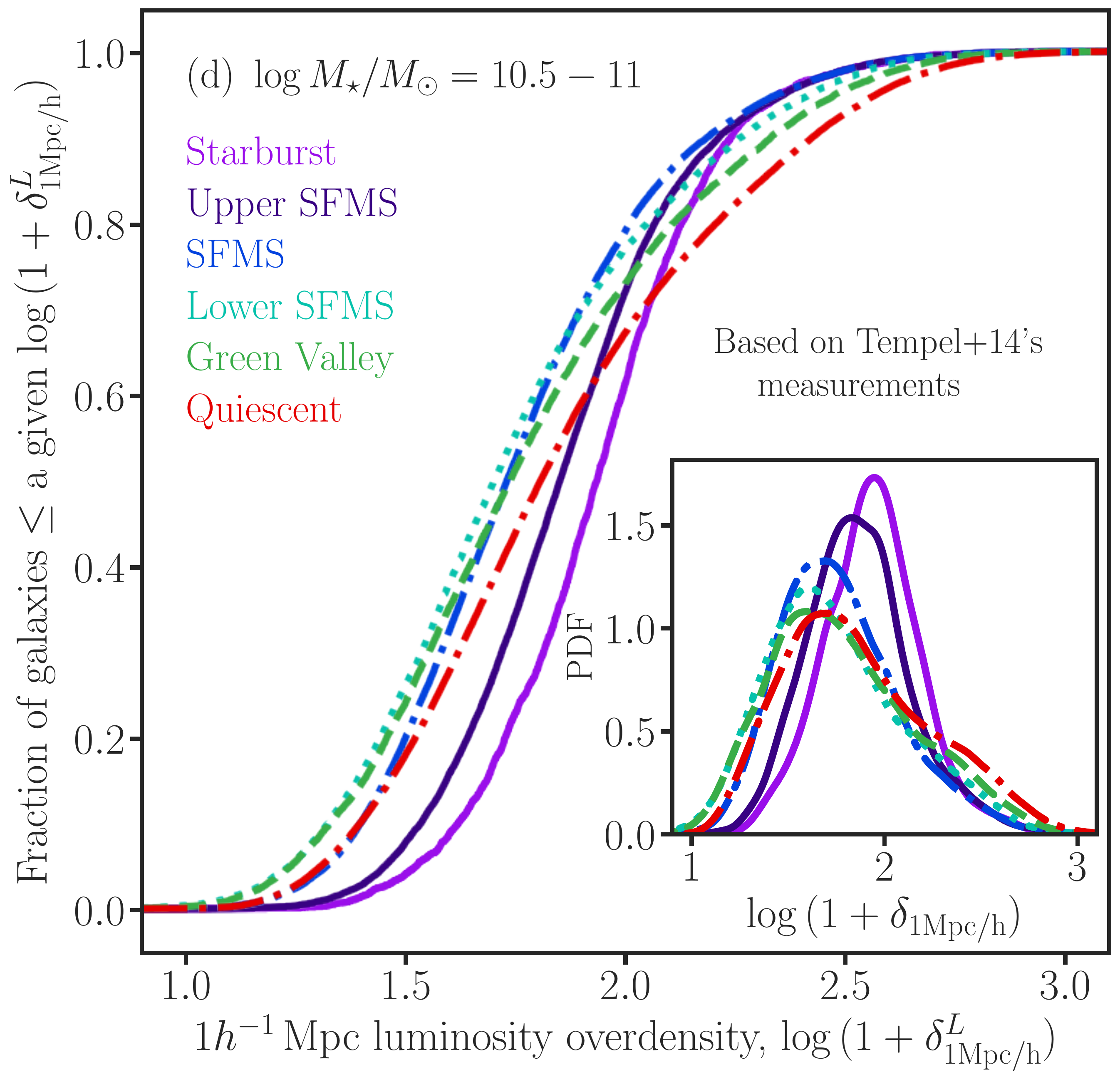}{0.48\textwidth}{}}
\caption{Comparing stellar mass overdensities with luminosity overdensities. The latter are computed by \citet{Tempel+14}. \label{fig:delTE_compare}}
\end{figure*}

Figure~\ref{fig:delTE_compare} compares $\deltaone$ and $\deltaeight$ mass overdensities measured by the author with the corresponding luminosity densities measured by \citet{Tempel+14}. Despite the difference on how the overdensities are normalized, $\deltaeight$ and $\delta^L_\mathrm{8Mpc}$ are consistent and lead to similar inference about the environments of starbursts and QPSBs. On the other hand,  there is a systematic difference between $\deltaone$ and $\delta^L_\mathrm{1Mpc}$; the mass density measurements within $1\,h^{-1}$Mpc indicates starbursts and upper SFMS galaxies have lower density than those of lower SFMS and SFMS galaxies, whereas \citet{Tempel+14}'s  $1\,h^{-1}$Mpc luminosity density measurements indicate the opposite. 



\section{Summary Statistics for Environments Upper SFMS, Lower SFMS, and Green-valley Galaxies} \label{sec:appB}
Table~\ref{tab:haloApp}, Table~\ref{tab:del18App}, and Table~\ref{tab:envApp} extend the comparison of environments of starbursts with those of upper SFMS, lower SFMS, green-valley galaxies. For ease of comparison, the same information presented in the main text for starbursts, SFMS galaxies, and QGs is repeated in these tables.

\begin{deluxetable*}{cccccccc}
\tablenum{4}
\tablecaption{Comparing the mean fractions of centrals and median (15\%, 85\%) halo masses of starbursts with other samples \label{tab:haloApp}}
\tablewidth{0pt}
\tablehead{
\colhead{Sample} & \colhead{Measurements} & \colhead{Starburst} & \colhead{Upper SFMS} &  \colhead{SFMS} & \colhead{Lower SFMS}
 & \colhead{Green Valley} & \colhead{Quiescent} 
}
\decimalcolnumbers
\startdata
& $f$ Central Tinker & 0.839 & 0.817 & 0.812 & 0.803 & 0.783 & 0.777\\
& $f$ Central Lim et al. & 0.832& 0.803 & 0.793 & 0.776 & 0.767 & 0.767 \\
M11 & $\log M_h$ Tinker & $12.3\,(11.8,13.2)$ & $12.4\,(11.8,13.4)$ &$12.3\,(11.7,13.4)$  & $12.4\,(11.7,13.5)$ & $12.6\,(11.8,13.7)$ & $12.9\,(11.8,13.8)$\\
& $\log M_h$ Lim et al.  & $12.5\,(12.3,12.8)$ & $12.5\,(12.3,13.0)$ & $12.7\,(12.4,13.2)$ & $12.8\,(12.5,13.3)$ & $12.9\,(12.6,13.5)$ & $13.1\,(12.7,13.6)$ \\ 
\hline
& $f$ Central Tinker  & 0.801 & 0.806 &0.797 & 0.756 & 0.726 & 0.708\\
& $f$ Central Lim et al.  & 0.799 & 0.779 & 0.757 & 0.708 & 0.673 & 0.647\\
M105 & $\log M_h$ Tinker  & $12.3\,(11.7,13.4)$ & $12.2\,(11.7,13.4)$ & $12.2\,(11.7,13.4)$ & $12.2\,(11.6,13.5)$ & $12.4\,(11.6,13.7)$ & $12.5\,(11.7,13.8)$ \\
& $\log M_h $ Lim et al.  & $12.1\,(11.9,12.5)$ & $12.1\,(11.9,12.6)$ & $12.2\,(12.0,12.7)$  &$12.3\,(12.0,13.0)$ & $12.4\,(12.1,13.3)$ & $12.5\,(12.2,13.4)$ \\
\hline
& $f$ Central Tinker  & 0.816 & 0.794 & 0.776 & 0.736 & 0.636 & 0.606 \\
& $f$ Central Lim et al.  & 0.787 & 0.746 & 0.727 & 0.653 & 0.563 & 0.526\\
M10 & $\log M_h$ Tinker  & $12.1\,(11.8,12.9)$ & $12.0\,(11.6,13.2)$ & $12.0\,(11.5,13.4)$ & $12.0\,(11.4,13.5)$ & $12.2\,(11.4,13.7)$ & $12.4\,(11.5,13.9)$ \\
 & $\log M_h$ Lim et al.  & $11.8\,(11.6,12.3)$ & $11.8\,(11.6,12.4)$ & $11.8\,(11.6,12.6)$& $11.9\,(11.7,13.1)$ & $12.0\,(11.8,13.4)$ & $12.0\,(11.8,13.7)$ \\
\hline 
& $f$ Central Tinker  & 0.791 &0.770  & 0.774 & 0.742 & 0.620 & 0.276\\
& $f$ Central Lim et al.  & 0.766 & 0.717 & 0.711 & 0.617 & 0.422 & 0.362\\
M95 & $\log M_h$ Tinker  & $12.0\,(11.7,12.4)$ & $11.9\,(11.5,13.4)$ & $11.8\,(11.3,13.4)$ & $12.0\,(11.3,13.5)$ & $12.3\,(11.3,13.8)$ & $13.2\,(11.3,14.3)$ \\
& $\log M_h$ Lim et al.  & $11.6\,(11.4,12.2)$ & $11.5\,(11.4,12.3)$ & $11.5\,(11.4,12.4)$ &$11.6\,(11.4,13.1)$ & $12.1\,(11.5,13.7)$ & $12.7\,(11.6,14.1)$ \\
\enddata
\tablecomments{The standard errors of mean fractions for all samples are $\sim 0.1-1.5\%$. In particular, the fractional errors for starbursts at the lowest and highest mass ranges are relatively higher due to smaller sample sizes (i.e, $\sim 1.5\%$ for Lim et al. data and $\lesssim 0.5\%$ for Tinker et al. data).}
\end{deluxetable*}

\begin{rotatetable*}
\movetableright=0.1in
\begin{deluxetable*}{llcccccc}
\tabletypesize{\footnotesize}
\tablenum{5}
\tablecaption{The Multiscale Environments of Starbursts and Comparison Samples \label{tab:del18App}}
\tablewidth{0pt}
\tablehead{
\colhead{Sample} & \colhead{Measurement} & \colhead{Starburst} & \colhead{Upper SFMS}  & \colhead{SFMS} & \colhead{Lower SFMS} & \colhead{GV} & \colhead{Quiescent}
}
\decimalcolnumbers
\startdata
 & $\log\,(1+\delta_\mathrm{0.5Mpc})$ & $-0.21\,(-0.42, 0.09)$ & $-0.15\,(-0.35,  0.17) $ & $-0.12\,(-0.33, 0.22)$ &$-0.06\,(-0.29, 0.28) $&$-0.01\,(-0.25, 0.35)$ & $0.07\,(-0.19,  0.43)$\\
 & $\log\,(1+\delta_\mathrm{1Mpc})$ & $-0.23\,(-0.53, 0.15)$ & $-0.15\,(-0.46, 0.24)$ & $-0.13\,(-0.44,  0.28)$ & $-0.06\,(-0.38,  0.35)$ & $0.00\,(-0.34, 0.41)$ & $0.07\,(-0.27, 0.49)$\\
& $\log\,(1+ \delta_\mathrm{2Mpc})$  & $-0.18\,(-0.62, 0.21)$ &  $-0.14\,(-0.57, 0.28)$ & $-0.11\,(-0.54, 0.31)$ &  $-0.05\,(-0.48,  0.36) $ & $0.00\,(0.42, 0.41)$ & $0.05\, (-0.35, 0.46) $\\
M11 & $\log\,(1+\delta_\mathrm{4Mpc})$  & $-0.14\,(-0.57, 0.24)$ & $-0.10\,(-0.55, 0.29)$ & $-0.08\,(-0.51,  0.30)$ & $-0.04\,(-0.47, 0.33)$ & $0.00\,(-0.41, 0.36)$ & $0.04\,(-0.36, 0.40)$\\
& $\log\,(1+\delta_\mathrm{8Mpc})$  & $-0.11\,(-0.51, 0.24)$ & $-0.06\,(-0.44, 0.25)$ & $-0.06\,(-0.43, 0.27)$ & $-0.03\,(-0.40, 0.28)$ & $0.00\,(-0.36, 0.31)$ & $0.03\,(-0.32, 0.33)$\\  
& $\log\,(1+\delta_5)$  & $-0.19\,(-0.69, 0.39)$ & $-0.15\,(-0.65, 0.48)$ & $-0.13\,(-0.64, 0.49)$ & $-0.08\,(-0.62,  0.58)$ & $-0.02\,(-0.56, 0.70)$ & $0.08\,(-0.48, 0.86)$\\  
& $\log\,Q_5$  & $-4.05\,(-5.30, -2.31)$ & $-4.06\,(-5.24, -2.38)$ & $-3.85\,(-5.09, -2.24)$ & $-3.70\,(-5.02, -2.06)$ & $-3.58\,(-5.04, -1.98)$ & $-3.41\,(-4.97, -1.77)$\\  
\hline
 & $\log\,(1+\delta_\mathrm{0.5Mpc})$ & $-0.22\,(-0.52, 0.26)$ & $-0.15\,(-0.42, 0.33)$ & $-0.11\,(-0.39, 0.39)$ & $-0.05\,(-0.36, 0.52)$ & $0.04\,(-0.29,  0.62)$ & $0.11\,(-0.20,  0.72)$\\
  & $\log\,(1+\delta_\mathrm{1Mpc})$ & $-0.23\,(-0.67, 0.36)$ & $-0.16\,(-0.59,  0.41)$ & $-0.13\,(-0.57,  0.44)$ & $-0.04\,(-0.52, 0.54)$ & $0.05\,(-0.45, 0.62)$ & $0.13\,(-0.37, 0.71)$\\
& $\log\,(1+ \delta_\mathrm{2Mpc})$  & $-0.18\,(-0.83, 0.32)$ &  $-0.12\,(-0.71, 0.39)$ & $-0.10\,(-0.67, 0.40$ &  $-0.03\,(-0.59, 0.47) $ & $0.04\,(-0.53, 0.53)$ & $0.10\, (-0.46, 0.59)$\\
M105 & $\log\,(1+\delta_\mathrm{4Mpc})$  & $-0.13\,(-0.68, 0.29)$ & $-0.08\,(-0.59, 0.33)$ & $-0.07\,(-0.56, 0.34)$ & $-0.03\,(-0.52, 0.39)$ & $0.02\,(0.46, 0.43)$ & $0.07\,(-0.41, 0.47)$\\
& $\log\,(1+\delta_\mathrm{8Mpc})$  & $-0.08\,(-0.50, 0.26)$ & $-0.05\,(-0.46, 0.28)$ & $-0.04\,(-0.43, 0.29)$ & $-0.02\,(-0.41, 0.31)$ & $0.02\,(-0.37, 0.35)$ & $0.04\,(-0.34, 0.37)$\\  
& $\log\,(1+\delta_5)$  & $-0.15\,(-0.65, 0.42)$ & $-0.11\,(-0.62, 0.50)$ & $-0.11\,(-0.63, 0.53)$ & $-0.03\,(-0.61,  0.68)$ & $0.03\,(-0.56, 0.81)$ & $0.12\,(-0.50, 0.96)$\\  
& $\log\,Q_5$  & $-4.14\,(-5.39, 2.40)$ & $-4.01\,(-5.21, -2.42)$ & $-3.86\,(-5.10, -2.28)$ & $-3.69\,(-5.06, -2.10)$ & $-3.65\,(-5.10, -2.02)$ & $-3.63\,(-5.18, -1.99)$\\  
\hline
 & $\log\,(1+\delta_\mathrm{0.5Mpc})$ & $-0.18\,(-0.59, 0.56)$ & $-0.26\,(-0.54, 0.23)$ & $-0.16\,(-0.40, 0.30)$ & $-0.02\,(-0.31, 0.51)$ & $-0.11\,(-0.53, 0.66)$ & $0.39\,(-0.28, 1.15)$\\
  & $\log\,(1+\delta_\mathrm{1Mpc})$ & $-0.20\,(-0.92, 0.55)$ & $-0.14\,(-0.83,  0.56)$ & $-0.12\,(-0.84,  0.53)$ & $0.01\,(-0.81, 0.67)$ & $0.18\,(-0.65,  0.80)$ & $0.29\,(-0.58, 0.93)$\\
  & $\log\,(1+ \delta_\mathrm{2Mpc})$  & $-0.13\,(-0.95, 0.42)$ &  $-0.09\,(-0.76, 0.45)$ & $-0.09\,(-0.76, 0.45) $ & $0.00\,(-0.67, 0.53) $ & $0.10\,(-0.55, 0.62)$ & $0.20\,(-0.48, 0.73)$\\
  M10 & $\log\,(1+\delta_\mathrm{4Mpc})$  & $-0.09\,(-0.68, 0.33)$ & $-0.06\,(-0.59, 0.38)$ & $-0.06\,(-0.57, 0.37)$ & $0.00\,(-0.51, 0.41)$ & $0.07\,(-0.44, 0.49)$ & $0.13\,(-0.39, 0.55)$\\
& $\log\,(1+\delta_\mathrm{8Mpc})$  & $-0.06\,(-0.50, 0.28)$ & $-0.03\,(-0.44, 0.32)$ & $-0.04\,(-0.44, 0.30)$ & $0.00\,(-0.39, 0.33)$ & $0.04\,(-0.34, 0.38)$ & $0.08\,(-0.31, 0.42)$\\  
& $\log\,(1+\delta_5)$  & $-0.11\,(-0.63, 0.47)$ & $-0.08\,(-0.62, 0.57)$ & $-0.10\,(-0.64, 0.57)$ & $-0.01\,(-0.62, 0.79)$ & $0.17\,(-0.53, 1.00)$ & $0.29\,(-0.47, 1.20)$\\  
& $\log\,Q_5$  & $-3.98\,(-5.18, -2.10)$ & $-3.73\,(-5.01, -2.05)$ & $-3.64\,(-4.92, -2.05)$ & $-3.46\,(-4.91, -1.88)$ & $-3.34\,(-5.01, -1.73)$ & $-3.40\,(-5.17, -1.77)$\\  
\hline
& $\log\,(1+\delta_\mathrm{0.5Mpc})$ & $-0.19\,(-0.68, 0.91)$ & $-0.10\,(-0.65, 0.98)$ & $-0.14\,(-0.76, 0.90)$ & $-0.03\,(-0.85, 0.93)$ & $0.54\,(-0.63, 1.21)$ & $0.77\,(-0.38, 1.43)$\\
 & $\log\,(1+\delta_\mathrm{1Mpc})$ & $-0.20\,(-1.29, 0.58)$ & $-0.09\,(-1.16,  0.63)$ & $-0.10\,(-1.13, 0.62)$ & $0.02\,(-1.11, 0.69)$ & $0.34\,(-0.63, 0.95)$ & $0.53\,(-0.33, 1.13)$\\
& $\log\,(1+ \delta_\mathrm{2Mpc})$  & $-0.18\,,(-0.92, 0.44)$ & $-0.06\,(-0.80, 0.51)$ & $-0.06\,(-0.79, 0.49) $ & $0.00\,(-0.70, 0.54) $ & $0.21\,(-0.45, 0.72)$ & $0.36\, (-0.27, 0.87)$\\
M95 & $\log\,(1+\delta_\mathrm{4Mpc})$  & $-0.13\,(-0.65, 0.35)$ & $-0.04\,(-0.56, 0.41)$ & $-0.04\,(-0.56, 0.39)$ & $0.00\,(-0.52, 0.41)$ & $0.12\,(-0.36, 0.54)$ & $0.23\,(-0.24, 0.64)$\\
& $\log\,(1+\delta_\mathrm{8Mpc})$  & $-0.10\,(-0.48, 0.27)$ & $-0.03\,(-0.42, 0.31)$ & $-0.02\,(-0.42, 0.31)$ & $0.00\,(-0.39, 0.32)$ & $0.09\,(-0.28, 0.42)$ & $0.15\,(-0.20, 0.48)$\\  
& $\log\,(1+\delta_5)$  & $-0.15\,(-0.65, 0.45)$ & $-0.06\,(-0.60, 0.62)$ & $-0.07\,(-0.64, 0.61)$ & $0.00\,(-0.65, 0.77)$ & $0.38\,(-0.45, 1.18)$ & $0.68\,(-0.26, 1.44)$\\  
& $\log\,Q_5$  & $-3.78\,(-5.12, -2.00)$ & $-3.48\,(-5.24, -1.39)$ & $-4.25\,(-5.24, -1.46)$ & $-4.21\,(-5.52, -2.13)$ & $-3.38\,(-4.76, -1.77)$ & $-2.64\,(-4.78, -1.18)$\\  
\enddata
\tablecomments{Column (1) : the $M_\star$ ranges of the samples binned by 0.5\,dex. The names of the samples indicate the minima of the $\log\, M_\star$ ranges. For example, M11 denotes $\log M_\star/M_\odot = 11-11.5$. We use the notation $X\,(Y, Z)$ to denote $X$ = median (50\%), $Y=$15\%, and $Z=85$\% of a distribution.}
\end{deluxetable*}
\end{rotatetable*}

\begin{rotatetable*}
\movetableright=0.1in
\begin{deluxetable*}{cccccccc}
\tablenum{6}
\tablecaption{The Mean Fractions (and 95\% Confidence Intervals) of Starbursts and Comparison Samples that are Isolated, Pairs, in Groups, or in Clusters. \label{tab:envApp}}
\tablewidth{0pt}
\tablehead{
\colhead{Sample} & \colhead{Measurements} & \colhead{Starburst} & \colhead{Upper SFMS} & \colhead{SFMS} & \colhead{Lower SFMS}
 & \colhead{Green Valley} & \colhead{Quiescent} 
}
\decimalcolnumbers
\startdata
M11 & Isolated & $0.632\,(0.576, 0.684)$ & $0.656\,(0.636, 0.676)$ & $0.605\,(0.593, 0.616)$ & $0.549\,(0.537, 0.562)$ & $0.498\,(0.487, 0.509)$ & $0.447\,(0.441, 0.453) $ \\
      & Pairs & $0.200\,(0.158, 0.248)$  & $0.187\,(0.171, 0.204)$ & $0.184\,(0.175, 0.194)$ & $183\,(0.173, 0.193)$ & $0.189\,(0.180, 0.198)$ & $0.188\,(0.183, 0.193)$ \\
      & $2-3$ neighbors & $0.108\,(0.078, 0.148)$ & $0.087\,(0.076, 0.100)$ &  $0.116\,(0.108, 0.124)$ & $0.133\,(0.125, 0.142)$ & $0.143\,(0.135, 0.151)$ & $0.156\,(0.152, 0.161)$\\ 
      & $4-5$ neighbors & $0.033\,(0.018, 0.060)$ & $0.029\,(0.023, 0.037)$ &  $0.035\,(0.031, 0.040)$ & $0.048\,(0.043, 0.054)$ & $0.058\,(0.053, 0.063)$ & $0.066\,(0.062, 0.069)$ \\
      & Rich group & $0.015\,(0.006, 0.036)$  & $0.030\,(0.024, 0.038)$ & $0.045\,(0.040, 0.050)$ & $0.067\,(0.061, 0.073)$ & $0.081\,(0.075, 0.087)$ & $0.100\,(0.096, 0.104)$ \\
      & Cluster & $0.013\,(0.005, 0.033)$ & $0.010\,(0.007, 0.015)$ & $0.015\,(0.012, 0.018)$ & $0.020\,(0.017, 0.024)$ & $0.032\,(0.028, 0.036)$ & $0.043\,(0.041, 0.046)$ \\
\hline
M105 & Isolated & $0.654\,(0.636, 0.673)$  & $0.647\,(0.638, 0.655)$ & $0.597\,(0.590, 0.603)$ & $0.520\,(0.511, 0.530)$ & $0.471\,(0.462, 0.480)$& $0.431\,(0.426, 0.436) $  \\
      & Pairs & $0.191\,(0.176, 0.207)$  &  $0.168\,(0.162, 0.175)$ & $0.169\,(0.164, 0.175)$ & $0.168\,(0.161, 0.175)$ & $0.167\,(0.161, 0.174)$ & $0.167\,(0.163, 0.171) $ \\
      & $2-3$ neighbors & $0.092\,(0.081, 0.103)$  & $0.103\,(0.098, 0.109)$  & $ 0.117\,(0.112, 0.121)$ &$ 0.131\,(0.124, 0.137)$ & $0.138\,(0.131, 0.144)$ & $0.139\,(0.136, 0.143)$ \\ 
      & $4-5$ neighbors & $0.028\,(0.022, 0.035)$  & $0.030\,(0.027, 0.033)$  & $0.040\,(0.038, 0.043)$ & $0.053\,(0.048, 0.057)$ & $0.057\,(0.053, 0.061)$ & $0.064\,(0.061, 0.066)$ \\
      & Rich group & $0.026\,(0.021, 0.033)$  & $0.040\,(0.036, 0.043)$ & $0.060\,(0.057, 0.064)$ & $ 0.098\,(0.093, 0.104)$ & $0.120\,(0.114, 0.126)$ & $0.131\,(0.128, 0.134)$ \\
      & Cluster & $0.008\,(0.006, 0.013)$ & $0.012\,(0.010, 0.014)$ & $0.017\,(0.016, 0.019)$ & $0.030\,(0.027, 0.033)$ & $ 0.047\,(0.043, 0.051)$ & $0.068\,(0.065, 0.070)$ \\
\hline
M10 & Isolated & $0.653\,(0.629, 0.676)$ & $0.578\,(0.569, 0.588)$ & $0.549\,(0.541, 0.556)$  & $0.471\,(0.458, 0.483)$ & $0.370\,(0.355, 0.384)$ & $0.332\,(0.323, 0.340) $ \\
      & Pairs & $0.176\,(0.158, 0.196)$ & $0.183\,(0.176, 0.190)$  & $0.171\,(0.165, 0.177)$ & $0.148\,(0.146, 0.164)$ & $0.149\,(0.138, 0.160)$ & $0.136\,(0.130, 0.142)$ \\
      & $2-3$ neighbors & $0.096\,(0.082, 0.112)$ & $0.116\,(0.110, 0.122)$  & $0.123\,(0.118, 0.128)$ & $0.124\,(0.116, 0.132)$ & $0.134\,(0.124, 0.144)$ & $0.123\,(0.118, 0.130)$  \\ 
      & $4-5$ neighbors & $0.026\,(0.020, 0.036)$  & $0.039\,(0.036, 0.043)$  & $0.045\,(0.043, 0.048)$ & $0.057\,(0.052, 0.063)$ & $0.070\,(0.063, 0.078)$ & $0.065\,(0.061, 0.070)$ \\
      & Rich group & $0.037\,(0.028, 0.047)$ & $0.063\,(0.059, 0.068)$ & $0.089\,(0.085, 0.093)$ & $0.148\,(0.140, 0.157)$ & $0.199\,(0.187, 0.211)$ & $0.216\,(0.208, 0.223)$ \\
      & Cluster &$0.012\,(0.008, 0.019)$ & $0.020\,(0.018, 0.023)$ & $0.024\,(0.021, 0.026)$ & $0.045\,(0.040, 0.051)$ &  $0.079\,(0.071, 0.088)$  & $0.128\,(0.122, 0.134)$ \\
\hline
M95 & Isolated & $0.623\,(0.576, 0.668)$  & $0.544\,(0.528,0.560)$ & $0.529\,(0.519, 0.539)$ & $0.447\,(0.432, 0.463)$ & $0.277\,(0.257, 0.299)$ & $0.216\,(0.196, 0.238)$  \\
      & Pairs & $0.179\,(0.145, 0.219)$  & $0.188\,(0.176, 0.201)$ & $0.175\,(0.167, 0.183)$ & $0.154\,(0.143, 0.166)$ & $0.110\,(0.096, 0.125)$ & $0.090\,(0.077, 0.106)$\\\
      & $2-3$ neighbors &  $0.103\,(0.077, 0.135)$ & $0.124\,(0.114, 0.135)$ & $0.123\,(0.116, 0.129)$ & $0.122\,(0.112, 0.133)$& $0.128\,(0.114, 0.145)$ & $0.091\,(0.077, 0.107)$  \\ 
      & $4-5$ neighbors & $0.027\,(0.016, 0.047)$  & $0.044\,(0.038, 0.051)$  & $0.050\,(0.046 0.055)$ & $0.059 \,(0.052, 0.067)$ & $0.061\,(0.050,  0.073)$ & $0.063\,(0.052, 0.077)$  \\
      & Rich group & $0.045\,(0.029, 0.070)$  & $0.076\,(0.068, 0.085)$ & $0.099\,(0.094, 0.106)$ & $0.173\,(0.161, 0.185)$& $0.302\,(0.281, 0.324)$ &  $0.336\,(0.312, 0.361)$ \\
      & Cluster & $0.022\,(0.012, 0.041)$ & $0.024\,(0.020, 0.029)$ & $0.024\,(0.021 0.027)$ & $0.045\,(0.039, 0.052)$ & $0.122\,(0.107, 0.138)$ & $0.203\,(0.184, 0.225)$\\
\enddata
\end{deluxetable*}
\end{rotatetable*}

\section{Correlation Between Environmental Overdensities and Halo Mass}\label{sec:appC}

Table~\ref{tab:rho_delMh} presents the the Spearman correlation coefficients between stellar mass overdensities and halo mass. The correlations between $M_h$ and $\deltahalf$ have $\rho  \approx 0.5-0.7$ depending on $M_\star$ and $\delSSFR$ while that of $M_h$ and $\deltaeight$ has $\rho \approx 0.1-0.3$.

\begin{deluxetable}{ccccccc}
\tablenum{7}
\tabletypesize{\footnotesize}
\tablecaption{The Spearman Correlation Coefficients Between Stellar Mass Overdensities and Halo Mass \label{tab:rho_delMh}}
\tablehead{
\colhead{Sample} & \colhead{Measurement} & \colhead{Starburst} & \colhead{SFMS} & \colhead{QG} & \colhead{All} & \colhead{Satellite} 
}
\decimalcolnumbers
\startdata
  & $\delta_\mathrm{0.5Mpc}$ vs. $M_h$ &0.69 & 0.74 & 0.72 & 0.78 & 0.61\\
  & $\delta_\mathrm{1Mpc}$ vs. $M_h$ &0.50 & 0.56 & 0.65 & 0.65 & 0.75 \\
All & $\delta_\mathrm{2Mpc}$ vs. $M_h$ &0.32 & 0.37 & 0.50 & 0.49 & 0.65\\
 &  $\delta_\mathrm{4Mpc}$ vs. $M_h$ &0.23 & 0.26 & 0.39 & 0.37 & 0.51\\
&   $\delta_\mathrm{8Mpc}$ vs. $M_h$ & 0.15  & 0.18 & 0.30 &0.28 & 0.40 \\
\hline
  & $\delta_\mathrm{0.5Mpc}$ vs. $M_h$ &0.61 & 0.66 & 0.71 & 0.72 & 0.39\\
& $\delta_\mathrm{1Mpc}$ vs. $M_h$ &0.45 & 0.57 & 0.64 & 0.65 & 0.64\\
M11 & $\delta_\mathrm{2Mpc}$ vs. $M_h$ &0.27 & 0.39 & 0.48 & 0.48 & 0.58\\
&  $\delta_\mathrm{4Mpc}$ vs. $M_h$ &0.14 & 0.26 & 0.36 &0.36 & 0.45 \\
&   $\delta_\mathrm{8Mpc}$ vs. $M_h$ & 0.12 & 0.19 & 0.27 &0.27 & 0.35\\
\hline
  & $\delta_\mathrm{0.5Mpc}$ vs. $M_h$ &0.69 & 0.70 & 0.74 & 0.75 & 0.56\\
  & $\delta_\mathrm{1Mpc}$ vs. $M_h$ &0.50 & 0.52 & 0.63 & 0.61 & 0.73\\
 M105 &  $\delta_\mathrm{2Mpc}$ vs. $M_h$ &0.32 & 0.35 &0.49 & 0.47 & 0.62\\
 &  $\delta_\mathrm{4Mpc}$ vs. $M_h$ &0.23 & 0.25 & 0.38 & 0.35 & 0.49\\
&   $\delta_\mathrm{8Mpc}$ vs. $M_h$ & 0.16 & 0.18 & 0.30 &0.27 & 0.38\\
\hline
  & $\delta_\mathrm{0.5Mpc}$ vs. $M_h$ & 0.54 & 0.64 & 0.74 & 0.69 & 0.67 \\
  & $\delta_\mathrm{1Mpc}$ vs. $M_h$ &0.39 & 0.47 & 0.65 & 0.56 & 0.75\\
 M10 & $\delta_\mathrm{2Mpc}$ vs. $M_h$ &0.26 & 0.51 & 0.53 & 0.43 & 0.64\\
 &  $\delta_\mathrm{4Mpc}$ vs. $M_h$ &0.18 & 0.24 & 0.43 & 0.33 & 0.51\\
&   $\delta_\mathrm{8Mpc}$ vs. $M_h$ & 0.13 & 0.17 & 0.35 &0.26 & 0.41\\
\hline
  & $\delta_\mathrm{0.5Mpc}$ vs. $M_h$ & 0.46 & 0.57 & 0.73 & 0.62 & 0.70\\
& $\delta_\mathrm{1Mpc}$ vs. $M_h$ &0.36 & 0.43 & 0.71 & 0.52 & 0.75\\
M95 &  $\delta_\mathrm{2Mpc}$ vs. $M_h$ &0.24 & 0.32 & 0.60 & 0.40 & 0.63\\ 
&  $\delta_\mathrm{4Mpc}$ vs. $M_h$ &0.19 & 0.24 & 0.48 & 0.31 & 0.50 \\
&  $\delta_\mathrm{8Mpc}$ vs. $M_h$ & 0.14  & 0.18 & 0.42 &0.25 & 0.40\\
\enddata
\tablecomments{Column (1) : the $M_\star$ ranges of the samples binned by 0.5\,dex or all galaxies with $\log\,M/M_\odot = 9.5-11.5$. Column (2) : gives the two parameters whose correlation coefficients are computed. All values are significant ($p \lesssim .001$).}
\end{deluxetable}

\section{$M_\star$ and $C$ Distributions Before Matching}\label{sec:appD}

\begin{figure}
\includegraphics[width=0.49\textwidth]{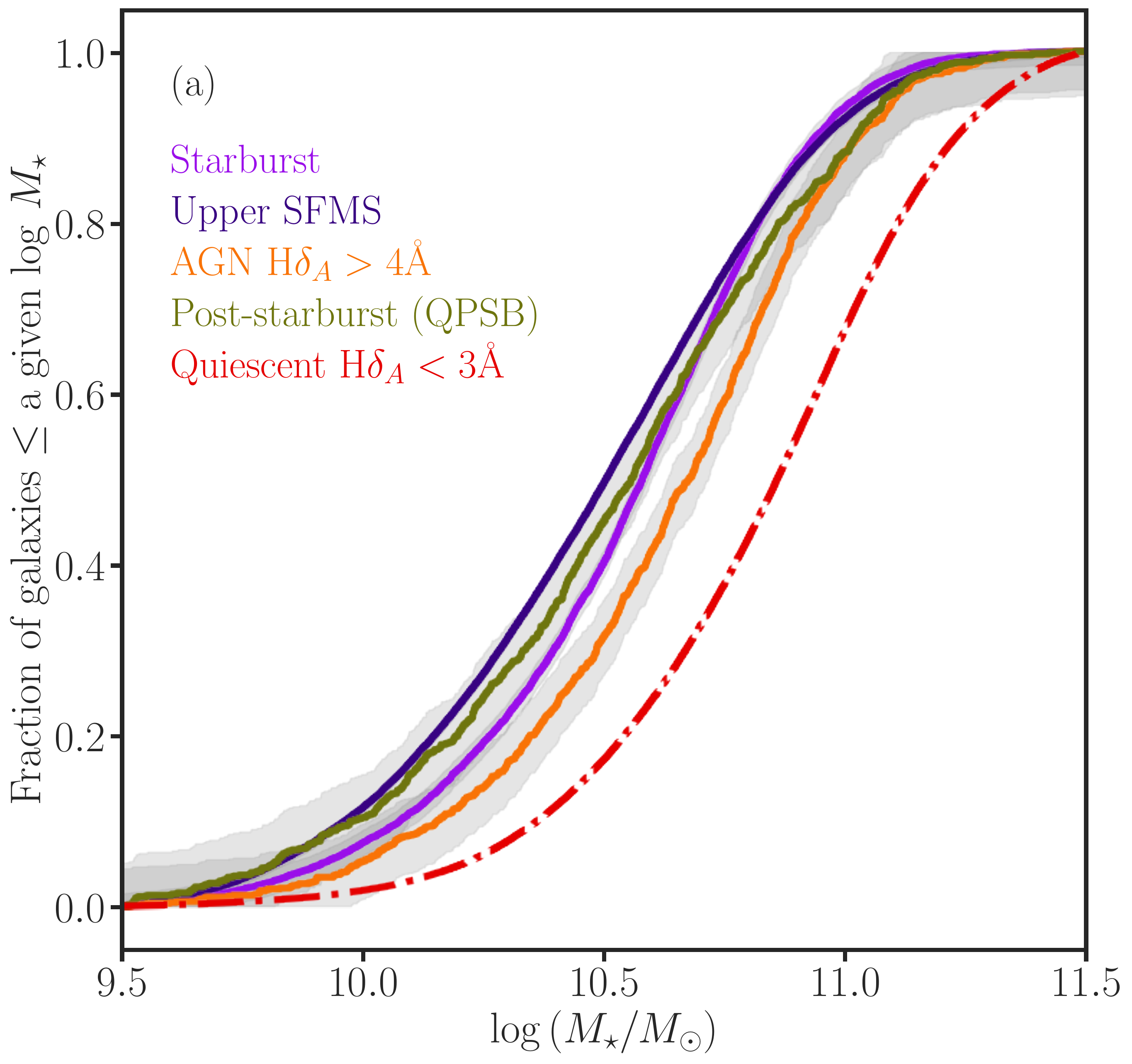}
\includegraphics[width=0.49\textwidth]{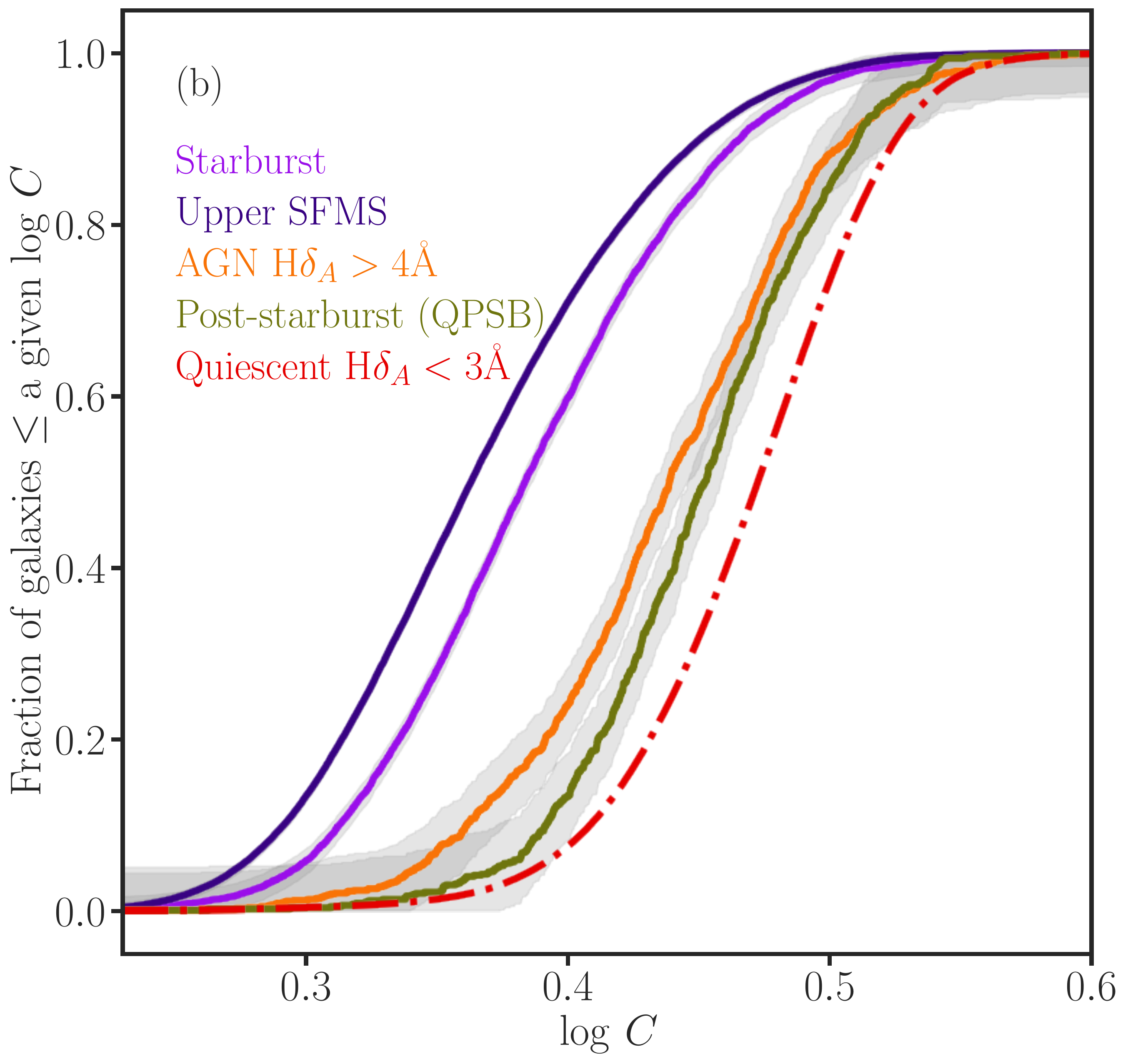}
\caption{The stellar mass and concentration index distributions of various samples before matching. The distributions of $C$ of QPSBs and strong-H$\delta_A$ AGNs are more similar to those QGs than to those of starbursts and upper SFMS galaxies while their $M_\star$ are significantly lower than those of QGs. \label{MC_beforematch}}
\end{figure}

Figure~\ref{MC_beforematch} shows the ECDFs of the stellar masses and concentration indices of QPSBs and of the parent samples of their possible progenitors or descendants. The distributions of $M_\star$ and $C$ of QPSBs are significantly different ($p \lesssim 0.001$) from those of the other samples before matching. The distribution of $C$ of QPSBs is more similar to those of QGs than to those of starbursts and upper SFMS galaxies while their $M_\star$ are significantly lower than those of QGs.

\end{document}